%% file: nbody-bbl.tex
\documentclass{aa}

\newcommand\dosingle[1]{#1}  \newcommand\dodouble[1]{ } 

\newcommand\nice[1]{#1}    \newcommand\subm[1]{}   
\subm{ \renewcommand\dosingle[1]{ }   \renewcommand\dodouble[1]{#1} }

\dodouble{\documentclass[usenatbib,referee]{mnras}}

\usepackage[T1]{fontenc}
\usepackage{aecompl}

\usepackage{graphicx}

\usepackage{amsfonts}
\usepackage{amsmath}

\usepackage{rotating} 

\usepackage[varg]{txfonts}
\usepackage{fixltx2e} 

\newcommand\mystamp[1]{{}}

\newcommand\mystamppreamble{
  \usepackage{eso-pic}
  \usepackage{color}
  \definecolor{redstamp}{rgb}{0.99,0.80,0.90} 
  \usepackage{datetime}
  \usepackage[normalem]{ulem}
}

\newcommand\mystampdothestamp{
  \AddToShipoutPicture{
    \AtTextLowerLeft{
      \makebox(550,300)[c]{\resizebox{\textwidth}{!}{
          \rotatebox{36}{\textsf{\textbf{\color{redstamp}brouillon~\frdutoday~\currenttime}}}}}
    }
  }
}

\mystamp{\mystamppreamble}

\usepackage{natbib}  
\usepackage{url}

\usepackage{etoolbox}

\newtoggle{postrefAA} \newtoggle{postrefBB} \newtoggle{postrefCC}

\newtoggle{prerefAA} \newtoggle{prerefBB} \newtoggle{prerefCC}
\newtoggle{prerefDD} \newtoggle{prerefEE} \newtoggle{prerefFF} \newtoggle{prerefGG}

\togglefalse{postrefAA} 

\togglefalse{postrefBB} 

\togglefalse{postrefCC} 

\togglefalse{prerefAA} \togglefalse{prerefBB} \togglefalse{prerefCC}
\togglefalse{prerefDD} \togglefalse{prerefEE} \togglefalse{prerefFF} \togglefalse{prerefGG}

\iftoggle{prerefAA}{
       \usepackage{color}  \definecolor{myred}{rgb}{0.7,0.0,0.2}
}{
      
}

\iftoggle{prerefBB}{
       \usepackage{color}  \definecolor{myred}{rgb}{0.7,0.0,0.2}
}{
      
}

\iftoggle{prerefCC}{
       \usepackage{color}  \definecolor{myred}{rgb}{0.7,0.0,0.2}
}{
      
}

\iftoggle{prerefDD}{
       \usepackage{color}  \definecolor{myred}{rgb}{0.7,0.0,0.2}
}{
      
}

\iftoggle{prerefEE}{
       \usepackage{color}  \definecolor{myred}{rgb}{0.7,0.0,0.2}
}{
      
}

\iftoggle{prerefFF}{
       \usepackage{color}  \definecolor{myred}{rgb}{0.7,0.0,0.2}
}{
      
}

\iftoggle{prerefGG}{
       \usepackage{color}  \definecolor{myred}{rgb}{0.7,0.0,0.2}
}{
      
}

\iftoggle{postrefAA}{
  \newcommand\postrefereeAAchanges[1]{{\bf \large \color{myred} #1}} \newcommand\postrefereeAAstart{ \bf \large \color{myred} }  \newcommand\postrefereeAAstop{ \rm \color{black} }  \usepackage{color}  \definecolor{myred}{rgb}{0.7,0.0,0.2}
}{
  \newcommand\postrefereeAAchanges[1]{#1}  \newcommand\postrefereeAAstart{  }  \newcommand\postrefereeAAstop{ }
}

\iftoggle{postrefBB}{
       \usepackage{color}  \definecolor{myred}{rgb}{0.7,0.0,0.2}
}{
      
}

\iftoggle{postrefCC}{
       \usepackage{color}  \definecolor{myred}{rgb}{0.7,0.0,0.2}
}{
      
}

\usepackage{color}
\definecolor{mygreen}{rgb}{0.11, 0.65, 0.02}

 \usepackage{hyperref}

\nice{

\providecommand{\url}[1]{\href{#1}{#1}}

}

\renewcommand{\eprint}[1]{\href{http://arxiv.org/abs/#1}{{\tt [arXiv:#1]}}}

\providecommand{\adsurl}[1]{}
\newcommand\SSS{Sect.~}
\newcommand\SSSS{Sections~}

\providecommand\apj{ApJ}                 
\providecommand\apjs{ApJSupp}                 
\providecommand\apjl{ApJL}                 
\providecommand\aap{A\&A}            
\providecommand\mnras{MNRAS}

\providecommand\PRL{Physical Review Letters}

\providecommand\PRL{PRL}
\providecommand\prd{PRD}

\providecommand\jcap{JCAP}

\providecommand\grg{Gen.~Rel.~Grav.}
\providecommand\annalesBruxelles{Ann. de la Soc. Sc. de Brux.}
\providecommand\annrevnucpartphys{Ann.~Rev.~Nucl.~Part.~Sci.}

\providecommand\pnas{Proc.~Nat.~Acad.~Sci.}
\providecommand\cqg{Class.~Quant.~Gra.}   %

\providecommand\newjphys{New Journ.~Phys.\/}

\nice{  }

\providecommand\kmsMpc{\mbox{\,$\mathrm{km/s/Mpc}$}}

\newcommand\gtapprox{\,\lower.6ex\hbox{$\buildrel >\over \sim$} \, }
\newcommand\ltapprox{\,\lower.6ex\hbox{$\buildrel <\over \sim$} \, }
\newcommand\propapprox{\,\lower.6ex\hbox{$\buildrel \propto\over \sim$} \, }

\newcommand\arcs{\ifmmode {'' }\else $'' $\fi}     
\newcommand\arcm{\ifmmode {' }\else $' $\fi}       
\newcommand\ddeg{\ifmmode^\circ\else$^\circ$\fi}    

\newcommand\diffd{\mathrm{d}}

\newcommand\frtoday{Le\space\number\day\space\ifcase\month\or
  janvier\or f\'evrier\or mars\or avril\or mai\or juin\or
  juillet\or ao\^ut\or septembre\or octobre\or novembre\or
d\'ecembre\fi\space \number\year}
\def\frdutoday{du\space\number\day\space\ifcase\month\or
  janvier\or f\'evrier\or mars\or avril\or mai\or juin\or
  juillet\or ao\^ut\or septembre\or octobre\or novembre\or
d\'ecembre\fi\space \number\year}

\newcommand\todayISO{\number\year-\ifnum\month<10 0\fi\number\month-\ifnum\day<10 0\fi\number\day}

\newcommand{\CD}{{\cal D}}

\newcommand{\CI}{{\cal I}}
\newcommand{\CQ}{{\cal Q}}
\newcommand{\CR}{{\cal R}}

\newcommand{\averageplain}[1]{\left\langle #1 \right\rangle}

\newcommand{\average}[1]{\left\langle #1 \right\rangle_\CD}

\newcommand{\averagebox}[1]{\left\langle #1 \right\rangle_{\mathrm{box}}}

\newcommand{\subbox}{_{\mathrm{box}}}

\newcommand{\initaverage}[1]{\left\langle #1 \right\rangle_{\CD_{\mathbf{i}}}}

\newcommand{\initaverageDunder}[1]{\left\langle #1 \right\rangle_{\CD^-_{\mathbf{i}}}}
\newcommand{\initaverageDover}[1]{\left\langle #1 \right\rangle_{\CD^+_{\mathbf{i}}}}
\newcommand{\initaverageriem}[1]{\left\langle #1 \right\rangle_{\CI}}

\newcommand{\initial}[1]{{{#1}_{\mathbf i}}}
\newcommand{\initialDunder}[1]{{{#1}_{\CD^+_{\mathbf i}}}}
\newcommand{\initialDover}[1]{{{#1}_{\CD^-_{\mathbf i}}}}
\newcommand\OmRDunder{\Omega_{\CR}^{\CD^-}}
\newcommand\OmRDover{\Omega_{\CR}^{\CD^+}}
\newcommand{\FLRW}{_{\mathrm{FL}}}

\newcommand\hzeroeffMpc{\mbox{Mpc/$\hzeroeff$}}
\newcommand\honebgMpc{\mbox{Mpc/[$\Honebg$/(100~km/s/Mpc)]}}

\newcommand\Omm{\Omega_{\mathrm{m}}}
\newcommand\Ommzero{\Omega_{\mathrm{m0}}}
\newcommand\OmLam{\Omega_{\Lambda}} 
\newcommand\OmLamzero{\Omega_{\Lambda0}} 
\newcommand\rhocrit{\rho_{\mathrm{crit}}}  

\newcommand\OmmD{\Omega_{\mathrm{m}}^{\CD}}

\newcommand\OmRD{\Omega_{\CR}^{\CD}}

\newcommand\OmQD{\Omega_{\CQ}^{\CD}}

\newcommand\CQTthree{{\cal Q}_{\mathrm{T}^3}}

\newcommand\aeff{a_{\mathrm{eff}}}

\newcommand\abg{a_{\mathrm{EdS}}}

\newcommand\tcollapse{t_{\mathrm{coll}}}
\newcommand\uncollapsed{_{\postrefereeAAchanges{\mathrm{uncoll}}}}
\newcommand\tturnaround{t_{\mathrm{turn}}}

\newcommand{\invI}{{\mathrm{I}}}
\newcommand{\invII}{{\mathrm{II}}}
\newcommand{\invIII}{{\mathrm{III}}}
\newcommand\bfv{\mathbf{v}}

\newcommand\sigmaSonexy{\widehat{\sigma}_{v^i_{,j}}^{j_1 \neq j \neq j_2}}
\newcommand\sigmaSonexyall{\widehat{\sigma}_{v^i_{,j}}^{j}}
\newcommand\sigmaSoneondiag{\widehat{\sigma}_{v^i_{,j}}^{\mathrm{diag}}}
\newcommand\sigmaSoneoffdiag{\widehat{\sigma}_{v^i_{,j}}^{\mathrm{off}}}
\newcommand\sigmameas{\sigma_{\mathrm{meas}}}
\newcommand\sigmaavISone{\widehat{\sigma}_{\average{\invI}}}
\newcommand\sigmaavIISone{\widehat{\sigma}_{\average{\invII}}}
\newcommand\sigmaavISoneglob{\widehat{\sigma}_{\averagebox{\invI}}}
\newcommand\sigmaavIISoneglob{\widehat{\sigma}_{\averagebox{\invII}}}
\newcommand\sigmaQSone{\widehat{\sigma}_{{\CQ}_{\CD}}}

\newcommand\vmodone{\mathbf{u}_1}
\newcommand\vmodtwo{\mathbf{u}_2}
\newcommand\nxyz{n_x}

\newcommand\dotabg{\dot{a}_{\mathrm{EdS}}}
\newcommand\Hzeroeff{H_{\mathrm{eff},0}}
\newcommand\hzeroeff{h_{\mathrm{eff}}}

\newcommand\Heff{H_{\mathrm{eff}}}
\newcommand\fvir{f_{\mathrm{vir}}}

\newcommand\Honebg{H_1^{\mathrm{EdS}}}
\newcommand\Hbg{H^{\mathrm{EdS}}}

\newcommand\LCDM{$\Lambda$CDM}

\newcommand\tzeroefffull{{t_{\aeff=1}}}
\newcommand\tzeroeff{{t_0}}

\newcommand\RAMSES{{\sc ramses}}
\newcommand\RAMSESscalav{{\sc ramses-scalav}}
\newcommand\DTFE{{\sc dtfe}}
\newcommand\DDTFE{{\sc Dtfe}}
\newcommand\inhomog{{\sc inhomog}}

\newcommand\Lbox{L_{\mathrm{box}}}
\newcommand\LD{L_{\CD}}
\newcommand\Lthirteeneight{L_{13.8}}
\newcommand\LDTFE{L_{\mathrm{DTFE}}}
\newcommand\LN{L_N}
\newcommand\Lsoft{L_{\mathrm{soft}}}
\newcommand\nD{n_{\CD}}
\newcommand\nDTFE{n_{\mathrm{DTFE}}}
\newcommand\nsoft{n_{\mathrm{soft}}}

\begin{document}

\title{Replacing dark energy by silent virialisation}
\titlerunning{Dark energy as silent virialisation}

\authorrunning{Roukema} 

\author{Boudewijn F. Roukema\inst{1,2}
}

\institute{
  Toru\'n Centre for Astronomy,
  Faculty of Physics, Astronomy and Informatics,
  Grudziadzka 5,
  Nicolaus Copernicus University,
  ul.~Gagarina 11, 87-100 Toru\'n, Poland
  \and
  Univ Lyon, Ens de Lyon, Univ Lyon1, CNRS, Centre de Recherche Astrophysique de
  Lyon UMR5574, F--69007, Lyon, France
}


\date{\frtoday}



\newcommand\Nchainsmain{16}
\newcommand\Npergroup{four}


\abstract
  {Standard cosmological $N$-body simulations have background scale
    factor evolution that is decoupled from non-linear structure
    formation.  Prior to gravitational collapse, kinematical
    backreaction ($\CQ_\CD$) justifies this approach in a Newtonian
    context.}
  {However, the final stages of a gravitational collapse event are
    sudden; a globally imposed smooth expansion rate forces at least
    one expanding region to suddenly and instantaneously decelerate in
    compensation for the virialisation event.  This is
    relativistically unrealistic.  A more conservative hypothesis is
    to allow non-collapsed domains to continue their volume evolution
    according to the $\CQ_\CD$ Zel'dovich approximation (QZA).  We aim to
    study the inferred average expansion under this `silent'
    virialisation hypothesis.}
  {We set standard ({\sc mpgrafic}) EdS 3-torus (T$^3$) cosmological
    $N$-body initial conditions.  Using \RAMSES{}, we partitioned the
    volume into domains and called the \DTFE{} library to estimate the
    per-domain initial values of the three invariants of the extrinsic
    curvature tensor that determine the QZA.  We integrated the
    Raychaudhuri equation in each domain using the \inhomog{} library,
    and adopted the stable clustering hypothesis to represent
    virialisation (VQZA).  We spatially averaged to obtain the
    effective global scale factor.  We adopted an
    early-epoch--normalised EdS reference-model Hubble constant
    $\Honebg = 37.7${\kmsMpc} and an effective Hubble constant
    $\Hzeroeff = 67.7${\kmsMpc}.}
  {From 2000 simulations at resolution $256^3$, we find that reaching
    a unity effective scale factor at 13.8~Gyr (16\% above EdS),
    occurs for an averaging scale of $\Lthirteeneight =
    2.5^{+0.1}_{-0.4}$~{\hzeroeffMpc}.  Relativistically interpreted,
    this corresponds to strong average negative curvature evolution,
    with the mean (median) curvature functional $\OmRD$ growing from
    zero to about 1.5--2 by the present. Over 100 realisations, the
    virialisation fraction and super-EdS expansion correlate strongly
    at fixed cosmological time.}
  {Thus, starting from EdS initial conditions and averaging on a
    typical non-linear structure formation scale, the VQZA
    dark-energy--free average expansion matches $\Lambda$CDM expansion
    to first order.  The software packages used here are
    free-licensed.}


\mystamp{\mystampdothestamp}

\maketitle

\dodouble{ \clearpage } 


\newcommand\FIGWIDTHONE{0.95\columnwidth}
\newcommand\FIGWIDTHTWO{1.05\columnwidth}

\newcommand\tversions{
\begin{table}
  \caption{Software versions for
    numerical uncertainty tests on $\invI$, $\invII$
    (\SSS\protect\ref{s-results-sigmaT3}), and for
    VQZA
    (\SSS\protect\ref{s-res-best-scales-VQZA}) and
    $N$-body--measured $\CQ_\CD$
    (\SSS\protect\ref{s-res-best-scales-Nbody}) simulations.
    \label{t-versions}}
  \begin{tabular}{lll}
    \hline\hline
    method & package & version and/or
    \rule{0ex}{2.5ex}\\ 
    && commit hash
    \\
    \hline
    \rule{0ex}{2.5ex} 
    $\invI$, $\invII$ errors & \DTFE{} & 1.1.1.Q, 1f2487e3 \\
    VQZA & {\sc mpgrafic} & 0.3.10 \\
    VQZA & \RAMSES{} (ramses/trunk) & 438f4659ae36 \\
    VQZA & \RAMSESscalav{} & 53257c54 \\
    VQZA & \DTFE{}  & 1.1.1.Q, 3d514657 \\
    VQZA & \inhomog{} & 0.0.67 \\
    $N$-body $\CQ_\CD$ & {\sc mpgrafic} & 0.3.10 \\
    $N$-body $\CQ_\CD$ & \RAMSES{} (ramses/trunk) & 438f4659ae36 \\
    $N$-body $\CQ_\CD$ & \RAMSESscalav{} & 1ba83730 \\
    $N$-body $\CQ_\CD$ & \DTFE{} & 1.1.1.Q, 3d514657 \\
    $N$-body $\CQ_\CD$ & \inhomog{} & 0.0.64 \\
    \hline
  \end{tabular} \\
\end{table}
}  

\newcommand\fbiscale{
  \begin{figure}
    \input{biscale_partition}
    \caption{Pre- and post-virialisation EdS-normalised scale factor evolution
      $\aeff/\abg$ for a biscale partition evolved using the VQZA
      as motivated in \protect\SSS\ref{s-newt-vs-GR} and defined
      in \protect\SSS\ref{s-results-VQZA}. From bottom to top at
      $t \gtapprox 7$~Gyr, the arbitrarily chosen dimensionless
      values of the average initial invariants
      (at initial scale factor $\initial{\abg} = 0.005$)
      in the expanding
      domain are $(\initaverageDunder{\invI}, \initaverageDunder{\invII},
      \initaverageDunder{\invIII}) = $
      $(0.01, 0, 0)$ (`planar' case);
      $(0.01, 10^{-4}/3, 10^{-6}/27)$ (`spherical' $\CD^-$ case);
      $(0.01, 10^{-4}, 0)$; and
      $(0.01, 0, 10^{-6})$; respectively.
      The sharp transitions from
      pre-virialisation to post-virialisation epochs are clearly visible
      and occur at roughly 7~Gyr to 2~Gyr, respectively. Prior to virialisation,
      the QZA is fully compatible with EdS global evolution (which can be thought of as
      Newtonian T$^3$ cosmology), especially in the planar case, in which the QZA
      is exact in the Newtonian case.
      \label{f-biscale}}
\end{figure}} 

\newcommand\fbiscaleunderdense{
  \begin{figure}
    \input{biscale_underdense}
    \caption{Expanding-domain scale factor evolution $a_{\CD^-}/\abg$
      for the biscale partition models illustrated in
      Fig.~\protect\ref{f-biscale}, comparing evolution of this domain
      according to the VQZA model (silent virialisation) and according to the standard
      $N$-body EdS `Newtonian' constraint (instantaneous feedback from virialisation).
      The VQZA scale factor
      evolution ($+$ symbols) correspond from bottom to top to
      those from bottom to top in Fig.~\protect\ref{f-biscale}.  The
      standard-model scale-factor
      $a_{\CD^-}^{\mathrm{Newt}}(t)/\abg(t)$
      [solid curves;
      Eq.~(\protect\eqref{e-biscale-aDminus-Newt})]
      are indistinguishable
      following virialisation of the overdense domain, since by
      assumption, the expanding domain suddenly switches from
      hyperbolic (super-EdS) evolution to very-nearly flat (EdS) evolution
      ($a_{\CD^-}^{\mathrm{Newt}}(t)/\abg(t) = 2^{1/3}$)
      when virialisation of the overdense domain occurs. A tiny difference
      from perfectly flat EdS evolution is present because the overdense
      domain has a small but non-zero fixed (stable clustering) volume.
      \label{f-biscale-underdense}}
\end{figure}
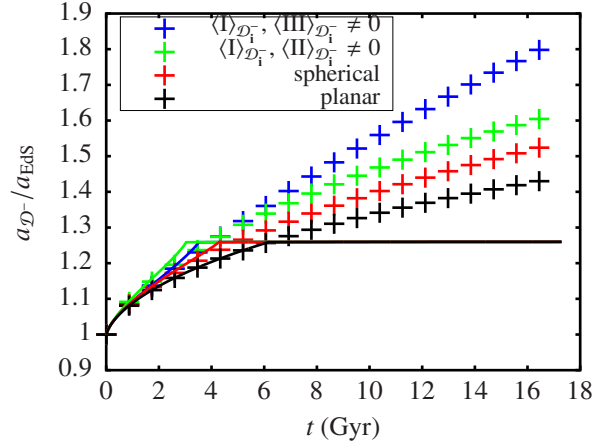} 

\newcommand\fbiscaleOmRD{
  \begin{figure}
    \input{biscale_OmRD}
    \caption{Mean curvature functional $\OmRDunder$ evolution of the
      expanding domain $\CD^-$ in the biscale partition models illustrated in
      Figs~\protect\ref{f-biscale}
      and \protect\ref{f-biscale-underdense}.
      VQZA curvature evolution is shown
      with $+$ symbols and the
      standard $N$-body EdS constraint
      (interpreted relativistically, not in Newtonian terms)
      is shown by curves. We hypothesise that the
      smooth curvature evolution ($+$) is relativistically more accurate than
      the sudden drop to flatness (solid curves) of the standard model.
      \label{f-biscale-OmRD}}
\end{figure}
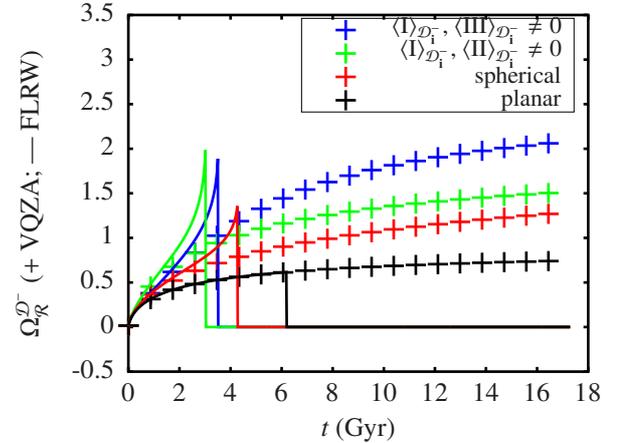} 

\newcommand\fSNinvariants{
  \begin{figure}
    \input{SN_invI}
    \input{SN_invII}
    \caption{Signal-to-noise ratios $S/N$ of $\invI$ (above) and $\invII$ (below)
      representing the ratio of rms signal to rms
      numerical noise $\sigmameas$, for analytical model
      $\vmodone$ (which has non-zero invariants) in
      Eq.~(\protect\ref{e-test-vmodels}), sampled by realisations of
      $N=128^3$ particles drawn from a uniform spatial random distribution
      for $8 \le \nDTFE \le 64$.
      The signal is negligibly small
      for $\nDTFE = 16, 128$ for
      $\nxyz = 8, 64$, respectively, since the test functions are sinusoidal and
      in phase with the box.
      \label{f-SN-invariants}}
\end{figure}
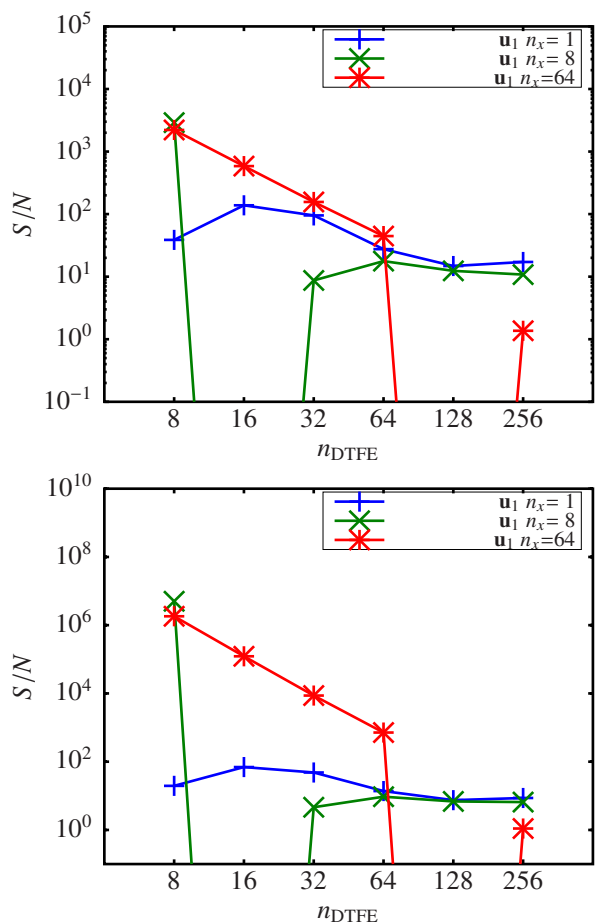} 

\newcommand\fSoneinvariants{
  \begin{figure}
    \input{S1_invI}
    \input{S1_invII}
    \caption{As in Fig.~\protect\ref{f-SN-invariants},
      ratio of measured noise $\sigmameas$ to
      rms error predicted by S$^1$-based error estimators
      for $\invI$ (above) and $\invII$ (below),
      for both models $\vmodone$ and $\vmodtwo$.
      These ratios are within two orders of
      magnitude of unity in both cases.
      \label{f-S1-invariants}}
\end{figure}
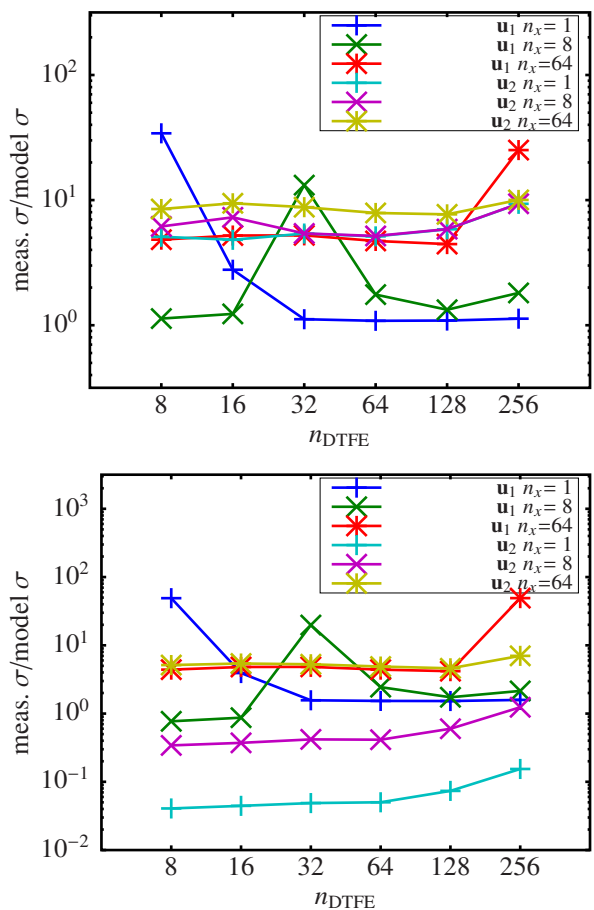} 

\newcommand\fSglobal{
  \begin{figure}
    \input{Sglobal_invI}
    \input{Sglobal_invII}
    \caption{As in Fig.~\protect\ref{f-S1-invariants},
      measured global
      $|\averagebox{\invI}|$ (above) and
      $|\averagebox{\invII}|$ (below),
      which should be zero in the absence of numerical error
      [by Stokes' theorem and Eq.~(\protect\ref{e-I-II-III-defns})],
      shown as symbols,
      compared to S$^1$-based error estimators
      under independent Gaussian error propagation,
      $\sigmaavISoneglob$,
      $\sigmaavIISoneglob$,
      shown as curves,
      for both models $\vmodone$ and $\vmodtwo$.
      \label{f-S1-global}}
\end{figure}} 

\newcommand\faD{
  \begin{figure}
    \input{aRZA_nsc8_ndtfe16_qdisable0}
    \input{aRZA_nsc8_ndtfe32_qdisable0}
    \input{aRZA_nsc8_ndtfe64_qdisable0}
    \caption{Full box cube-root-of-total-volume
      scale factor $\aeff(t)$
      evolution
      for
      {\em (top:)} $L_{\mathrm{box}}  = 8 \LD = 2 L_{\mathrm{DTFE}} = 16 L_N$,
      {\em (middle:)} $L_{\mathrm{box}}  = 8 \LD = 4 L_{\mathrm{DTFE}} = 8 L_N$,
      {\em(bottom:)} $L_{\mathrm{box}}  = 8 \LD = 8 L_{\mathrm{DTFE}} = 4 L_N$,
      for $N=256^3$ particles; where $\hzeroeff$ is written as $h$
      to reduce clutter. The $\LD$ scales are labelled, increasing in length
      from top to bottom within any panel. Cosmic variance is visible as non-uniformity
      in the $\LD$ dependence of the separation of the curves.
      {\em Online:} a fixed
      scale $\LD$ is shown with the same colour in all panels
      in this and later figures
      (for example, $\LD = 2.0 \,\hzeroeffMpc$ is purple).
      \label{f-aD}}
  \end{figure}} 

\newcommand\faDQdisabled{
  \begin{figure}
    \input{aRZA_ndtfe32_qdisable1}
    \input{aRZA_ndtfe64_qdisable1}
    \input{aRZA_ndtfe128_qdisable1}
    \caption{As in Fig.~\ref{f-aD}, full box cube-root-of-total-volume
      scale factor $\aeff(t)$, but ignoring kinematical
      backreaction (setting $\CQ_{\CD}=0 \; \forall \CD\; \forall t$),
      for
      {\em (top:)} $L_{\mathrm{box}}  = 8 \LD = 4 L_{\mathrm{DTFE}} = 8 L_N$,
      {\em (middle:)} $L_{\mathrm{box}}  = 8 \LD = 8 L_{\mathrm{DTFE}} = 4 L_N$,
      {\em (bottom:)} $L_{\mathrm{box}}  = 8 \LD = 16 L_{\mathrm{DTFE}} = 2 L_N$.
      \label{f-aD-Qdisabled}}
\end{figure}} 

\newcommand\faDQbigbox{
  \begin{figure}
    \input{aRZA_nsc64_ndtfe128_qdisable0}
    \caption{As in Fig.~\ref{f-aD}, full box cube-root-of-total-volume
      scale factor $\aeff(t)$,
      for $\Lbox  = 64 \LD = 2 \LDTFE = 2 \LN$,
      for $N=256^3$ particles: the $\LD = 2$~Mpc$/\hzeroeff$ scale
      corresponds to a 128~Mpc$/\hzeroeff$ comoving box size.
      \label{f-aD-bigbox}}
\end{figure}} 

\newcommand\fQD{
  \begin{figure}
    \input{OmQmean}
    \input{OmQmedian}
    \caption{Evolution of the volume-weighted mean (above) and median
      (below) globally normalised kinematical backreaction functional
      $\OmQD$ of uncollapsed domains, corresponding to the same scales
      as in the middle panel of Fig.~\protect\ref{f-aD}.
      \label{f-QD}}
\end{figure}} 

\newcommand\fRD{
  \begin{figure}
    \input{OmRmean}
    \input{OmRmedian}
    \caption{Evolution of the volume-weighted mean (above) and median
      (below) globally normalised curvature functional $\OmRD$ of
      uncollapsed domains, corresponding to the same scales as in the
      middle panel of Fig.~\protect\ref{f-aD}.
      \label{f-RD}}
\end{figure}} 

\newcommand\ffvir{
  \begin{figure}
    \input{superEdS_fvir_nsc8_ndtfe32_qdisable0}
    \caption{Super-EdS ratio $\aeff/\abg$
      versus virialisation fraction $\fvir$
      for 100 VQZA simulations with $\LD= 2$~Mpc$/\hzeroeff$ and
      $L_{\mathrm{box}}  = 8 \LD = 4 L_{\mathrm{DTFE}} = 8 L_N$,
      at an early ($\circ$), middle ($\times$), and late ($+$) time,
      as indicated.
      At any fixed time, $\aeff/\abg$ correlates strongly with $\fvir$.
      \label{f-fvir}}
\end{figure}} 

\newcommand\fLscalavnbodyLag{
  \begin{figure}
    \input{a_nbody_Lag_128_F}
    \caption{As in Fig.~\protect\ref{f-aD} {\em (middle)}, for
      $N$-body evolution and
      Lagrangian tracing of initial domains, for
      $L_{\mathrm{box}}  = 8 \LD = 4 L_{\mathrm{DTFE}} = 4 L_N$,
      with $N=128^3$ particles,
      and  $\nsoft/(N^{1/3}) = 32$. 
      The curves' behaviour is not monotonic with respect to $\LD$ scales;
      the top curve is for
      $\LD = 2.0 $~Mpc$/\hzeroeff$,
      the middle pair of curves are for
      $\LD = 1.0 $~Mpc$/\hzeroeff$ (upper) and
      $\LD = 4.0 $~Mpc$/\hzeroeff$ (lower), and the bottom
      pair are for
      $\LD = 8.0 $~Mpc$/\hzeroeff$ (upper) and
      $\LD = 0.5 $~Mpc$/\hzeroeff$ (lower).
      \label{f-Lscalav-nbody-Lag}}
\end{figure}} 

\newcommand\fLscalavnbodyLagQdisabled{
  \begin{figure}
    \input{a_nbody_Lag_128_T}
    \caption{As in Fig.~\protect\ref{f-aD-Qdisabled} {\em (top)}, for
      $N$-body evolution and
      Lagrangian tracing of initial domains, for
      $L_{\mathrm{box}}  = 8 \LD = 4 L_{\mathrm{DTFE}} = 4 L_N$,
      with $N=128^3$ particles,
      \postrefereeAAchanges{(artificially
        setting $\CQ_{\CD}=0 \; \forall \CD\; \forall t$)}.
      The curves' behaviour is monotonic with respect to $\LD$ scales,
      as labelled.
      \label{f-Lscalav-nbody-Lag-Qdisabled}}
\end{figure}} 

\section{Introduction} \label{s-intro}

Cosmological $N$-body simulations normally assume that cosmological
expansion is gravitationally decoupled from structure formation: scale
factor evolution is inserted into the simulation independently of the
non-linear structure growth calculated in the simulation.  This is a
simplification primarily due to the limitations of both analytical and
numerical methods of calculation. However, in practice, this
decoupling assumption is usually considered to be a corollary of the
homogeneity and isotropy assumption underlying the
Friedmann--Lema\^{\i}tre--Robertson--Walker (FLRW;
\citealt{Fried23,Fried24,Lemaitre31ell,Lemaitre31elltrans,Rob35})
family of cosmological solutions of the Einstein equation, which
includes the present standard model of cosmology, the $\Lambda$CDM
(cosmological-constant--dominated cold dark matter) model.  The
validity of the decoupling assumption has long been debated
\citep[e.g.][and references therein]{EllisStoeger87,Buchert11Towards}.

Here, our aim is to take into account the role of virialisation
and calculate the
effective scale factor evolution in an $N$-body
simulation context, guided by
general-relativistic constraints defined by scalar averaging
\citep*{Buch00scalav,Buch01scalav,Buchert11Towards} that
are closely analogous to Newtonian equivalents but are justified
relativistically
\citep*{BKS00,BuchRZA1,BuchRZA2}.
By providing tools for calculating effective background expansion from
structure formation itself, we also encourage the community to explore
alternatives to the conservative approach adopted in this paper.
In \SSS\ref{s-newt-vs-GR}, we explain why the standard treatment
of virialisation appears to imply, via an externally imposed background
expansion rate,
a relativistically unrealistic constraint on spatial expansion.
Specific calculations
to illustrate the argument are given in
\SSS\ref{s-results-VQZA}.

Given that at
early redshifts $z \gg 10$, the observational distinction between the
$\Lambda$CDM model and the Einstein-de~Sitter (EdS) model (a flat FLRW
model with a zero cosmological constant) is very weak, we followed
Occam's razor and adopted an EdS model at early times, setting up
standard ({\sc mpgrafic}) cosmological $N$-body initial conditions.
We evolved the initial conditions by a primarily analytical approach --
which we refer to here as the
kinematical-backreaction
($\CQ_\CD$)
Zel'dovich approximation (QZA).
This provides an
analytical expression for kinematical backreaction $\CQ_\CD$
evolution in the Newtonian \citep{BKS00}
and general-relativistic contexts
(\citealt{Kasai95,Kasai98} in the form
\citealt{BuchRZA1,BuchRZA2,BuchRZA3}; see also
\citealt{MatarreseTerranova96,VillaMatarrese11}), with
closely related but differing definitions and meaning.

It has long been accepted observationally that non-linear
structure, that is, statistical patterns of overdensities and
underdensities with respect to an `average' (mean) density, can be
characterised by a length scale of several megaparsecs
\citep{Peebles74,PeeblesHauser74}.
If the average expansion history in an initially EdS simulation is distinct from
EdS evolution, then it can reasonably be
expected that the difference will depend on the choice of length
scale corresponding to non-linear structure. Moreover, if the
treatment of volume evolution proposed here is a fair approximation
of the evolution of the real Universe, then the length scale
that yields the appropriate amount of super-EdS
scale factor evolution to match a $\Lambda$CDM approximation of observational
data should be reasonably close to the several-megaparsec scale.

\postrefereeAAchanges{The approach presented here}
adds detail to previous work in modelling
inhomogeneous curvature and expansion
\citep{Rasanen04,Buchert08status,WiegBuch10,BuchRas12} that is
constrained by the Raychaudhuri equation and the Hamiltonian
constraint (whose spatial averages generalise the Friedmannian expansion and acceleration laws).
The recent emergence of average
negative scalar curvature in cosmic history (\citealt{Buchert05CQGLetter, BuchCarf08,
  Buchert08status}; see also \citealt{Wiltshire07clocks,
  Wiltshire07exact}), in which curvature inhomogeneity evolves
together with kinematical backreaction (according to a combined
conservation law; \citealt{Buch00scalav}),
is found by several of these groups
to push the effective
scale factor to evolve in a way that potentially mimics `dark energy' on
large scales \citep{RoyFOCUS}. Observationally, the non-negligible
peculiar expansion rate of voids \citep*{ROB13} and the 10\%-level
environmental dependence of the baryon acoustic oscillation (BAO) peak
location \citep{RBOF15,RBFO15} are consistent with a structure
formation alternative to dark energy \citep{RMBO16Hbg1}.  However,
precise observational analysis in a more accurate universe model than
the FLRW model requires precise modelling.  Specific implementations
of emerging negative curvature models have been developed using many
different approximations
(\citealt{Rasanen06negemergcurv,
  NambuTanimoto05,KaiNambu07,
  Rasanen08peakmodel,
  Larena09template,Chiesa14Larenatest,
  WiegBuch10,BuchRas12,
  Wiltshire09timescape,DuleyWilt13,NazerW15CMB,
  ROB13,
  Barbosa16viscos,
  BolejkoC10SwissSzek,
  LRasSzybka13};
see also
\citealt{Sussman15LTBpertGIC,Chirinos16LTBav,
  Bolejko17negcurv,Bolejko17styro,
  Kras81kevolving,Kras82kevolving,Kras83kevolving,Stichel16,
  Coley10scalars,KasparSvitek14Cartan,RaczDobos16}).

There have been very recent attempts to run ``fully'' relativistic
cosmological simulations, which nevertheless require numerical
shortcuts and strategies in order to obtain results in reasonable
amounts of computing time, and, in some methods, adopt standard
flat-space tools such as Fourier transforms.  However, these have so
far failed to show compatibility with the simpler derivations of
emerging average negative curvature models cited above.
Two possible reasons for this may be that these codes
are too tightly coupled to a background whose
average curvature is not allowed to evolve,
and they do not appear to take virialisation into
account. As explained below in \SSS\ref{s-newt-vs-GR},
we adopt the stable clustering hypothesis
\citep{Peebles1980}
for virialisation,
and the QZA for non-virialised domains, without tightly constraining
the average curvature.
This approach thus allows the emergence of average negative curvature
associated with super-EdS expansion rates.

In the terminology of
\citet{AdamekDDK15,AdamekDDK16code},
the approach we present here aims to model the growth of what these authors
call ``homogeneous modes'' in
their hybrid particle--mesh simulations that aim towards consistency
with the Einstein equation.  Pure mesh-based cosmological simulation
codes aiming at consistency with the Einstein equation include those
of \citet{BentivegnaBruni15} and \cite{Macpherson17} using the free-licensed
Einstein Toolkit, as well as those of
\citet*{GiblinMS15departFLRW,GiblinMS16raytrace}.
All of these codes assume a T$^3$
spatial section of their universe model, with the aim being
numerical convenience, even though this choice of topology is
physically reasonable in terms of the Einstein equation
\citep[][and references therein]{
  AslanMan11, BielewPogosJaffePlanck13, RFKB13, Aurich15T3,
  Steiner17}.
Our approach only imposes the dynamical constraint
of a T$^3$ uniformly flat spatial section in the initial conditions.
The mesh-based codes do not yet claim to resolve
gravitational collapse (virialisation). The \citet{AdamekDDK16code}
code aims to do this, but is currently tied closely to an FLRW
background, and is complementary to the approach presented here
(see also \citealt*{Daverio17}).
\postrefereeAAchanges{For approximations related to virialisation and/or
  gradients of effective pressure in
  the context of inhomogeneous cosmology, see
  \citet{Buch01scalav},
  \citet{BolejkoLasky08},
  \citet{BolejkoFerreira12},
  and \citet{AdamekCDK15}.}

The method adopted here to some degree resembles the
\citet{Rasanen06negemergcurv} approach, in the sense that collapsing
and expanding regions are evolved individually,
and the \citet{RaczDobos16} approach of partitioning
$N$-body realisations into small domains and evolving these in order
to generate an effective scale factor evolution.
However, both of those approaches assume spherical
symmetry to calculate the dynamical evolution of their domains, and
neglect spatial continuity by
assuming that the kinematical backreaction $\CQ_\CD$ is zero, which is only
strictly correct in special cases,
such as that of spherically symmetric
collapse against a spatially flat background
[see also \citet{Bolejko08Szekaverage} regarding
  \citet{SzekeresP75} models].
We do not assume spherical symmetry of domains, we have spatial
continuity, and we include the role of $\CQ_\CD$.
\citet{Rasanen06negemergcurv} does not
consider domains whose disjoint union constitutes the global spatial
section.  \citet{RaczDobos16} do consider the density of domains whose
union gives the full spatial section, but the domains are in reality
non-spherical; the authors' assumption of spherical symmetry is an
approximation that assumes kinematical backreaction to be zero.
Their domains appear to be non-Lagrangian: shifting of matter across domains
does not seem to be taken into account in their Equations (3) and (4).

Thus, the method of this paper
differs from
\citet{Rasanen06negemergcurv} and \citet{RaczDobos16}
in that we use generic initial condition
realisations over a contiguous spatial section,
we calculate kinematical backreaction explicitly and use it in the Raychaudhuri equation for
evolving each domain (thus bypassing any need to assume or impose spherical symmetry), and
we adopt the stable clustering hypothesis for virialisation.
After presentation of definitions and method, the method proposed in
this paper is summarised in
Definition~1 (\SSS\ref{defn-VQZA}). 
Future development of our approach will be to make it
free from an assumed reference model,
as in the \cite{WiegBuch10} volume-partitioning approach
or the \citet{Wiltshire07clocks,Wiltshire07exact} biscale
partition. The latter divides the spatial section contiguously,
though not smoothly, into negatively curved void domains and
spatially flat collapsing and virialised structures (`walls'),
with a statistical justification for matching
walls to the expansion history.

The method is presented in \SSS\ref{s-method}, starting
with EdS reference model parameters in \SSS\ref{s-method-H1bg}
and with terminology for the different scales present in $N$-body simulations
in general, and for the method presented here in particular,
in \SSS\ref{s-method-scales}.
The generation of initial
conditions on a mesh covering the
fundamental domain of the T$^3$ spatial
section
is presented in \SSS\ref{s-method-ic}.
This is done using {\sc mpgrafic}
(this package and the others used in this work
are free-licensed\footnote{\url{https://www.gnu.org/licenses/licenses.html}}
software).
In \SSS\ref{s-method-inv-Q}
we explain how to numerically measure the
invariants of what in Newtonian terms is the peculiar velocity
gradient tensor.
This is only done in the initial conditions for our main model.
However, these invariants can also be measured at later time steps
for full $N$-body calculations.
To the degree that
the peculiar gradient tensor
approximates the
(general-relativistic) extrinsic curvature
tensor, our approach should not deviate by much from a relativistic approach, even
though the initial conditions are defined in Newtonian terms.
The library form of the Delaunay Tessellation Field Estimator {\sc dtfe},
slightly extended by the addition of a small number of functions and corrected for
minor bugs, is used to estimate these invariants.
The {\sc inhomog} library is introduced in
\SSS\ref{s-method-RZA}. This is used
to evolve individual cubical domains using the QZA,
each of comoving sidelength $\LD$.
For convenience and in order to ease the use of this approach in full
$N$-body simulations, this is done using \RAMSESscalav{}, an extension to
\RAMSES, as a front end to read in the initial conditions
and call the \DTFE{} and \inhomog{} libraries.
In \SSS\ref{s-method-double-averaging} we
define how to relate the domain-based average scale factors $a_{\CD}(t)$
to a global effective scale factor $\aeff(t)$.

As a complement to the main model of this work,
a method of evolving initially cubical domains $\CD$ in a full $N$-body
simulation, in which $\CQ_\CD$ is estimated numerically at each major time
step instead of evolved using the QZA, is presented
in \SSS\ref{s-method-nbody}.
In this case, the Raychaudhuri equation is integrated by following
the Lagrangian volume defined by the particles of each initial
domain 
and again calculating
an effective global expansion rate $\aeff(t)$. This approach again uses
\RAMSESscalav{} as a front end for reading in initial conditions,
and \DTFE{} as the library for calculating the peculiar velocity
gradient invariants.
The \inhomog{} library is not used in this case.

Since the estimation of $\CQ_\CD$ is crucial to this work,
the accuracy of the \DTFE{}
numerical estimation of the peculiar velocity gradient invariants
(and thus $\CQ_\CD$) is presented in \SSS\ref{s-results-sigmaT3}.
The main results of this work, the comparison of
effective scale factor evolution to EdS evolution and
\LCDM{} evolution, are presented in
\SSS\ref{s-res-best-scales-VQZA}.
In \SSS\ref{s-results-fvir}, the close relation between
virialisation and super-EdS global effective expansion
is shown.
In \SSS\ref{s-res-best-scales-Nbody},
we briefly compare our main results with those of
calculating $\CQ_\CD$ numerically in
our $N$-body comparison method.
The relation between Newtonian and relativistic reasoning
is discussed again in the light of these results, qualitatively presenting a
perfect fluid effective model of virialisation, in \SSS\ref{s-disc}.
A summary and conclusions are given in \SSS\ref{s-conclu}.
We set the spacetime unit conversion constant
\citep{TaylorWheeler92} $c := 1$ unless otherwise specified.
\postrefereeAAchanges{Appendix~\ref{app-invariant-formulae}
  provides a computer algebra script
  for confirming the equivalence of expressions for
  invariants of the peculiar velocity gradient
  tensor (Eq.~\eqref{e-I-II-III-defns}).
  Appendix~\ref{s-appendix-six-point} provides the
  correct expression for the variance of
  the third invariant, which is wrong in
  Eqs~(C20)--(C22) of \citet{BKS00}.}

\section{Virialisation versus spatial expansion}
\label{s-newt-vs-GR}

Although the aim here, as in several related projects, is general-relativistic cosmology
simulations, it is useful to consider Newtonian reasoning.
However, defining a mathematically solid Newtonian cosmological model is not
easy. \citet{BuchertEhlers97}
showed that a T$^3$ Newtonian cosmological model defined
in terms of a dust fluid and gravitational potential, which avoids the divergent infinite sum problem in
the pointwise two-point attraction approach,
can provide a self-consistent solution
(see also
\citealt{EhlersBuchert97};
\citealt{GibbEllis13Newt} for a recent review
and an interesting family of solutions;
and recent reminders in \citealt{Kaiser17noQ,Buchert17Newt}).
In this case, if $\CQ_\CD$ evolution in local domains is
calculated within the constraints of the same
(expanding) T$^3$ `absolute' space, then
the global scale factor evolution of an initially EdS model
retains EdS behaviour (proportionality to $t^{2/3}$),
even when scale factor evolution is calculated in the
local domains $\CD$ using $\CQ_\CD$ and
averaged.
In terms of kinematical backreaction,
the global $\CQTthree = 0$
(see \SSS\ref{s-method-sigmaT3}).
[Equation (61) of \citet{BKS00} is incorrect; the middle
  element of the double equation has to be omitted to make the equation
  correct.]
However, this justification for exact EdS global expansion is only valid
until gravitational
collapse and virialisation (or more generally, shell-crossing).
Before discussing virialisation, let us first consider
the QZA, that is,
the algebraic expression for evaluating $\CQ_\CD$ time evolution
as presented in \citet{BKS00} and \citet{BuchRZA2} in the spirit
of the Zel'dovich approximation \citep{ZA}.

The QZA evolution of the kinematical backreaction $\CQ_\CD$
has the same algebraic structure in the Newtonian
[\citealt{BKS00}, eq.~(42); \citealt{BuchRZA2}, eq.~(57); NZA]
and relativistic [\citealt{BuchRZA2}, eq.~(50)]
cases, but with differing spatial definitions and
interpretations \citep[][IV.A]{BuchRZA2};
we cite this expression in Eqs~\eqref{e-resultQ2}, \eqref{eq:GammaDef}.
This expression for $\CQ_\CD$ evolution is exact for some particular cases
in the Newtonian context
\citep[][III.D, III.E]{BKS00}.
In the relativistic context,
it is exact for plane symmetric collapse [\citealt{BuchRZA2}, eq.~(75)]
and provides exact solutions for certain cases of
spherically symmetric collapse
\citep[][LTB]{Lemaitre33,Tolman34,Bondi47}
models \citep[][V.B.1]{BuchRZA2}.
The QZA requires knowing the growth function for linear perturbations
in a reference model (in our case, the EdS reference model), and
of the initial invariants
of the extrinsic curvature tensor. Here, we
estimate the latter using the Newtonian peculiar velocity gradients,
generated for a standard T$^3$ cosmological
$N$-body simulation.
The initial invariants in non-global domains should
collectively satisfy a partitioning formula
[\citealt{BuchCarf08}, \SSS{}3.2; \citealt{WiegBuch10}, eqs~(16), (17)]
that yields globally zero mean invariants due to the T$^3$ constraint; see
\SSS\ref{s-method-sigmaT3} for details.


How are virialisation, Newtonian and relativistic reasoning related?
The assumptions of the Newtonian collisionless,
compressible fluid approach for exact
spherically symmetric collapse, interpreted strictly, imply
collapse to a singularity in a finite time. A Lagrangian
model is easier to relate to
a relativistic model than an Eulerian model in the sense that the
Lagrangian approach ties matter and space together directly.
However, in this idealised case of spherical collapse,
a positive Lagrangian volume drops suddenly to zero at the final
moment of collapse.
In other words,
the validity of the approach is limited to times before
gravitational collapse events occur. More
generically, the assumptions fail when shell crossing first occurs. In
$N$-body simulations or with physical particles, the usual way to avoid
these Newtonian singularities is to adopt a hybrid fluid--particle
model:
the fluid model allows the model to be defined globally
as shown by \citet{BuchertEhlers97}
(or approximated regionally
in an adaptive mesh refinement method), while
virialisation of a set of particles is expected (and modelled) to
occur instead of
singularity formation.
At a small enough scale, baryon pressure also opposes gravitational
collapse.
Disregarding baryon pressure,
overdensities in a T$^3$ $N$-body simulation
cannot be strictly spherically symmetric because they are composed of
particles; and they cannot evolve
in an exact spherically symmetric way because
the global spatial section is T$^3$. Moreover, Gaussian initial conditions
will generically lead to at least moderately anisotropic collapse.
Finally, excessively high two-particle
close-encounter accelerations are avoided by a small-length-scale
softening parameter. Thus, the limitations of the Newtonian
fluid approach are dealt with
in $N$-body simulations by considering singularity formation
to be unrealistic (unless a simulation is detailed enough to include
active galactic nucleus formation and star formation through to
supermassive and stellar black hole formation).

This avoidance of singularity formation is generally modelled
without any overt compensation in volume evolution at the global
T$^3$ volume evolution level. Below we argue that in the standard
approach (whether analytical or $N$-body), there is, in fact,
{\em hidden} compensation in volume evolution at the global level.

In this work, we are not interested in modelling the details of
virialisation. What is relevant for a relativistic interpretation is
whether or not compensation between volume loss and gain in collapsing
and expanding (respectively) Lagrangian domains is included in the
model.  It is clear that, when interpreting observations in terms
of FLRW models, the virialisation of high-mass collapsed objects
correlates closely with the
cosmologically recent appearance of a non-negligible value of the dark
energy parameter ($\OmLam$).  This observational clue is quantified
in \citet{ROB13}. Relativistically, the assumption of no sudden compensation
between virialising and expanding domains can be described as a
`silent virialisation' approximation
(following \citealt{Matarr94silentPRL,Matarr94silentMN}
\postrefereeAAchanges{for the somewhat different ``silent universe'' approximation};
see also \citealt{EllisTsagas02,Bolejko17styro}).

In \SSS\ref{s-results-VQZA},
we present a two-domain
partition of the T$^3$ volume that illustrates our claim that
the standard approach of passing from pre- to post-virialisation
is relativistically unrealistic.
In \SSS\ref{s-res-VQZA-previr}, we define the biscale partitions
and study volume evolution prior to the collapse and virialisation event.
In \SSS\ref{s-res-VQZA-postvir}, we present the dilemma between
choosing the standard approach versus using
the QZA for the expanding domain
and the stable clustering approximation for virialised domains
(presented in more detail for our main multi-domain partition
modelling in \SSS\ref{s-method-virialisation}). We thus define
the Virialisation $\CQ_\CD$ Zel'dovich Approximation.

\section{Method} \label{s-method}

\subsection{EdS reference model extrapolated from early epochs}
\label{s-method-H1bg}
The use of an
Einstein--de~Sitter (EdS) reference model at early times
(that we previously called a
``background'' EdS model; \citealt{ROB13,RMBO16Hbg1})
or on
large spatial scales in inhomogeneous cosmological models that
aim to be more accurate than the FLRW model risks leading
to confusion when referring to `the' Hubble constant,
since several different values compete for this name.
Here, as in \citet{RMBO16Hbg1}, we set up the EdS reference
model in terms of a reference-model scale factor
$\abg$
and a reference-model expansion rate $\Hbg$ given by
\begin{align}
  \abg := (3 \Honebg t /2)^{2/3} \,, \quad
  \Hbg := \dotabg/\abg = 2/(3t) \,,
  \label{e-defn-Hbg1}
\end{align}
where $\Honebg := \Hbg(\abg =1)$.
This EdS model is intended to approximately fit early epochs.
In order to derive an observationally realistic value of
$\Honebg$, an effective scale factor $\aeff$
which is nearly identical to the reference model scale factor
at early times, that is, $\aeff \approx \abg$ for small $t$,
is adopted.
This is done by normalising to
$\aeff = 1$ at the present time $\tzeroeff \equiv \tzeroefffull$,
where the Planck Surveyor
\citep[Table 4, sixth data column,][]{Planck2015cosmparam}
parametrisation of the $\Lambda$CDM model is used as a proxy for
cosmological observations.
As explained in
\citet{RMBO16Hbg1},
this requires that
we adopt an early-epoch--normalised EdS--reference-model Hubble constant
$\Honebg = 37.7${\kmsMpc}
(\citealt{RaczDobos16} adopt this value too)
and an effective Hubble constant -- the limiting value that should
be observed in the local few hundred Mpc according to an FLRW
fit of the data -- $\Hzeroeff = 67.74${\kmsMpc}.

\subsection{Scales} \label{s-method-scales}

A standard T$^3$ $N$-body simulation has three built-in length scales:
\begin{list}{(\roman{enumi})}{\usecounter{enumi}}
  \item $\Lbox$ -- the side length of the fundamental
    domain, often called the box size,
    \addtocounter{enumi}{2}
  \item $\LN$ -- the mean interparticle separation
    along one of the three fundamental directions
    between `adjacent' particles among a set of $N^{1/3}$
    particles sorted along that axis,
    where the simulation contains $N$
    particles; that is,
    $\LN := \Lbox /N^{1/3}$, and
  \item $\Lsoft$ -- the softening length of Newtonian two-point
    instantaneous attraction; in the case of \RAMSES{} (\SSS\ref{s-method-nbody}),
    this can be considered to be the maximum level of resolution for calculating the Newtonian gravitational
    potential in an adaptively defined cell, $\Lsoft := 2^{-{\mathrm{levelmax}}} \Lbox$.
\end{list}
We insert two additional scales between $\Lbox$ and $\LN$
(we order the definitions from (i) to (v) to monotonically match
length scales).
Non-linear structure formation has characteristic scales
\citep{Peebles74,PeeblesHauser74},
so as in earlier work, we need to set a scale at which
we
estimate the kinematical backreaction $\CQ_\CD$ and
apply the scalar averaged evolution equations
(\SSS\ref{s-method-scalav}).
Since $\CQ_\CD$ is an average of the peculiar velocity gradient tensor invariants,
the latter should be estimated at a smaller scale.
However, to reduce the Poisson noise effects
of using a finite number of particles, the scale at which these invariants are
estimated
(\SSS\ref{s-method-inv-Q})
should be greater than $\LN$. Thus, our intermediate scales are
\begin{list}{(\roman{enumi})}{\usecounter{enumi}}
  \addtocounter{enumi}{1}
\item
  $\LD$ -- the scalar averaging initial domain size in comoving units
  (\SSS\ref{s-method-scalav}), and
\item
  $\LDTFE$ -- the \DTFE{} mesh size
  (\SSS\ref{s-method-inv-Q}).
\end{list}
We correspondingly define
  \begin{align}
    \nD := \Lbox/\LD \,, \quad
    \nDTFE := \Lbox/\LDTFE \,, \quad
    \nsoft := \Lbox/\Lsoft\,.
  \end{align}
Using this terminology, numerical scalar-averaging modelling of
cosmological expansion requires, in general,
\begin{align}
  \Lbox \gg & \LD \gg \LDTFE \gg \LN \gg \Lsoft \nonumber \\
  \Leftrightarrow \;
  & 1 \ll \nD \ll \nDTFE \ll N^{1/3} \ll \nsoft
  \label{e-big-numbers-assumption}
\end{align}
for an $N$-particle simulation,
where the validity of the minimal ratios in this multiple inequality
needs to be studied both numerically and analytically. Setting
the first three ratios of these values to $8$ or $16$ would require $N=512^3$ or $4096^3$,
respectively.
Given these heavy requirements in computer resources,
we limit this initial study to numerical exploration of the modest
value $N=256^3$. For our main calculations, in which the QZA uses
only the initial values of the invariants of the peculiar velocity
gradient (in a Newtonian interpretation), $\nsoft$ is irrelevant.
For the $N$-body comparison, we use $N=128^3$ and set $\nsoft/(N^{1/3}) = 32$.

\subsection{Initial conditions} \label{s-method-ic}

We use the free-licensed
(GNU General Public License, version~2 or later; GPL-2$+$)
package {\sc mpgrafic-0.3.10}
\citep{Bert01grafic,PrunetPichon08mpgrafic}\footnote{\url{http://www2.iap.fr/users/pichon/mpgrafic.html},
  \url{https://bitbucket.org/broukema/mpgrafic}} to generate cosmological initial conditions for an EdS model, with
$\Honebg = 37.7${\kmsMpc} -- which,
in the absence of structure formation,
would give a unity scale factor (for the reference model)
at the unrealistically late foliation time
of 17.3~Gyr [see Eq.~\eqref{e-defn-Hbg1}].

The comoving fundamental domain size $\Lbox$
(see \SSS\ref{s-method-scales})
needs to be expressed
in units of $\honebgMpc$, since the simulation starts with the EdS
reference model. However, in order for length scales to be interpretable
in terms of standard descriptions of low-redshift observations,
it is more useful to choose $\Lbox$ based on a given
low-redshift length scale $\LDTFE$, so
as above, we adopt the \citet{Planck2015cosmparam} estimate
of $\Hzeroeff = 67.74${\kmsMpc}. Thus, we have
\begin{align}
  \frac{\Lbox}{\honebgMpc} &=
  \frac{\Lbox}{\hzeroeffMpc} \frac{\Honebg}{\Hzeroeff} \nonumber \\
  &=
  \frac{\Lbox}{\LDTFE} \frac{\LDTFE}{\hzeroeffMpc} \frac{\Honebg}{\Hzeroeff},
\end{align}
where $\hzeroeff := \Hzeroeff/$100{\kmsMpc}.

We match the \citet{Planck2015cosmparam} power spectrum
normalisation of $\sigma_8^{\Lambda\mathrm{CDM}} = 0.8159$ to our
EdS reference model. We evolve
the effective Planck-parametrised $\Lambda$CDM model backwards
to $z=1000$, that is, to $t=547$~Myr, and then forwards using our
EdS model, obtaining $\sigma_8 = 1.0395$ when $\abg = 1$, which
we use for {\sc mpgrafic} initial conditions.

The value of $\LN$ is set in {\sc mpgrafic}, and later read in
automatically by \RAMSES{}.

\subsection{Estimating `velocity gradient' invariants}
\label{s-method-inv-Q}

In order to model what from a general-relativistic point of view
is the extrinsic curvature tensor, we numerically estimate what
in Newtonian terms is the peculiar velocity
gradient tensor.
That is, we make the standard assumption that
world lines of particles can have their time derivatives separated into
`peculiar velocities' and a refence model expansion $\abg(t)$.

Delaunay and Voronoi tessellation are two methods that \citet{BernardeauWeygaert96} argued
provide close to optimal tracing of the (Newtonian) peculiar velocity field.
Here, we use the former, as implemented in the Delaunay Tessellation Field Estimator
(\DTFE{})\footnote{DTFE-1.1.1.Q: \url{https://bitbucket.org/broukema/dtfe};\\
  DTFE-1.1.1:\\\url{https://www.astro.rug.nl/\%7Evoronoi/DTFE/dtfe.html}}
 which is a free-licensed code that
has been found to be highly competitive in comparison
with other codes that aim to extract peculiar velocity fields from
cosmological $N$-body simulations
\citep{SchaapvdWeygaert00,vdWeygaertSchaap07,CW11,Kennel04kdtreeDTFE}, with good tracing of the details
of the cosmic web of virialised structure
\citep{AragonCalvo10CosmicSpine,Sousbie11cweb,AragonCalvo10Delaunay,
  ParkPranav13,PranavEdelsb16}.
\DDTFE{} uses the free-licensed library
{\sc cgal} (Computational Geometry Algorithms Library\footnote{\url{https://www.cgal.org/}})
to make Delaunay tessellations.

A Delaunay tessellation of an $N$-body simulation snapshot is a
division of the T$^3$ volume into tetrahedra using the particles'
positions, in such a way that none of the particles lies inside the
circum-2-sphere of any of the tetrahedra. Given one such tetrahedron
with vertices $k=0,1,2,3$ and simulation peculiar velocity vectors at these vertices
$v^i_k$, $i=1,2,3$, a linear interpolation of the peculiar velocity gradient
$v^i_{,j}$
relates velocity component changes from the 0-th to the $k$-th vertex for $k\not=0$
\begin{align}
  v^i_k - v^i_0 &= v^i_{,j} \, (x^j_k - x^j_0)\,,
\end{align}
for the usual Einstein summation notation over corresponding upper/lower indices
and a comma in the subscript to indicate a partial derivative.
If the matrix $x^j_k - x^j_0$ is invertible (which, in numerical practice, should
almost always be the case), the velocity gradient
can be estimated as
\begin{align}
  v^i_{,j} &= (x^j_k - x^j_0)^{-1} \, (v^i_k - v^i_0)\,.
\end{align}
We use \DTFE{}
to carry out this linear interpolation
using \DTFE{} ``method 1''. This consists of Monte Carlo sampling
with a quasi-random spatial distribution within a Delaunay tetrahedron,
giving interpolated values that accumulate in
cubical cells of a regular grid of ${\nDTFE}^3$
cells within the simulation box, yielding an averaged interpolated
estimate of $v^i_{,j}$ in each DTFE grid cell.

\subsubsection{S$^1$-based estimates of velocity gradient uncertainties}
\label{s-method-sigmaS1}

We introduce a method of estimating the \DTFE{} grid estimates
of $v^i_{,j}$ that does not seem to have been used previously. This new method
is based on the numerical estimate of the
integral of $v^i_{,j}$ over any closed straight loop S$^1$
passing through the $\nDTFE$ points ${\mathbf x}_k = \{x^j_k\}_j$
in
an ${\nDTFE}^3$ cubical DTFE grid of velocity gradient estimates
in a time slice of a simulation,
where $k \in \{0,\ldots,\nDTFE-1\}$,
and $x^{j_1}$, $x^{j_2}$ are two fixed values
in the $j_1$-th, $j_2$-th directions, respectively,
with $j_1 \neq j \neq j_2$.
Analytically,
\begin{align}
  \int_{S^1} v^i_{,j} \,\diffd x^{j} &= 0
\end{align}
should
hold by the second fundamental theorem of calculus (Stokes' theorem)
if $v^i_{,j}$ is continuous and real-valued. Provided that we
allow the usual S-shaped peculiar velocity profiles across filaments
and virialised objects, this should be satisfied by a standard interpretation
of the information numerically represented by the simulation particles.

Numerically, we make the assumption that the numerical uncertainty
$\sigma_{v^i_{,j}}$ is Gaussian and identically and independently distributed
at each of the \DTFE{} grid points along a given S$^1$ loop.
This is obviously an oversimplification. Collapsing or expanding structures
on scales greater than $\LDTFE$ could reasonably lead to correlations
in numerical errors in the estimates of $v^i_{,j}$ across
close pairs of \DTFE{} grid points, and tensor invariants
of $v^i_{,j}$ have represent
symmetries that may exacerbate these correlations. It is also
quite realistic for the per-grid-point errors to be non-gaussian.
Nevertheless, this order of magnitude estimate of the error
depends on very few assumptions, and provides a consistency check
that can be interpreted in terms of expected failures of these minimal
assumptions.

Thus, we introduce
\begin{align}
  \sigmaSonexy
  &:= \frac{\left\vert \sum_{k \in \{0,\ldots,\nDTFE-1\}}  v^i_{,j} \right\vert}
      {\sqrt{\nDTFE-1}},
      \label{e-sigmaS1-basic}
\end{align}
where we use $\nDTFE-1$ to increase the error estimate slightly at low $\nDTFE$, and the absolute value becomes irrelevant when
this formula is used in Eq.~\eqref{e-sigmaS1-from-var} below.
There is no constraint $i = j$: the velocity components' derivatives
need to correspond to the direction of integration, but the
velocity components themselves do not need to.
This provides $\nDTFE^2$ estimates for
a fixed direction $j$. We define the rms (root mean square)
of these $\nDTFE^2$ error estimates of the $i$-th component summed in the $j$-th
direction:
\begin{align}
  \sigmaSonexyall &:=
  \sqrt{\averageplain{
      \left(\sigmaSonexy\right)^2}_{0 \le k_1 < \nDTFE, 0 \le k_2 < \nDTFE }
  },
  \label{e-sigmaS1-from-var}
\end{align}
where $k_1, k_2$ label the grid positions in the $j_1$-th, $j_2$-th
directions, respectively.
We introduce this into \DTFE{} with the function
{\sc estimateSigmaGradientOneDirec}. For propagating these uncertainties
to estimates in uncertainties in the scalar invariants of $v^i_{,j}$
(\SSS\ref{s-method-sigmaT3}),
we estimate the rms over the diagonal and off-diagonal components
of $v^i_{,j}$ separately:
\begin{align}
  \sigmaSoneondiag
  &:= \sqrt{\averageplain{\left(\widehat{\sigma}_{v^i_{,i}}^{i} \right)^2}}_i \;, 
  \quad\quad
  \sigmaSoneoffdiag
  := \sqrt{\averageplain{\left(\sigmaSonexyall\right)^2}_{i,j,i\neq j} } .
    \label{e-sigmaS1-on-off-diag}
\end{align}
These two error estimators summarise and simplify the uncertainty
information from the S$^1$ constraint from all velocity
gradient components in all three orthogonal directions
(of the simulation interpreted as a Newtonian simulation)
over the whole fundamental domain.
The motivation for separating diagonal from off-diagonal
components is that in a physically realistic simulation,
this might help to separate different types of error.

\subsubsection{S$^1$ and T$^3$-based uncertainties in
  $\invI$ and $\invII$}
\label{s-method-sigmaT3}

\postrefereeAAchanges{In flat space,
the} tensor invariants of the peculiar velocity gradient can
be written
(\citealt{Buchert94,EhlersBuchert97};
see App.~\ref{app-invariant-formulae} for a derivation)
\begin{align}
  \invI(v^i_{,j})  :=&\; \mathrm{tr}(v^i_{,j}) = v^i_{,i} = \nabla \cdot \bfv \nonumber \\
  \invII(v^i_{,j})  :=&\;
  \tfrac{1}{2}
  \left\{
  \left[\mathrm{tr} \left(v^i_{,j}\right) \right]^2 -
  \mathrm{tr} \left[ \left(v^i_{,j}\right)^2 \right]
  \right\} \nonumber \\
   =&\; \tfrac{1}{2}\left((v^i_{,i})^2 - v^i_{,j}\, v^j_{,i}\right) \nonumber \\
   =&\; \tfrac{1}{2}\nabla\cdot
  \Big(\bfv (\nabla\cdot\bfv) - (\bfv\cdot\nabla)\bfv \Big) \nonumber
  \\
  \invIII(v^i_{,j})
   :=&\; \mathrm{det} (v^i_{,j}) \nonumber \\
   =&\; \tfrac{1}{3} v^i_{,j}\, v^j_{,k}\, v^k_{,i}
  - \tfrac{1}{2} v^i_{,i} \, (v^i_{,j}\, v^j_{,i})
  + \tfrac{1}{6} (v^i_{,i})^3 \nonumber \\
   =&\; \tfrac{1}{3}
  \nabla\cdot\Bigg\{ \tfrac{1}{2}
  \; \left[ \nabla\cdot
  \Big( \bfv(\nabla\cdot\bfv) - (\bfv\cdot\nabla)\bfv \Big) \; \right] \; \bfv
  \nonumber \\
  &\;-
  \left[ \; \Big(\bfv(\nabla\cdot\bfv) - (\bfv\cdot\nabla)\bfv\Big)
  \cdot\nabla\right] \;\bfv
  \Bigg\},
  \label{e-I-II-III-defns}
\end{align}
where $\bfv := (v^0,v^1,v^2)$.
Thus, all three invariants can be expressed as divergences, so
that Stokes' theorem again applies, this time over the full T$^3$.
Again, sums of $\invI$, $\invII$, or $\invIII$ over
the full simulation volume should
analytically be zero, but numerically should
indicate the level of numerical
noise. However, assuming that the \DTFE{} grid over the full box
gives statistically independent and identically distributed errors
is likely to be a worse approximation than assuming this within
any S$^1$ straight loop.
Nevertheless, global means of the invariants and of
kinematical backreaction [see Eq.~\eqref{e-defn-Q} below] can
be checked for consistency with zero based on the S$^1$ method of error
estimation.

Again assuming statistically independent gaussian errors
and using Eqs~\eqref{e-I-II-III-defns} and
\eqref{e-defn-Q}, we obtain an S$^1$-based
uncertainty for $\average{\invI}, \average{\invII},$ and
${\CQ}_{\CD}$ of
\begin{align}
  \sigmaavISone &=
  \sqrt{3} \,{\sigmaSoneondiag}
  \nonumber \\
  \sigmaavIISone &=
  \sqrt{ \averageplain{2 \, \left(\sigmaSoneondiag\, v^i_{,i}\right)^2 +
       \left[\sigmaSoneoffdiag\, v^i_{,j} (1-\delta^j_i)\right]^2}_{\mathbf{x}}
       }
  \nonumber \\
  \sigmaQSone &=
  \sqrt{ 4 \, {\sigmaavIISone}^2
    + \frac{16}{9} \, {\sigmaavISone}^2
    {\averageplain{v^i_{,i}}_{\mathbf{x}}}^2 }\;,
  % ...
  \label{e-sigmaS1-I-II}
\end{align}
where
$\mathbf{x}$ represents all $\nDTFE^3$ \DTFE{} grid points and
$\delta^j_i \in \{0,1\}$ is the usual Kronecker delta.

To test \DTFE{}'s numerical accuracy in estimating velocity field
gradients, we define analytical velocity fields
\begin{align}
  \vmodone &:= \left(
  \sin\frac{\pi \nxyz x^0 }{\Lbox},\,   \sin\frac{\pi \nxyz x^1 }{\Lbox},\,   \sin\frac{\pi \nxyz x^2 }{\Lbox} \right)\,,
  \nonumber \\
  \vmodtwo &:= \left(
  \sin\frac{\pi \nxyz x^1 }{\Lbox},\,  \sin\frac{\pi \nxyz x^2 }{\Lbox},\,   \sin\frac{\pi \nxyz x^0 }{\Lbox} \right)\,,
  \label{e-test-vmodels}
\end{align}
for which $\invI$ and $\invII$ are straightforward to calculate
analytically; the only non-zero gradient components are aligned
with the sinusoidal directions in $\vmodone$, and orthogonal
to them in $\vmodtwo$ (giving
$\invI(\vmodtwo) = 0= \invII(\vmodtwo)$).
We study the two following questions:
\begin{list}{(\roman{enumi})}{\usecounter{enumi}}
\item
  What is the signal-to-noise ratio ($S/N$) of the rms of
  $\vmodone$ or $\vmodtwo$, where the signal is defined by
  Eq.~\eqref{e-test-vmodels} and the noise
  $\sigmameas$ is the rms of the difference
  between the analytical expression for $\invI$ or $\invII$ and the
  numerical estimate?
\item
  What is the ratio of $\sigmameas$ to the rms error predicted
  by the S$^1$-based error estimators indicated above
  [Eqs~\eqref{e-sigmaS1-basic}
    --\eqref{e-sigmaS1-on-off-diag},
    \eqref{e-sigmaS1-I-II}]?
\end{list}
The results of these calculations using
{\sc test\_vgrad\_DTFE.cpp} in \DTFE{} are presented
in \SSS\ref{s-results-sigmaT3}.

\subsection{$\CQ_\CD$ Zel'dovich approximation (QZA)}
\label{s-method-scalav}
\label{s-method-RZA} 

Here we present our method of integrating the
averaged Raychaudhuri evolution
at the $\LD$ scale, used in our method (i),
that we implement in the \inhomog{} free-licensed
(GPL-2$+$) library\footnote{\url{https://bitbucket.org/broukema/inhomog},
  \url{https://tracker.debian.org/inhomog}}.
Averaging of $a_{\CD}$ within the full volume, that is,
at the box scale $\Lbox$, is discussed below in
\SSS\ref{s-method-double-averaging}.
For the reader's convenience,
we make this subsection slightly more general than is needed for the specific
method adopted here: we include the case of
a non-zero cosmological constant $\Lambda$.

\subsubsection{Scalar averaging equations}
\label{s-scalav-basic}

The averaged Raychaudhuri equation in the relativistic case
[\citealt{BuchRZA2}, (9)]
is:
\begin{equation}
  3\frac{{\ddot{a}}_{\CD}}{a_{\CD}}
  +4\pi  G\frac{M_{\initial{\CD}}\, a_{\initial{\CD}}^3}{V_{\initial{\CD}} \,a_{\CD}^{3}}
  -\Lambda=\CQ_{\CD}\;,
\label{eq:expansion-law-GR}
\end{equation}
where
$a_{\CD}$ is defined in the usual way [\citealt{BuchRZA2},
\postrefereeAAchanges{eqs~(2), (3)}],
we write the per-domain scale factor and volume evolution in terms
of their initial ($\mathbf{i}$) values
$a_{\CD}^3 / a_{\initial{\CD}}^3 = V_{\CD}(t)/V_{\initial{\CD}}$,
$M_{\initial{\CD}}$ is the mass in the domain $\CD$,
and the averaged Hamiltonian constraint [\citealt{BuchRZA2}, eq.~(10)] is:
\begin{equation}
  \left( \frac{{\dot{a}}_{\CD}}{a_{\CD}} \right)^2
  -\frac{8\pi G}{3} \frac{M_{\initial{\CD}} \, a_{\initial{\CD}}^3}{V_{\initial{\CD}} \,a_{\CD}^{3}}
  + \frac{\average\CR}{6}
  -\frac{\Lambda}{3} = - \frac{\CQ_{\CD}}{6}
\label{eq:adot-GR}
\end{equation}
\citep{Buch00Hiroshima,Buch00scalav,Buch01scalav},
where the kinematical backreaction
is defined in terms of the invariants of the extrinsic curvature tensor,
approximated in terms of the invariants of the velocity gradient tensor
on a domain $\CD$:
\begin{align}
  {\CQ}_{\CD} &:= 2\, \average{\invII} -
  \frac{2}{3} {\average{\invI}}^2.
  \label{e-defn-Q}
\end{align}
For a more physically motivated definition in the Newtonian case,
see \citet[][II.B., Eq.~(5), App.~A]{BKS00}.

We use \DTFE{} to numerically estimate
the initial values of these invariants on a regular cubical mesh
of ${\nDTFE}^3$ comoving positions in the EdS reference model.
Thus, each domain $\CD$ (again a cubical cell) contains $(\nDTFE/\nD)^3$ estimates of
${\invI}$ and ${\invII}$.
We average these to obtain
$\average{\invI}$ and $\average{\invII}$.
As stated above [Eq.~\eqref{e-big-numbers-assumption}],
it should be preferable to have $\nDTFE/\nD \gg 1$.

We only consider the $\Lambda = 0$ (dark-energy--free) case here,
since Occam's razor favours first calculating a DE-free model with
expansion generated consistently with structure formation, which is the main
theme of this paper.
Non-flat FLRW backgrounds are more difficult to model correctly, since Fourier analysis
is invalid in these cases; the power spectrum
needs to be expressed in terms of an orthonormal basis for a constant
non-zero curvature 3-manifold of the appropriate topology.

\subsubsection{Initial invariants and other initial conditions}

The \inhomog{} software
allows the reference model scale factor $a(t)$ to be normalised at unity
either at the initial time or at the present, setting
(cf \citealt{BKS00}~C.1 last paragraph)
\begin{equation}
\initial a :=  a\FLRW(\initial t)
  \label{e-a-FLRW-initial}
\end{equation}
so that
\begin{equation}
  \varrho\FLRW(t)=\frac{ \varrho\FLRW(\initial{t}) \,
    \postrefereeAAchanges{\initial{a}^3}}{a(t)^3},
  \label{e-homog-density-evolution}
\end{equation}
where the subscript `FL' indicates the FLRW reference model.
In this paper,
the reference model is the EdS model described in \SSS\ref{s-method-H1bg},
and $a\FLRW(\initial t)$ is set as described in \SSS\ref{s-method-ic}.
In \citet{BuchRZA2}, the reference model is termed a background.


\citet{BKS00} internally alternates between defining the invariants
of the extrinsic cuvature tensor and the
(normalised and zeroed) growth function $\xi$
[\eqref{eq:def-xi} below] to have a time dimension ($[\invI] = \mathrm{T}^{-1}$,
$[\invII] = \mathrm{T}^{-2}$, $[\invIII] = \mathrm{T}^{-3}$,
$[\xi] = \mathrm{T}$, where $[x]$ denotes the dimension of $x$ and $\mathrm T$ is
the time dimension) or to be dimensionless. \citet{BuchRZA2} uses the dimensional definitions
throughout, except in Section VI.A.
Here we adopt the dimensionless definitions. This
requires the replacements
\begin{eqnarray}
  \xi \rightarrow \frac{\initial{\dot{q}}}{\initial{q}} \xi, \nonumber \nonumber &\;\;\;\;\;\;&
  \invI \rightarrow \left(\frac{\initial{{q}}}{\initial{\dot{q}}}\right) \invI, \nonumber \\
  \invII \rightarrow \left(\frac{\initial{{q}}}{\initial{\dot{q}}}\right)^2 \invII, &\;\;\;\;\;\;&
  \invIII \rightarrow \left(\frac{\initial{{q}}}{\initial{\dot{q}}}\right)^3 \invIII
  \label{e-dimensionless-isation}
\end{eqnarray}
everywhere in the text of \citet{BuchRZA2} except for Section VI.A,
where the growing mode $q\FLRW(t)$ is given in
Eqs~(\ref{e-growth-EdS}) or (\ref{e-growth-flat}).
In many formulae, the replacements
cancel so that the published formulae are unaffected by this change of convention.

Numerical evaluation of these invariants for
the RZA version of the QZA formalism developed
in \cite{BKS00,BuchRZA1,BuchRZA2} requires using a
power spectrum in the reference model,
so the spatial section of the reference model must be
$\mathbb{R}^3$ or $T^3$ (some other flat FLRW spatial sections also
allow Fourier analysis). Here we adopt a standard $\mathbb{R}^3$ power spectrum
\citep{EisenHu98Tofk}.

\citet{BuchRZA2}~VI.A gives the initial conditions for
$^{{\rm RZA}}a_{\CD}$ (for brevity we drop the `RZA' pre-superscript),
following from (A3) and (49) in \citet{BKS00},
where here we use
the dimensionless definition of the growth rate $\xi$ and the invariants,
and a bold {$\mathbf i$} subscript to indicate that the invariants
are calculated at the initial time
on the domain $\CD$ in Lagrangian coordinates ($\initial{\CD}$),
\begin{align}
  a_{\initial{\CD}} &= \initial{a} \nonumber \\
  {\dot{a}_{\CD}}\left(\initial t\right) & =
  {\dot{a}\FLRW}\left(\initial t\right)\left(1+\frac{1}{3}\initaverage{\initial{\invI}}\right).
  \label{e-init-conds-RZA2}
\end{align}
An alternative choice would be to use
the Hamiltonian constraint \eqref{eq:adot-GR} and assume
that the initial epoch is early enough that turnaround has not yet been reached,
$\dot{a}_{\CD}(\initial t) > 0$,
in order to select
the positive square root for $\dot{a}_{\CD}(\initial t)$,
yielding
\begin{align}
  a_{\initial{\CD}} &= \initial{a} \\
  {\dot{a}_{\CD}}\left(\initial t\right) & =
  \initial{a}
  \sqrt{
    \frac{8\pi G}{3} \frac{M_{\initial{\CD}}}{V_{\initial{\CD}}\initial{a}^{3}}
    - \frac{\average\CR \left(\initial t\right)}{6}
    - \frac{\CQ_{\CD} \left(\initial t\right)}{6}
    +\frac{\Lambda}{3} }\;,
  \label{e-init-conds-RZA2-better}
\end{align}
where we postpone expresssions
for $\average\CR \left(\initial t\right)$ and
$\CQ_{\CD} \left(\initial t\right)$ to
\eqref{e-Q-init} and
\eqref{e-R-init} below.

\subsubsection{Evolution equation}

Writing the per-domain density
$\rho_{\CD} = \left(M_{\initial{\CD}}/V_{\initial{\CD}}\right)\,
a_{\initial{\CD}}^3/a_{\CD}^3$,
equation \eqref{eq:expansion-law-GR} becomes
\begin{equation}
  3\frac{{\ddot{a}}_{\CD}}{a_{\CD}} +
  {4\pi G \average{\varrho}}
  = \CQ_{\CD} + \Lambda\;,
  \label{eq:expansion-law-GR-II}
\end{equation}
and with \citet[][eq.~(92)]{BuchRZA2} (with a
dimensionless definition of $\xi$ and the invariants) can be
expressed as
\begin{equation}
  3\frac{{\ddot{a}}_{\CD}}{a_{\CD}}+4\pi
  G \frac{a\FLRW^{3}}{a_{\CD}^{3}}\varrho\FLRW\left(1-\initaverage{\initial{\invI}}\right)
  = \CQ_{\CD}  + \Lambda\;
  \label{eq:expansion-law-GR-III}
\end{equation}
and thus, using \eqref{e-homog-density-evolution},
\begin{equation}
  3\frac{{\ddot{a}}_{\CD}}{a_{\CD}}+ \frac{A}{a_{\CD}^{3}}  = \CQ_{\CD}  + \Lambda\;,
  \label{eq:expansion-law-GR-IV}
\end{equation}
where the constant $A$ is defined
\begin{eqnarray}
  A &:=& 4\pi G {\varrho\FLRW}(\initial{t}) a_{\initial{\CD}}^3
  \left(1-\initaverage{\initial{\invI}}\right)
  \nonumber \\
  &=& \frac{3}{2} {H}\FLRW^2(\initial{t}) \Omm(\initial{t}) a_{\initial{\CD}}^3
  \left(1-\initaverage{\initial{\invI}}\right),
  \label{e-defn-Aconst}
\end{eqnarray}
with $H\FLRW := \dot{a}\FLRW/a\FLRW$
and $\Omm := 8\pi G \rho\FLRW/(3H\FLRW^2)$ the usual FLRW matter density parameter.
For the EdS reference model we have
\begin{equation}
  A = \frac{3}{2} {H}\FLRW^2(\initial{t}) a_{\initial{\CD}}^3
  \left(1-\initaverage{\initial{\invI}}\right).
  \label{e-EdS-Aconst}
\end{equation}
This allows \eqref{e-init-conds-RZA2-better} to be rewritten
\begin{align}
  {\dot{a}_{\CD}}\left(\initial t\right) & =
  \initial{a}
  \sqrt{
    \frac{2}{3}\frac{A}{\initial{a}^3}
    - \frac{\average\CR \left(\initial t\right)}{6}
    - \frac{\CQ_{\CD} \left(\initial t\right)}{6}
    +\frac{\Lambda}{3} }\;
  \label{e-init-conds-RZA2-ham}
\end{align}
and the Hamiltonian evolution equation becomes
\begin{align}
  {\dot{a}_{\CD}}\left( t\right) & =
  \pm a_{\CD}
  \sqrt{
    \frac{2}{3}\frac{A}{{a}^3}
    - \frac{\average\CR \left( t\right)}{6}
    - \frac{\CQ_{\CD} \left( t\right)}{6}
    +\frac{\Lambda}{3} }\;.
  \label{e-RZA2-ham-sqrt-integrator}
\end{align}
Typical collapsing solutions start with $\dot{a}_{\CD} > 0 $,
the positive square root. Finding the first local minimum
of $\dot{a}_{\CD}^2$ in the positive square root case yields
an estimate of the turnaround epoch $\tturnaround$
(when $\dot{a}_{\CD}$ drops to zero), enabling
a switch to the negative square root for an initial approximate
solution that continues through to collapse. Iterative improvement
of the estimates of $\tturnaround$ and $a_{\CD}(t), \dot{a}_{\CD}(t)$
yields improved accuracy.

\subsubsection{Auxiliary equations}

To numerically solve
Eq.~\eqref{eq:adot-GR} would require estimating
the evolution of both the
kinematical backreaction
$\CQ_{\CD}$ and the mean curvature $\average{\CR}$.
The kinematical backreaction is given by \citet{BuchRZA2}~(50)
(or its Newtonian equivalent):
\begin{eqnarray}
  \CQ_{\CD}\;= & {\displaystyle
    \frac{\dot{\xi}^{2}\left(\gamma_{1}+\xi\gamma_{2}+\xi^{2}\gamma_{3}\right)}{\left(1+\xi\initaverageriem{{\rm
          I}_{{\rm {\bf i}}}}+\xi^{2}\initaverageriem{{\rm
          II}_{{\rm {\bf i}}}}+\xi^{3}\initaverageriem{{\rm
          III}_{{\rm {\bf i}}}}\right)^{2}}\;,} \nonumber \\
  \label{e-resultQ2}
\end{eqnarray}
where
\begin{align}
  \begin{cases}
    \gamma_{1}:=2\initaverageriem{{\rm II}_{{\rm {\bf i}}}}-\frac{2}{3}\initaverageriem{{\rm I}_{{\rm {\bf i}}}}^{2}\\
    \gamma_{2}:=6\initaverageriem{{\rm III}_{{\rm {\bf i}}}}-\frac{2}{3}\initaverageriem{{\rm II}_{{\rm {\bf i}}}}\initaverageriem{{\rm I}_{{\rm {\bf i}}}}\\
    \gamma_{3}:=2\initaverageriem{{\rm I}_{{\rm {\bf i}}}}\initaverageriem{{\rm III}_{{\rm {\bf i}}}}-\frac{2}{3}\initaverageriem{{\rm II}_{{\rm {\bf
          i}}}}^{2},\end{cases}
\label{eq:GammaDef}
\end{align}
and the subscript
$\CI$ indicates that a fully relativistic calculation
would integrate the curved, Riemannian volume even at
this early time when perturbations are weak,
and the dimensionless growth rate $\xi$ is given in Eq.~\eqref{eq:def-xi}.
The mean curvature is given by \citet{BuchRZA2}~(13), (54) for the flat FLRW background
case:
\begin{eqnarray}
  \average{\CR}\;= & {\displaystyle
    \frac{\dot{\xi}^{2}\left(\tilde\gamma_{1}+\xi\tilde\gamma_{2}+\xi^{2}\tilde\gamma_{3}\right)}{\left(1+\xi\initaverageriem{{\rm
          I}_{{\rm {\bf i}}}}+\xi^{2}\initaverageriem{{\rm
          II}_{{\rm {\bf i}}}}+\xi^{3}\initaverageriem{{\rm
          III}_{{\rm {\bf i}}}}\right)^{2}}\;,} \nonumber \\
  \label{e-resultR2}
\end{eqnarray}
where
\begin{align}
  \begin{cases}
    \tilde\gamma_{1}:=
    -2\initaverageriem{{\rm II}_{{\rm {\bf i}}}}
    -12\initaverageriem{{\rm I}_{{\rm {\bf i}}}} \frac{H}{\dot\xi}
    -4\initaverageriem{{\rm I}_{{\rm {\bf i}}}} \frac{\ddot\xi}{\dot\xi^2}\\
    \tilde\gamma_{2}:=
    -6\initaverageriem{{\rm III}_{{\rm {\bf i}}}}
    -24\initaverageriem{{\rm II}_{{\rm {\bf i}}}} \frac{H}{\dot\xi}
    -8 \initaverageriem{{\rm II}_{{\rm {\bf i}}}} \frac{\ddot\xi}{\dot\xi^2}\\
    \tilde\gamma_{3}:=
    -36\initaverageriem{{\rm III}_{{\rm {\bf i}}}}\frac{H}{\dot\xi}
    -12\initaverageriem{{\rm III}_{{\rm {\bf i}}}} \frac{\ddot\xi}{\dot\xi^2} .
  \end{cases}
\label{eq:GammaTildeDef}
\end{align}
As detailed below in \SSS\ref{s-rza-num-strategy},
we bypass the direct dependence on $\average{\CR}$
by integrating
Eq.~\eqref{eq:expansion-law-GR-IV} instead.
Evaluation of these expressions requires the growth rate
$\xi(t)$, for which we use the dimensionless form of
\citet{BuchRZA2}~(32),
\begin{equation}
  \xi(t):= \frac{\lbrack q\FLRW(t)-q\FLRW(\initial t)\rbrack}{{{q}\FLRW(\initial t)}} \;.
  \label{eq:def-xi}
\end{equation}
For the EdS reference model, the growing mode is
\begin{equation}
  q\FLRW(t) = a\FLRW(t)
  \label{e-growth-EdS}
\end{equation}
(for example, Chapter 15, \citealt{Wein72};
eq.~(21) \citealt{Bildhauer92LambdaPancake}).
For completeness, the growing mode for low-density flat FLRW backgrounds
(such as \LCDM) with
present density parameter $\Ommzero<1$ and
$\OmLamzero := 1 -\Ommzero$ is:
\begin{align}
  q\FLRW(t) =& \frac{5\Ommzero}{2\OmLamzero}
  \left\{
  \frac{3 \Ommzero + \OmLamzero a\FLRW}{\OmLamzero a\FLRW} \; - \right. \nonumber \\
  &\left.    \frac{ 3\Ommzero \sqrt{\Ommzero +\OmLamzero a\FLRW} }{
    (\OmLamzero a\FLRW)^{3/2} }
  \,\mathrm{asinh} \left(\sqrt{\frac{\OmLamzero a\FLRW}{\Ommzero}}\right)
  \right\}
       \nonumber \\
  \label{e-growth-flat}
\end{align}
[\citealt{Bildhauer92LambdaPancake}, (21)].

Since $\initial\xi = 0$ by definition (see \eqref{eq:def-xi}),
the initial backreaction terms
needed in \eqref{e-init-conds-RZA2-ham}
follow from \eqref{e-resultQ2}--\eqref{eq:GammaTildeDef}:
\begin{align}
  \CQ_{\CD}(\initial t) &=
  \dot{\initial \xi}^{2}
  \left( 2\initaverageriem{{\rm II}_{{\rm {\bf i}}}}-\frac{2}{3}\initaverageriem{{\rm I}_{{\rm {\bf i}}}}^{2} \right)
  \label{e-Q-init}
\end{align}
and
\begin{align}
  \CR_{\CD}(\initial t) &=
  \dot{\initial\xi}^{2}
    \left(
    -2\initaverageriem{{\rm II}_{{\rm {\bf i}}}}
    -12\initaverageriem{{\rm I}_{{\rm {\bf i}}}} \frac{H}{\dot\xi}
    -4\initaverageriem{{\rm I}_{{\rm {\bf i}}}} \frac{\ddot\xi}{\dot\xi^2}
    \right)\;.
    \label{e-R-init}
\end{align}

For EdS, we have $a\FLRW(t) = \initial{a} (t/\initial{t})^{2/3}$, so from
\eqref{e-a-FLRW-initial} and \eqref{e-EdS-Aconst} we have
\begin{eqnarray}
\frac{ \dot{a}\FLRW(\initial t)}{\initial{a} } &=&  H\FLRW(\initial t) = \frac{2 }{3\initial t} \nonumber \\
  A & = &  \frac{2 \initial{a}  }{3\initial t^2} \left(1-\initaverage{\initial{\invI}}\right).
  \label{e-EdS-IC-RZA2}
\end{eqnarray}
The standard FLRW expressions can be used for the low-density flat background case.

\subsubsection{Numerical strategy} \label{s-rza-num-strategy}

The Raychaudhuri equation \eqref{eq:expansion-law-GR-IV}
is a second-order ordinary differential equation (ODE),
which can be
reduced to two first-order equations
\begin{eqnarray}
  \dot{a}_{\CD} &=& y \nonumber \\
  \dot{y} &=& \frac{1}{3} \left[(\CQ_{\CD}+\Lambda) a_{\CD} - \frac{A}{a_{\CD}^2} \right]
  \label{e-first-order-ODEs}
\end{eqnarray}
where $y \equiv \dot{a}_{\CD}$ is defined to clarify the numerical strategy.
This pair of first-order equations can be solved numerically by calculating
$\CQ_\CD(t)$ from \eqref{e-resultQ2} and \eqref{eq:GammaDef} and
the initial values of the invariants
$\initaverage{\initial{\invI}},
\initaverage{\initial{\invII}},$
and $\initaverage{\initial{\invIII}}$,
and setting the scale factor initial conditions
from Eq.~\eqref{e-init-conds-RZA2}
and the constant $A$ from
Eqs~\eqref{e-EdS-Aconst} or
\eqref{e-EdS-IC-RZA2},
and the growth function from
Eq.~\eqref{e-growth-EdS}.
In principle,
as in \citet{OstrowskiBuchR16},
typical values of
$\initaverage{\initial{\invI}},
\initaverage{\initial{\invII}},$
and $\initaverage{\initial{\invIII}}$,
could be evaluated
under the assumption of Gaussianity of the initial invariants,
using eqs~(C5), (C14) and (C18) of \citet{BKS00} and
the corrected equivalent of eqs~(C20)--(C22) of \citet{BKS00}, which we provide
here in Eq.~\eqref{e-BKS00-C20-corrected} in App.~\ref{s-appendix-six-point},
since it has not been published previously.
However, in this work, we estimate
$\initaverage{\initial{\invI}},
\initaverage{\initial{\invII}},$
and $\initaverage{\initial{\invIII}}$ from the initial conditions
generated by {\sc mpgrafic}, which only assumes Gaussianity of
the density fluctuations. Thus, the distributions of the second and third invariants
of the peculiar velocity gradients are
not constrained to be Gaussian.

Numerical solvers for ODEs often require Jacobians. In the
case of Eq.~\eqref{e-first-order-ODEs}, these are
\begin{eqnarray}
  \frac{\partial\dot{a}_{\CD}}{\partial a_{\CD}} = 0 &\;\;\;\;\;\;&
  \frac{\partial \dot{y}}{\partial a_{\CD}} =
  \frac{1}{3} (\CQ_{\CD}+\Lambda)  +   \frac{2}{3} {A}{a_{\CD}^{-3}}
  \nonumber \\
  \frac{\partial\dot{a}_{\CD}}{\partial y} = 1
  &\;\;\;\;\;\;&
  \frac{\partial \dot{y}}{\partial y} =
  0 \nonumber \\
  \frac{\partial \dot{a}_{\CD}}{\partial t} = 0 &\;\;\;\;\;\;&
  \frac{\partial \dot{y}}{\partial t} =
   \frac{1}{3} a_{\CD} \frac{\diffd (\CQ_{\CD}+\Lambda)}{\diffd t} .
  \label{e-first-order-ODEs-Jacobian}
\end{eqnarray}
We solve these equations using the embedded Runge--Kutta--Fehlberg $(4, 5)$ method
of the {\sc GNU Scientific Library (GSL)} routine {\sc gsl\_odeiv\_step\_rkf45}.

The averaged density parameters then follow from (18) and (19) in
\citet{BuchRZA2},
\begin{align}
  \OmmD =
  \frac{8\pi G \average{\varrho} }{3 {H_{\CD}}^2}
  &= \frac{ H\FLRW^2(\initial{t}) \,\Omm(\initial{t}) a_{\initial{\CD}}^3}{\dot{a}_{\CD}^2 a_{\CD}}
  \left( 1 - \initaverage{\initial{\invI}}\right) \nonumber \\
  &= \frac{2}{3} \frac{A \,\Omm(\initial{t})}{\dot{a}_{\CD}^2 a_{\CD}}
  \nonumber \\
  \OmQD &= -\frac{\CQ_{\CD}}{6 H_{\CD}^2} \nonumber \\
  \OmLam^{\CD} &= \frac{\Lambda}{3 H_{\CD}^2} = {\OmLam}\FLRW \frac{H\FLRW^2}{H_{\CD}^2} \nonumber \\
  \OmRD &= 1 - \OmmD - \OmLam^{\CD} - \OmQD.
  \label{e-Omegas-a-ODE}
\end{align}

\subsubsection{Virialisation}
\label{s-method-virialisation}

After turnaround of a given domain $\CD$, that is,
when $\dot{a}_{\CD}$ becomes negative,
we continue integration
towards numerical collapse
($a_{\CD} \ll 1$)
at foliation time $\tcollapse$.
The initial solution $\{ a_{\CD}(t), \dot{a}_\CD(t) : t < \tcollapse \} $ is used as an
approximation that is improved iteratively.
We assume virialisation -- we set
 \begin{align}
   a_{\CD} (t > \tcollapse) := a_{\mathrm{EdS}}(\tcollapse)\;
   / (18\pi^2)^{1/3},
   \label{e-vir-EdS}
 \end{align}
 where $18\pi^2$ is the usual EdS
 stable clustering \citep{Peebles1980}
 virialisation overdensity
\citep[][eq.~(A16)]{LaceyCole93MN}. For completeness, the \inhomog{}
package provides the corresponding value for the flat-$\Lambda$ case
if that is chosen, using the fitting formula of
\citet[][(6)]{BryanNorman98} to \citet{EkeColeFrenk96}'s
calculation.

As motivated below in \SSS\ref{s-results-VQZA},
we do not introduce any behaviour in other
domains that compensates for virialisation events.
Indeed, it is not obvious how such compensating volume
evolution could reasonably be approximated, apart from imposing
the reference model expansion rate and abandoning the aim
of estimating structure-generated effective expansion.

\subsection{Regional versus global averaging}
\label{s-method-double-averaging}

As in \citet[][3.1]{Rasanen06negemergcurv} and \citet{RaczDobos16}, we cubically average
the domain-wise scale factors,
\begin{align}
  \aeff(t) : = \left(\frac{\sum_{\CD} a_{\CD}^3(t)}{\sum_{\CD} 1}\right)^{1/3}
  \; = \; \frac{\left(\sum_{\CD} a_{\CD}^3(t)\right)^{1/3}}{\nD}\,,
  \label{e-double-averaging}
\end{align}
where the volumes of collapsed domains are bound below by their
virialisation volumes at the time of collapse,
as defined in Eq.~\eqref{e-vir-EdS}.
\postrefereeAAchanges{Although this represents volume averaging at
  the global level, virialisation is not represented in $\CQ_\CD$,
  so unless $\CD$ is chosen large enough for no virialisation events to occur,
  Eq.~\eqref{eq:expansion-law-GR} is expected to fail at the
  global level, since the single fluid stream assumption fails at
  virialisation. Thus,
  $\aeff(t) = \abg(t)$ is not expected in the presence of virialisation.
  See \SSS\ref{s-disc-flat-T3-zeroQ-violation} for more discussion of this
  point.}

\subsection{\RAMSES{} front end and $N$-body measurement of $\CQ_\CD$}
\label{s-method-nbody}

The two main classes of faster simulation techniques are tree
codes and particle-mesh (PM) based methods \citep[e.g., ][]{Bagla05,Dehnen11}.
In this work, we provide a patch to the adaptive mesh refinement (AMR)
\RAMSES{} code\footnote{\url{https://bitbucket.org/rteyssie/ramses}}
\citep{Teyssier02,GuilletTeyssier11}, available under
CeCILL (GNU GPL compatible) as Fortran 90 code, that we
tentatively refer to as \RAMSESscalav{}\footnote{\url{https://bitbucket.org/broukema/ramses-scalav}}
This serves as a front end to read in the initial conditions files
(\SSS\ref{s-method-ic})
and call the \DTFE{} and \inhomog{} libraries to
estimate the per-domain initial invariants
$\initaverage{\invI}$,
$\initaverage{\invII}$, and
$\initaverage{\invIII}$, and to
carry out VQZA evolution
and volume averaging
as described in
\SSSS\ref{s-method-RZA} and
\ref{s-method-double-averaging}.

In an initial exploration of more numerical approaches,
the \RAMSESscalav{} front end also allows $\CQ_\CD$ to be calculated
numerically from the invariants
at each major time step
using Eq.~\eqref{e-defn-Q} instead of using the QZA analytical calculation of
\eqref{e-resultQ2} and \eqref{eq:GammaDef}, which is based on the initial
values of the invariants and the growth rate of the reference model.
We then integrate Eq.~\eqref{eq:expansion-law-GR-IV},
again rewriting it in the form of Eq.~\eqref{e-first-order-ODEs}.
In this case, we integrate using the classical (non-adaptive) Runge--Kutta (fourth-order) algorithm.
As in \SSS\ref{s-method-double-averaging}, we average the domain-level
scale factors $a_{\CD}$ to obtain $\aeff$ using
Eq.~\eqref{e-double-averaging}.
For any given domain $\CD$, we store the list of particles
initially present in $\CD$, and at time $t$ we
calculate $\average{\invI}$ and $\average{\invII}$ by
weighting $\invI$ and $\invII$ by the volume element
$\diffd \mu_i$ of the $i$-th particle.
All parameters ${\invI}_i$, ${\invII}_i$ and $\diffd \mu_i$ are
calculated as (Euclidean) linear interpolations
of the ${\invI}$, ${\invII}$ and $1/\rho$ values
of the eight
neighbouring \DTFE{} cells surrounding the particle. The
global normalisation between $\diffd \mu_i$ and
$1/\rho_i$ is arbitrary.

  \begin{figure}
    \input{biscale_partition}
    \caption{Pre- and post-virialisation EdS-normalised scale factor evolution
      $\aeff/\abg$ for a biscale partition evolved using the VQZA
      as motivated in \protect\SSS\ref{s-newt-vs-GR} and defined
      in \protect\SSS\ref{s-results-VQZA}. From bottom to top at
      $t \gtapprox 7$~Gyr, the arbitrarily chosen dimensionless
      values of the average initial invariants
      (at initial scale factor $\initial{\abg} = 0.005$)
      in the expanding
      domain are $(\initaverageDunder{\invI}, \initaverageDunder{\invII},
      \initaverageDunder{\invIII}) = $
      $(0.01, 0, 0)$ (`planar' case);
      $(0.01, 10^{-4}/3, 10^{-6}/27)$ (`spherical' $\CD^-$ case);
      $(0.01, 10^{-4}, 0)$; and
      $(0.01, 0, 10^{-6})$; respectively.
      The sharp transitions from
      pre-virialisation to post-virialisation epochs are clearly visible
      and occur at roughly 7~Gyr to 2~Gyr, respectively. Prior to virialisation,
      the QZA is fully compatible with EdS global evolution (which can be thought of as
      Newtonian T$^3$ cosmology), especially in the planar case, in which the QZA
      is exact in the Newtonian case.
      \label{f-biscale}}
\end{figure}

  \begin{figure}
    \input{biscale_underdense}
    \caption{Expanding-domain scale factor evolution $a_{\CD^-}/\abg$
      for the biscale partition models illustrated in
      Fig.~\protect\ref{f-biscale}, comparing evolution of this domain
      according to the VQZA model (silent virialisation) and according to the standard
      $N$-body EdS `Newtonian' constraint (instantaneous feedback from virialisation).
      The VQZA scale factor
      evolution ($+$ symbols) correspond from bottom to top to
      those from bottom to top in Fig.~\protect\ref{f-biscale}.  The
      standard-model scale-factor
      $a_{\CD^-}^{\mathrm{Newt}}(t)/\abg(t)$
      [solid curves;
      Eq.~(\protect\eqref{e-biscale-aDminus-Newt})]
      are indistinguishable
      following virialisation of the overdense domain, since by
      assumption, the expanding domain suddenly switches from
      hyperbolic (super-EdS) evolution to very-nearly flat (EdS) evolution
      ($a_{\CD^-}^{\mathrm{Newt}}(t)/\abg(t) = 2^{1/3}$)
      when virialisation of the overdense domain occurs. A tiny difference
      from perfectly flat EdS evolution is present because the overdense
      domain has a small but non-zero fixed (stable clustering) volume.
      \label{f-biscale-underdense}}
\end{figure}

Since we do not change the list of
particles in a domain between time steps, the domain shape,
which we can conceptually imagine as the union of Voronoi tessellation
tetrahedra surrounding the particles,
is unlikely to remain cubical. It is
likely to become irregular and disconnected. The domain at late times
will contain holes that exclude particles that have entered from
neighbouring domains, and will include small islands of space around
particles that have escaped the main
body of the domain. In principle, this is not a problem, since
Eq.~\eqref{eq:expansion-law-GR-IV} is based on Riemannian (or Lebesgue
in the case of a global T$^3$ domain) integration, which is additive.

By default, a \RAMSES{} cosmological simulation runs like a typical
$N$-body simulation, with the reference model expansion inserted at each time
step from a pre-calculated model, decoupled from structure formation.
For future work, our \RAMSESscalav{} front end makes it easy to
adopt a more consistent approach, that is,
to use the cubically averaged
$\aeff(t)$ [Eq.~\eqref{e-double-averaging}] when calculating
the Newtonian gravitational potential that is used for inferring
accelerations of particles. In the present work,
we limit calculations to those using the reference model EdS scale
factor evolution, in order to compare with
the VQZA approach.

\section{VQZA super-EdS expansion: biscale partition}
\label{s-results-VQZA}

  \begin{figure}
    \input{biscale_OmRD}
    \caption{Mean curvature functional $\OmRDunder$ evolution of the
      expanding domain $\CD^-$ in the biscale partition models illustrated in
      Figs~\protect\ref{f-biscale}
      and \protect\ref{f-biscale-underdense}.
      VQZA curvature evolution is shown
      with $+$ symbols and the
      standard $N$-body EdS constraint
      (interpreted relativistically, not in Newtonian terms)
      is shown by curves. We hypothesise that the
      smooth curvature evolution ($+$) is relativistically more accurate than
      the sudden drop to flatness (solid curves) of the standard model.
      \label{f-biscale-OmRD}}
\end{figure}

\citet{Rasanen06negemergcurv} proposed that the volume average
of expanding and contracting domains defined against an FLRW reference model
may evolve differently from, and in particular, at later times, evolve faster
than the background. As stated in \SSS\ref{s-newt-vs-GR}, prior to virialisation,
this claim fails in a strictly Newtonian T$^3$
case for contiguous space: the kinematical backreaction $\CQ_\CD$
in the expanding and contracting domains
is a non-linear effect that
compensates for what otherwise appears to be dynamical asymmetry in non-linear evolution
of initially small density perturbations. This was established in
\citep{BuchertEhlers97}.

\subsection{Newtonian conservation of EdS expansion prior to virialisation}
\label{s-res-VQZA-previr}

We illustrate this here by dividing a T$^3$ volume of an EdS model
into exactly two domains
$\CD^+$ and $\CD^-$ of initially equal volume:
T$^3 = \CD^+ \cup \CD^-$; $\CD^+ \cap \CD^- = \emptyset$;
$\initialDover{V} = \initialDunder{V}$.
We set $\CD^-$ to be the expanding domain and $\CD^+$ the contracting domain.
Thanks to Stokes' theorem,
the global {(Newtonian)}
averages of the three invariants of the
peculiar velocity gradient (approximating
{what in the relativistic case
  is} the peculiar expansion tensor;
see III.B.1 in \citealt{BuchRZA2}),
where the invariants are given in \SSS\ref{s-method-sigmaT3},
are each zero on this compact spatial 3-manifold.
With volume partitioning [see \citealt{WiegBuch10}, eqs~(16), (17)] and our
choice to give the two domains initially equal volumes, we have per-domain average
initial invariants related by
\begin{align}
  \initaverageDover{I} = -\initaverageDunder{I},\;\;
  \initaverageDover{II} = -\initaverageDunder{II},\;\;
  \initaverageDover{III} = -\initaverageDunder{III}\;.
  \label{e-biscale-complements}
\end{align}
Using Eqs~\eqref{e-resultQ2}, \eqref{eq:GammaDef}, \eqref{eq:def-xi}, and
\eqref{e-growth-EdS}, the QZA on this biscale volume split should provide
approximate confirmation that the EdS scale factor evolution is conserved during
a Newtonian pre-virialisation period in cases in which the QZA is approximate,
and exact (within numerical accuracy) confirmation in cases in which it is exact.

We consider several subcases --
a class that includes planar collapse:
\begin{align}
  \initaverageDunder{\invII}= 0, \initaverageDunder{\invIII} = 0 \;;
  \label{e-planar-collapse}
\end{align}
a class that includes spherically symmetric expansion or collapse:
\begin{align}
  \initaverageDunder{\invII}=
  \initaverageDunder{\invI}^2/3  \,,\;\;
  \initaverageDunder{\invIII} =
  \initaverageDunder{\invI}^3/27\;;
  \label{e-spherical-expansion}
\end{align}
and cases where either
$\initaverageDunder{\invII}$ or
$\initaverageDunder{\invIII}$ is set to zero and the non-zero invariants
are set to values that give terms in
Eqs~\eqref{e-resultQ2} and \eqref{eq:GammaDef} of roughly similar orders
of magnitude. Specific values are indicated in the caption of
Fig.~\ref{f-biscale}.

The planar collapse class of solutions \citep[][III.D]{BKS00} constitute an exact solution family,
which is clearly supported numerically in Fig.~\ref{f-biscale} prior
to the collapse of the collapsing domain at about 6.2~Gyr.
Prior to 6.2~Gyr, the global super-EdS scale factor ratio is unity
to within a precision of $|1 - \aeff(t)/\abg| < 2.3 \times 10^{-5}$,
where $\aeff$ is defined in Eq.~\eqref{e-double-averaging}.
In other words, prior to virialisation, the QZA in the planar class of solutions
is fully consistent with
the Newtonian $T^3$ cosmology
that is normally used to calculate structure formation in the flat FLRW
models (EdS and $\Lambda$CDM).

The spherical symmetric class of solutions is also exact
\citep[][III.E]{BKS00}, with $\gamma_1 = \gamma_2 = \gamma_3 = 0$.
However, if Eq.~\eqref{e-spherical-expansion} is satisfied so that
the expanding domain $\CD^-$ is a member of this class, then
Eq.~\eqref{e-biscale-complements} prevents
the contracting domain from satisfying
Eq.~\eqref{e-spherical-expansion} (with $\CD^-$ replaced by $\CD^+$),
since the squared second invariant is necessarily positive in both
cases. Thus, a small deviation from global EdS evolution
is visible in Fig.~\ref{f-biscale}: the global super-EdS scale factor ratio
is close to unity but drops a few percent below unity just prior to the collapse
at 4.3~Gyr.
The other two cases show similar levels of deviation from an exact global
EdS expansion just prior to collapse of the collapsing domain.

\subsection{After virialisation: the VQZA}
\label{s-res-VQZA-postvir}

In Fig.~\ref{f-biscale}, the global scale factor evolution
follows from assuming the stable clustering hypothesis [Eq.~\eqref{e-vir-EdS}]
for the collapsed domain and by assuming that the expanding domain continues its expansion
as given in the QZA in Eqs~\eqref{e-resultQ2}, \eqref{eq:GammaDef}, \eqref{eq:def-xi}, and
\eqref{e-growth-EdS}. This conservative assumption motivates the following definition.
\begin{definition}\label{defn-VQZA}
  {\em Virialisation $\CQ_\CD$ Zel'dovich approximation (VQZA)}:
  \begin{list}{(\roman{enumi})}{\usecounter{enumi}}
  \item
    Divide a large volume to be studied into contiguous domains $\CD$;
  \item
    evolve the scale factor $a_{\CD}(t)$ in each domain according to
    the Raychaudhuri equation
    [Eqs~\eqref{eq:expansion-law-GR-IV}, \eqref{e-defn-Aconst}],
    where the kinematical backreaction $\CQ_\CD$ is approximated by the
    $\CQ_\CD$ Zel'dovich approximation (QZA)
    [Eqs~\eqref{e-I-II-III-defns}, \eqref{e-resultQ2}, \eqref{eq:GammaDef}],
    unless gravitational collapse and virialisation
    (iii) occur at some time $\tcollapse$;
  \item
    if virialisation [\SSS\ref{s-method-virialisation}] occurs in a domain $\CD$,
    QZA evolution ceases
    at $t = \tcollapse$
    and the stable clustering approximation
    [Eq.~\eqref{e-vir-EdS} in the EdS case]
    for $a_{\CD}(t)$ is adopted
    for $t \ge \tcollapse$ for the domain $\CD$;
    \label{e-virialisation-condition}
  \item
    the global scale factor evolution is estimated by cubical averaging
    of the per-domain scale factors $a_{\CD}(t)$ [Eq.~\eqref{e-double-averaging}].
  \end{list}
\end{definition}

Is this model Newtonian or relativistic? Our aim is to model spatial
expansion in a close approximation to general-relativistic
behaviour. As discussed above, a fully Newtonian cosmology is
difficult to define. The \citet{BuchertEhlers97} fluid model
justification of T$^3$ Newtonian cosmology does not apply past
shell-crossing. When one domain collapses and the others continue
expanding, should there be a sudden compensation in the behaviour of
the expanding domain(s) in response to the virialisation of the
collapsed domain? We hypothesise that this particular feedback effect
would not occur in a relativistic model. In other words, we
hypothesise silent virialisation: that the details of collapse have a
negligible effect at large distances from the collapsing domain
\citep{Matarr94silentPRL,Matarr94silentMN,EllisTsagas02,Bolejko17styro}.
In this sense, the VQZA is a relativistic approximation.  It is not
{\em exactly} relativistic. For example, the sum of rest masses is
conserved via the constants $A$ used in per domain Raychaudhuri
integration [Eqs.~\eqref{eq:expansion-law-GR-IV},
  \eqref{e-defn-Aconst}]; peculiar velocities of the order of
$10^{-2.5}$ would imply that the system mass (in the Minkowski
spacetime sense) should be about $10^{-5}$ higher than the sum of the
rest masses.  In work closely analogous to the present paper,
\citet{Bolejko17styro,Bolejko17styroLCDM} uses a silent universe
family of cosmological solutions of the Einstein equation based on
four scalar quantities (density, expansion rate, shear and Weyl
curvature).  That model numerically evolves an ensemble of worldlines
starting with \LCDM{} initial conditions, making similar assumptions
to those of the VQZA: stable clustering and silent virialisation.  The
emergence of strong negative mean curvature is inferred in
\postrefereeAAchanges{those works}.

The dynamical difference between the VQZA and the standard approach is
also shown in Figure~\ref{f-biscale-underdense} in the biscale case.  This figure shows
the evolution of the expanding domain $\CD^-$ in the VQZA ($+$
symbols) compared with evolution that follows the standard (implicit)
dynamical assumption made about expanding domains in cosmological
$N$-body simulations and typical analytical calculations (solid
curves). The latter is defined by
\begin{align}
  a_{\CD^-}^{\mathrm{Newt}} := \left( 2 {\abg}^3 - a_{\CD^+}^3 \right)^{1/3},
  \label{e-biscale-aDminus-Newt}
\end{align}
where the factor of two represents division of the spatial section
into two domains.
The expanding domain has smooth VQZA behaviour independently
of the collapse event.
In contrast to the smooth volume evolution under the VQZA, the
standard model implicitly requires the expanding domain(s) to be
suddenly pulled more strongly by the virialised object once collapse
and virialisation have finished, in order to preserve a globally EdS
expansion rate: $ \ddot{a}_{\CD^-}^{\mathrm{Newt}}$ has to suddenly
become less positive or more negative when $\dot{a}_{\CD^+}$ very
rapidly switches from a strong negative value to zero.

Interpreted relativistically, the expanding domain prior to collapse
corresponds to negatively curved
space that is below the critical density.  This is shown in
Fig.~\ref{f-biscale-OmRD} for the same biscale examples. The curvature functional is given in
Eq.~\eqref{e-Omegas-a-ODE} in the VQZA case.
In the EdS constraint case, we again interpret $\OmRD$ relativistically
(an EdS model is intended as a relativistic, not Newtonian, model), using
the partitioning formula, eq.~(10) of
\citet{WiegBuch10} applied to a globally flat spatial section,
\begin{align}
  {\OmRDunder} := - \frac{a_{\CD^+}^3}{a_{\CD^-}^3} \OmRDover \frac{H_{\CD^+}^2}{H{\FLRW}^2},
\end{align}
where the ${H_{\CD^+}^2}/{H{\FLRW}^2}$ factor converts between
globally normalised and per-domain normalised definitions of $\Omega$.
{Following collapse,
the stable clustering assumption gives $H_{\CD^+} := 0$,
and both domains are interpreted as flat.}
Thus, in the EdS constraint case,
the expanding domain in the biscale case needs to suddenly switch
from smooth $\OmRD$ growth to $\OmRD \approx 0$. Interpreted from a Newtonian point
of view, this sudden switch is allowed;
here, we hypothesise that it is relativistically inaccurate.

Ironically, it is the {\em standard} model that assumes
a form of feedback (backreaction) in this situation; by not adopting this
virialisation-induced feedback, the VQZA is more conservative than
the standard approach.

\tversions

  \begin{figure}
    \input{SN_invI}
    \input{SN_invII}
    \caption{Signal-to-noise ratios $S/N$ of $\invI$ (above) and $\invII$ (below)
      representing the ratio of rms signal to rms
      numerical noise $\sigmameas$, for analytical model
      $\vmodone$ (which has non-zero invariants) in
      Eq.~(\protect\ref{e-test-vmodels}), sampled by realisations of
      $N=128^3$ particles drawn from a uniform spatial random distribution
      for $8 \le \nDTFE \le 64$.
      The signal is negligibly small
      for $\nDTFE = 16, 128$ for
      $\nxyz = 8, 64$, respectively, since the test functions are sinusoidal and
      in phase with the box.
      \label{f-SN-invariants}}
\end{figure}

  \begin{figure}
    \input{S1_invI}
    \input{S1_invII}
    \caption{As in Fig.~\protect\ref{f-SN-invariants},
      ratio of measured noise $\sigmameas$ to
      rms error predicted by S$^1$-based error estimators
      for $\invI$ (above) and $\invII$ (below),
      for both models $\vmodone$ and $\vmodtwo$.
      These ratios are within two orders of
      magnitude of unity in both cases.
      \label{f-S1-invariants}}
\end{figure}

  \begin{figure}
    \input{Sglobal_invI}
    \input{Sglobal_invII}
    \caption{As in Fig.~\protect\ref{f-S1-invariants},
      measured global
      $|\averagebox{\invI}|$ (above) and
      $|\averagebox{\invII}|$ (below),
      which should be zero in the absence of numerical error
      [by Stokes' theorem and Eq.~(\protect\ref{e-I-II-III-defns})],
      shown as symbols,
      compared to S$^1$-based error estimators
      under independent Gaussian error propagation,
      $\sigmaavISoneglob$,
      $\sigmaavIISoneglob$,
      shown as curves,
      for both models $\vmodone$ and $\vmodtwo$.
      \label{f-S1-global}}
\end{figure}

\section{Results} \label{s-results}

In studying the question of whether dark energy can be replaced
by a non-linear structure formation scale at a first level of approximation,
we treat the \LCDM{} model as an observational proxy,
as in \citet{RMBO16Hbg1}. In principle,
given the microlensed Galactic Bulge star
likely upper age limit of $T \approx 14.7^{+0.3}_{-0.7}$~Gyr
\citep{Bensby13Bulgetzero,RMBO16Hbg1}
and EdS CMB fits indicating an age of the Universe slightly higher than
this \citep{BlanchSarkar03,HuntSarkar07glitches,NadSarkar11LTB},
it could reasonably argued that we should consider the structure formation
scale that yields an effective scale factor $\aeff$ that
reaches unity just a little earlier than $t=15$~Gyr, rather than
the \LCDM{} current age of the Universe estimate of $t_0 = 13.80$~Gyr.
Nevertheless, it is premature to make detailed observational tests
using the VQZA approach, so in this work we consider the \LCDM{}
proxy estimate of the age of the Universe as a reasonable first-order approximation.

  \begin{figure}
    \input{aRZA_nsc8_ndtfe16_qdisable0}
    \input{aRZA_nsc8_ndtfe32_qdisable0}
    \input{aRZA_nsc8_ndtfe64_qdisable0}
    \caption{Full box cube-root-of-total-volume
      scale factor $\aeff(t)$
      evolution
      for
      {\em (top:)} $L_{\mathrm{box}}  = 8 \LD = 2 L_{\mathrm{DTFE}} = 16 L_N$,
      {\em (middle:)} $L_{\mathrm{box}}  = 8 \LD = 4 L_{\mathrm{DTFE}} = 8 L_N$,
      {\em(bottom:)} $L_{\mathrm{box}}  = 8 \LD = 8 L_{\mathrm{DTFE}} = 4 L_N$,
      for $N=256^3$ particles; where $\hzeroeff$ is written as $h$
      to reduce clutter. The $\LD$ scales are labelled, increasing in length
      from top to bottom within any panel. Cosmic variance is visible as non-uniformity
      in the $\LD$ dependence of the separation of the curves.
      {\em Online:} a fixed
      scale $\LD$ is shown with the same colour in all panels
      in this and later figures
      (for example, $\LD = 2.0 \,\hzeroeffMpc$ is purple).
      \label{f-aD}}
  \end{figure}

  \begin{figure}
    \input{aRZA_nsc64_ndtfe128_qdisable0}
    \caption{As in Fig.~\ref{f-aD}, full box cube-root-of-total-volume
      scale factor $\aeff(t)$,
      for $\Lbox  = 64 \LD = 2 \LDTFE = 2 \LN$,
      for $N=256^3$ particles: the $\LD = 2$~Mpc$/\hzeroeff$ scale
      corresponds to a 128~Mpc$/\hzeroeff$ comoving box size.
      \label{f-aD-bigbox}}
\end{figure}

  \begin{figure}
    \input{aRZA_ndtfe32_qdisable1}
    \input{aRZA_ndtfe64_qdisable1}
    \input{aRZA_ndtfe128_qdisable1}
    \caption{As in Fig.~\ref{f-aD}, full box cube-root-of-total-volume
      scale factor $\aeff(t)$, but ignoring kinematical
      backreaction (setting $\CQ_{\CD}=0 \; \forall \CD\; \forall t$),
      for
      {\em (top:)} $L_{\mathrm{box}}  = 8 \LD = 4 L_{\mathrm{DTFE}} = 8 L_N$,
      {\em (middle:)} $L_{\mathrm{box}}  = 8 \LD = 8 L_{\mathrm{DTFE}} = 4 L_N$,
      {\em (bottom:)} $L_{\mathrm{box}}  = 8 \LD = 16 L_{\mathrm{DTFE}} = 2 L_N$.
      \label{f-aD-Qdisabled}}
\end{figure}

  \begin{figure}
    \input{OmQmean}
    \input{OmQmedian}
    \caption{Evolution of the volume-weighted mean (above) and median
      (below) globally normalised kinematical backreaction functional
      $\OmQD$ of uncollapsed domains, corresponding to the same scales
      as in the middle panel of Fig.~\protect\ref{f-aD}.
      \label{f-QD}}
\end{figure}

  \begin{figure}
    \input{OmRmean}
    \input{OmRmedian}
    \caption{Evolution of the volume-weighted mean (above) and median
      (below) globally normalised curvature functional $\OmRD$ of
      uncollapsed domains, corresponding to the same scales as in the
      middle panel of Fig.~\protect\ref{f-aD}.
      \label{f-RD}}
\end{figure}

  \begin{figure}
    \input{superEdS_fvir_nsc8_ndtfe32_qdisable0}
    \caption{Super-EdS ratio $\aeff/\abg$
      versus virialisation fraction $\fvir$
      for 100 VQZA simulations with $\LD= 2$~Mpc$/\hzeroeff$ and
      $L_{\mathrm{box}}  = 8 \LD = 4 L_{\mathrm{DTFE}} = 8 L_N$,
      at an early ($\circ$), middle ($\times$), and late ($+$) time,
      as indicated.
      At any fixed time, $\aeff/\abg$ correlates strongly with $\fvir$.
      \label{f-fvir}}
\end{figure}

\subsection{Uncertainties in peculiar velocity gradient invariants $\invI$ and $\invII$}
\label{s-results-sigmaT3}

Measurement of $\CQ_\CD$ is crucial to calculating volume evolution,
so we first consider numerical uncertainties.
Application of the error estimation methods described in
\SSSS\ref{s-method-sigmaS1} and~\ref{s-method-sigmaT3} to the
test functions [Eq.~\eqref{e-test-vmodels}] are illustrated in
Figs~\ref{f-SN-invariants}--\ref{f-S1-global}, using the \DTFE{}
version indicated in Table~\ref{t-versions}.

Figure~\ref{f-SN-invariants} shows values mostly greater than unity.
In other words, given that we know the true test-function velocity field
$\vmodone$ [Eq.~\eqref{e-test-vmodels}],
numerical \DTFE{} estimates of the first and second tensor invariants of
the peculiar velocity gradient tensor typically have errors of smaller amplitude
than the signal. This is consistent with the claims in the literature regarding
\DTFE's accuracy \citep{SchaapvdWeygaert00,vdWeygaertSchaap07,CW11}.

The $\nxyz=64, \nDTFE=256$ points for both $\invI$ and $\invII$ (upper and lower plots,
respectively) are for \DTFE{} sampling at a higher resolution than the $N=128^3$ particle
resolution.
That is, one full sinusoidal wavelength of $\vmodone$ in one direction
is sampled by about two particles, and \DTFE{} estimates the velocity gradient
at four grid points along this direction. To obtain any reasonable velocity gradient
in this situation would obviously push the extremes of the available information
content, so obtaining $S/N \sim 1$, as shown in both panels of the figure, is clearly
reasonable.

However, the error estimates in Fig.~\ref{f-SN-invariants} are based
on prior knowledge of the true velocity field, which, in general,
is not the case. The S$^1$-based error estimation method introduced
in \SSS\ref{s-method-sigmaS1} bypasses this requirement.
Figure~\ref{f-S1-invariants} shows that
this new method appears to be viable, in the sense that the
ratios of rms measured noise to rms predicted noise are within at most
two orders of magnitude of unity.
The assumptions of statistical independence and Gaussianity used in the
derivation of Eq.~(\protect\ref{e-sigmaS1-basic}) are unlikely to be fully
satisfied, so better agreement between modelled and measured noise
would be surprising.
Structures sized just twice that of
the particle resolution ($\nxyz = 64 = N/2$, asterisks) tend to give
the worst (highest) ratios. Most of the ratios worsen when the \DTFE{}
resolution is higher than the particle resolution ($\nDTFE = 2 N$).
These behaviours are reasonable.

The measured-to-predicted noise ratios for the second invariant
$\invII(\vmodtwo)$ (Fig.~\ref{f-S1-invariants}, lower panel, lower curves with $+$ and $\times$ symbols)
mostly are {lower} than unity. That is, the measured errors are lower
than that given by Eq.~(\protect\ref{e-sigmaS1-basic}).
This is presumably related to the fact that
$\invII(\vmodtwo) = 0 $, and possibly also to symmetries in sampling
this function which cancel each other very effectively, so that
the assumption of statistically independent distributions overestimates the noise level.

Global (T$^3$) application of Stokes' theorem gives consistent error propagation
results to those for the S$^1$-based estimates, to within a few orders of magnitude.
Figure~\ref{f-S1-global} shows that $\averagebox{\invI}$ (global first invariant) errors are bounded above
by the S$^1$-based errors with Gaussian error propagation. In contrast,
$\averagebox{\invII}$ errors are only bounded above by the S$^1$-based errors
for large structures ($\nxyz =1$ or $8$). Structures whose wavelength is twice that
of the particle resolution ($\nxyz=64 = N/2$, asterisks) numerically fail the
analytically required $\averagebox{\invII} = 0$ test at many $\nDTFE$ resolutions.
The equivalent plot for $\CQ\subbox$ is not displayed, because
the low amplitude of $|\averagebox{\invI}|$ estimates and
Eq.~(\protect\ref{e-defn-Q}) make the $\averagebox{\invI}^2$ contribution negligible,
giving a plot visually identical to that
for $\averagebox{\invII}$, but raised by a factor of two.

\subsection{VQZA super-EdS expansion}
\label{s-res-best-scales-VQZA}

Figure~\ref{f-aD} shows effective scale factors for VQZA simulations with
$\Lbox  = 8 \LD$
and an {\sc mpgrafic} grid of $N=256^3$ particles,
with $\LD$ increasing within any of the three panels as labelled,
and $\LD/\LDTFE$ increasing from two in the top panel to eight in the
bottom panel.
(See Table~\ref{t-versions} for software versions.)
The scale factor growth is very close to EdS for $\LD = 8\, \hzeroeffMpc$
and $\LD = 16 \,\hzeroeffMpc$. This is reasonable, because typical overdensities
on these scales do not have time to collapse by the present.

On smaller averaging scales $\LD$, silent virialisation (the VQZA,
Definition~1, \SSS\ref{defn-VQZA})
starts to play a role in volume expansion.
This can be seen in Figure~\ref{f-aD}, where
the scale factor growth is stronger as $\LD$ decreases.
The \LCDM{} proxy scale factor evolution ($\times$ symbols)
mostly lies between the $\LD = 2$ and $4\,${\hzeroeffMpc} curves.
In other words, these curves provide the required super-EdS expansion
required to reach a unity effective scale factor $\aeff(t)$ by
$t = 13.8$~Gyr.

Figure~\ref{f-aD} also indicates that the preferred $\LD$ scale for
providing sufficient super-EdS expansion is only weakly sensitive to
the choice of $\Lbox$, $\LDTFE$, and $\LN$. The spacing between the
curves in any individual panel is not uniform: this is very likely a
sign of cosmic variance: the $\Lbox$ sizes here are small.  The box
size $\Lbox$ can be increased further, but at the expense of leaving
only small ratios $\LD/\LDTFE$ and $\LDTFE/\LN$.
Figure~\ref{f-aD-bigbox} shows that $2\,{\hzeroeffMpc} \ltapprox \LD
\ltapprox 4\,{\hzeroeffMpc}$ is still favoured in this case.

By running a large number of these simulations, we can
estimate $\Lthirteeneight$, the value of $\LD$ which gives
$\aeff(13.8~\mathrm{Gyr}) = 1$.
This is done here as follows.
For a set of five simulations with different $\LD$ values, as in one of
the panels of Fig.~\ref{f-aD}, estimate
$t_{\aeff=1}$ by a spline of $t$ against $\aeff(t)$ for each simulation.
Least-squares fit a quadratic to $t_{\aeff=1} (\LD)$ and solve
$t_{\aeff=1}(\LD) = 13.8$~Gyr to obtain $\LD$.
We find that for $\Lbox = 8 \LD = 4 \LDTFE = 8 \LN$
(as in Fig.~\ref{f-aD}, middle panel),
100 sets of 5 simulations yield an averaging
scale of
$\Lthirteeneight = 2.51\pm0.22$~{\hzeroeffMpc}. This is the scale
at which $\aeff(t)$ reaches a unity scale factor at
13.8~Gyr, which is 16\% above $\abg(13.8~\mathrm{Gyr})$.
The error estimate is the standard deviation of the 100 estimates.
This error estimate represents the effect of cosmic variance for
$4\,\hzeroeffMpc \le \Lbox  \le 64\,\hzeroeffMpc$.

Corresponding values for smaller and bigger box sizes,
$\Lbox = 8 \LD = 2 \LDTFE = 16 \LN$
(Fig.~\ref{f-aD}, top panel)
and
$\Lbox = 8 \LD = 8 \LDTFE = 4 \LN$
(Fig.~\ref{f-aD}, bottom panel),
are
$\Lthirteeneight = 2.14\pm0.26$~{\hzeroeffMpc} and
$\Lthirteeneight = 2.59\pm0.25$~{\hzeroeffMpc},
respectively, for 100 sets of 5 simulations of $N=256^3$ particles
in each case.
Thus, the uncertainty in changing the box
size and resolution parameters is of about the same magnitude
as the uncertainty associated with cosmic variance, and about ten
times the standard error in the mean of $\Lthirteeneight$ for any fixed set of size and
resolution parameters. We summarise these by writing the (dominating) systematic
error and taking the median of the three sets of calculations
($\Lbox = 8 \LD = 4 \LDTFE = 8 \LN$, as in Fig.~\ref{f-aD}, middle panel,
for one quintuple of realisations):
$\Lthirteeneight = 2.5^{+0.1}_{-0.4}$~{\hzeroeffMpc}.

We also carried out 100 sets of five simulations on still
bigger box scales
(as in Fig.~\ref{f-aD-bigbox}),
at the risk of greater numerical error due to
the small factors between the shorter length scales
($ \LD/\LDTFE = \LDTFE/\LN = 2$).
This set of simulations gives a much more
precise estimate: $\Lthirteeneight = 2.56 \pm 0.01$~{\hzeroeffMpc}, most likely
because of the reduced cosmic variance.
However, since there is a greater risk of numerical error,
this estimate is not likely to be more accurate
(the resolution-related systematic error is likely to be higher). Thus
we retain $\Lthirteeneight = 2.5^{+0.1}_{-0.4}$~{\hzeroeffMpc} as a
conservative summary of our main quantitative result.

\citet{Rasanen06negemergcurv}'s early model was only a toy model,
without a proposal for a specific characteristic length scale $\LD$.
Using the FLRW critical density $\rhocrit = 1.52 \times 10^{11}
M_{\odot}/\mathrm{Mpc}^3$, \citet{RaczDobos16}'s preferred mass scale
for matching the observational distance-modulus--redshift relation,
$2.03 \times 10^{11}M_{\sun}$, corresponds to
$1.10$~{\hzeroeffMpc}, which is a little below half our scale. We do
not expect our main result to closely match that of
\citet{RaczDobos16}, since our methods differ.  If we artificially
suppress kinematical backreaction in order to compare with
\citet{RaczDobos16}, that is, if we set $\CQ_{\CD}(t)=0$ in all
domains $\CD$ at all times, then the super-EdS expansion rates should
increase for any fixed $\LD$.  Figure~\ref{f-aD-Qdisabled} shows that
this is indeed the case. Since kinematical backreaction is ignored,
stronger super-EdS global volume evolution results, giving
$\Lthirteeneight(\CQ_{\CD} := 0) \approx 8~\hzeroeffMpc$, that is, an
overestimate by a factor of about three.  This is a factor of about
seven greater than \citet{RaczDobos16}'s estimate: the latter's model
does not quite match our zero-$\CQ_\CD$ simulations.

Figure~\ref{f-QD} helps to show the role of kinematical backreaction
in our calculations,
\postrefereeAAchanges{where the volume-weighted mean and median values for the uncollapsed
  domains are defined
  \begin{align}
    \left<\OmQD\right>\uncollapsed &:= -\frac{\left<\CQ_{\CD}\right>\uncollapsed}{6 {\Heff}^2} \nonumber \\
    \left<\OmRD\right>\uncollapsed &:= -\frac{\left<\CR_{\CD}\right>\uncollapsed}{6 {\Heff}^2} \nonumber \\
    \mu\left(\OmQD\right)\uncollapsed &:= -\frac{\CQ_{{\CD}_k}}{6 {\Heff}^2} \nonumber \\
    \mu\left(\OmRD\right)\uncollapsed &:= -\frac{\CR_{{\CD}_k}}{6 {\Heff}^2} \,,
  \end{align}
  where ${\CD}_k$ is the $k$-th domain among the sorted $i$-th scale factors
  of uncollapsed domains satisfying
  \begin{align}
    \sum_{i\le k-1} {a_{\CD}}_i^3 \le \frac{1}{2} \sum_i {a_{\CD}}_i^3 \; &\quad\mathrm{and}\quad
    \sum_{i\ge k+1} {a_{\CD}}_i^3 \le \frac{1}{2} \sum_i {a_{\CD}}_i^3 ,
  \end{align}
  and Eqs~\eqref{e-resultQ2}, \eqref{eq:GammaDef}, \eqref{e-resultR2}, \eqref{eq:GammaTildeDef} are
  used to evaluate $\CQ$ and $\CR$ (ties in the median, if they occur, are upper weighted).
The $\OmQD$} volume-weighted mean and median values
for uncollapsed domains \postrefereeAAchanges{are} positive,
corresponding to negative $\CQ_\CD$, \postrefereeAAchanges{which tends} to cause
\postrefereeAAchanges{earlier}
collapse of overdensities
and slower expansion of voids.
Thus, if
\postrefereeAAchanges{kinematical backreaction
  were artificially set to zero in all domains,
  as appears to be the case in \citet{RaczDobos16}
  and is shown here in Fig.~\ref{f-aD-Qdisabled},
  then it would be reasonable for the faster expansion
  of voids to contribute to much stronger
  scale factor growth. In that case,}
larger scale (lower amplitude) initial overdensities
\postrefereeAAchanges{would be}
sufficient to obtain $\aeff(13.8~{\mathrm{Gyr}}) = 1$,
\postrefereeAAchanges{as can be seen in Fig.~\ref{f-aD-Qdisabled}.}

The values of the mean and median $\OmQD$ values shown
in Fig.~\ref{f-QD} only sample uncollapsed domains, and are
normalised globally [\citealt{WiegBuch10}, eq.~(10)],
rather than per-domain
[Eq.~\eqref{e-Omegas-a-ODE}]. This implies that
prior to virialisation, the mean should
be constrained to be close to zero
by the Stokes' theorem T$^3$ constraint on the global
kinematical backreaction $\CQ_{T^3}$.
This does appear to be the case in the upper panel of
Fig.~\ref{f-QD} at very early times,
but the first virialisation events happen quickly, so
$\averageplain{\OmQD}\uncollapsed$ quickly grows away from zero.

The spikiness of Fig.~\ref{f-QD} (the vertical axis is logarithmic)
is physically realistic: it emphasises the discontinuous nature
of virialisation. Each time a domain collapses, it no longer contributes
to $\averageplain{\OmQD}\uncollapsed$ as defined for this figure.
Forcing a smooth $\abg(t)$ (or $a_{\Lambda\mathrm{CDM}}(t)$) global volume
evolution artificially hides this spikiness.
Unsurprisingly, the volume-weighted medians $\mu(\OmQD)\uncollapsed$ evolve
much more smoothly, since the collapse of a domain shifts the median
within the central part of the $\OmQD$ distribution, which is denser than
the tails.
For $\LD = 2\,${\hzeroeffMpc}, these medians
peak rapidly at about $0.03$ and then gradually decay
to the present.

As expected by \citeauthor{Buchert05CQGLetter}
(\citeyear{Buchert05CQGLetter}; see \SSS\ref{s-intro} above for more
literature), the super-EdS $\aeff(t)$ is accompanied by emerging
negative average curvature, as shown in Fig.~\ref{f-RD}.  At the $\LD
= 2\,${\hzeroeffMpc} scale (middle curve at early times), the
volume-weighted mean $\OmRD$ for uncollapsed domains grows from zero
to about $1$ by about half the age of the Universe and then grows more
slowly to about $1.5$ by the present.  The volume-weighted median
$\OmRD$ also grows quickly to about $1.5$ by $t \sim 4$~Gyr, the same
epoch, and then grows slowly to about $2$ by the present.  New
observational tests will be needed to constrain recent emerging
curvature such as that modelled here, since constraints that assume
comoving {\em non}-evolving {\em uniform} curvature do not constrain {\em evolving}
{\em non}-uniform mean curvature.
Moreover, in this work, no distinction is made between volume-average (``bare'')
$\Omega$ parameters and those likely to be observed from our Galaxy
(``dressed''; \citealt{BuchCarf02,BuchCarf03nudity}), so more work will be
needed to calculate parameters such as the detailed estimations for
the Timescape model given in \citet[][Table II]{NazerW15CMB}.

  \begin{figure}
    \input{a_nbody_Lag_128_F}
    \caption{As in Fig.~\protect\ref{f-aD} {\em (middle)}, for
      $N$-body evolution and
      Lagrangian tracing of initial domains, for
      $L_{\mathrm{box}}  = 8 \LD = 4 L_{\mathrm{DTFE}} = 4 L_N$,
      with $N=128^3$ particles,
      and  $\nsoft/(N^{1/3}) = 32$. 
      The curves' behaviour is not monotonic with respect to $\LD$ scales;
      the top curve is for
      $\LD = 2.0 $~Mpc$/\hzeroeff$,
      the middle pair of curves are for
      $\LD = 1.0 $~Mpc$/\hzeroeff$ (upper) and
      $\LD = 4.0 $~Mpc$/\hzeroeff$ (lower), and the bottom
      pair are for
      $\LD = 8.0 $~Mpc$/\hzeroeff$ (upper) and
      $\LD = 0.5 $~Mpc$/\hzeroeff$ (lower).
      \label{f-Lscalav-nbody-Lag}}
\end{figure}

  \begin{figure}
    \input{a_nbody_Lag_128_T}
    \caption{As in Fig.~\protect\ref{f-aD-Qdisabled} {\em (top)}, for
      $N$-body evolution and
      Lagrangian tracing of initial domains, for
      $L_{\mathrm{box}}  = 8 \LD = 4 L_{\mathrm{DTFE}} = 4 L_N$,
      with $N=128^3$ particles,
      \postrefereeAAchanges{(artificially
        setting $\CQ_{\CD}=0 \; \forall \CD\; \forall t$)}.
      The curves' behaviour is monotonic with respect to $\LD$ scales,
      as labelled.
      \label{f-Lscalav-nbody-Lag-Qdisabled}}
\end{figure}

\subsection{Dependence of super-EdS expansion on virialisation fraction $\fvir$}
\label{s-results-fvir}

\citet{ROB13} argued that virialisation and non-rigidity of
voids constitute a key element to an emerging negative average curvature
alternative to dark energy. The VQZA provides an {\em ab initio}
model of this argument, so
it should be possible to verify the relationship between super-EdS
expansion and virialisation.
The virialisation fraction (fraction of mass
in virialised objects) can be estimated by counting the number
$N_{\mathrm{coll}}$ of
domains $\CD$ that have gravitationally collapsed at any given time
and normalising:
\begin{equation}
  \fvir := \frac{N_{\mathrm{coll}}}{\nD^3}.
\end{equation}
Figure~\ref{f-fvir} shows that
at any fixed time,
VQZA simulations run from independent realisations of initial conditions
yield super-EdS expansion $\aeff/\abg$
that correlate strongly with $\fvir$.
Spearman's rank correlation $\rho$ tests overwhelmingly confirm this, giving
a probability $p = 3\times 10^{-9}, 10^{-9},$ and $10^{-6}$, of
no correlation between $\fvir$ and $\aeff/\abg$
at the early, middle, and late times (as listed in
Fig.~\ref{f-fvir}), respectively.

Thus, unsurprisingly, silent virialisation leads to
a strong correlation
between the degree of virialisation $\fvir(t)$ that results
from a given realisation of initial conditions at a given time $t$
and the amount of super-EdS expansion $\aeff(t)/\abg(t)$ at that
same epoch. Enforcement of Newtonian global expansion, as in standard
cosmological $N$-body simulations, would destroy this correlation, since
in that case $\aeff(t)/\abg(t) := 1$ by definition.

\subsection{$N$-body measurement of $\CQ_\CD$}
\label{s-res-best-scales-Nbody}

What happens when $\CQ_\CD$ is calculated from the Lagrangian domains
in the evolving $N$-body simulation, as detailed in \SSS\ref{s-method-nbody} above, where numerical noise plays a stronger role?
Figures~\ref{f-Lscalav-nbody-Lag} and~\ref{f-Lscalav-nbody-Lag-Qdisabled} show
the equivalent of Figs~\ref{f-aD} (middle) and \ref{f-aD-Qdisabled} (top), respectively.
(See Table~\ref{t-versions} for software versions.)

Figures~\ref{f-aD-Qdisabled} (top) and \ref{f-Lscalav-nbody-Lag-Qdisabled},
in which $\CQ_\CD$ is \postrefereeAAchanges{artificially} set to zero,
appear, by
inspection, to be almost identical. This close correspondence provides a check on the numerical integration
and coordination between software modules, since
the former integration is carried out by calls from the \inhomog{} library to the {\sc GSL}
routine {\sc gsl\_odeiv\_step\_rkf45},
while the latter is carried out by
the \RAMSESscalav{} routine {\sc update\_av\_scalars\_global\_module} with a hardwired version of
the classical (non-adaptive) Runge--Kutta algorithm.
The initial conditions
were not set to be identical between these two panels, so the small differences indicate both cosmic variance
and numerical error.

For more realistic calculations, when kinematical backreaction is not set to zero,
Fig.~\ref{f-Lscalav-nbody-Lag} shows more complex behaviour than that in
Fig.~\ref{f-aD} (middle).
The sequence of bigger scales $\LD = 2, 4, 8$~Mpc$/\hzeroeff$
has slightly stronger $\aeff(t)$
than for the VQZA case.
In contrast, for the
sequence of smaller scales ($\LD = 2, 1, 0.5$~Mpc$/\hzeroeff$)
the $\aeff(t)$ growth is successively weaker.
Cosmic variance is strong on these scales, so seeking an explanation
from a small number of small simulations is not strongly justified:
many competing numerical effects such as close-encounter
scattering and softening may be involved. More importantly
for the present work,
a value of $1 \ltapprox \Lthirteeneight/\hzeroeffMpc \ltapprox 4$
indicates consistency with the VQZA calculations. Future work on higher
numbers of higher $N$ simulations will
enable a statistically
more significant comparison between the analytical and numerical
models of $\CQ_\CD$ and their effects on $\aeff(t)$.

\section{Discussion} \label{s-disc}

\postrefereeAAstart

\subsection{Should global backreaction from T$^3$ initial conditions be zero?}
\label{s-disc-flat-T3-zeroQ-violation}

The scalar averaging equations, given in \SSS\ref{s-scalav-basic} together
with Eq.~\eqref{e-double-averaging}, can be applied on any domain size,
including the global volume,
which in this work is T$^3$, provided that the assumptions
underlying the equations are satisfied.
For flat space, given the definition of $\CQ_\CD$ in Eq.~\eqref{e-defn-Q}
and the flat space expression of the peculiar velocity gradient invariants
as exact divergences in Eq.~\eqref{e-I-II-III-defns},
it would appear that $\CQ_{\mathrm{T}^3}$ must be zero.
Thus, Eq.~\eqref{eq:expansion-law-GR} should
yield $\aeff(t) \equiv \abg(t)$, no matter what scale of $\CD$ is chosen,
provided that the evolution of $\CQ_\CD$ is calculated exactly.
More realistically, in the present case
$\CQ_\CD$ is calculated using the QZA [Eqs~\eqref{e-resultQ2},
\eqref{eq:GammaDef}],
so $\aeff(t) \approx \abg(t)$ should still be a good approximation, to
the degree that Eqs~\eqref{e-resultQ2},
\eqref{eq:GammaDef}
remain accurate at $t$,
modulo Poisson noise contributed by combining many domains.
Indeed, Fig.~\ref{f-biscale} shows that prior to virialisation, this
approximation holds reasonably well in illustrative cases.

However, Eq.~\eqref{eq:expansion-law-GR} and the definition of $\CQ_\CD$
normally used in the literature are derived for a pressureless
perfect fluid, that is, `dust', with no allowance for shell crossings
or avoidance of collapse to a singularity. This is not a problem
for studying non-linear gravitational collapse times of overdense objects,
in which case these equations yield realistic estimates in a more elegant
way than the usual alternatives
\citep{Ostrowski16MG14MF,Ostrowski2016PhD,Ostrowski17AcPPS,OstrowskiBuchR16}.
On the other hand,
Eq.~\eqref{eq:expansion-law-GR} (in its form stated above)
is no longer valid once the final
stages of gravitational collapse and virialisation have taken place.
As $a_{\CD} \rightarrow 0^+$ in Eq.~\eqref{eq:expansion-law-GR}, $\CQ_\CD$ via
Eqs~\eqref{e-resultQ2} and \eqref{eq:GammaDef} is not able to cancel
self-gravity, so
a literal interpretation of Eq.~\eqref{eq:expansion-law-GR}
would imply that $\dot{a}_{\CD} \rightarrow -\infty$.

In this sense, the global volume-averaging expressed in
Eq.~\eqref{e-double-averaging} is used in this work as an extension
beyond the usual set of scalar averaging equations; the metric volumes are
summed, but the choice of the $\CD$ scale affects the result via
virialisation.

Obviously, if the {\em aim} of calculations of this sort were to study
EdS cosmology and to substitute a perfect fluid by a set of
collisionless particles in which shell-crossings were allowed, then it
would have to be possible to modify Eq.~\eqref{eq:expansion-law-GR} by
adding a virialisation term to the kinematical backreaction in order
to prevent collapse to a singularity. For example, this could take the
form of the dynamical backreaction proposed in eqs~(13a), (13d) of
\citet{Buch01scalav} (see also \citealt{Yodzis74};
\citealt[][footnote~7]{Buchert17Newt}).  To retain a
global EdS expansion history, the new term would not only have to
prevent collapsing domains from forming singularities, but would also
have to correspondingly slow down the expansion rate of expanding domains,
in order to retain a net EdS expansion rate during the post-virialisation
phase.

The main aim of this work is a step towards relativistic cosmology
rather than testing the consistency of Newtonian cosmology.
The latter requires instant reaction at a distance, but the former
is constrained by causality, making instantaneous reaction to
virialisation unphysical.
Thus, the approach defined in Definition~1 (\SSS\ref{defn-VQZA}) is
retained for the present work.
In other words, here, the QZA formulae
for expanding regions are not modified nor supplemented in reaction to
virialisation, and the resulting global volume evolution is not
constrained to be Newtonian.

Moreover, since stable clustering is adopted for virialised domains,
if we consider their volume-weighted curvature contributions
to be negligible (either because the volumes are tiny, the curvature
is weak, or both), then the negative curvature of the expanding domains
will tend to yield an overall negative mean curvature, invalidating
the derivation of the divergence expressions in Eq.~\eqref{e-I-II-III-defns}
at late times, since a flat manifold is no longer a good approximation.

\subsection{How can virialisation provide an order-unity dark-energy expansion?}

How is it possible that stable clustering, which is typically thought
of as a Newtonian statistical gravitational effect, can provide a
reasonable relativistic approximation in this context?  The reasoning
above can be reworded as follows.
Prior to virialisation, the QZA formulae
provide a good cancellation between expansion and collapse compared to
EdS evolution of the scale factor.  At a virialisation event,
the spatially contracting contribution of a virialised domain to the
general cosmological volume evolution is cancelled, either by the
stable clustering assumption or (in future work) using more detailed
analytical models
\citep{Buch01scalav,BolejkoLasky08,BolejkoFerreira12,Buchert17Newt}.
To better than an order of magnitude, about half the
mass of any large volume today is located in virialised dark matter
haloes. Thus, among the ensemble of
domains on the nonlinear, few-megaparsec scale, the
domains that contribute about half the total mass have their
gravitational roles cancelled, in the sense that the
averaged Raychaudhuri equation, Eq.~\eqref{eq:expansion-law-GR}, is
replaced by
\begin{align}
  {\ddot{a}}_{\CD}& \approx 0
  \label{e-Ray-virialised}
\end{align}
for those domains.
Under an EdS constraint, expanding regions are forced to slow their
expansion down 
in response to this virialisation. Relativistically interpreted,
this would mean that the expanding, negatively curved regions are forced
to slow down and flatten. Without the EdS constraint, that is, without
modifying the QZA formula for expanding domains (or in other words, with
the hypothesis that silent virialisation is a fair approximation), about half
the mass that contributes to deceleration of the global, effective
scale factor is
effectively cancelled, in terms of its
contribution to average deceleration. Thus,
the cancelling of about half the deceleration results in
volume acceleration. This
effect (related to that of ``finite infinity'' regions:
\citealt{Ellis84relcosmo,Wiltshire07clocks,Cox07finiteinfinity})
is overlooked in the standard
approach because Eq.~\eqref{e-double-averaging} (cubical averaging of
domain-wise scale factors) is not normally used.

In this sense, the `new physics' that contributes significantly to
dark energy is old physics -- Newtonian virialisation -- that is
assumed to not have instantaneous effects at a distance. This could
be described as non-compensation for virialisation.
The global acceleration effect emerges indirectly from a
statistical Newtonian particle effect via the combination of
Eq.~\eqref{eq:expansion-law-GR} for pre-virialisation average volume evolution (with $\Lambda \equiv
0$), Eq.~\eqref{e-Ray-virialised} for virialised domains, and
Eq.~\eqref{e-double-averaging} to silently prevent the latter from
affecting the former.

\subsection{Caveats of the VQZA, the fluid model and vorticity}

The accuracy of the VQZA depends on the
accuracy of the assumptions stated in Definition~1, \SSS\ref{defn-VQZA}.
For example, if propagation of relativistic constraints
across domain boundaries implied that virialisation in one domain
induced inaccuracy in the QZA formulae
[Eqs~\eqref{e-I-II-III-defns}, \eqref{e-resultQ2}, \eqref{eq:GammaDef}]
in one or more neighbouring domains, then this would lead to systematic error
due to violation of the silent virialisation
assumption of (ii) and (iv) of Definition~1.
Other simplifying assumptions either stated above or implicit to the VQZA
approach presented here include:
\begin{list}{(\roman{enumi})}{\usecounter{enumi}}
\item
  a spacetime foliation that matches the EdS foliation at
  early times is assumed (proper time and coordinate time are
  not distinguished);
\item
  the difference between the sum of rest masses and the system mass
  is assumed to  be negligible
  \cite[see][ for an example of modelling Lorentzian effects]{AdamekDDK15,AdamekDDK16code};
\item
  equations \eqref{eq:expansion-law-GR},
  \eqref{eq:adot-GR}, and
  \eqref{e-defn-Q} are justified in the relativistic case under
  the assumption of zero vorticity;
\item
  there is assumed to be no shell-crossing prior to virialisation;
\item
  although the scale found to be of most interest is found to
  be $\Lthirteeneight = 2.5^{+0.1}_{-0.4}$~{\hzeroeffMpc}, the
  calculations were performed (in particular see Fig.~\ref{f-aD})
  down to $\LD = 0.25$~{\hzeroeffMpc}, where it is clear that
  the zero vorticity and lack of shell-crossing assumptions could
  potentially lead to non-negligible correction terms.
\end{list}
Thus, as with each of the other approaches towards
general-relativistic cosmological simulations published so far (see
\SSS\ref{s-intro}), the approximations and simplifying assumptions of
this approach have to be studied further in order to judge how much
they affect the results.

\postrefereeAAstop

\section{Conclusion} \label{s-conclu}

This {work}
presents the first in a series of cosmological expansion
modelling improvements.
{These} aim to incorporate successively more
elements towards obtaining general-relativistically valid $N$-body
simulations in which the effective expansion rate is calculated from
local structure formation rather than inserted arbitrarily.  A key
element of the present work in obtaining super-EdS expansion is
described in \SSS\ref{s-newt-vs-GR} and illustrated quantitatively
in \SSS\ref{s-results-VQZA}. This is the virialisation--$\CQ_\CD$
Zel'dovich approximation (Definition~1, \SSS\ref{defn-VQZA}), in which
locally Newtonian expansion
modelled by the QZA in combination with virialisation modelled
under the stable clustering hypothesis together imply super-EdS
evolution of the effective expansion rate.

The standard FLRW structure formation alternative to the VQZA can be
summarised as follows:
\begin{list}{(\roman{enumi})}{\usecounter{enumi}}
\item
  prior to a collapse/virialisation event, the Lagrangian per-domain
  average scale factor evolution is implicitly modelled to be smooth
  for both the collapsing domain and the expanding domain;
\item
  the final stages of a gravitational collapse and virialisation event
  are very sudden in comparison with cosmological time scales;
\item
  the global scale factor evolution is assumed to be smooth (FLRW)
  passing from times prior to the collapse/virialisation event through
  to times following the event;
\item
  assumptions (i) + (ii) + (iii) together imply that at least one
  expanding domain $\CD^-$ must undergo a sudden drop in acceleration
  $\ddot{a}_{\CD^-}$ when the collapse/virialisation event occurs, in
  order to compensate for the latter; smooth expansion would violate
  either (ii) or (iii); in the case of a biscale model, the mean spatial
  3-Ricci curvature (in a relativistic interpretation)
  of the expanding domain must suddenly switch from
  negative curvature to zero curvature.
\end{list}
The hypothesis adopted here is that (iv) is relativistically unrealistic
(as shown by the solid curves in Fig.~\ref{f-biscale-underdense} for biscale examples).
This motivates the VQZA, in which (i) and (ii) hold, but (iii) is violated
to the degree that any collapse/virialisation process is accepted to be a sudden
event, and to be silent --- without a significant sudden effect
on expanding domains. Thus, the VQZA avoids (iv) the non-differentiability of
$\dot{a}_{\CD^-}$ for expanding domains $\CD^-$.  In other words, we postulate that smoothness in
volume growth of the expanding domain, as approximated in the QZA, is
relativistically more accurate than smoothness in global mean volume
growth in situations in which collapse terminates suddenly in a small,
already greatly contracted domain.  In \SSS\ref{s-results-VQZA}, this
argument is illustrated analytically and numerically with a biscale
model that shows the sudden change required to happen in the expanding
domain in the standard model.

The consequences of the VQZA were examined here for a more generic case,
by partitioning general $N$-body simulation
initial conditions into initially cubical domains.
\begin{list}{(\roman{enumi})}{\usecounter{enumi}}
\item
  We introduced
  (\SSSS\ref{s-method-sigmaS1}, \ref{s-method-sigmaT3})
  a method of estimating
  numerical error in velocity gradient numerical estimates by applying
  Stokes' theorem over S$^1$.
  We found
  (\SSS\ref{s-results-sigmaT3})
  that for a simulation with $\Lbox/\LN =128$, structure scales of
  $\Lbox/64$ or greater had $S/N$ ratios in the accuracy of
  \DTFE{} determination of the velocity gradient invariants $\invI$
  and $\invII$ of about ten or higher, except when attempting
  sub--mean-interparticle-separation sampling at $\LDTFE = \LN/2$.
\item
  Based on our VQZA simulations (Figs~\ref{f-aD}, \ref{f-aD-bigbox}),
  we consider $\Lthirteeneight = 2.5^{+0.1}_{-0.4}$~{\hzeroeffMpc} to
  be a reasonable representation of the averaging scale $\CD$ which
  provides the optimal amount of super-EdS scale factor evolution to
  match observations as represented by an \LCDM{} proxy. The error
  represents our estimate of systematic error associated with changing
  the box (T$^3$ fundamental domain), \DTFE{} and particle resolution
  scales, and running 100 VQZA simulations of size $N=256^3$ at each
  relevant set of scales. Our largest box size simulations yield
  $\Lthirteeneight = 2.56 \pm 0.01~{\hzeroeffMpc}$, where the error is
  one standard deviation (not the standard error in the mean), but
  since these have low ratios $\LD/\LDTFE = \LDTFE/\LN = 2$, it is likely
  that systematic, resolution-related error is greater than this
  random error.
\item
  This super-EdS scale factor evolution is associated with
  kinematical backreaction and curvature normalised functionals
  $\OmQD$ and $\OmRD$, respectively, whose volume-weighted
  means and medians for uncollapsed domains are
  mostly positive throughout cosmic history
  (Figs~\ref{f-QD}, \ref{f-RD}). Kinematical backreaction
  is strongest at early times, with $\averageplain{\OmQD}\uncollapsed$
  reaching a median value of about 0.03 at $t \approx 2$~Gyr
  and then dropping gradually towards zero. The recent emergence of
  average negative curvature is confirmed in these calculations,
  with the $\OmRD$ mean and median
  \postrefereeAAchanges{values for uncollapsed domains}
  growing to about 1.5 to 2,
  respectively, by the present.
\item
  Cosmic variance, as represented by independent realisations
  of initial conditions, gives different numbers of collapsed
  objects at a given cosmological time, and thus virialisation
  fractions $\fvir$, for a given choice of $\LD$. The super-EdS
  scale factor ratio $\aeff/\abg$ is very strongly correlated with
  $\fvir$ (\SSS\ref{s-results-fvir}, Fig.~\ref{f-fvir}).
\item
  Artificially setting $\CQ_\CD=0$
  \postrefereeAAchanges{gives}
  $\Lthirteeneight(\CQ_\CD := 0) \approx 8$~{\hzeroeffMpc}
  (Figs~\ref{f-aD-Qdisabled}, \ref{f-Lscalav-nbody-Lag-Qdisabled}).
  This
  is the result that we would expect from \citet{RaczDobos16}'s approach.
\item
  Small numbers of small exploratory simulations with numerical
  estimation of $\CQ_\CD$ at every major time step give rough
  agreement with the VQZA $\Lthirteeneight$ scale
  (Fig.~\ref{f-Lscalav-nbody-Lag}).
\end{list}

Thus, from EdS initial conditions, dropping the decoupling hypothesis
and estimating spatial expansion using the VQZA provides sufficient
dark-energy--free average expansion by the standard age of the
Universe estimate, for an averaging $\CQ_\CD$ scale of
$\Lthirteeneight = 2.5^{+0.1}_{-0.4}$~{\hzeroeffMpc}.
This estimate is just a factor of about three
lower than (more non-linear than) the $8\hzeroeffMpc$ scale at which the
$\sigma_8$ normalisation of the
initial power spectrum of density perturbations in flat space $P_k$ is
normally defined. The historically important non-linearity scale,
the $r_0$ scale in the two-point spatial auto-correlation
function of galaxies,
at which the excess probability of finding a pair of galaxies is unity,
is close in value. This was estimated over four decades ago
\citep{Peebles74} as $5 \hzeroeffMpc \ltapprox r_0 \ltapprox 10 \hzeroeffMpc$
\citep[][Eq.~(48)]{PeeblesHauser74}.
In this presentation of the VQZA, we set $\sigma_8$ appropriate
for an early-epoch Planck-parametrised \LCDM{} model
(\SSS\ref{s-method-ic}). Since silent virialisation is the
key driver of the super-EdS expansion that we find here, it is unsurprising
that (i) $\Lthirteeneight$, (ii) the 8$\hzeroeffMpc$ scale at which $\sigma_8$ is not far
from unity, and (iii) the galaxy two-point auto-correlation function zero point
$r_0$ are close in value to one another.

This work provides a step towards correcting
standard $N$-body simulations for the general-relativistic constraints
imposed by scalar averaging.
\postrefereeAAchanges{This work is} too preliminary to provide observational
predictions at comparable precision to that of the \LCDM{} model.
Future work will need to include:
inhomogeneous spatial curvature for calculating averages;
curvature effects of particle movement;
feedback of the global expansion rate on the effective background model growth rate $\xi(t)$;
improvement beyond the QZA, since the QZA is only a first-order approximation
(\citealt{BuchRZA2}, \SSSS{}IV, V; see \citealt{BuchRZA3}),
in particular by taking into account an effective background with
evolving average curvature;
and ray-tracing for
calculating luminosity distances and angular diameter distances
through inhomogeneously curved spatial sections
(see, for example, \citealt{SikoraGlod17},
\postrefereeAAchanges{in which the effective scale factor
  of the scalar-averaging averaging approach is found to be a good
  match to detailed ray-tracing calculations in a specific cosmological
  inhomogeneous metric}).

Nevertheless, \postrefereeAAchanges{it seems reasonable to consider}
this work to be more accurate than the
standard model: sudden, \postrefereeAAchanges{acausal}
switches in curvature and expansion rates
in expanding regions that contain no singular spacetime events
\postrefereeAAchanges{would be relativistically surprising}.


\section*{Acknowledgments}
Thank you to
J\'er\'emy Blaizot,
Krzysztof Bolejko,
Justyna Borkowska,
Thomas Buchert,
\postrefereeAAchanges{Miko{\l}aj Korzy{\'n}ski},
Bartosz Lew,
Jan Ostrowski,
Pierre Mourier,
Yann Rasera,
Joakim Rosdahl,
Quentin Vigneron,
David Wiltshire,
\postrefereeAAchanges{and an anonymous referee}
for useful comments.
A part of this project was funded by the National
Science Centre, Poland, under grant 2014/13/B/ST9/00845.  Part of this
work consists of research conducted within the scope of the HECOLS
International Associated Laboratory, supported in part by the Polish
NCN grant DEC-2013/08/M/ST9/00664. A part of this project has made use
of computations made under grant 314 of the Pozna\'n
Supercomputing and Networking Center (PSNC).  This work has used the
free-licensed {\sc GNU Octave} package \citep{Eaton14}.

\subm{ \clearpage }

%



\appendix

\section{Newtonian velocity gradient invariants}
\label{app-invariant-formulae}

Here we provide a script
\citep[App.~A][]{Ostrowski2016PhD}
that can be run using the free-licensed
computer algebra system {\sc maxima}, version
v.5.32.1,\footnote{\url{http://maxima.sourceforge.net}} in order to
confirm the equivalence of the expressions in
Eq.~\eqref{e-I-II-III-defns}.  The three {\sc printf} statements at the
end should yield ``0'' to show algebraic equivalences.

\begin{verbatim}
/* Maxima script for checking Euclidean
   velocity gradient invariant relations

 (C) 2016,2017 J.J. Ostrowski, B.F. Roukema,
 GPL-3+

 This program is free software: you can
 redistribute it and/or modify it under the
 terms of the GNU General Public License as
 published by the Free Software Foundation,
 either version 3 of the License, or (at your
 option) any later version.

 This program is distributed in the hope that
 it will be useful, but WITHOUT ANY WARRANTY;
 without even the implied warranty of
 MERCHANTABILITY or FITNESS FOR A PARTICULAR
 PURPOSE.  See the GNU General Public License
 for more details.

 You should have received a copy of the GNU
 General Public License along with this
 program.  If not, see
 https://www.gnu.org/licenses/gpl-3.0.html
*/

v: [v_x,v_y,v_z]; /* declare velocity fld */

depends(v,[x,y,z]); /* each component may
 depend on any coordinate */

derivabbrev : true; /* select subscript
 style of showing derivatives */

/* 1st order partial derivs of velocity */
dvdx: diff(v,x,1);
dvdy: diff(v,y,1);
dvdz: diff(v,z,1);

/* gradient of the velocity field */
vgrad: transpose(matrix(dvdx,dvdy,dvdz));

/* packages for matrix manipulations */
load("nchrpl");
load("eigen");

/* divergence and curl of vector field */
mydiv(v) := diff(v[1],x,1) + diff(v[2],y,1)
     + diff(v[3],z,1);
mycurl(v) := [diff(v[3],y,1)-diff(v[2],z,1),
     diff(v[1],z,1)-diff(v[3],x,1),
     diff(v[2],x,1)-diff(v[1],y,1)];

/* gradient of scalar field */
mygrad(phi) := [diff(phi,x,1), diff(phi,y,1),
     diff(phi,z,1)];

/* directional derivative of vector w in
   direction v */
my_v_dot_del_w(v,w) :=
    [mygrad(w[1]).transpose(v),
     mygrad(w[2]).transpose(v),
     mygrad(w[3]).transpose(v)];

/*******************************************/
/* divergence formula for second invariant */
/*******************************************/

/* Invariant II by definition: */
IIdefn : 1/2*(mattrace(vgrad)^2 -
         mattrace(vgrad.vgrad)),expand;

/* Divergence formula */
IItmp : v * mydiv(v) - my_v_dot_del_w(v,v),
        expand;
IIb : 1/2 * mydiv(IItmp), expand;

/* expansion rate */
theta : mydiv(v);

/* for Levi-Civita symbol */
load(itensor);

/* vorticity */
angular_velocity : 1/2 * mycurl(v);
/* declare as a matrix */
vorticity_tensor : zeromatrix(3,3);
for i : 1 thru 3 do
   for j : 1 thru 3 do
      for k : 1 thru 3 do
        vorticity_tensor[i,j] :
             vorticity_tensor[i,j] -
             levi_civita([i,j,k]) *
             angular_velocity[k];

vorticity_sq : 1/2 *
     mattrace( vorticity_tensor .
     transpose(vorticity_tensor) ), expand;

/* shear */
/* declare as a matrix */
shear_tensor : zeromatrix(3,3);
for i : 1 thru 3 do
   for j : 1 thru 3 do
      shear_tensor[i,j] : vgrad[i,j] -
           1/3 * kdelta([i],[j]) * theta -
           vorticity_tensor[i,j], expand;

/* verify that shear is symmetric,
   i.e. zero antisymmetric part */
shear_tensor - transpose(shear_tensor);

shear_sq : 1/2 * mattrace( shear_tensor .
     transpose(shear_tensor) ),
     expand;

/* Vorticity-shear-expansion formula */
IIc : vorticity_sq - shear_sq
      + 1/3 * theta^2, expand;

/******************************************/
/* divergence formula for third invariant */
/******************************************/

/* Invariant III by definition: */
IIIdefn: determinant(vgrad);

/* Invariant III as sum of sums over
                                indices: */
vtr : mattrace(vgrad);
IIIa : 1/3 * mattrace( vgrad.vgrad.vgrad )
     - 1/2 * vtr* mattrace(vgrad.vgrad)
     + 1/6 * (vtr)^3, expand;

/* Divergence formula */
IIItmp : IIb*v, expand;
IIIb : 1/3 *
   mydiv( IIItmp - my_v_dot_del_w(IItmp,v)),
   expand;

/* Vorticity-shear-expansion formula */
IIIc : 1/3 * (1/9 * theta^3 +
     theta *(vorticity_sq - shear_sq) +
     mattrace(shear_tensor.shear_tensor.
              shear_tensor)) +
     angular_velocity.shear_tensor.
              transpose(angular_velocity),
     expand;

/****************************************/
/* Check that definitions match
   alternative expressions */
/****************************************/

/* Defn of invariant vs
                     divergence formula */
printf(false,"IIdefn - IIb: ~a",
       expand(IIdefn-IIb));

/* Defn of invariant vs
              vorticity-shear-expansion */
printf(false,"IIdefn - IIc: ~a",
       expand(IIdefn-IIc));

/* Defn of invariant vs sum of sums */
printf(false,"IIIdefn - IIIa: ~a",
       expand(IIIdefn-IIIa));

/* Defn of invariant vs
                     divergence formula */
printf(false,"IIIdefn - IIIb: ~a",
       expand(IIIdefn-IIIb));

/* Defn of invariant vs
              vorticity-shear-expansion */
printf(false,"IIIdefn - IIIc: ~a",
       expand(IIIdefn-IIIc));
\end{verbatim}

\section{Variance of the third invariant
  $\mathbb{E}[\left<\mathbf{III}_{\mathbf{i}}\right>^2_{{\cal B}_R} ]$}
\label{s-appendix-six-point}


As in App.~B of BKS, for a Gaussian density field of
power spectrum $P_{\mathbf{i}}(k)$ for
a Fourier mode $\mathbf{k}$ of amplitude $k := |\mathbf{k}|$,
we label independent modes $\mathbf{k}_i$, for $1 \le i \le 6$,
and estimate the ensemble mean $\mathbb{E}[.]$ by using the
spatial mean.
The six-point correlation function can be factorised
by finding all distinct triples of pairs, similarly to (C17) of
BKS00,
\begin{align}
\mathbb{E}&[ \tilde{\delta}_{\mathbf{i}}(\mathbf{k}_1)
  \tilde{\delta}_{\mathbf{i}}(\mathbf{k}_2)
  \tilde{\delta}_{\mathbf{i}}(\mathbf{k}_3)
  \tilde{\delta}_{\mathbf{i}}(\mathbf{k}_4)
  \tilde{\delta}_{\mathbf{i}}(\mathbf{k}_5)
  \tilde{\delta}_{\mathbf{i}}(\mathbf{k}_6) ]
\nonumber \\
&= \; P_{\mathbf{i}}(k_1) P_{\mathbf{i}}(k_3) P_{\mathbf{i}}(k_5) \,\delta^D(\mathbf{k}_1 + \mathbf{k}_2) \, \delta^D(\mathbf{k}_3 + \mathbf{k}_4) \, \delta^D(\mathbf{k}_5 + \mathbf{k}_6) \, \nonumber \\
&\;+ P_{\mathbf{i}}(k_1) P_{\mathbf{i}}(k_3) P_{\mathbf{i}}(k_4) \,\delta^D(\mathbf{k}_1 + \mathbf{k}_2) \, \delta^D(\mathbf{k}_3 + \mathbf{k}_5) \, \delta^D(\mathbf{k}_4 + \mathbf{k}_6) \, \nonumber \\
&\;+ P_{\mathbf{i}}(k_1) P_{\mathbf{i}}(k_3) P_{\mathbf{i}}(k_4) \,\delta^D(\mathbf{k}_1 + \mathbf{k}_2) \, \delta^D(\mathbf{k}_3 + \mathbf{k}_6) \, \delta^D(\mathbf{k}_4 + \mathbf{k}_5) \, \nonumber \\
&\;+ P_{\mathbf{i}}(k_1) P_{\mathbf{i}}(k_2) P_{\mathbf{i}}(k_5) \,\delta^D(\mathbf{k}_1 + \mathbf{k}_3) \, \delta^D(\mathbf{k}_2 + \mathbf{k}_4) \, \delta^D(\mathbf{k}_5 + \mathbf{k}_6) \, \nonumber \\
&\;+ P_{\mathbf{i}}(k_1) P_{\mathbf{i}}(k_2) P_{\mathbf{i}}(k_4) \,\delta^D(\mathbf{k}_1 + \mathbf{k}_3) \, \delta^D(\mathbf{k}_2 + \mathbf{k}_5) \, \delta^D(\mathbf{k}_4 + \mathbf{k}_6) \, \nonumber \\
&\;+ P_{\mathbf{i}}(k_1) P_{\mathbf{i}}(k_2) P_{\mathbf{i}}(k_4) \,\delta^D(\mathbf{k}_1 + \mathbf{k}_3) \, \delta^D(\mathbf{k}_2 + \mathbf{k}_6) \, \delta^D(\mathbf{k}_4 + \mathbf{k}_5) \, \nonumber \\
&\;+ P_{\mathbf{i}}(k_1) P_{\mathbf{i}}(k_2) P_{\mathbf{i}}(k_5) \,\delta^D(\mathbf{k}_1 + \mathbf{k}_4) \, \delta^D(\mathbf{k}_2 + \mathbf{k}_3) \, \delta^D(\mathbf{k}_5 + \mathbf{k}_6) \, \nonumber \\
&\;+ P_{\mathbf{i}}(k_1) P_{\mathbf{i}}(k_2) P_{\mathbf{i}}(k_3) \,\delta^D(\mathbf{k}_1 + \mathbf{k}_4) \, \delta^D(\mathbf{k}_2 + \mathbf{k}_5) \, \delta^D(\mathbf{k}_3 + \mathbf{k}_6) \, \nonumber \\
&\;+ P_{\mathbf{i}}(k_1) P_{\mathbf{i}}(k_2) P_{\mathbf{i}}(k_3) \,\delta^D(\mathbf{k}_1 + \mathbf{k}_4) \, \delta^D(\mathbf{k}_2 + \mathbf{k}_6) \, \delta^D(\mathbf{k}_3 + \mathbf{k}_5) \, \nonumber \\
&\;+ P_{\mathbf{i}}(k_1) P_{\mathbf{i}}(k_2) P_{\mathbf{i}}(k_4) \,\delta^D(\mathbf{k}_1 + \mathbf{k}_5) \, \delta^D(\mathbf{k}_2 + \mathbf{k}_3) \, \delta^D(\mathbf{k}_4 + \mathbf{k}_6) \, \nonumber \\
&\;+ P_{\mathbf{i}}(k_1) P_{\mathbf{i}}(k_2) P_{\mathbf{i}}(k_3) \,\delta^D(\mathbf{k}_1 + \mathbf{k}_5) \, \delta^D(\mathbf{k}_2 + \mathbf{k}_4) \, \delta^D(\mathbf{k}_3 + \mathbf{k}_6) \, \nonumber \\
&\;+ P_{\mathbf{i}}(k_1) P_{\mathbf{i}}(k_2) P_{\mathbf{i}}(k_3) \,\delta^D(\mathbf{k}_1 + \mathbf{k}_5) \, \delta^D(\mathbf{k}_2 + \mathbf{k}_6) \, \delta^D(\mathbf{k}_3 + \mathbf{k}_4) \, \nonumber \\
&\;+ P_{\mathbf{i}}(k_1) P_{\mathbf{i}}(k_2) P_{\mathbf{i}}(k_4) \,\delta^D(\mathbf{k}_1 + \mathbf{k}_6) \, \delta^D(\mathbf{k}_2 + \mathbf{k}_3) \, \delta^D(\mathbf{k}_4 + \mathbf{k}_5) \, \nonumber \\
&\;+ P_{\mathbf{i}}(k_1) P_{\mathbf{i}}(k_2) P_{\mathbf{i}}(k_3) \,\delta^D(\mathbf{k}_1 + \mathbf{k}_6) \, \delta^D(\mathbf{k}_2 + \mathbf{k}_4) \, \delta^D(\mathbf{k}_3 + \mathbf{k}_5) \, \nonumber \\
&\;+ P_{\mathbf{i}}(k_1) P_{\mathbf{i}}(k_2) P_{\mathbf{i}}(k_3) \,\delta^D(\mathbf{k}_1 + \mathbf{k}_6) \, \delta^D(\mathbf{k}_2 + \mathbf{k}_5) \, \delta^D(\mathbf{k}_3 + \mathbf{k}_4) \, ,
  \label{e-BKS00-C17-sixpoint}
\end{align}
where
$\delta^D$ is the Dirac delta function.
Use of \eqref{e-BKS00-C17-sixpoint} together with (C9) and (C10) of BKS,
where ${\cal B}_R$ is a ball of radius $R$ and $\widetilde{W}_R$ is the Fourier
transform of the normalised top-hat window function $W_R$,
leads to
\begin{align}
  \mathbb{E}&[\left<\mathbf{III}_{\mathbf{i}}\right>^2_{{\cal B}_R} ]
  \nonumber \\
  &=      {32\, \pi^6}
          \int_{\mathbb{R}^3} \diffd^3 k_1
          \int_{\mathbb{R}^3} \diffd^3 k_2
          \int_{\mathbb{R}^3} \diffd^3 k_3 \nonumber \\
          &\;
              { \frac{{ \widetilde{W}_R\left(| {\mathbf{k}}_1 + {\mathbf{k}}_ 2 + {\mathbf{k}}_3 |\right)}^2\,
                  P_{\mathbf{i}}(k_1) P_{\mathbf{i}}(k_2) P_{\mathbf{i}}(k_3)}
                {
                  3\, { {k}_1}^4\, { {k}_2}^4\, { {k}_3}^4} } \;\times \nonumber \\
              &\; \left[{ {k}_1}^2\, { {k}_2}^2\, { {k}_3}^2-3\,
              { ({\mathbf k}_2 \cdot {\mathbf k}_3)}^2\, { {k}_1}^2+2\, { ({\mathbf k}_1 \cdot {\mathbf k}_2)}\,
              { ({\mathbf k}_1 \cdot {\mathbf k}_3)}\, { ({\mathbf k}_2 \cdot {\mathbf k}_3)}\right]
              \;\times
              \nonumber \\
              &\; \left[{ {k}_1}^2\,
                        { {k}_2}^2\, { {k}_3}^2-{ ({\mathbf k}_1 \cdot {\mathbf k}_2)}^2\, { {k}_3}^2-
                        { ({\mathbf k}_1 \cdot {\mathbf k}_3)}^2\, { {k}_2}^2-{ ({\mathbf k}_2 \cdot {\mathbf k}_3)}^2\, { {k}_1}^2  \right. \nonumber \\
                        &\;\;\; +\left. 2 \vphantom{{{\mathbf k}_9}^9}
                        \, { ({\mathbf k}_1 \cdot {\mathbf k}_2)}\, { ({\mathbf k}_1 \cdot {\mathbf k}_3)}\, { ({\mathbf k}_2 \cdot {\mathbf k}_3)}\right]
                        \nonumber \\
                        \label{e-BKS00-C20-corrected}
\end{align}
rather than the original (C20)--(C22).
Terms induced by \eqref{e-BKS00-C17-sixpoint} that
involve $\tilde{W}_R({{k}}_1)^2$ cancel exactly.

These expressions have been derived with the help of
{\sc GNU Octave} and the free-licensed computer algebra system {\sc maxima}
and tested numerically using numerical 18-dimensional integration with
two sample points per dimension, performed with the GNU Scientific Library.

\newcommand\ctwentytwentytwodoublecheck{
  {\bf JUST TO CHECK: cf less edited version directly from maxima: }

  $$
  32\,\pi^6\,\left({\it k_1}^2\,{\it k_2}^2\,{\it k_3}^2-3\,
  {\it dot\_k_{23}}^2\,{\it k_1}^2+2\,{\it dot\_k_{12}}\,
  {\it dot\_k_{13}}\,{\it dot\_k_{23}}\right)
  $$

  $$
  \,\left({\it k_1}^2\,
         {\it k_2}^2\,{\it k_3}^2-{\it dot\_k_{12}}^2\,{\it k_3}^2-
         {\it dot\_k_{13}}^2\,{\it k_2}^2-{\it dot\_k_{23}}^2\,{\it k_1}^2+2
         \,{\it dot\_k_{12}}\,{\it dot\_k_{13}}\,{\it dot\_k_{23}}\right)
         $$

         $${ \frac{
             {\it W\_k1\_k2\_k_3}^2\,{\it P_1}\,{\it P_2}\,{\it P_3}}
           {
             3\,{\it k_1}^4\,{\it k_2}^4\,{\it k_3}^4}}$$
} 


\end{document}

%% file: biscale_partition.tex
\begingroup
  \makeatletter
  \providecommand\color[2][]{%
    \GenericError{(gnuplot) \space\space\space\@spaces}{%
      Package color not loaded in conjunction with
      terminal option `colourtext'%
    }{See the gnuplot documentation for explanation.%
    }{Either use 'blacktext' in gnuplot or load the package
      color.sty in LaTeX.}%
    \renewcommand\color[2][]{}%
  }%
  \providecommand\includegraphics[2][]{%
    \GenericError{(gnuplot) \space\space\space\@spaces}{%
      Package graphicx or graphics not loaded%
    }{See the gnuplot documentation for explanation.%
    }{The gnuplot epslatex terminal needs graphicx.sty or graphics.sty.}%
    \renewcommand\includegraphics[2][]{}%
  }%
  \providecommand\rotatebox[2]{#2}%
  \@ifundefined{ifGPcolor}{%
    \newif\ifGPcolor
    \GPcolorfalse
  }{}%
  \@ifundefined{ifGPblacktext}{%
    \newif\ifGPblacktext
    \GPblacktexttrue
  }{}%
  \let\gplgaddtomacro\g@addto@macro
  \gdef\gplbacktext{}%
  \gdef\gplfronttext{}%
  \makeatother
  \ifGPblacktext
    \def\colorrgb#1{}%
    \def\colorgray#1{}%
  \else
    \ifGPcolor
      \def\colorrgb#1{\color[rgb]{#1}}%
      \def\colorgray#1{\color[gray]{#1}}%
      \expandafter\def\csname LTw\endcsname{\color{white}}%
      \expandafter\def\csname LTb\endcsname{\color{black}}%
      \expandafter\def\csname LTa\endcsname{\color{black}}%
      \expandafter\def\csname LT0\endcsname{\color[rgb]{1,0,0}}%
      \expandafter\def\csname LT1\endcsname{\color[rgb]{0,1,0}}%
      \expandafter\def\csname LT2\endcsname{\color[rgb]{0,0,1}}%
      \expandafter\def\csname LT3\endcsname{\color[rgb]{1,0,1}}%
      \expandafter\def\csname LT4\endcsname{\color[rgb]{0,1,1}}%
      \expandafter\def\csname LT5\endcsname{\color[rgb]{1,1,0}}%
      \expandafter\def\csname LT6\endcsname{\color[rgb]{0,0,0}}%
      \expandafter\def\csname LT7\endcsname{\color[rgb]{1,0.3,0}}%
      \expandafter\def\csname LT8\endcsname{\color[rgb]{0.5,0.5,0.5}}%
    \else
      \def\colorrgb#1{\color{black}}%
      \def\colorgray#1{\color[gray]{#1}}%
      \expandafter\def\csname LTw\endcsname{\color{white}}%
      \expandafter\def\csname LTb\endcsname{\color{black}}%
      \expandafter\def\csname LTa\endcsname{\color{black}}%
      \expandafter\def\csname LT0\endcsname{\color{black}}%
      \expandafter\def\csname LT1\endcsname{\color{black}}%
      \expandafter\def\csname LT2\endcsname{\color{black}}%
      \expandafter\def\csname LT3\endcsname{\color{black}}%
      \expandafter\def\csname LT4\endcsname{\color{black}}%
      \expandafter\def\csname LT5\endcsname{\color{black}}%
      \expandafter\def\csname LT6\endcsname{\color{black}}%
      \expandafter\def\csname LT7\endcsname{\color{black}}%
      \expandafter\def\csname LT8\endcsname{\color{black}}%
    \fi
  \fi
  \setlength{\unitlength}{0.0200bp}%
  \begin{picture}(11520.00,8640.00)%
    \gplgaddtomacro\gplbacktext{%
      \colorrgb{0.00,0.00,0.00}%
      \put(1628,1408){\makebox(0,0)[r]{\strut{}0.9}}%
      \colorrgb{0.00,0.00,0.00}%
      \put(1628,2525){\makebox(0,0)[r]{\strut{}1}}%
      \colorrgb{0.00,0.00,0.00}%
      \put(1628,3642){\makebox(0,0)[r]{\strut{}1.1}}%
      \colorrgb{0.00,0.00,0.00}%
      \put(1628,4760){\makebox(0,0)[r]{\strut{}1.2}}%
      \colorrgb{0.00,0.00,0.00}%
      \put(1628,5877){\makebox(0,0)[r]{\strut{}1.3}}%
      \colorrgb{0.00,0.00,0.00}%
      \put(1628,6994){\makebox(0,0)[r]{\strut{}1.4}}%
      \colorrgb{0.00,0.00,0.00}%
      \put(1628,8111){\makebox(0,0)[r]{\strut{}1.5}}%
      \colorrgb{0.00,0.00,0.00}%
      \put(1892,968){\makebox(0,0){\strut{}0}}%
      \colorrgb{0.00,0.00,0.00}%
      \put(2874,968){\makebox(0,0){\strut{}2}}%
      \colorrgb{0.00,0.00,0.00}%
      \put(3855,968){\makebox(0,0){\strut{}4}}%
      \colorrgb{0.00,0.00,0.00}%
      \put(4837,968){\makebox(0,0){\strut{}6}}%
      \colorrgb{0.00,0.00,0.00}%
      \put(5819,968){\makebox(0,0){\strut{}8}}%
      \colorrgb{0.00,0.00,0.00}%
      \put(6800,968){\makebox(0,0){\strut{}10}}%
      \colorrgb{0.00,0.00,0.00}%
      \put(7782,968){\makebox(0,0){\strut{}12}}%
      \colorrgb{0.00,0.00,0.00}%
      \put(8764,968){\makebox(0,0){\strut{}14}}%
      \colorrgb{0.00,0.00,0.00}%
      \put(9745,968){\makebox(0,0){\strut{}16}}%
      \colorrgb{0.00,0.00,0.00}%
      \put(10727,968){\makebox(0,0){\strut{}18}}%
      \colorrgb{0.00,0.00,0.00}%
      \put(352,4759){\rotatebox{90}{\makebox(0,0){\strut{}$\aeff/\abg$}}}%
      \colorrgb{0.00,0.00,0.00}%
      \put(6309,308){\makebox(0,0){\strut{}$t$ (Gyr)}}%
    }%
    \gplgaddtomacro\gplfronttext{%
      \colorrgb{0.00,0.00,0.00}%
      \put(7235,7828){\makebox(0,0)[r]{\footnotesize $\initaverageDunder{\invI}$, $\initaverageDunder{\invIII} \neq0$~~}}%
      \colorrgb{0.00,0.00,0.00}%
      \put(7235,7388){\makebox(0,0)[r]{\footnotesize $\initaverageDunder{\invI}$, $\initaverageDunder{\invII} \neq0$~~}}%
      \colorrgb{0.00,0.00,0.00}%
      \put(7235,6948){\makebox(0,0)[r]{\footnotesize spherical~~}}%
      \colorrgb{0.00,0.00,0.00}%
      \put(7235,6508){\makebox(0,0)[r]{\footnotesize planar~~}}%
    }%
    \gplbacktext
    \put(0,0){\includegraphics[scale=0.4]{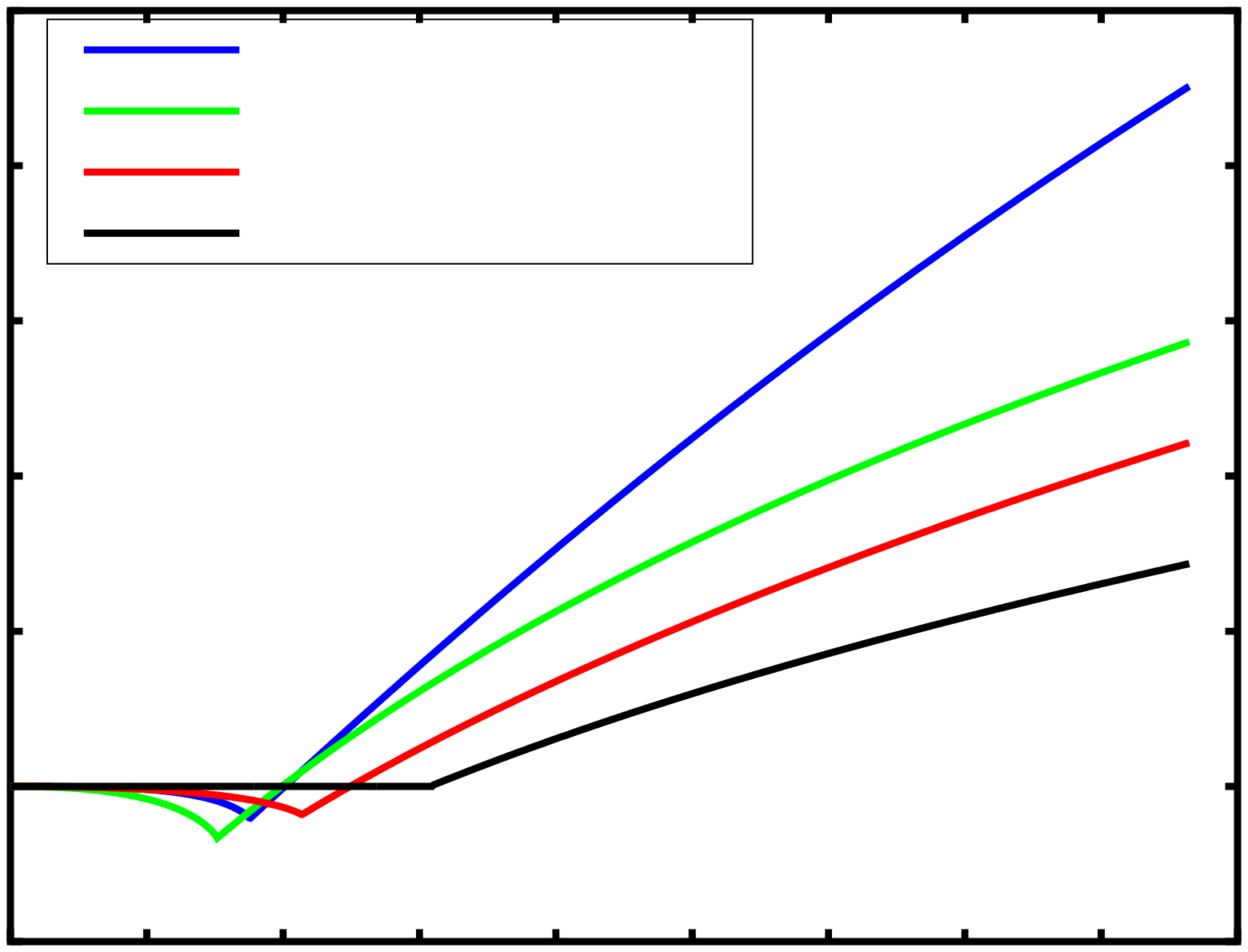}}%
    \gplfronttext
  \end{picture}%
\endgroup

%% file: biscale_underdense.tex
\begingroup
  \makeatletter
  \providecommand\color[2][]{%
    \GenericError{(gnuplot) \space\space\space\@spaces}{%
      Package color not loaded in conjunction with
      terminal option `colourtext'%
    }{See the gnuplot documentation for explanation.%
    }{Either use 'blacktext' in gnuplot or load the package
      color.sty in LaTeX.}%
    \renewcommand\color[2][]{}%
  }%
  \providecommand\includegraphics[2][]{%
    \GenericError{(gnuplot) \space\space\space\@spaces}{%
      Package graphicx or graphics not loaded%
    }{See the gnuplot documentation for explanation.%
    }{The gnuplot epslatex terminal needs graphicx.sty or graphics.sty.}%
    \renewcommand\includegraphics[2][]{}%
  }%
  \providecommand\rotatebox[2]{#2}%
  \@ifundefined{ifGPcolor}{%
    \newif\ifGPcolor
    \GPcolorfalse
  }{}%
  \@ifundefined{ifGPblacktext}{%
    \newif\ifGPblacktext
    \GPblacktexttrue
  }{}%
  \let\gplgaddtomacro\g@addto@macro
  \gdef\gplbacktext{}%
  \gdef\gplfronttext{}%
  \makeatother
  \ifGPblacktext
    \def\colorrgb#1{}%
    \def\colorgray#1{}%
  \else
    \ifGPcolor
      \def\colorrgb#1{\color[rgb]{#1}}%
      \def\colorgray#1{\color[gray]{#1}}%
      \expandafter\def\csname LTw\endcsname{\color{white}}%
      \expandafter\def\csname LTb\endcsname{\color{black}}%
      \expandafter\def\csname LTa\endcsname{\color{black}}%
      \expandafter\def\csname LT0\endcsname{\color[rgb]{1,0,0}}%
      \expandafter\def\csname LT1\endcsname{\color[rgb]{0,1,0}}%
      \expandafter\def\csname LT2\endcsname{\color[rgb]{0,0,1}}%
      \expandafter\def\csname LT3\endcsname{\color[rgb]{1,0,1}}%
      \expandafter\def\csname LT4\endcsname{\color[rgb]{0,1,1}}%
      \expandafter\def\csname LT5\endcsname{\color[rgb]{1,1,0}}%
      \expandafter\def\csname LT6\endcsname{\color[rgb]{0,0,0}}%
      \expandafter\def\csname LT7\endcsname{\color[rgb]{1,0.3,0}}%
      \expandafter\def\csname LT8\endcsname{\color[rgb]{0.5,0.5,0.5}}%
    \else
      \def\colorrgb#1{\color{black}}%
      \def\colorgray#1{\color[gray]{#1}}%
      \expandafter\def\csname LTw\endcsname{\color{white}}%
      \expandafter\def\csname LTb\endcsname{\color{black}}%
      \expandafter\def\csname LTa\endcsname{\color{black}}%
      \expandafter\def\csname LT0\endcsname{\color{black}}%
      \expandafter\def\csname LT1\endcsname{\color{black}}%
      \expandafter\def\csname LT2\endcsname{\color{black}}%
      \expandafter\def\csname LT3\endcsname{\color{black}}%
      \expandafter\def\csname LT4\endcsname{\color{black}}%
      \expandafter\def\csname LT5\endcsname{\color{black}}%
      \expandafter\def\csname LT6\endcsname{\color{black}}%
      \expandafter\def\csname LT7\endcsname{\color{black}}%
      \expandafter\def\csname LT8\endcsname{\color{black}}%
    \fi
  \fi
  \setlength{\unitlength}{0.0200bp}%
  \begin{picture}(11520.00,8640.00)%
    \gplgaddtomacro\gplbacktext{%
      \colorrgb{0.00,0.00,0.00}%
      \put(1628,1408){\makebox(0,0)[r]{\strut{}0.9}}%
      \colorrgb{0.00,0.00,0.00}%
      \put(1628,2078){\makebox(0,0)[r]{\strut{}1}}%
      \colorrgb{0.00,0.00,0.00}%
      \put(1628,2749){\makebox(0,0)[r]{\strut{}1.1}}%
      \colorrgb{0.00,0.00,0.00}%
      \put(1628,3419){\makebox(0,0)[r]{\strut{}1.2}}%
      \colorrgb{0.00,0.00,0.00}%
      \put(1628,4089){\makebox(0,0)[r]{\strut{}1.3}}%
      \colorrgb{0.00,0.00,0.00}%
      \put(1628,4760){\makebox(0,0)[r]{\strut{}1.4}}%
      \colorrgb{0.00,0.00,0.00}%
      \put(1628,5430){\makebox(0,0)[r]{\strut{}1.5}}%
      \colorrgb{0.00,0.00,0.00}%
      \put(1628,6100){\makebox(0,0)[r]{\strut{}1.6}}%
      \colorrgb{0.00,0.00,0.00}%
      \put(1628,6770){\makebox(0,0)[r]{\strut{}1.7}}%
      \colorrgb{0.00,0.00,0.00}%
      \put(1628,7441){\makebox(0,0)[r]{\strut{}1.8}}%
      \colorrgb{0.00,0.00,0.00}%
      \put(1628,8111){\makebox(0,0)[r]{\strut{}1.9}}%
      \colorrgb{0.00,0.00,0.00}%
      \put(1892,968){\makebox(0,0){\strut{}0}}%
      \colorrgb{0.00,0.00,0.00}%
      \put(2874,968){\makebox(0,0){\strut{}2}}%
      \colorrgb{0.00,0.00,0.00}%
      \put(3855,968){\makebox(0,0){\strut{}4}}%
      \colorrgb{0.00,0.00,0.00}%
      \put(4837,968){\makebox(0,0){\strut{}6}}%
      \colorrgb{0.00,0.00,0.00}%
      \put(5819,968){\makebox(0,0){\strut{}8}}%
      \colorrgb{0.00,0.00,0.00}%
      \put(6800,968){\makebox(0,0){\strut{}10}}%
      \colorrgb{0.00,0.00,0.00}%
      \put(7782,968){\makebox(0,0){\strut{}12}}%
      \colorrgb{0.00,0.00,0.00}%
      \put(8764,968){\makebox(0,0){\strut{}14}}%
      \colorrgb{0.00,0.00,0.00}%
      \put(9745,968){\makebox(0,0){\strut{}16}}%
      \colorrgb{0.00,0.00,0.00}%
      \put(10727,968){\makebox(0,0){\strut{}18}}%
      \colorrgb{0.00,0.00,0.00}%
      \put(352,4759){\rotatebox{90}{\makebox(0,0){\strut{}$a_{\CD^-}/\abg$}}}%
      \colorrgb{0.00,0.00,0.00}%
      \put(6309,308){\makebox(0,0){\strut{}$t$ (Gyr)}}%
    }%
    \gplgaddtomacro\gplfronttext{%
      \colorrgb{0.00,0.00,0.00}%
      \put(7235,7828){\makebox(0,0)[r]{\footnotesize $\initaverageDunder{\invI}$, $\initaverageDunder{\invIII} \neq0$~~}}%
      \colorrgb{0.00,0.00,0.00}%
      \put(7235,7388){\makebox(0,0)[r]{\footnotesize $\initaverageDunder{\invI}$, $\initaverageDunder{\invII} \neq0$~~}}%
      \colorrgb{0.00,0.00,0.00}%
      \put(7235,6948){\makebox(0,0)[r]{\footnotesize spherical~~}}%
      \colorrgb{0.00,0.00,0.00}%
      \put(7235,6508){\makebox(0,0)[r]{\footnotesize planar~~}}%
    }%
    \gplbacktext
    \put(0,0){\includegraphics[scale=0.4]{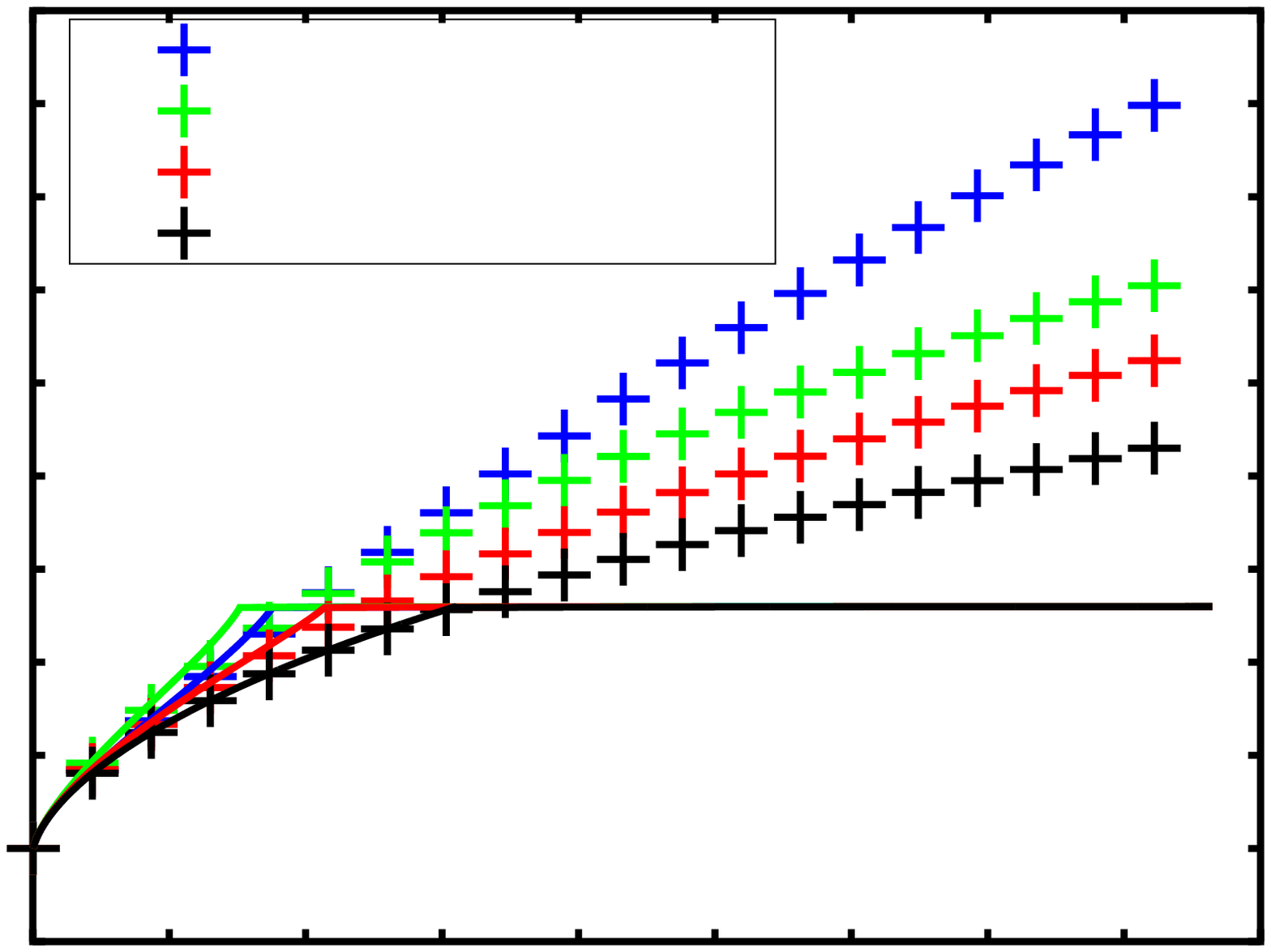}}%
    \gplfronttext
  \end{picture}%
\endgroup

%% file: biscale_OmRD.tex
\begingroup
  \makeatletter
  \providecommand\color[2][]{%
    \GenericError{(gnuplot) \space\space\space\@spaces}{%
      Package color not loaded in conjunction with
      terminal option `colourtext'%
    }{See the gnuplot documentation for explanation.%
    }{Either use 'blacktext' in gnuplot or load the package
      color.sty in LaTeX.}%
    \renewcommand\color[2][]{}%
  }%
  \providecommand\includegraphics[2][]{%
    \GenericError{(gnuplot) \space\space\space\@spaces}{%
      Package graphicx or graphics not loaded%
    }{See the gnuplot documentation for explanation.%
    }{The gnuplot epslatex terminal needs graphicx.sty or graphics.sty.}%
    \renewcommand\includegraphics[2][]{}%
  }%
  \providecommand\rotatebox[2]{#2}%
  \@ifundefined{ifGPcolor}{%
    \newif\ifGPcolor
    \GPcolorfalse
  }{}%
  \@ifundefined{ifGPblacktext}{%
    \newif\ifGPblacktext
    \GPblacktexttrue
  }{}%
  \let\gplgaddtomacro\g@addto@macro
  \gdef\gplbacktext{}%
  \gdef\gplfronttext{}%
  \makeatother
  \ifGPblacktext
    \def\colorrgb#1{}%
    \def\colorgray#1{}%
  \else
    \ifGPcolor
      \def\colorrgb#1{\color[rgb]{#1}}%
      \def\colorgray#1{\color[gray]{#1}}%
      \expandafter\def\csname LTw\endcsname{\color{white}}%
      \expandafter\def\csname LTb\endcsname{\color{black}}%
      \expandafter\def\csname LTa\endcsname{\color{black}}%
      \expandafter\def\csname LT0\endcsname{\color[rgb]{1,0,0}}%
      \expandafter\def\csname LT1\endcsname{\color[rgb]{0,1,0}}%
      \expandafter\def\csname LT2\endcsname{\color[rgb]{0,0,1}}%
      \expandafter\def\csname LT3\endcsname{\color[rgb]{1,0,1}}%
      \expandafter\def\csname LT4\endcsname{\color[rgb]{0,1,1}}%
      \expandafter\def\csname LT5\endcsname{\color[rgb]{1,1,0}}%
      \expandafter\def\csname LT6\endcsname{\color[rgb]{0,0,0}}%
      \expandafter\def\csname LT7\endcsname{\color[rgb]{1,0.3,0}}%
      \expandafter\def\csname LT8\endcsname{\color[rgb]{0.5,0.5,0.5}}%
    \else
      \def\colorrgb#1{\color{black}}%
      \def\colorgray#1{\color[gray]{#1}}%
      \expandafter\def\csname LTw\endcsname{\color{white}}%
      \expandafter\def\csname LTb\endcsname{\color{black}}%
      \expandafter\def\csname LTa\endcsname{\color{black}}%
      \expandafter\def\csname LT0\endcsname{\color{black}}%
      \expandafter\def\csname LT1\endcsname{\color{black}}%
      \expandafter\def\csname LT2\endcsname{\color{black}}%
      \expandafter\def\csname LT3\endcsname{\color{black}}%
      \expandafter\def\csname LT4\endcsname{\color{black}}%
      \expandafter\def\csname LT5\endcsname{\color{black}}%
      \expandafter\def\csname LT6\endcsname{\color{black}}%
      \expandafter\def\csname LT7\endcsname{\color{black}}%
      \expandafter\def\csname LT8\endcsname{\color{black}}%
    \fi
  \fi
  \setlength{\unitlength}{0.0200bp}%
  \begin{picture}(11520.00,8640.00)%
    \gplgaddtomacro\gplbacktext{%
      \colorrgb{0.00,0.00,0.00}%
      \put(1892,1408){\makebox(0,0)[r]{\strut{}-0.5}}%
      \colorrgb{0.00,0.00,0.00}%
      \put(1892,2246){\makebox(0,0)[r]{\strut{}0}}%
      \colorrgb{0.00,0.00,0.00}%
      \put(1892,3084){\makebox(0,0)[r]{\strut{}0.5}}%
      \colorrgb{0.00,0.00,0.00}%
      \put(1892,3922){\makebox(0,0)[r]{\strut{}1}}%
      \colorrgb{0.00,0.00,0.00}%
      \put(1892,4760){\makebox(0,0)[r]{\strut{}1.5}}%
      \colorrgb{0.00,0.00,0.00}%
      \put(1892,5597){\makebox(0,0)[r]{\strut{}2}}%
      \colorrgb{0.00,0.00,0.00}%
      \put(1892,6435){\makebox(0,0)[r]{\strut{}2.5}}%
      \colorrgb{0.00,0.00,0.00}%
      \put(1892,7273){\makebox(0,0)[r]{\strut{}3}}%
      \colorrgb{0.00,0.00,0.00}%
      \put(1892,8111){\makebox(0,0)[r]{\strut{}3.5}}%
      \colorrgb{0.00,0.00,0.00}%
      \put(2156,968){\makebox(0,0){\strut{}0}}%
      \colorrgb{0.00,0.00,0.00}%
      \put(3108,968){\makebox(0,0){\strut{}2}}%
      \colorrgb{0.00,0.00,0.00}%
      \put(4061,968){\makebox(0,0){\strut{}4}}%
      \colorrgb{0.00,0.00,0.00}%
      \put(5013,968){\makebox(0,0){\strut{}6}}%
      \colorrgb{0.00,0.00,0.00}%
      \put(5965,968){\makebox(0,0){\strut{}8}}%
      \colorrgb{0.00,0.00,0.00}%
      \put(6918,968){\makebox(0,0){\strut{}10}}%
      \colorrgb{0.00,0.00,0.00}%
      \put(7870,968){\makebox(0,0){\strut{}12}}%
      \colorrgb{0.00,0.00,0.00}%
      \put(8822,968){\makebox(0,0){\strut{}14}}%
      \colorrgb{0.00,0.00,0.00}%
      \put(9775,968){\makebox(0,0){\strut{}16}}%
      \colorrgb{0.00,0.00,0.00}%
      \put(10727,968){\makebox(0,0){\strut{}18}}%
      \colorrgb{0.00,0.00,0.00}%
      \put(352,4759){\rotatebox{90}{\makebox(0,0){\strut{}$\OmRDunder$ (+ VQZA; --- FLRW)}}}%
      \colorrgb{0.00,0.00,0.00}%
      \put(6441,308){\makebox(0,0){\strut{}$t$ (Gyr)}}%
    }%
    \gplgaddtomacro\gplfronttext{%
      \colorrgb{0.00,0.00,0.00}%
      \put(10463,7828){\makebox(0,0)[r]{\footnotesize $\initaverageDunder{\invI}$, $\initaverageDunder{\invIII} \neq0$~~}}%
      \colorrgb{0.00,0.00,0.00}%
      \put(10463,7388){\makebox(0,0)[r]{\footnotesize $\initaverageDunder{\invI}$, $\initaverageDunder{\invII} \neq0$~~}}%
      \colorrgb{0.00,0.00,0.00}%
      \put(10463,6948){\makebox(0,0)[r]{\footnotesize spherical~~}}%
      \colorrgb{0.00,0.00,0.00}%
      \put(10463,6508){\makebox(0,0)[r]{\footnotesize planar~~}}%
    }%
    \gplbacktext
    \put(0,0){\includegraphics[scale=0.4]{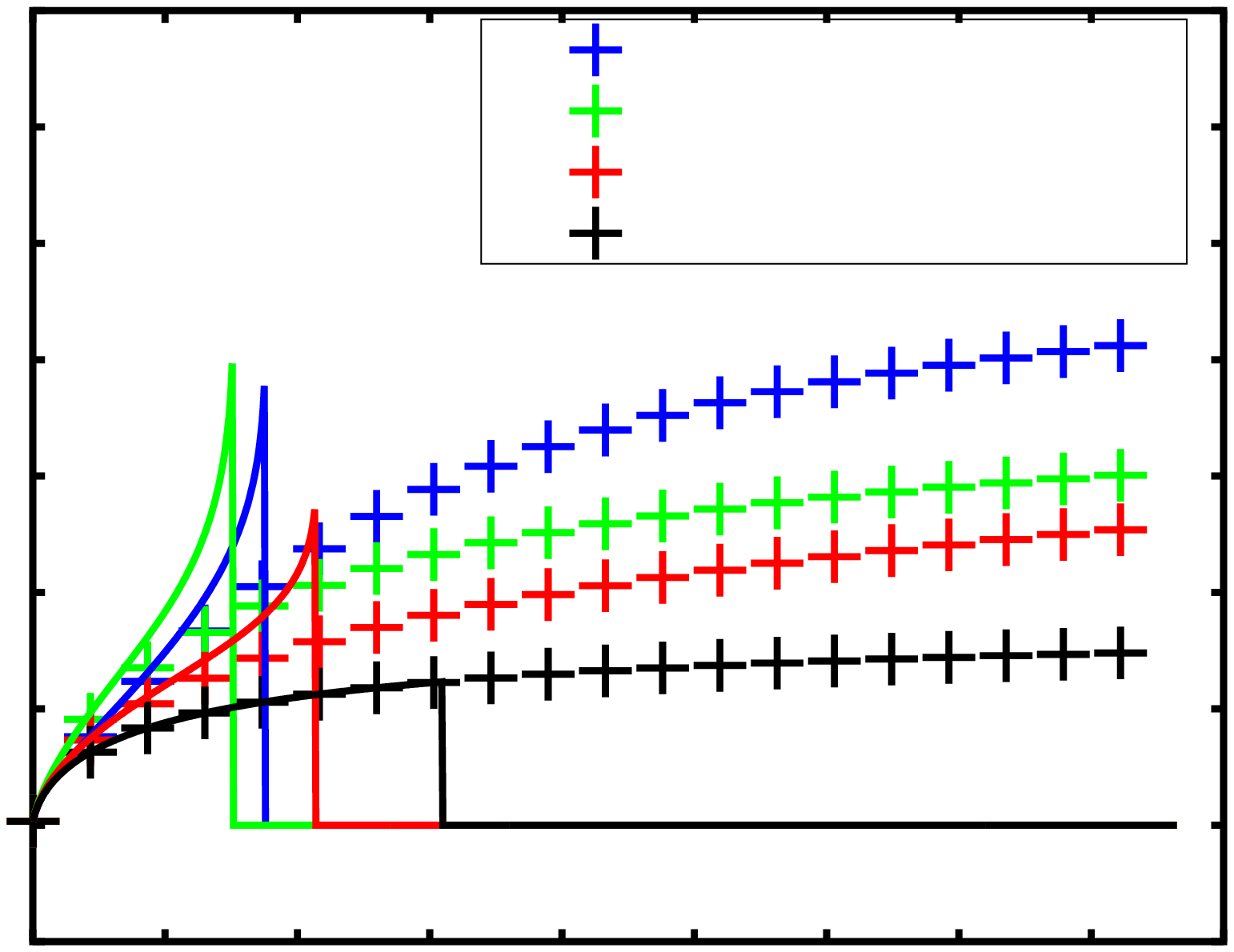}}%
    \gplfronttext
  \end{picture}%
\endgroup

%% file: SN_invI.tex
\begingroup
  \makeatletter
  \providecommand\color[2][]{%
    \GenericError{(gnuplot) \space\space\space\@spaces}{%
      Package color not loaded in conjunction with
      terminal option `colourtext'%
    }{See the gnuplot documentation for explanation.%
    }{Either use 'blacktext' in gnuplot or load the package
      color.sty in LaTeX.}%
    \renewcommand\color[2][]{}%
  }%
  \providecommand\includegraphics[2][]{%
    \GenericError{(gnuplot) \space\space\space\@spaces}{%
      Package graphicx or graphics not loaded%
    }{See the gnuplot documentation for explanation.%
    }{The gnuplot epslatex terminal needs graphicx.sty or graphics.sty.}%
    \renewcommand\includegraphics[2][]{}%
  }%
  \providecommand\rotatebox[2]{#2}%
  \@ifundefined{ifGPcolor}{%
    \newif\ifGPcolor
    \GPcolorfalse
  }{}%
  \@ifundefined{ifGPblacktext}{%
    \newif\ifGPblacktext
    \GPblacktexttrue
  }{}%
  \let\gplgaddtomacro\g@addto@macro
  \gdef\gplbacktext{}%
  \gdef\gplfronttext{}%
  \makeatother
  \ifGPblacktext
    \def\colorrgb#1{}%
    \def\colorgray#1{}%
  \else
    \ifGPcolor
      \def\colorrgb#1{\color[rgb]{#1}}%
      \def\colorgray#1{\color[gray]{#1}}%
      \expandafter\def\csname LTw\endcsname{\color{white}}%
      \expandafter\def\csname LTb\endcsname{\color{black}}%
      \expandafter\def\csname LTa\endcsname{\color{black}}%
      \expandafter\def\csname LT0\endcsname{\color[rgb]{1,0,0}}%
      \expandafter\def\csname LT1\endcsname{\color[rgb]{0,1,0}}%
      \expandafter\def\csname LT2\endcsname{\color[rgb]{0,0,1}}%
      \expandafter\def\csname LT3\endcsname{\color[rgb]{1,0,1}}%
      \expandafter\def\csname LT4\endcsname{\color[rgb]{0,1,1}}%
      \expandafter\def\csname LT5\endcsname{\color[rgb]{1,1,0}}%
      \expandafter\def\csname LT6\endcsname{\color[rgb]{0,0,0}}%
      \expandafter\def\csname LT7\endcsname{\color[rgb]{1,0.3,0}}%
      \expandafter\def\csname LT8\endcsname{\color[rgb]{0.5,0.5,0.5}}%
    \else
      \def\colorrgb#1{\color{black}}%
      \def\colorgray#1{\color[gray]{#1}}%
      \expandafter\def\csname LTw\endcsname{\color{white}}%
      \expandafter\def\csname LTb\endcsname{\color{black}}%
      \expandafter\def\csname LTa\endcsname{\color{black}}%
      \expandafter\def\csname LT0\endcsname{\color{black}}%
      \expandafter\def\csname LT1\endcsname{\color{black}}%
      \expandafter\def\csname LT2\endcsname{\color{black}}%
      \expandafter\def\csname LT3\endcsname{\color{black}}%
      \expandafter\def\csname LT4\endcsname{\color{black}}%
      \expandafter\def\csname LT5\endcsname{\color{black}}%
      \expandafter\def\csname LT6\endcsname{\color{black}}%
      \expandafter\def\csname LT7\endcsname{\color{black}}%
      \expandafter\def\csname LT8\endcsname{\color{black}}%
    \fi
  \fi
  \setlength{\unitlength}{0.0200bp}%
  \begin{picture}(11520.00,8640.00)%
    \gplgaddtomacro\gplbacktext{%
      \colorrgb{0.00,0.00,0.00}%
      \put(1548,1152){\makebox(0,0)[r]{\strut{}$10^{-1}$}}%
      \colorrgb{0.00,0.00,0.00}%
      \put(1548,2328){\makebox(0,0)[r]{\strut{}$10^{0}$}}%
      \colorrgb{0.00,0.00,0.00}%
      \put(1548,3504){\makebox(0,0)[r]{\strut{}$10^{1}$}}%
      \colorrgb{0.00,0.00,0.00}%
      \put(1548,4680){\makebox(0,0)[r]{\strut{}$10^{2}$}}%
      \colorrgb{0.00,0.00,0.00}%
      \put(1548,5855){\makebox(0,0)[r]{\strut{}$10^{3}$}}%
      \colorrgb{0.00,0.00,0.00}%
      \put(1548,7031){\makebox(0,0)[r]{\strut{}$10^{4}$}}%
      \colorrgb{0.00,0.00,0.00}%
      \put(1548,8207){\makebox(0,0)[r]{\strut{}$10^{5}$}}%
      \colorrgb{0.00,0.00,0.00}%
      \put(3065,792){\makebox(0,0){\strut{}8}}%
      \colorrgb{0.00,0.00,0.00}%
      \put(4366,792){\makebox(0,0){\strut{}16}}%
      \colorrgb{0.00,0.00,0.00}%
      \put(5667,792){\makebox(0,0){\strut{}32}}%
      \colorrgb{0.00,0.00,0.00}%
      \put(6968,792){\makebox(0,0){\strut{}64}}%
      \colorrgb{0.00,0.00,0.00}%
      \put(8269,792){\makebox(0,0){\strut{}128}}%
      \colorrgb{0.00,0.00,0.00}%
      \put(9570,792){\makebox(0,0){\strut{}256}}%
      \colorrgb{0.00,0.00,0.00}%
      \put(288,4679){\rotatebox{90}{\makebox(0,0){\strut{}$S/N$}}}%
      \colorrgb{0.00,0.00,0.00}%
      \put(6317,252){\makebox(0,0){\strut{}$n_{\mathrm{DTFE}}$}}%
    }%
    \gplgaddtomacro\gplfronttext{%
      \colorrgb{0.00,0.00,0.00}%
      \put(10655,7964){\makebox(0,0)[r]{\strut{} {\tiny$\vmodone$ $\nxyz$= 1  }}}%
      \colorrgb{0.00,0.00,0.00}%
      \put(10655,7604){\makebox(0,0)[r]{\strut{} {\tiny$\vmodone$ $\nxyz$= 8  }}}%
      \colorrgb{0.00,0.00,0.00}%
      \put(10655,7244){\makebox(0,0)[r]{\strut{} {\tiny$\vmodone$ $\nxyz$=64  }}}%
    }%
    \gplbacktext
    \put(0,0){\includegraphics[scale=0.4]{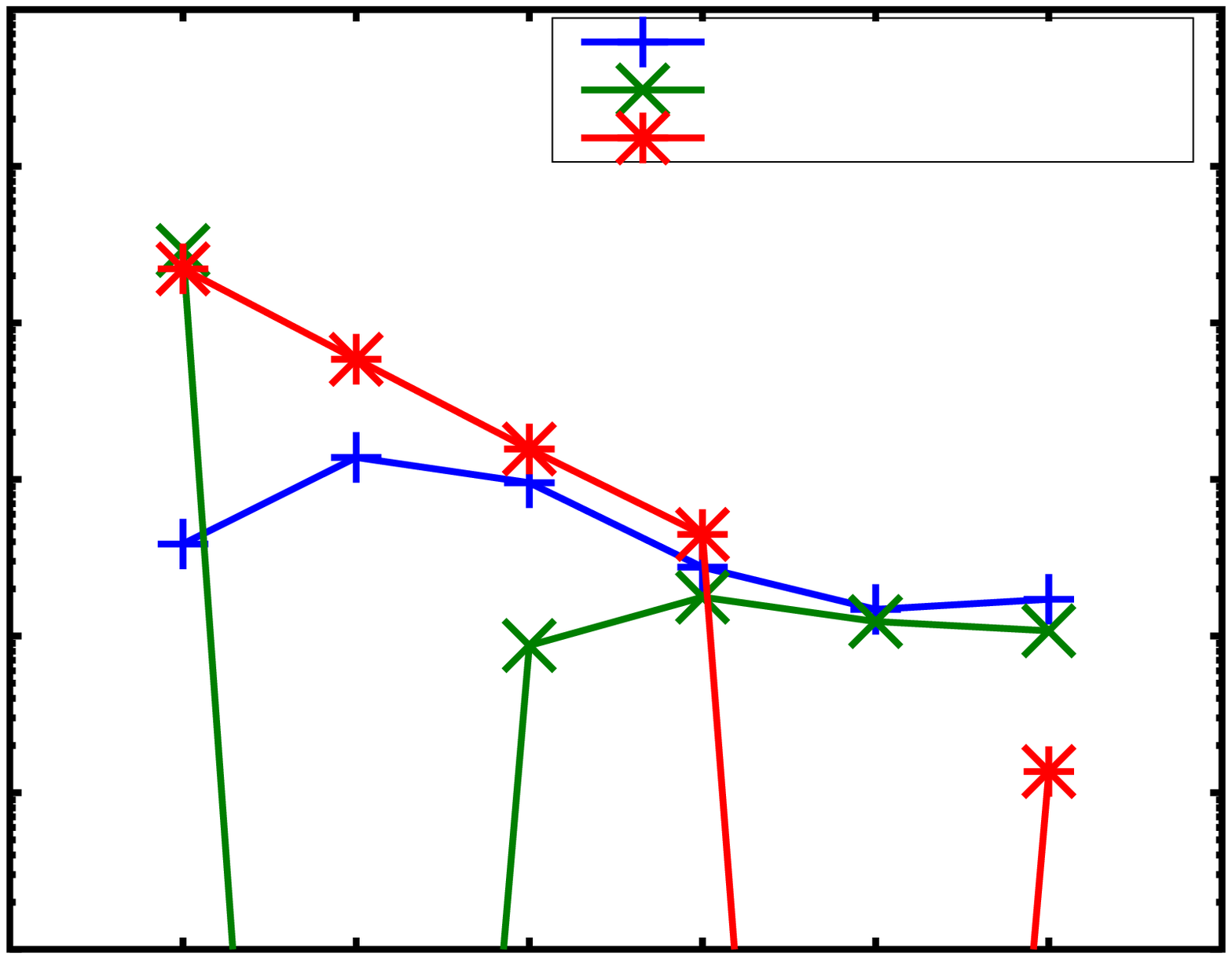}}%
    \gplfronttext
  \end{picture}%
\endgroup

%% file: SN_invII.tex
\begingroup
  \makeatletter
  \providecommand\color[2][]{%
    \GenericError{(gnuplot) \space\space\space\@spaces}{%
      Package color not loaded in conjunction with
      terminal option `colourtext'%
    }{See the gnuplot documentation for explanation.%
    }{Either use 'blacktext' in gnuplot or load the package
      color.sty in LaTeX.}%
    \renewcommand\color[2][]{}%
  }%
  \providecommand\includegraphics[2][]{%
    \GenericError{(gnuplot) \space\space\space\@spaces}{%
      Package graphicx or graphics not loaded%
    }{See the gnuplot documentation for explanation.%
    }{The gnuplot epslatex terminal needs graphicx.sty or graphics.sty.}%
    \renewcommand\includegraphics[2][]{}%
  }%
  \providecommand\rotatebox[2]{#2}%
  \@ifundefined{ifGPcolor}{%
    \newif\ifGPcolor
    \GPcolorfalse
  }{}%
  \@ifundefined{ifGPblacktext}{%
    \newif\ifGPblacktext
    \GPblacktexttrue
  }{}%
  \let\gplgaddtomacro\g@addto@macro
  \gdef\gplbacktext{}%
  \gdef\gplfronttext{}%
  \makeatother
  \ifGPblacktext
    \def\colorrgb#1{}%
    \def\colorgray#1{}%
  \else
    \ifGPcolor
      \def\colorrgb#1{\color[rgb]{#1}}%
      \def\colorgray#1{\color[gray]{#1}}%
      \expandafter\def\csname LTw\endcsname{\color{white}}%
      \expandafter\def\csname LTb\endcsname{\color{black}}%
      \expandafter\def\csname LTa\endcsname{\color{black}}%
      \expandafter\def\csname LT0\endcsname{\color[rgb]{1,0,0}}%
      \expandafter\def\csname LT1\endcsname{\color[rgb]{0,1,0}}%
      \expandafter\def\csname LT2\endcsname{\color[rgb]{0,0,1}}%
      \expandafter\def\csname LT3\endcsname{\color[rgb]{1,0,1}}%
      \expandafter\def\csname LT4\endcsname{\color[rgb]{0,1,1}}%
      \expandafter\def\csname LT5\endcsname{\color[rgb]{1,1,0}}%
      \expandafter\def\csname LT6\endcsname{\color[rgb]{0,0,0}}%
      \expandafter\def\csname LT7\endcsname{\color[rgb]{1,0.3,0}}%
      \expandafter\def\csname LT8\endcsname{\color[rgb]{0.5,0.5,0.5}}%
    \else
      \def\colorrgb#1{\color{black}}%
      \def\colorgray#1{\color[gray]{#1}}%
      \expandafter\def\csname LTw\endcsname{\color{white}}%
      \expandafter\def\csname LTb\endcsname{\color{black}}%
      \expandafter\def\csname LTa\endcsname{\color{black}}%
      \expandafter\def\csname LT0\endcsname{\color{black}}%
      \expandafter\def\csname LT1\endcsname{\color{black}}%
      \expandafter\def\csname LT2\endcsname{\color{black}}%
      \expandafter\def\csname LT3\endcsname{\color{black}}%
      \expandafter\def\csname LT4\endcsname{\color{black}}%
      \expandafter\def\csname LT5\endcsname{\color{black}}%
      \expandafter\def\csname LT6\endcsname{\color{black}}%
      \expandafter\def\csname LT7\endcsname{\color{black}}%
      \expandafter\def\csname LT8\endcsname{\color{black}}%
    \fi
  \fi
  \setlength{\unitlength}{0.0200bp}%
  \begin{picture}(11520.00,8640.00)%
    \gplgaddtomacro\gplbacktext{%
      \colorrgb{0.00,0.00,0.00}%
      \put(1548,1793){\makebox(0,0)[r]{\strut{}$10^{0}$}}%
      \colorrgb{0.00,0.00,0.00}%
      \put(1548,3076){\makebox(0,0)[r]{\strut{}$10^{2}$}}%
      \colorrgb{0.00,0.00,0.00}%
      \put(1548,4359){\makebox(0,0)[r]{\strut{}$10^{4}$}}%
      \colorrgb{0.00,0.00,0.00}%
      \put(1548,5642){\makebox(0,0)[r]{\strut{}$10^{6}$}}%
      \colorrgb{0.00,0.00,0.00}%
      \put(1548,6924){\makebox(0,0)[r]{\strut{}$10^{8}$}}%
      \colorrgb{0.00,0.00,0.00}%
      \put(1548,8207){\makebox(0,0)[r]{\strut{}$10^{10}$}}%
      \colorrgb{0.00,0.00,0.00}%
      \put(3065,792){\makebox(0,0){\strut{}8}}%
      \colorrgb{0.00,0.00,0.00}%
      \put(4366,792){\makebox(0,0){\strut{}16}}%
      \colorrgb{0.00,0.00,0.00}%
      \put(5667,792){\makebox(0,0){\strut{}32}}%
      \colorrgb{0.00,0.00,0.00}%
      \put(6968,792){\makebox(0,0){\strut{}64}}%
      \colorrgb{0.00,0.00,0.00}%
      \put(8269,792){\makebox(0,0){\strut{}128}}%
      \colorrgb{0.00,0.00,0.00}%
      \put(9570,792){\makebox(0,0){\strut{}256}}%
      \colorrgb{0.00,0.00,0.00}%
      \put(288,4679){\rotatebox{90}{\makebox(0,0){\strut{}$S/N$}}}%
      \colorrgb{0.00,0.00,0.00}%
      \put(6317,252){\makebox(0,0){\strut{}$n_{\mathrm{DTFE}}$}}%
    }%
    \gplgaddtomacro\gplfronttext{%
      \colorrgb{0.00,0.00,0.00}%
      \put(10655,7964){\makebox(0,0)[r]{\strut{} {\tiny$\vmodone$ $\nxyz$= 1  }}}%
      \colorrgb{0.00,0.00,0.00}%
      \put(10655,7604){\makebox(0,0)[r]{\strut{} {\tiny$\vmodone$ $\nxyz$= 8  }}}%
      \colorrgb{0.00,0.00,0.00}%
      \put(10655,7244){\makebox(0,0)[r]{\strut{} {\tiny$\vmodone$ $\nxyz$=64  }}}%
    }%
    \gplbacktext
    \put(0,0){\includegraphics[scale=0.4]{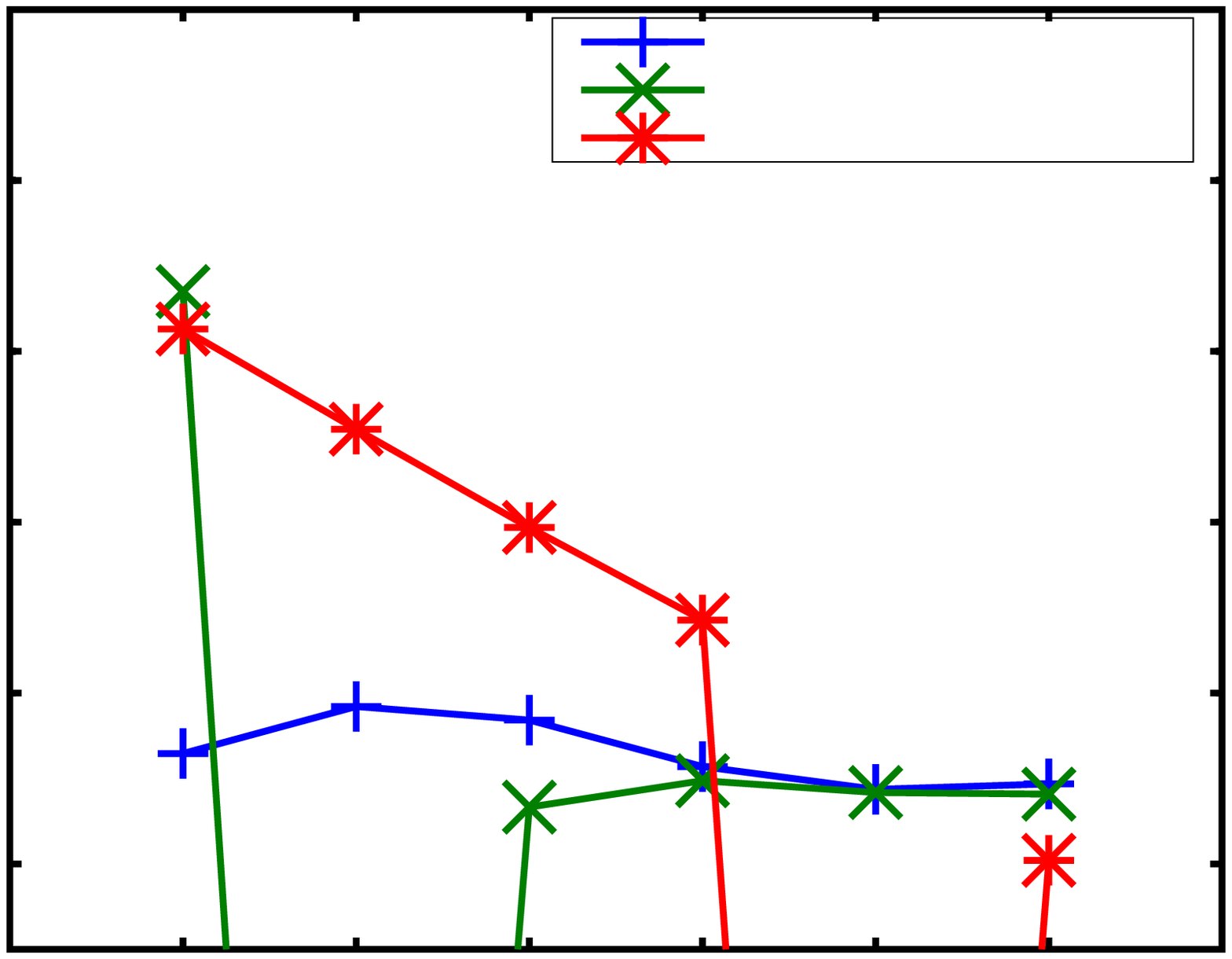}}%
    \gplfronttext
  \end{picture}%
\endgroup

%% file: S1_invI.tex
\begingroup
  \makeatletter
  \providecommand\color[2][]{%
    \GenericError{(gnuplot) \space\space\space\@spaces}{%
      Package color not loaded in conjunction with
      terminal option `colourtext'%
    }{See the gnuplot documentation for explanation.%
    }{Either use 'blacktext' in gnuplot or load the package
      color.sty in LaTeX.}%
    \renewcommand\color[2][]{}%
  }%
  \providecommand\includegraphics[2][]{%
    \GenericError{(gnuplot) \space\space\space\@spaces}{%
      Package graphicx or graphics not loaded%
    }{See the gnuplot documentation for explanation.%
    }{The gnuplot epslatex terminal needs graphicx.sty or graphics.sty.}%
    \renewcommand\includegraphics[2][]{}%
  }%
  \providecommand\rotatebox[2]{#2}%
  \@ifundefined{ifGPcolor}{%
    \newif\ifGPcolor
    \GPcolorfalse
  }{}%
  \@ifundefined{ifGPblacktext}{%
    \newif\ifGPblacktext
    \GPblacktexttrue
  }{}%
  \let\gplgaddtomacro\g@addto@macro
  \gdef\gplbacktext{}%
  \gdef\gplfronttext{}%
  \makeatother
  \ifGPblacktext
    \def\colorrgb#1{}%
    \def\colorgray#1{}%
  \else
    \ifGPcolor
      \def\colorrgb#1{\color[rgb]{#1}}%
      \def\colorgray#1{\color[gray]{#1}}%
      \expandafter\def\csname LTw\endcsname{\color{white}}%
      \expandafter\def\csname LTb\endcsname{\color{black}}%
      \expandafter\def\csname LTa\endcsname{\color{black}}%
      \expandafter\def\csname LT0\endcsname{\color[rgb]{1,0,0}}%
      \expandafter\def\csname LT1\endcsname{\color[rgb]{0,1,0}}%
      \expandafter\def\csname LT2\endcsname{\color[rgb]{0,0,1}}%
      \expandafter\def\csname LT3\endcsname{\color[rgb]{1,0,1}}%
      \expandafter\def\csname LT4\endcsname{\color[rgb]{0,1,1}}%
      \expandafter\def\csname LT5\endcsname{\color[rgb]{1,1,0}}%
      \expandafter\def\csname LT6\endcsname{\color[rgb]{0,0,0}}%
      \expandafter\def\csname LT7\endcsname{\color[rgb]{1,0.3,0}}%
      \expandafter\def\csname LT8\endcsname{\color[rgb]{0.5,0.5,0.5}}%
    \else
      \def\colorrgb#1{\color{black}}%
      \def\colorgray#1{\color[gray]{#1}}%
      \expandafter\def\csname LTw\endcsname{\color{white}}%
      \expandafter\def\csname LTb\endcsname{\color{black}}%
      \expandafter\def\csname LTa\endcsname{\color{black}}%
      \expandafter\def\csname LT0\endcsname{\color{black}}%
      \expandafter\def\csname LT1\endcsname{\color{black}}%
      \expandafter\def\csname LT2\endcsname{\color{black}}%
      \expandafter\def\csname LT3\endcsname{\color{black}}%
      \expandafter\def\csname LT4\endcsname{\color{black}}%
      \expandafter\def\csname LT5\endcsname{\color{black}}%
      \expandafter\def\csname LT6\endcsname{\color{black}}%
      \expandafter\def\csname LT7\endcsname{\color{black}}%
      \expandafter\def\csname LT8\endcsname{\color{black}}%
    \fi
  \fi
  \setlength{\unitlength}{0.0200bp}%
  \begin{picture}(11520.00,8640.00)%
    \gplgaddtomacro\gplbacktext{%
      \colorrgb{0.00,0.00,0.00}%
      \put(1332,2328){\makebox(0,0)[r]{\strut{}$10^{0}$}}%
      \colorrgb{0.00,0.00,0.00}%
      \put(1332,4680){\makebox(0,0)[r]{\strut{}$10^{1}$}}%
      \colorrgb{0.00,0.00,0.00}%
      \put(1332,7031){\makebox(0,0)[r]{\strut{}$10^{2}$}}%
      \colorrgb{0.00,0.00,0.00}%
      \put(2880,792){\makebox(0,0){\strut{}8}}%
      \colorrgb{0.00,0.00,0.00}%
      \put(4212,792){\makebox(0,0){\strut{}16}}%
      \colorrgb{0.00,0.00,0.00}%
      \put(5544,792){\makebox(0,0){\strut{}32}}%
      \colorrgb{0.00,0.00,0.00}%
      \put(6875,792){\makebox(0,0){\strut{}64}}%
      \colorrgb{0.00,0.00,0.00}%
      \put(8207,792){\makebox(0,0){\strut{}128}}%
      \colorrgb{0.00,0.00,0.00}%
      \put(9539,792){\makebox(0,0){\strut{}256}}%
      \colorrgb{0.00,0.00,0.00}%
      \put(288,4679){\rotatebox{90}{\makebox(0,0){\strut{}meas. $\sigma$/model $\sigma$}}}%
      \colorrgb{0.00,0.00,0.00}%
      \put(6209,252){\makebox(0,0){\strut{}$n_{\mathrm{DTFE}}$}}%
    }%
    \gplgaddtomacro\gplfronttext{%
      \colorrgb{0.00,0.00,0.00}%
      \put(10655,7964){\makebox(0,0)[r]{\strut{} {\tiny$\vmodone$ $\nxyz$= 1  }}}%
      \colorrgb{0.00,0.00,0.00}%
      \put(10655,7604){\makebox(0,0)[r]{\strut{} {\tiny$\vmodone$ $\nxyz$= 8  }}}%
      \colorrgb{0.00,0.00,0.00}%
      \put(10655,7244){\makebox(0,0)[r]{\strut{} {\tiny$\vmodone$ $\nxyz$=64  }}}%
      \colorrgb{0.00,0.00,0.00}%
      \put(10655,6884){\makebox(0,0)[r]{\strut{} {\tiny$\vmodtwo$ $\nxyz$= 1  }}}%
      \colorrgb{0.00,0.00,0.00}%
      \put(10655,6524){\makebox(0,0)[r]{\strut{} {\tiny$\vmodtwo$ $\nxyz$= 8  }}}%
      \colorrgb{0.00,0.00,0.00}%
      \put(10655,6164){\makebox(0,0)[r]{\strut{} {\tiny$\vmodtwo$ $\nxyz$=64  }}}%
    }%
    \gplbacktext
    \put(0,0){\includegraphics[scale=0.4]{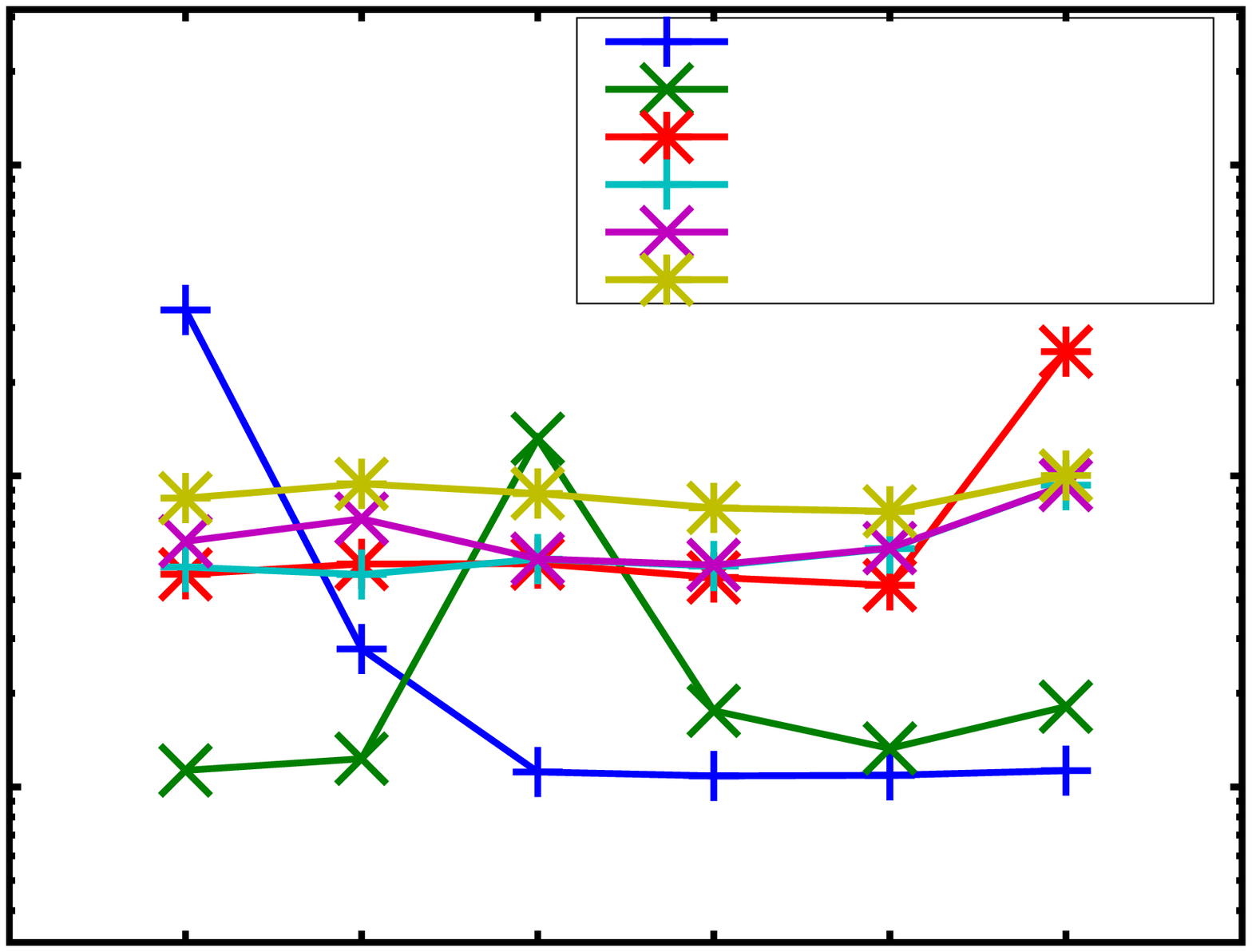}}%
    \gplfronttext
  \end{picture}%
\endgroup

%% file: S1_invII.tex
\begingroup
  \makeatletter
  \providecommand\color[2][]{%
    \GenericError{(gnuplot) \space\space\space\@spaces}{%
      Package color not loaded in conjunction with
      terminal option `colourtext'%
    }{See the gnuplot documentation for explanation.%
    }{Either use 'blacktext' in gnuplot or load the package
      color.sty in LaTeX.}%
    \renewcommand\color[2][]{}%
  }%
  \providecommand\includegraphics[2][]{%
    \GenericError{(gnuplot) \space\space\space\@spaces}{%
      Package graphicx or graphics not loaded%
    }{See the gnuplot documentation for explanation.%
    }{The gnuplot epslatex terminal needs graphicx.sty or graphics.sty.}%
    \renewcommand\includegraphics[2][]{}%
  }%
  \providecommand\rotatebox[2]{#2}%
  \@ifundefined{ifGPcolor}{%
    \newif\ifGPcolor
    \GPcolorfalse
  }{}%
  \@ifundefined{ifGPblacktext}{%
    \newif\ifGPblacktext
    \GPblacktexttrue
  }{}%
  \let\gplgaddtomacro\g@addto@macro
  \gdef\gplbacktext{}%
  \gdef\gplfronttext{}%
  \makeatother
  \ifGPblacktext
    \def\colorrgb#1{}%
    \def\colorgray#1{}%
  \else
    \ifGPcolor
      \def\colorrgb#1{\color[rgb]{#1}}%
      \def\colorgray#1{\color[gray]{#1}}%
      \expandafter\def\csname LTw\endcsname{\color{white}}%
      \expandafter\def\csname LTb\endcsname{\color{black}}%
      \expandafter\def\csname LTa\endcsname{\color{black}}%
      \expandafter\def\csname LT0\endcsname{\color[rgb]{1,0,0}}%
      \expandafter\def\csname LT1\endcsname{\color[rgb]{0,1,0}}%
      \expandafter\def\csname LT2\endcsname{\color[rgb]{0,0,1}}%
      \expandafter\def\csname LT3\endcsname{\color[rgb]{1,0,1}}%
      \expandafter\def\csname LT4\endcsname{\color[rgb]{0,1,1}}%
      \expandafter\def\csname LT5\endcsname{\color[rgb]{1,1,0}}%
      \expandafter\def\csname LT6\endcsname{\color[rgb]{0,0,0}}%
      \expandafter\def\csname LT7\endcsname{\color[rgb]{1,0.3,0}}%
      \expandafter\def\csname LT8\endcsname{\color[rgb]{0.5,0.5,0.5}}%
    \else
      \def\colorrgb#1{\color{black}}%
      \def\colorgray#1{\color[gray]{#1}}%
      \expandafter\def\csname LTw\endcsname{\color{white}}%
      \expandafter\def\csname LTb\endcsname{\color{black}}%
      \expandafter\def\csname LTa\endcsname{\color{black}}%
      \expandafter\def\csname LT0\endcsname{\color{black}}%
      \expandafter\def\csname LT1\endcsname{\color{black}}%
      \expandafter\def\csname LT2\endcsname{\color{black}}%
      \expandafter\def\csname LT3\endcsname{\color{black}}%
      \expandafter\def\csname LT4\endcsname{\color{black}}%
      \expandafter\def\csname LT5\endcsname{\color{black}}%
      \expandafter\def\csname LT6\endcsname{\color{black}}%
      \expandafter\def\csname LT7\endcsname{\color{black}}%
      \expandafter\def\csname LT8\endcsname{\color{black}}%
    \fi
  \fi
  \setlength{\unitlength}{0.0200bp}%
  \begin{picture}(11520.00,8640.00)%
    \gplgaddtomacro\gplbacktext{%
      \colorrgb{0.00,0.00,0.00}%
      \put(1548,1152){\makebox(0,0)[r]{\strut{}$10^{-2}$}}%
      \colorrgb{0.00,0.00,0.00}%
      \put(1548,2435){\makebox(0,0)[r]{\strut{}$10^{-1}$}}%
      \colorrgb{0.00,0.00,0.00}%
      \put(1548,3717){\makebox(0,0)[r]{\strut{}$10^{0}$}}%
      \colorrgb{0.00,0.00,0.00}%
      \put(1548,5000){\makebox(0,0)[r]{\strut{}$10^{1}$}}%
      \colorrgb{0.00,0.00,0.00}%
      \put(1548,6283){\makebox(0,0)[r]{\strut{}$10^{2}$}}%
      \colorrgb{0.00,0.00,0.00}%
      \put(1548,7566){\makebox(0,0)[r]{\strut{}$10^{3}$}}%
      \colorrgb{0.00,0.00,0.00}%
      \put(3065,792){\makebox(0,0){\strut{}8}}%
      \colorrgb{0.00,0.00,0.00}%
      \put(4366,792){\makebox(0,0){\strut{}16}}%
      \colorrgb{0.00,0.00,0.00}%
      \put(5667,792){\makebox(0,0){\strut{}32}}%
      \colorrgb{0.00,0.00,0.00}%
      \put(6968,792){\makebox(0,0){\strut{}64}}%
      \colorrgb{0.00,0.00,0.00}%
      \put(8269,792){\makebox(0,0){\strut{}128}}%
      \colorrgb{0.00,0.00,0.00}%
      \put(9570,792){\makebox(0,0){\strut{}256}}%
      \colorrgb{0.00,0.00,0.00}%
      \put(288,4679){\rotatebox{90}{\makebox(0,0){\strut{}meas. $\sigma$/model $\sigma$}}}%
      \colorrgb{0.00,0.00,0.00}%
      \put(6317,252){\makebox(0,0){\strut{}$n_{\mathrm{DTFE}}$}}%
    }%
    \gplgaddtomacro\gplfronttext{%
      \colorrgb{0.00,0.00,0.00}%
      \put(10655,7964){\makebox(0,0)[r]{\strut{} {\tiny$\vmodone$ $\nxyz$= 1  }}}%
      \colorrgb{0.00,0.00,0.00}%
      \put(10655,7604){\makebox(0,0)[r]{\strut{} {\tiny$\vmodone$ $\nxyz$= 8  }}}%
      \colorrgb{0.00,0.00,0.00}%
      \put(10655,7244){\makebox(0,0)[r]{\strut{} {\tiny$\vmodone$ $\nxyz$=64  }}}%
      \colorrgb{0.00,0.00,0.00}%
      \put(10655,6884){\makebox(0,0)[r]{\strut{} {\tiny$\vmodtwo$ $\nxyz$= 1  }}}%
      \colorrgb{0.00,0.00,0.00}%
      \put(10655,6524){\makebox(0,0)[r]{\strut{} {\tiny$\vmodtwo$ $\nxyz$= 8  }}}%
      \colorrgb{0.00,0.00,0.00}%
      \put(10655,6164){\makebox(0,0)[r]{\strut{} {\tiny$\vmodtwo$ $\nxyz$=64  }}}%
    }%
    \gplbacktext
    \put(0,0){\includegraphics[scale=0.4]{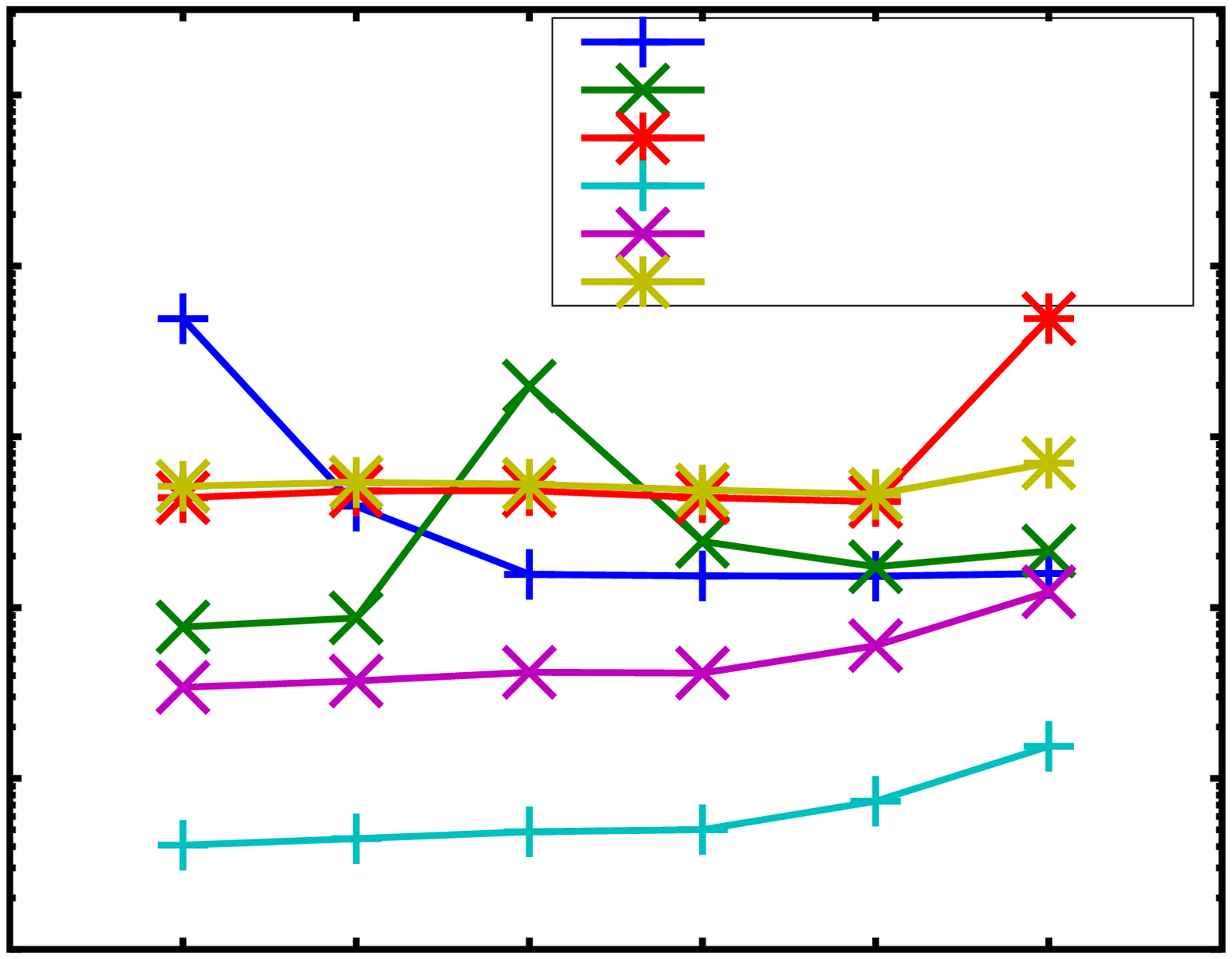}}%
    \gplfronttext
  \end{picture}%
\endgroup

%% file: Sglobal_invI.tex
\begingroup
  \makeatletter
  \providecommand\color[2][]{%
    \GenericError{(gnuplot) \space\space\space\@spaces}{%
      Package color not loaded in conjunction with
      terminal option `colourtext'%
    }{See the gnuplot documentation for explanation.%
    }{Either use 'blacktext' in gnuplot or load the package
      color.sty in LaTeX.}%
    \renewcommand\color[2][]{}%
  }%
  \providecommand\includegraphics[2][]{%
    \GenericError{(gnuplot) \space\space\space\@spaces}{%
      Package graphicx or graphics not loaded%
    }{See the gnuplot documentation for explanation.%
    }{The gnuplot epslatex terminal needs graphicx.sty or graphics.sty.}%
    \renewcommand\includegraphics[2][]{}%
  }%
  \providecommand\rotatebox[2]{#2}%
  \@ifundefined{ifGPcolor}{%
    \newif\ifGPcolor
    \GPcolorfalse
  }{}%
  \@ifundefined{ifGPblacktext}{%
    \newif\ifGPblacktext
    \GPblacktexttrue
  }{}%
  \let\gplgaddtomacro\g@addto@macro
  \gdef\gplbacktext{}%
  \gdef\gplfronttext{}%
  \makeatother
  \ifGPblacktext
    \def\colorrgb#1{}%
    \def\colorgray#1{}%
  \else
    \ifGPcolor
      \def\colorrgb#1{\color[rgb]{#1}}%
      \def\colorgray#1{\color[gray]{#1}}%
      \expandafter\def\csname LTw\endcsname{\color{white}}%
      \expandafter\def\csname LTb\endcsname{\color{black}}%
      \expandafter\def\csname LTa\endcsname{\color{black}}%
      \expandafter\def\csname LT0\endcsname{\color[rgb]{1,0,0}}%
      \expandafter\def\csname LT1\endcsname{\color[rgb]{0,1,0}}%
      \expandafter\def\csname LT2\endcsname{\color[rgb]{0,0,1}}%
      \expandafter\def\csname LT3\endcsname{\color[rgb]{1,0,1}}%
      \expandafter\def\csname LT4\endcsname{\color[rgb]{0,1,1}}%
      \expandafter\def\csname LT5\endcsname{\color[rgb]{1,1,0}}%
      \expandafter\def\csname LT6\endcsname{\color[rgb]{0,0,0}}%
      \expandafter\def\csname LT7\endcsname{\color[rgb]{1,0.3,0}}%
      \expandafter\def\csname LT8\endcsname{\color[rgb]{0.5,0.5,0.5}}%
    \else
      \def\colorrgb#1{\color{black}}%
      \def\colorgray#1{\color[gray]{#1}}%
      \expandafter\def\csname LTw\endcsname{\color{white}}%
      \expandafter\def\csname LTb\endcsname{\color{black}}%
      \expandafter\def\csname LTa\endcsname{\color{black}}%
      \expandafter\def\csname LT0\endcsname{\color{black}}%
      \expandafter\def\csname LT1\endcsname{\color{black}}%
      \expandafter\def\csname LT2\endcsname{\color{black}}%
      \expandafter\def\csname LT3\endcsname{\color{black}}%
      \expandafter\def\csname LT4\endcsname{\color{black}}%
      \expandafter\def\csname LT5\endcsname{\color{black}}%
      \expandafter\def\csname LT6\endcsname{\color{black}}%
      \expandafter\def\csname LT7\endcsname{\color{black}}%
      \expandafter\def\csname LT8\endcsname{\color{black}}%
    \fi
  \fi
  \setlength{\unitlength}{0.0200bp}%
  \begin{picture}(11520.00,8640.00)%
    \gplgaddtomacro\gplbacktext{%
      \colorrgb{0.00,0.00,0.00}%
      \put(1764,1152){\makebox(0,0)[r]{\strut{}$10^{-16}$}}%
      \colorrgb{0.00,0.00,0.00}%
      \put(1764,2563){\makebox(0,0)[r]{\strut{}$10^{-14}$}}%
      \colorrgb{0.00,0.00,0.00}%
      \put(1764,3974){\makebox(0,0)[r]{\strut{}$10^{-12}$}}%
      \colorrgb{0.00,0.00,0.00}%
      \put(1764,5385){\makebox(0,0)[r]{\strut{}$10^{-10}$}}%
      \colorrgb{0.00,0.00,0.00}%
      \put(1764,6796){\makebox(0,0)[r]{\strut{}$10^{-8}$}}%
      \colorrgb{0.00,0.00,0.00}%
      \put(1764,8207){\makebox(0,0)[r]{\strut{}$10^{-6}$}}%
      \colorrgb{0.00,0.00,0.00}%
      \put(3250,792){\makebox(0,0){\strut{}8}}%
      \colorrgb{0.00,0.00,0.00}%
      \put(4520,792){\makebox(0,0){\strut{}16}}%
      \colorrgb{0.00,0.00,0.00}%
      \put(5790,792){\makebox(0,0){\strut{}32}}%
      \colorrgb{0.00,0.00,0.00}%
      \put(7061,792){\makebox(0,0){\strut{}64}}%
      \colorrgb{0.00,0.00,0.00}%
      \put(8331,792){\makebox(0,0){\strut{}128}}%
      \colorrgb{0.00,0.00,0.00}%
      \put(9601,792){\makebox(0,0){\strut{}256}}%
      \colorrgb{0.00,0.00,0.00}%
      \put(288,4679){\rotatebox{90}{\makebox(0,0){\strut{}meas. $\averagebox{\invI}$, model $\averagebox{\invI}$}}}%
      \colorrgb{0.00,0.00,0.00}%
      \put(6425,252){\makebox(0,0){\strut{}$n_{\mathrm{DTFE}}$}}%
    }%
    \gplgaddtomacro\gplfronttext{%
      \colorrgb{0.00,0.00,0.00}%
      \put(7011,3195){\makebox(0,0)[r]{\strut{} {\tiny$\vmodone$ $\nxyz$= 1  }}}%
      \colorrgb{0.00,0.00,0.00}%
      \put(7011,2835){\makebox(0,0)[r]{\strut{} {\tiny$\vmodone$ $\nxyz$= 8  }}}%
      \colorrgb{0.00,0.00,0.00}%
      \put(7011,2475){\makebox(0,0)[r]{\strut{} {\tiny$\vmodone$ $\nxyz$=64  }}}%
      \colorrgb{0.00,0.00,0.00}%
      \put(7011,2115){\makebox(0,0)[r]{\strut{} {\tiny$\vmodtwo$ $\nxyz$= 1  }}}%
      \colorrgb{0.00,0.00,0.00}%
      \put(7011,1755){\makebox(0,0)[r]{\strut{} {\tiny$\vmodtwo$ $\nxyz$= 8  }}}%
      \colorrgb{0.00,0.00,0.00}%
      \put(7011,1395){\makebox(0,0)[r]{\strut{} {\tiny$\vmodtwo$ $\nxyz$=64  }}}%
    }%
    \gplbacktext
    \put(0,0){\includegraphics[scale=0.4]{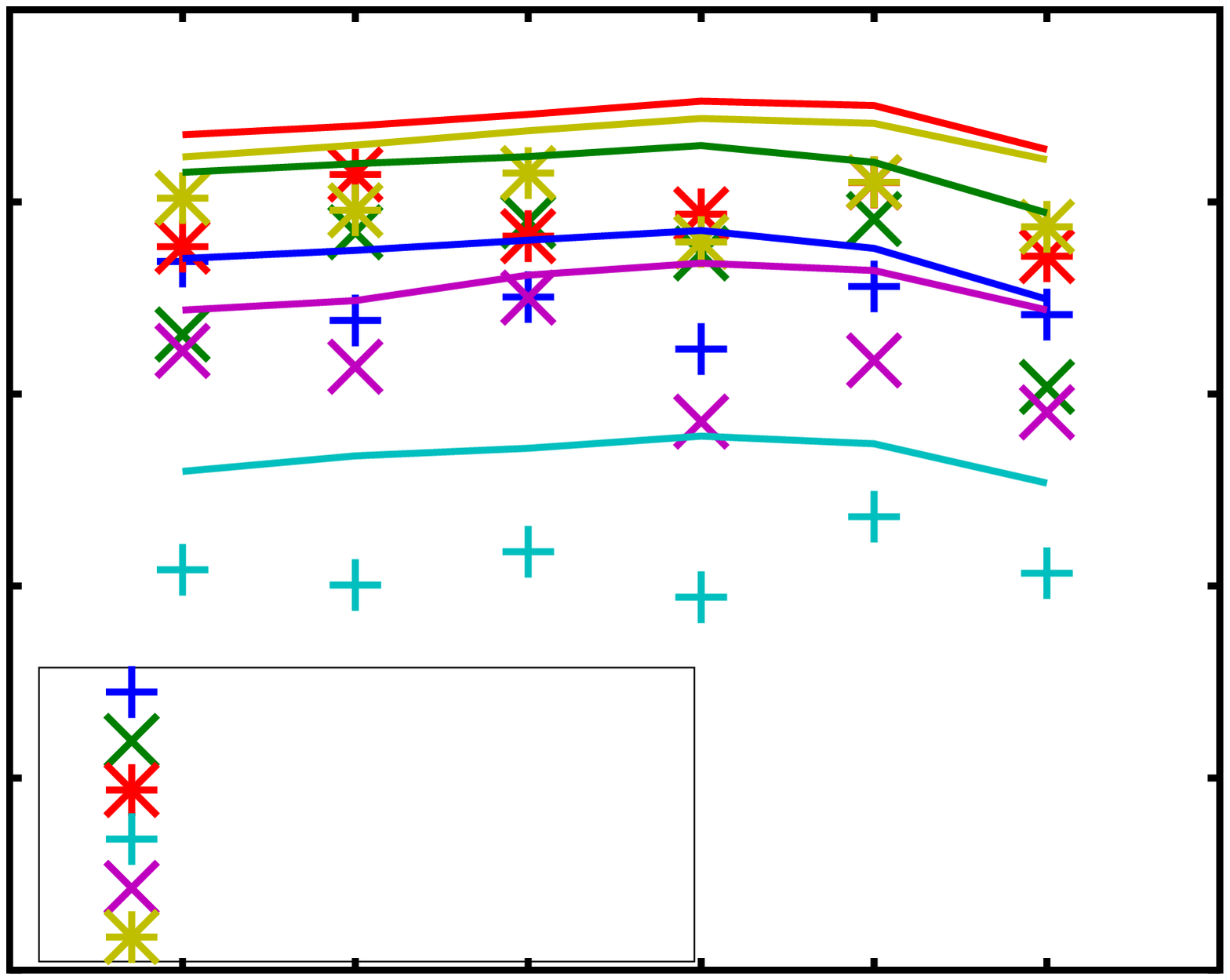}}%
    \gplfronttext
  \end{picture}%
\endgroup

%% file: Sglobal_invII.tex
\begingroup
  \makeatletter
  \providecommand\color[2][]{%
    \GenericError{(gnuplot) \space\space\space\@spaces}{%
      Package color not loaded in conjunction with
      terminal option `colourtext'%
    }{See the gnuplot documentation for explanation.%
    }{Either use 'blacktext' in gnuplot or load the package
      color.sty in LaTeX.}%
    \renewcommand\color[2][]{}%
  }%
  \providecommand\includegraphics[2][]{%
    \GenericError{(gnuplot) \space\space\space\@spaces}{%
      Package graphicx or graphics not loaded%
    }{See the gnuplot documentation for explanation.%
    }{The gnuplot epslatex terminal needs graphicx.sty or graphics.sty.}%
    \renewcommand\includegraphics[2][]{}%
  }%
  \providecommand\rotatebox[2]{#2}%
  \@ifundefined{ifGPcolor}{%
    \newif\ifGPcolor
    \GPcolorfalse
  }{}%
  \@ifundefined{ifGPblacktext}{%
    \newif\ifGPblacktext
    \GPblacktexttrue
  }{}%
  \let\gplgaddtomacro\g@addto@macro
  \gdef\gplbacktext{}%
  \gdef\gplfronttext{}%
  \makeatother
  \ifGPblacktext
    \def\colorrgb#1{}%
    \def\colorgray#1{}%
  \else
    \ifGPcolor
      \def\colorrgb#1{\color[rgb]{#1}}%
      \def\colorgray#1{\color[gray]{#1}}%
      \expandafter\def\csname LTw\endcsname{\color{white}}%
      \expandafter\def\csname LTb\endcsname{\color{black}}%
      \expandafter\def\csname LTa\endcsname{\color{black}}%
      \expandafter\def\csname LT0\endcsname{\color[rgb]{1,0,0}}%
      \expandafter\def\csname LT1\endcsname{\color[rgb]{0,1,0}}%
      \expandafter\def\csname LT2\endcsname{\color[rgb]{0,0,1}}%
      \expandafter\def\csname LT3\endcsname{\color[rgb]{1,0,1}}%
      \expandafter\def\csname LT4\endcsname{\color[rgb]{0,1,1}}%
      \expandafter\def\csname LT5\endcsname{\color[rgb]{1,1,0}}%
      \expandafter\def\csname LT6\endcsname{\color[rgb]{0,0,0}}%
      \expandafter\def\csname LT7\endcsname{\color[rgb]{1,0.3,0}}%
      \expandafter\def\csname LT8\endcsname{\color[rgb]{0.5,0.5,0.5}}%
    \else
      \def\colorrgb#1{\color{black}}%
      \def\colorgray#1{\color[gray]{#1}}%
      \expandafter\def\csname LTw\endcsname{\color{white}}%
      \expandafter\def\csname LTb\endcsname{\color{black}}%
      \expandafter\def\csname LTa\endcsname{\color{black}}%
      \expandafter\def\csname LT0\endcsname{\color{black}}%
      \expandafter\def\csname LT1\endcsname{\color{black}}%
      \expandafter\def\csname LT2\endcsname{\color{black}}%
      \expandafter\def\csname LT3\endcsname{\color{black}}%
      \expandafter\def\csname LT4\endcsname{\color{black}}%
      \expandafter\def\csname LT5\endcsname{\color{black}}%
      \expandafter\def\csname LT6\endcsname{\color{black}}%
      \expandafter\def\csname LT7\endcsname{\color{black}}%
      \expandafter\def\csname LT8\endcsname{\color{black}}%
    \fi
  \fi
  \setlength{\unitlength}{0.0200bp}%
  \begin{picture}(11520.00,8640.00)%
    \gplgaddtomacro\gplbacktext{%
      \colorrgb{0.00,0.00,0.00}%
      \put(1764,1152){\makebox(0,0)[r]{\strut{}$10^{-18}$}}%
      \colorrgb{0.00,0.00,0.00}%
      \put(1764,2160){\makebox(0,0)[r]{\strut{}$10^{-16}$}}%
      \colorrgb{0.00,0.00,0.00}%
      \put(1764,3168){\makebox(0,0)[r]{\strut{}$10^{-14}$}}%
      \colorrgb{0.00,0.00,0.00}%
      \put(1764,4176){\makebox(0,0)[r]{\strut{}$10^{-12}$}}%
      \colorrgb{0.00,0.00,0.00}%
      \put(1764,5183){\makebox(0,0)[r]{\strut{}$10^{-10}$}}%
      \colorrgb{0.00,0.00,0.00}%
      \put(1764,6191){\makebox(0,0)[r]{\strut{}$10^{-8}$}}%
      \colorrgb{0.00,0.00,0.00}%
      \put(1764,7199){\makebox(0,0)[r]{\strut{}$10^{-6}$}}%
      \colorrgb{0.00,0.00,0.00}%
      \put(1764,8207){\makebox(0,0)[r]{\strut{}$10^{-4}$}}%
      \colorrgb{0.00,0.00,0.00}%
      \put(3250,792){\makebox(0,0){\strut{}8}}%
      \colorrgb{0.00,0.00,0.00}%
      \put(4520,792){\makebox(0,0){\strut{}16}}%
      \colorrgb{0.00,0.00,0.00}%
      \put(5790,792){\makebox(0,0){\strut{}32}}%
      \colorrgb{0.00,0.00,0.00}%
      \put(7061,792){\makebox(0,0){\strut{}64}}%
      \colorrgb{0.00,0.00,0.00}%
      \put(8331,792){\makebox(0,0){\strut{}128}}%
      \colorrgb{0.00,0.00,0.00}%
      \put(9601,792){\makebox(0,0){\strut{}256}}%
      \colorrgb{0.00,0.00,0.00}%
      \put(288,4679){\rotatebox{90}{\makebox(0,0){\strut{}meas. $\averagebox{\invII}$, model $\averagebox{\invII}$}}}%
      \colorrgb{0.00,0.00,0.00}%
      \put(6425,252){\makebox(0,0){\strut{}$n_{\mathrm{DTFE}}$}}%
    }%
    \gplgaddtomacro\gplfronttext{%
      \colorrgb{0.00,0.00,0.00}%
      \put(7011,7964){\makebox(0,0)[r]{\strut{} {\tiny$\vmodone$ $\nxyz$= 1  }}}%
      \colorrgb{0.00,0.00,0.00}%
      \put(7011,7604){\makebox(0,0)[r]{\strut{} {\tiny$\vmodone$ $\nxyz$= 8  }}}%
      \colorrgb{0.00,0.00,0.00}%
      \put(7011,7244){\makebox(0,0)[r]{\strut{} {\tiny$\vmodone$ $\nxyz$=64  }}}%
      \colorrgb{0.00,0.00,0.00}%
      \put(7011,6884){\makebox(0,0)[r]{\strut{} {\tiny$\vmodtwo$ $\nxyz$= 1  }}}%
      \colorrgb{0.00,0.00,0.00}%
      \put(7011,6524){\makebox(0,0)[r]{\strut{} {\tiny$\vmodtwo$ $\nxyz$= 8  }}}%
      \colorrgb{0.00,0.00,0.00}%
      \put(7011,6164){\makebox(0,0)[r]{\strut{} {\tiny$\vmodtwo$ $\nxyz$=64  }}}%
    }%
    \gplbacktext
    \put(0,0){\includegraphics[scale=0.4]{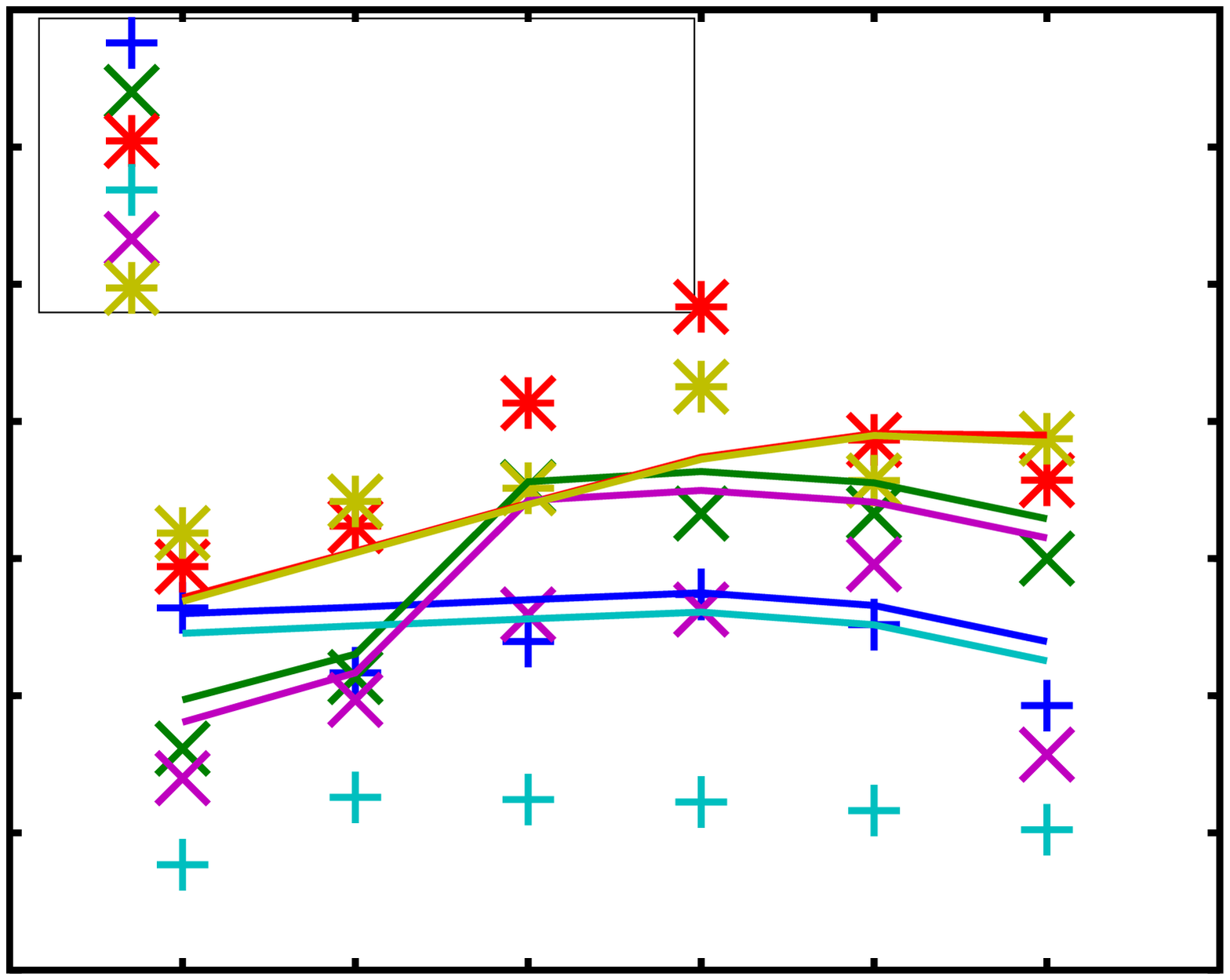}}%
    \gplfronttext
  \end{picture}%
\endgroup

%% file: aRZA_nsc8_ndtfe16_qdisable0.tex
\begingroup
  \makeatletter
  \providecommand\color[2][]{%
    \GenericError{(gnuplot) \space\space\space\@spaces}{%
      Package color not loaded in conjunction with
      terminal option `colourtext'%
    }{See the gnuplot documentation for explanation.%
    }{Either use 'blacktext' in gnuplot or load the package
      color.sty in LaTeX.}%
    \renewcommand\color[2][]{}%
  }%
  \providecommand\includegraphics[2][]{%
    \GenericError{(gnuplot) \space\space\space\@spaces}{%
      Package graphicx or graphics not loaded%
    }{See the gnuplot documentation for explanation.%
    }{The gnuplot epslatex terminal needs graphicx.sty or graphics.sty.}%
    \renewcommand\includegraphics[2][]{}%
  }%
  \providecommand\rotatebox[2]{#2}%
  \@ifundefined{ifGPcolor}{%
    \newif\ifGPcolor
    \GPcolorfalse
  }{}%
  \@ifundefined{ifGPblacktext}{%
    \newif\ifGPblacktext
    \GPblacktexttrue
  }{}%
  \let\gplgaddtomacro\g@addto@macro
  \gdef\gplbacktext{}%
  \gdef\gplfronttext{}%
  \makeatother
  \ifGPblacktext
    \def\colorrgb#1{}%
    \def\colorgray#1{}%
  \else
    \ifGPcolor
      \def\colorrgb#1{\color[rgb]{#1}}%
      \def\colorgray#1{\color[gray]{#1}}%
      \expandafter\def\csname LTw\endcsname{\color{white}}%
      \expandafter\def\csname LTb\endcsname{\color{black}}%
      \expandafter\def\csname LTa\endcsname{\color{black}}%
      \expandafter\def\csname LT0\endcsname{\color[rgb]{1,0,0}}%
      \expandafter\def\csname LT1\endcsname{\color[rgb]{0,1,0}}%
      \expandafter\def\csname LT2\endcsname{\color[rgb]{0,0,1}}%
      \expandafter\def\csname LT3\endcsname{\color[rgb]{1,0,1}}%
      \expandafter\def\csname LT4\endcsname{\color[rgb]{0,1,1}}%
      \expandafter\def\csname LT5\endcsname{\color[rgb]{1,1,0}}%
      \expandafter\def\csname LT6\endcsname{\color[rgb]{0,0,0}}%
      \expandafter\def\csname LT7\endcsname{\color[rgb]{1,0.3,0}}%
      \expandafter\def\csname LT8\endcsname{\color[rgb]{0.5,0.5,0.5}}%
    \else
      \def\colorrgb#1{\color{black}}%
      \def\colorgray#1{\color[gray]{#1}}%
      \expandafter\def\csname LTw\endcsname{\color{white}}%
      \expandafter\def\csname LTb\endcsname{\color{black}}%
      \expandafter\def\csname LTa\endcsname{\color{black}}%
      \expandafter\def\csname LT0\endcsname{\color{black}}%
      \expandafter\def\csname LT1\endcsname{\color{black}}%
      \expandafter\def\csname LT2\endcsname{\color{black}}%
      \expandafter\def\csname LT3\endcsname{\color{black}}%
      \expandafter\def\csname LT4\endcsname{\color{black}}%
      \expandafter\def\csname LT5\endcsname{\color{black}}%
      \expandafter\def\csname LT6\endcsname{\color{black}}%
      \expandafter\def\csname LT7\endcsname{\color{black}}%
      \expandafter\def\csname LT8\endcsname{\color{black}}%
    \fi
  \fi
  \setlength{\unitlength}{0.0200bp}%
  \begin{picture}(11520.00,8640.00)%
    \gplgaddtomacro\gplbacktext{%
      \colorrgb{0.00,0.00,0.00}%
      \put(1406,1216){\makebox(0,0)[r]{\strut{}0}}%
      \colorrgb{0.00,0.00,0.00}%
      \put(1406,1990){\makebox(0,0)[r]{\strut{}0.2}}%
      \colorrgb{0.00,0.00,0.00}%
      \put(1406,2764){\makebox(0,0)[r]{\strut{}0.4}}%
      \colorrgb{0.00,0.00,0.00}%
      \put(1406,3538){\makebox(0,0)[r]{\strut{}0.6}}%
      \colorrgb{0.00,0.00,0.00}%
      \put(1406,4312){\makebox(0,0)[r]{\strut{}0.8}}%
      \colorrgb{0.00,0.00,0.00}%
      \put(1406,5087){\makebox(0,0)[r]{\strut{}1}}%
      \colorrgb{0.00,0.00,0.00}%
      \put(1406,5861){\makebox(0,0)[r]{\strut{}1.2}}%
      \colorrgb{0.00,0.00,0.00}%
      \put(1406,6635){\makebox(0,0)[r]{\strut{}1.4}}%
      \colorrgb{0.00,0.00,0.00}%
      \put(1406,7409){\makebox(0,0)[r]{\strut{}1.6}}%
      \colorrgb{0.00,0.00,0.00}%
      \put(1406,8183){\makebox(0,0)[r]{\strut{}1.8}}%
      \colorrgb{0.00,0.00,0.00}%
      \put(1634,836){\makebox(0,0){\strut{}0}}%
      \colorrgb{0.00,0.00,0.00}%
      \put(2686,836){\makebox(0,0){\strut{}2}}%
      \colorrgb{0.00,0.00,0.00}%
      \put(3737,836){\makebox(0,0){\strut{}4}}%
      \colorrgb{0.00,0.00,0.00}%
      \put(4789,836){\makebox(0,0){\strut{}6}}%
      \colorrgb{0.00,0.00,0.00}%
      \put(5840,836){\makebox(0,0){\strut{}8}}%
      \colorrgb{0.00,0.00,0.00}%
      \put(6892,836){\makebox(0,0){\strut{}10}}%
      \colorrgb{0.00,0.00,0.00}%
      \put(7943,836){\makebox(0,0){\strut{}12}}%
      \colorrgb{0.00,0.00,0.00}%
      \put(8995,836){\makebox(0,0){\strut{}14}}%
      \colorrgb{0.00,0.00,0.00}%
      \put(10046,836){\makebox(0,0){\strut{}16}}%
      \colorrgb{0.00,0.00,0.00}%
      \put(304,4699){\rotatebox{90}{\makebox(0,0){\strut{}scale factor $a$}}}%
      \colorrgb{0.00,0.00,0.00}%
      \put(6234,266){\makebox(0,0){\strut{}$t$ (Gyr)}}%
    }%
    \gplgaddtomacro\gplfronttext{%
      \colorrgb{0.00,0.00,0.00}%
      \put(6029,7930){\makebox(0,0)[r]{\footnotesize EdS   }}%
      \colorrgb{0.00,0.00,0.00}%
      \put(6029,7550){\makebox(0,0)[r]{\footnotesize 0.25 Mpc/$h$ }}%
      \colorrgb{0.00,0.00,0.00}%
      \put(6029,7170){\makebox(0,0)[r]{\footnotesize 0.5 Mpc/$h$ }}%
      \colorrgb{0.00,0.00,0.00}%
      \put(6029,6790){\makebox(0,0)[r]{\footnotesize 1.0 Mpc/$h$ }}%
      \colorrgb{0.00,0.00,0.00}%
      \put(6029,6410){\makebox(0,0)[r]{\footnotesize 2.0 Mpc/$h$ }}%
      \colorrgb{0.00,0.00,0.00}%
      \put(6029,6030){\makebox(0,0)[r]{\footnotesize $\LD$=4.0 Mpc/$h$ }}%
      \colorrgb{0.00,0.00,0.00}%
      \put(6029,5650){\makebox(0,0)[r]{\footnotesize  {\LCDM}   }}%
    }%
    \gplbacktext
    \put(0,0){\includegraphics[scale=0.4]{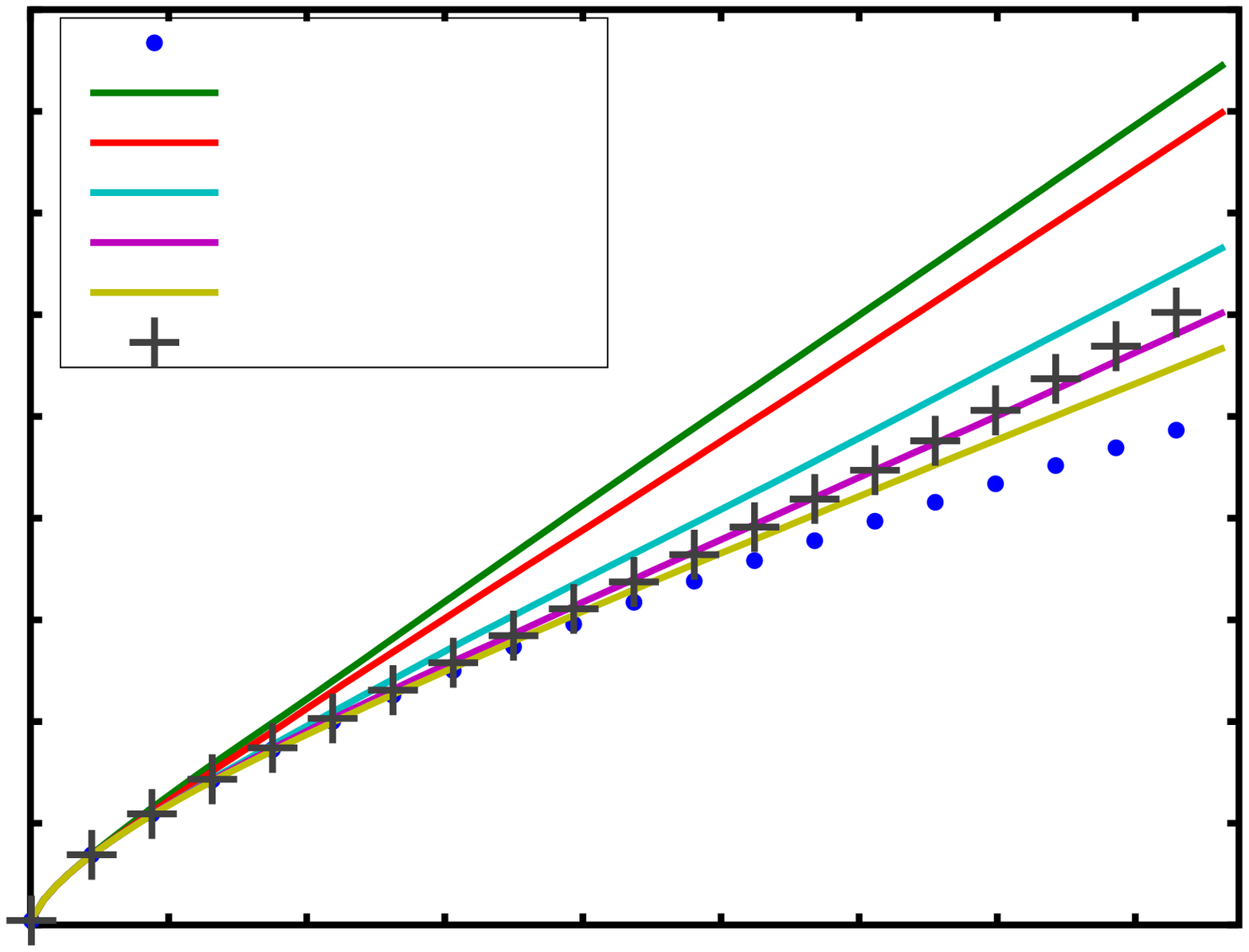}}%
    \gplfronttext
  \end{picture}%
\endgroup

%% file: aRZA_nsc8_ndtfe32_qdisable0.tex
\begingroup
  \makeatletter
  \providecommand\color[2][]{%
    \GenericError{(gnuplot) \space\space\space\@spaces}{%
      Package color not loaded in conjunction with
      terminal option `colourtext'%
    }{See the gnuplot documentation for explanation.%
    }{Either use 'blacktext' in gnuplot or load the package
      color.sty in LaTeX.}%
    \renewcommand\color[2][]{}%
  }%
  \providecommand\includegraphics[2][]{%
    \GenericError{(gnuplot) \space\space\space\@spaces}{%
      Package graphicx or graphics not loaded%
    }{See the gnuplot documentation for explanation.%
    }{The gnuplot epslatex terminal needs graphicx.sty or graphics.sty.}%
    \renewcommand\includegraphics[2][]{}%
  }%
  \providecommand\rotatebox[2]{#2}%
  \@ifundefined{ifGPcolor}{%
    \newif\ifGPcolor
    \GPcolorfalse
  }{}%
  \@ifundefined{ifGPblacktext}{%
    \newif\ifGPblacktext
    \GPblacktexttrue
  }{}%
  \let\gplgaddtomacro\g@addto@macro
  \gdef\gplbacktext{}%
  \gdef\gplfronttext{}%
  \makeatother
  \ifGPblacktext
    \def\colorrgb#1{}%
    \def\colorgray#1{}%
  \else
    \ifGPcolor
      \def\colorrgb#1{\color[rgb]{#1}}%
      \def\colorgray#1{\color[gray]{#1}}%
      \expandafter\def\csname LTw\endcsname{\color{white}}%
      \expandafter\def\csname LTb\endcsname{\color{black}}%
      \expandafter\def\csname LTa\endcsname{\color{black}}%
      \expandafter\def\csname LT0\endcsname{\color[rgb]{1,0,0}}%
      \expandafter\def\csname LT1\endcsname{\color[rgb]{0,1,0}}%
      \expandafter\def\csname LT2\endcsname{\color[rgb]{0,0,1}}%
      \expandafter\def\csname LT3\endcsname{\color[rgb]{1,0,1}}%
      \expandafter\def\csname LT4\endcsname{\color[rgb]{0,1,1}}%
      \expandafter\def\csname LT5\endcsname{\color[rgb]{1,1,0}}%
      \expandafter\def\csname LT6\endcsname{\color[rgb]{0,0,0}}%
      \expandafter\def\csname LT7\endcsname{\color[rgb]{1,0.3,0}}%
      \expandafter\def\csname LT8\endcsname{\color[rgb]{0.5,0.5,0.5}}%
    \else
      \def\colorrgb#1{\color{black}}%
      \def\colorgray#1{\color[gray]{#1}}%
      \expandafter\def\csname LTw\endcsname{\color{white}}%
      \expandafter\def\csname LTb\endcsname{\color{black}}%
      \expandafter\def\csname LTa\endcsname{\color{black}}%
      \expandafter\def\csname LT0\endcsname{\color{black}}%
      \expandafter\def\csname LT1\endcsname{\color{black}}%
      \expandafter\def\csname LT2\endcsname{\color{black}}%
      \expandafter\def\csname LT3\endcsname{\color{black}}%
      \expandafter\def\csname LT4\endcsname{\color{black}}%
      \expandafter\def\csname LT5\endcsname{\color{black}}%
      \expandafter\def\csname LT6\endcsname{\color{black}}%
      \expandafter\def\csname LT7\endcsname{\color{black}}%
      \expandafter\def\csname LT8\endcsname{\color{black}}%
    \fi
  \fi
  \setlength{\unitlength}{0.0200bp}%
  \begin{picture}(11520.00,8640.00)%
    \gplgaddtomacro\gplbacktext{%
      \colorrgb{0.00,0.00,0.00}%
      \put(1406,1216){\makebox(0,0)[r]{\strut{}0}}%
      \colorrgb{0.00,0.00,0.00}%
      \put(1406,1990){\makebox(0,0)[r]{\strut{}0.2}}%
      \colorrgb{0.00,0.00,0.00}%
      \put(1406,2764){\makebox(0,0)[r]{\strut{}0.4}}%
      \colorrgb{0.00,0.00,0.00}%
      \put(1406,3538){\makebox(0,0)[r]{\strut{}0.6}}%
      \colorrgb{0.00,0.00,0.00}%
      \put(1406,4312){\makebox(0,0)[r]{\strut{}0.8}}%
      \colorrgb{0.00,0.00,0.00}%
      \put(1406,5087){\makebox(0,0)[r]{\strut{}1}}%
      \colorrgb{0.00,0.00,0.00}%
      \put(1406,5861){\makebox(0,0)[r]{\strut{}1.2}}%
      \colorrgb{0.00,0.00,0.00}%
      \put(1406,6635){\makebox(0,0)[r]{\strut{}1.4}}%
      \colorrgb{0.00,0.00,0.00}%
      \put(1406,7409){\makebox(0,0)[r]{\strut{}1.6}}%
      \colorrgb{0.00,0.00,0.00}%
      \put(1406,8183){\makebox(0,0)[r]{\strut{}1.8}}%
      \colorrgb{0.00,0.00,0.00}%
      \put(1634,836){\makebox(0,0){\strut{}0}}%
      \colorrgb{0.00,0.00,0.00}%
      \put(2686,836){\makebox(0,0){\strut{}2}}%
      \colorrgb{0.00,0.00,0.00}%
      \put(3737,836){\makebox(0,0){\strut{}4}}%
      \colorrgb{0.00,0.00,0.00}%
      \put(4789,836){\makebox(0,0){\strut{}6}}%
      \colorrgb{0.00,0.00,0.00}%
      \put(5840,836){\makebox(0,0){\strut{}8}}%
      \colorrgb{0.00,0.00,0.00}%
      \put(6892,836){\makebox(0,0){\strut{}10}}%
      \colorrgb{0.00,0.00,0.00}%
      \put(7943,836){\makebox(0,0){\strut{}12}}%
      \colorrgb{0.00,0.00,0.00}%
      \put(8995,836){\makebox(0,0){\strut{}14}}%
      \colorrgb{0.00,0.00,0.00}%
      \put(10046,836){\makebox(0,0){\strut{}16}}%
      \colorrgb{0.00,0.00,0.00}%
      \put(304,4699){\rotatebox{90}{\makebox(0,0){\strut{}scale factor $a$}}}%
      \colorrgb{0.00,0.00,0.00}%
      \put(6234,266){\makebox(0,0){\strut{}$t$ (Gyr)}}%
    }%
    \gplgaddtomacro\gplfronttext{%
      \colorrgb{0.00,0.00,0.00}%
      \put(6029,7930){\makebox(0,0)[r]{\footnotesize EdS   }}%
      \colorrgb{0.00,0.00,0.00}%
      \put(6029,7550){\makebox(0,0)[r]{\footnotesize 0.5 Mpc/$h$ }}%
      \colorrgb{0.00,0.00,0.00}%
      \put(6029,7170){\makebox(0,0)[r]{\footnotesize 1.0 Mpc/$h$ }}%
      \colorrgb{0.00,0.00,0.00}%
      \put(6029,6790){\makebox(0,0)[r]{\footnotesize 2.0 Mpc/$h$ }}%
      \colorrgb{0.00,0.00,0.00}%
      \put(6029,6410){\makebox(0,0)[r]{\footnotesize 4.0 Mpc/$h$ }}%
      \colorrgb{0.00,0.00,0.00}%
      \put(6029,6030){\makebox(0,0)[r]{\footnotesize 8.0 Mpc/$h$ }}%
      \colorrgb{0.00,0.00,0.00}%
      \put(6029,5650){\makebox(0,0)[r]{\footnotesize  {\LCDM}   }}%
    }%
    \gplbacktext
    \put(0,0){\includegraphics[scale=0.4]{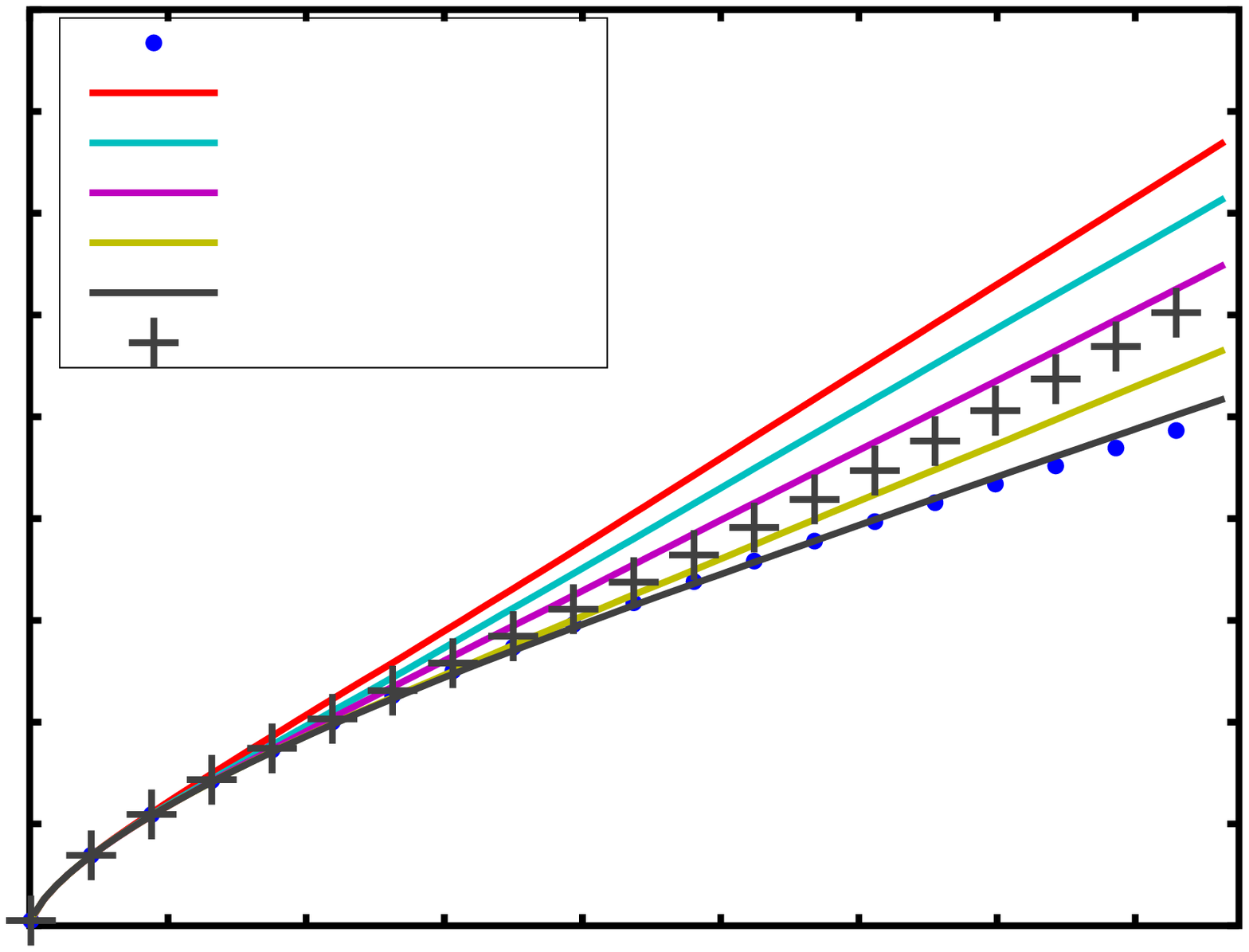}}%
    \gplfronttext
  \end{picture}%
\endgroup

%% file: aRZA_nsc8_ndtfe64_qdisable0.tex
\begingroup
  \makeatletter
  \providecommand\color[2][]{%
    \GenericError{(gnuplot) \space\space\space\@spaces}{%
      Package color not loaded in conjunction with
      terminal option `colourtext'%
    }{See the gnuplot documentation for explanation.%
    }{Either use 'blacktext' in gnuplot or load the package
      color.sty in LaTeX.}%
    \renewcommand\color[2][]{}%
  }%
  \providecommand\includegraphics[2][]{%
    \GenericError{(gnuplot) \space\space\space\@spaces}{%
      Package graphicx or graphics not loaded%
    }{See the gnuplot documentation for explanation.%
    }{The gnuplot epslatex terminal needs graphicx.sty or graphics.sty.}%
    \renewcommand\includegraphics[2][]{}%
  }%
  \providecommand\rotatebox[2]{#2}%
  \@ifundefined{ifGPcolor}{%
    \newif\ifGPcolor
    \GPcolorfalse
  }{}%
  \@ifundefined{ifGPblacktext}{%
    \newif\ifGPblacktext
    \GPblacktexttrue
  }{}%
  \let\gplgaddtomacro\g@addto@macro
  \gdef\gplbacktext{}%
  \gdef\gplfronttext{}%
  \makeatother
  \ifGPblacktext
    \def\colorrgb#1{}%
    \def\colorgray#1{}%
  \else
    \ifGPcolor
      \def\colorrgb#1{\color[rgb]{#1}}%
      \def\colorgray#1{\color[gray]{#1}}%
      \expandafter\def\csname LTw\endcsname{\color{white}}%
      \expandafter\def\csname LTb\endcsname{\color{black}}%
      \expandafter\def\csname LTa\endcsname{\color{black}}%
      \expandafter\def\csname LT0\endcsname{\color[rgb]{1,0,0}}%
      \expandafter\def\csname LT1\endcsname{\color[rgb]{0,1,0}}%
      \expandafter\def\csname LT2\endcsname{\color[rgb]{0,0,1}}%
      \expandafter\def\csname LT3\endcsname{\color[rgb]{1,0,1}}%
      \expandafter\def\csname LT4\endcsname{\color[rgb]{0,1,1}}%
      \expandafter\def\csname LT5\endcsname{\color[rgb]{1,1,0}}%
      \expandafter\def\csname LT6\endcsname{\color[rgb]{0,0,0}}%
      \expandafter\def\csname LT7\endcsname{\color[rgb]{1,0.3,0}}%
      \expandafter\def\csname LT8\endcsname{\color[rgb]{0.5,0.5,0.5}}%
    \else
      \def\colorrgb#1{\color{black}}%
      \def\colorgray#1{\color[gray]{#1}}%
      \expandafter\def\csname LTw\endcsname{\color{white}}%
      \expandafter\def\csname LTb\endcsname{\color{black}}%
      \expandafter\def\csname LTa\endcsname{\color{black}}%
      \expandafter\def\csname LT0\endcsname{\color{black}}%
      \expandafter\def\csname LT1\endcsname{\color{black}}%
      \expandafter\def\csname LT2\endcsname{\color{black}}%
      \expandafter\def\csname LT3\endcsname{\color{black}}%
      \expandafter\def\csname LT4\endcsname{\color{black}}%
      \expandafter\def\csname LT5\endcsname{\color{black}}%
      \expandafter\def\csname LT6\endcsname{\color{black}}%
      \expandafter\def\csname LT7\endcsname{\color{black}}%
      \expandafter\def\csname LT8\endcsname{\color{black}}%
    \fi
  \fi
  \setlength{\unitlength}{0.0200bp}%
  \begin{picture}(11520.00,8640.00)%
    \gplgaddtomacro\gplbacktext{%
      \colorrgb{0.00,0.00,0.00}%
      \put(1406,1216){\makebox(0,0)[r]{\strut{}0}}%
      \colorrgb{0.00,0.00,0.00}%
      \put(1406,1990){\makebox(0,0)[r]{\strut{}0.2}}%
      \colorrgb{0.00,0.00,0.00}%
      \put(1406,2764){\makebox(0,0)[r]{\strut{}0.4}}%
      \colorrgb{0.00,0.00,0.00}%
      \put(1406,3538){\makebox(0,0)[r]{\strut{}0.6}}%
      \colorrgb{0.00,0.00,0.00}%
      \put(1406,4312){\makebox(0,0)[r]{\strut{}0.8}}%
      \colorrgb{0.00,0.00,0.00}%
      \put(1406,5087){\makebox(0,0)[r]{\strut{}1}}%
      \colorrgb{0.00,0.00,0.00}%
      \put(1406,5861){\makebox(0,0)[r]{\strut{}1.2}}%
      \colorrgb{0.00,0.00,0.00}%
      \put(1406,6635){\makebox(0,0)[r]{\strut{}1.4}}%
      \colorrgb{0.00,0.00,0.00}%
      \put(1406,7409){\makebox(0,0)[r]{\strut{}1.6}}%
      \colorrgb{0.00,0.00,0.00}%
      \put(1406,8183){\makebox(0,0)[r]{\strut{}1.8}}%
      \colorrgb{0.00,0.00,0.00}%
      \put(1634,836){\makebox(0,0){\strut{}0}}%
      \colorrgb{0.00,0.00,0.00}%
      \put(2686,836){\makebox(0,0){\strut{}2}}%
      \colorrgb{0.00,0.00,0.00}%
      \put(3737,836){\makebox(0,0){\strut{}4}}%
      \colorrgb{0.00,0.00,0.00}%
      \put(4789,836){\makebox(0,0){\strut{}6}}%
      \colorrgb{0.00,0.00,0.00}%
      \put(5840,836){\makebox(0,0){\strut{}8}}%
      \colorrgb{0.00,0.00,0.00}%
      \put(6892,836){\makebox(0,0){\strut{}10}}%
      \colorrgb{0.00,0.00,0.00}%
      \put(7943,836){\makebox(0,0){\strut{}12}}%
      \colorrgb{0.00,0.00,0.00}%
      \put(8995,836){\makebox(0,0){\strut{}14}}%
      \colorrgb{0.00,0.00,0.00}%
      \put(10046,836){\makebox(0,0){\strut{}16}}%
      \colorrgb{0.00,0.00,0.00}%
      \put(304,4699){\rotatebox{90}{\makebox(0,0){\strut{}scale factor $a$}}}%
      \colorrgb{0.00,0.00,0.00}%
      \put(6234,266){\makebox(0,0){\strut{}$t$ (Gyr)}}%
    }%
    \gplgaddtomacro\gplfronttext{%
      \colorrgb{0.00,0.00,0.00}%
      \put(6257,7930){\makebox(0,0)[r]{\footnotesize EdS   }}%
      \colorrgb{0.00,0.00,0.00}%
      \put(6257,7550){\makebox(0,0)[r]{\footnotesize 1.0 Mpc/$h$ }}%
      \colorrgb{0.00,0.00,0.00}%
      \put(6257,7170){\makebox(0,0)[r]{\footnotesize 2.0 Mpc/$h$ }}%
      \colorrgb{0.00,0.00,0.00}%
      \put(6257,6790){\makebox(0,0)[r]{\footnotesize 4.0 Mpc/$h$ }}%
      \colorrgb{0.00,0.00,0.00}%
      \put(6257,6410){\makebox(0,0)[r]{\footnotesize 8.0 Mpc/$h$ }}%
      \colorrgb{0.00,0.00,0.00}%
      \put(6257,6030){\makebox(0,0)[r]{\footnotesize 16.0 Mpc/$h$ }}%
      \colorrgb{0.00,0.00,0.00}%
      \put(6257,5650){\makebox(0,0)[r]{\footnotesize  {\LCDM}   }}%
    }%
    \gplbacktext
    \put(0,0){\includegraphics[scale=0.4]{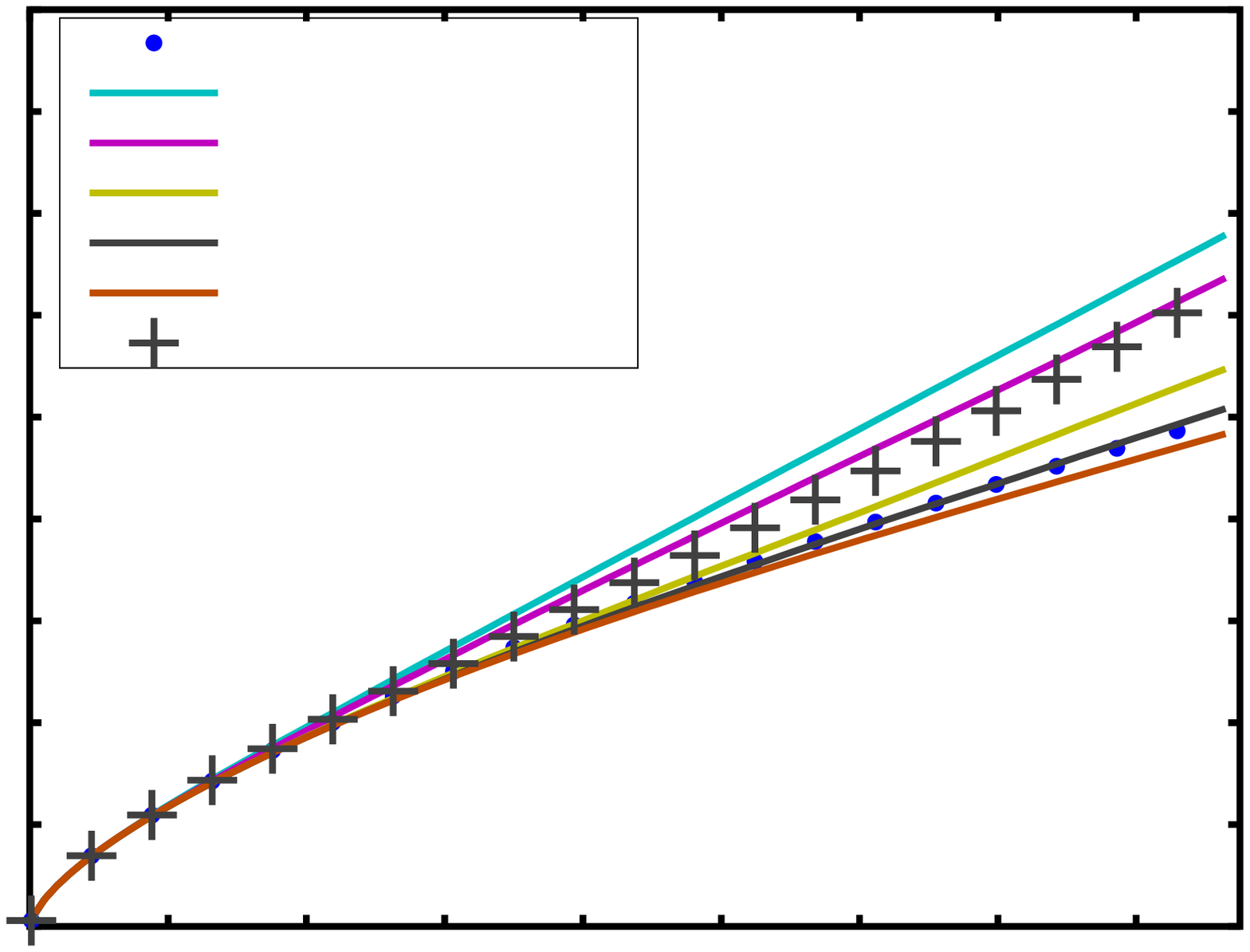}}%
    \gplfronttext
  \end{picture}%
\endgroup

%% file: aRZA_ndtfe32_qdisable1.tex
\begingroup
  \makeatletter
  \providecommand\color[2][]{%
    \GenericError{(gnuplot) \space\space\space\@spaces}{%
      Package color not loaded in conjunction with
      terminal option `colourtext'%
    }{See the gnuplot documentation for explanation.%
    }{Either use 'blacktext' in gnuplot or load the package
      color.sty in LaTeX.}%
    \renewcommand\color[2][]{}%
  }%
  \providecommand\includegraphics[2][]{%
    \GenericError{(gnuplot) \space\space\space\@spaces}{%
      Package graphicx or graphics not loaded%
    }{See the gnuplot documentation for explanation.%
    }{The gnuplot epslatex terminal needs graphicx.sty or graphics.sty.}%
    \renewcommand\includegraphics[2][]{}%
  }%
  \providecommand\rotatebox[2]{#2}%
  \@ifundefined{ifGPcolor}{%
    \newif\ifGPcolor
    \GPcolorfalse
  }{}%
  \@ifundefined{ifGPblacktext}{%
    \newif\ifGPblacktext
    \GPblacktexttrue
  }{}%
  \let\gplgaddtomacro\g@addto@macro
  \gdef\gplbacktext{}%
  \gdef\gplfronttext{}%
  \makeatother
  \ifGPblacktext
    \def\colorrgb#1{}%
    \def\colorgray#1{}%
  \else
    \ifGPcolor
      \def\colorrgb#1{\color[rgb]{#1}}%
      \def\colorgray#1{\color[gray]{#1}}%
      \expandafter\def\csname LTw\endcsname{\color{white}}%
      \expandafter\def\csname LTb\endcsname{\color{black}}%
      \expandafter\def\csname LTa\endcsname{\color{black}}%
      \expandafter\def\csname LT0\endcsname{\color[rgb]{1,0,0}}%
      \expandafter\def\csname LT1\endcsname{\color[rgb]{0,1,0}}%
      \expandafter\def\csname LT2\endcsname{\color[rgb]{0,0,1}}%
      \expandafter\def\csname LT3\endcsname{\color[rgb]{1,0,1}}%
      \expandafter\def\csname LT4\endcsname{\color[rgb]{0,1,1}}%
      \expandafter\def\csname LT5\endcsname{\color[rgb]{1,1,0}}%
      \expandafter\def\csname LT6\endcsname{\color[rgb]{0,0,0}}%
      \expandafter\def\csname LT7\endcsname{\color[rgb]{1,0.3,0}}%
      \expandafter\def\csname LT8\endcsname{\color[rgb]{0.5,0.5,0.5}}%
    \else
      \def\colorrgb#1{\color{black}}%
      \def\colorgray#1{\color[gray]{#1}}%
      \expandafter\def\csname LTw\endcsname{\color{white}}%
      \expandafter\def\csname LTb\endcsname{\color{black}}%
      \expandafter\def\csname LTa\endcsname{\color{black}}%
      \expandafter\def\csname LT0\endcsname{\color{black}}%
      \expandafter\def\csname LT1\endcsname{\color{black}}%
      \expandafter\def\csname LT2\endcsname{\color{black}}%
      \expandafter\def\csname LT3\endcsname{\color{black}}%
      \expandafter\def\csname LT4\endcsname{\color{black}}%
      \expandafter\def\csname LT5\endcsname{\color{black}}%
      \expandafter\def\csname LT6\endcsname{\color{black}}%
      \expandafter\def\csname LT7\endcsname{\color{black}}%
      \expandafter\def\csname LT8\endcsname{\color{black}}%
    \fi
  \fi
  \setlength{\unitlength}{0.0200bp}%
  \begin{picture}(11520.00,8640.00)%
    \gplgaddtomacro\gplbacktext{%
      \colorrgb{0.00,0.00,0.00}%
      \put(1406,1216){\makebox(0,0)[r]{\strut{}0}}%
      \colorrgb{0.00,0.00,0.00}%
      \put(1406,1990){\makebox(0,0)[r]{\strut{}0.2}}%
      \colorrgb{0.00,0.00,0.00}%
      \put(1406,2764){\makebox(0,0)[r]{\strut{}0.4}}%
      \colorrgb{0.00,0.00,0.00}%
      \put(1406,3538){\makebox(0,0)[r]{\strut{}0.6}}%
      \colorrgb{0.00,0.00,0.00}%
      \put(1406,4312){\makebox(0,0)[r]{\strut{}0.8}}%
      \colorrgb{0.00,0.00,0.00}%
      \put(1406,5087){\makebox(0,0)[r]{\strut{}1}}%
      \colorrgb{0.00,0.00,0.00}%
      \put(1406,5861){\makebox(0,0)[r]{\strut{}1.2}}%
      \colorrgb{0.00,0.00,0.00}%
      \put(1406,6635){\makebox(0,0)[r]{\strut{}1.4}}%
      \colorrgb{0.00,0.00,0.00}%
      \put(1406,7409){\makebox(0,0)[r]{\strut{}1.6}}%
      \colorrgb{0.00,0.00,0.00}%
      \put(1406,8183){\makebox(0,0)[r]{\strut{}1.8}}%
      \colorrgb{0.00,0.00,0.00}%
      \put(1634,836){\makebox(0,0){\strut{}0}}%
      \colorrgb{0.00,0.00,0.00}%
      \put(2686,836){\makebox(0,0){\strut{}2}}%
      \colorrgb{0.00,0.00,0.00}%
      \put(3737,836){\makebox(0,0){\strut{}4}}%
      \colorrgb{0.00,0.00,0.00}%
      \put(4789,836){\makebox(0,0){\strut{}6}}%
      \colorrgb{0.00,0.00,0.00}%
      \put(5840,836){\makebox(0,0){\strut{}8}}%
      \colorrgb{0.00,0.00,0.00}%
      \put(6892,836){\makebox(0,0){\strut{}10}}%
      \colorrgb{0.00,0.00,0.00}%
      \put(7943,836){\makebox(0,0){\strut{}12}}%
      \colorrgb{0.00,0.00,0.00}%
      \put(8995,836){\makebox(0,0){\strut{}14}}%
      \colorrgb{0.00,0.00,0.00}%
      \put(10046,836){\makebox(0,0){\strut{}16}}%
      \colorrgb{0.00,0.00,0.00}%
      \put(304,4699){\rotatebox{90}{\makebox(0,0){\strut{}scale factor $a$}}}%
      \colorrgb{0.00,0.00,0.00}%
      \put(6234,266){\makebox(0,0){\strut{}$t$ (Gyr)}}%
    }%
    \gplgaddtomacro\gplfronttext{%
      \colorrgb{0.00,0.00,0.00}%
      \put(6029,7930){\makebox(0,0)[r]{\footnotesize EdS   }}%
      \colorrgb{0.00,0.00,0.00}%
      \put(6029,7550){\makebox(0,0)[r]{\footnotesize 0.5 Mpc/$h$ }}%
      \colorrgb{0.00,0.00,0.00}%
      \put(6029,7170){\makebox(0,0)[r]{\footnotesize 1.0 Mpc/$h$ }}%
      \colorrgb{0.00,0.00,0.00}%
      \put(6029,6790){\makebox(0,0)[r]{\footnotesize 2.0 Mpc/$h$ }}%
      \colorrgb{0.00,0.00,0.00}%
      \put(6029,6410){\makebox(0,0)[r]{\footnotesize 4.0 Mpc/$h$ }}%
      \colorrgb{0.00,0.00,0.00}%
      \put(6029,6030){\makebox(0,0)[r]{\footnotesize 8.0 Mpc/$h$ }}%
      \colorrgb{0.00,0.00,0.00}%
      \put(6029,5650){\makebox(0,0)[r]{\footnotesize  {\LCDM}   }}%
    }%
    \gplbacktext
    \put(0,0){\includegraphics[scale=0.4]{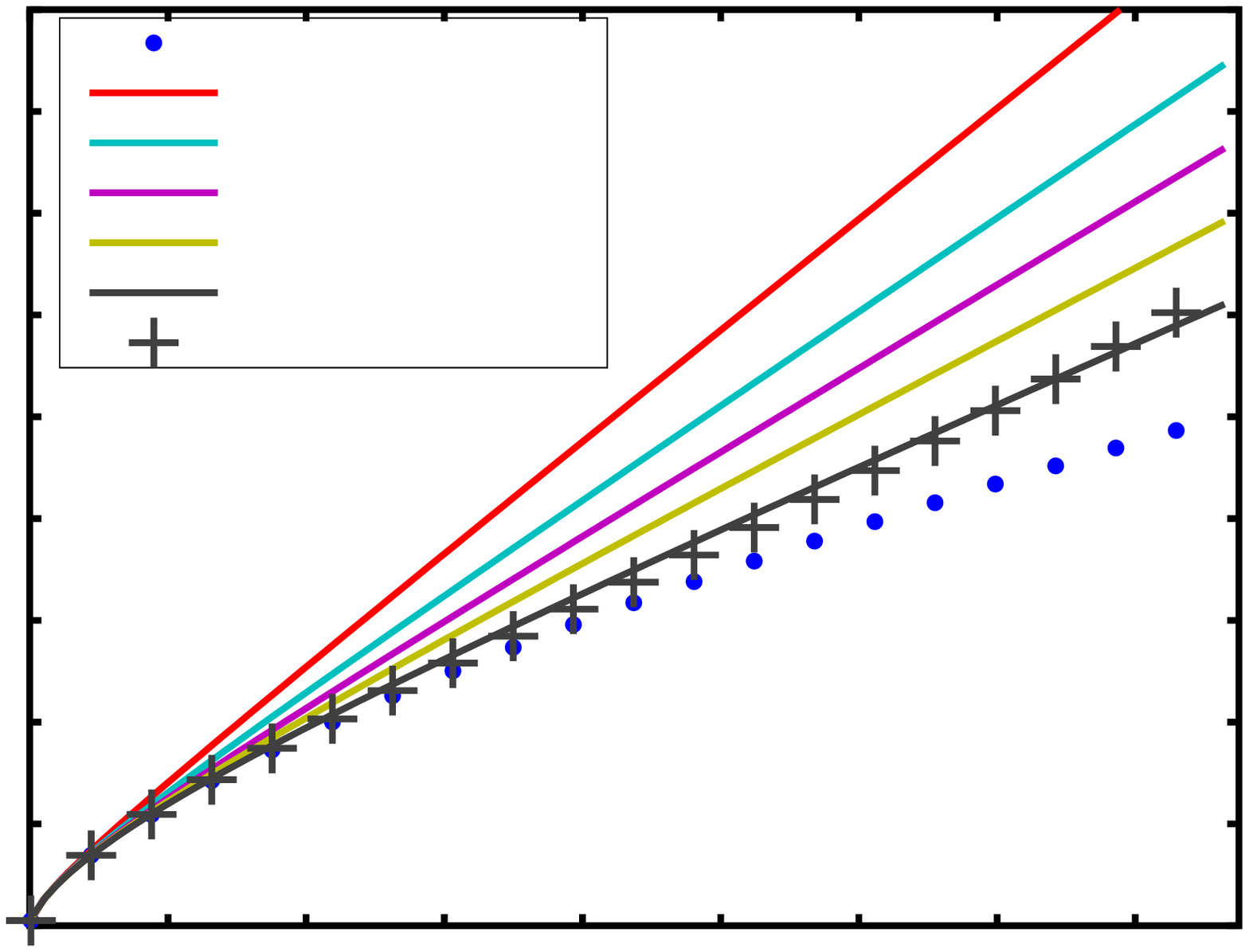}}%
    \gplfronttext
  \end{picture}%
\endgroup

%% file: aRZA_ndtfe64_qdisable1.tex
\begingroup
  \makeatletter
  \providecommand\color[2][]{%
    \GenericError{(gnuplot) \space\space\space\@spaces}{%
      Package color not loaded in conjunction with
      terminal option `colourtext'%
    }{See the gnuplot documentation for explanation.%
    }{Either use 'blacktext' in gnuplot or load the package
      color.sty in LaTeX.}%
    \renewcommand\color[2][]{}%
  }%
  \providecommand\includegraphics[2][]{%
    \GenericError{(gnuplot) \space\space\space\@spaces}{%
      Package graphicx or graphics not loaded%
    }{See the gnuplot documentation for explanation.%
    }{The gnuplot epslatex terminal needs graphicx.sty or graphics.sty.}%
    \renewcommand\includegraphics[2][]{}%
  }%
  \providecommand\rotatebox[2]{#2}%
  \@ifundefined{ifGPcolor}{%
    \newif\ifGPcolor
    \GPcolorfalse
  }{}%
  \@ifundefined{ifGPblacktext}{%
    \newif\ifGPblacktext
    \GPblacktexttrue
  }{}%
  \let\gplgaddtomacro\g@addto@macro
  \gdef\gplbacktext{}%
  \gdef\gplfronttext{}%
  \makeatother
  \ifGPblacktext
    \def\colorrgb#1{}%
    \def\colorgray#1{}%
  \else
    \ifGPcolor
      \def\colorrgb#1{\color[rgb]{#1}}%
      \def\colorgray#1{\color[gray]{#1}}%
      \expandafter\def\csname LTw\endcsname{\color{white}}%
      \expandafter\def\csname LTb\endcsname{\color{black}}%
      \expandafter\def\csname LTa\endcsname{\color{black}}%
      \expandafter\def\csname LT0\endcsname{\color[rgb]{1,0,0}}%
      \expandafter\def\csname LT1\endcsname{\color[rgb]{0,1,0}}%
      \expandafter\def\csname LT2\endcsname{\color[rgb]{0,0,1}}%
      \expandafter\def\csname LT3\endcsname{\color[rgb]{1,0,1}}%
      \expandafter\def\csname LT4\endcsname{\color[rgb]{0,1,1}}%
      \expandafter\def\csname LT5\endcsname{\color[rgb]{1,1,0}}%
      \expandafter\def\csname LT6\endcsname{\color[rgb]{0,0,0}}%
      \expandafter\def\csname LT7\endcsname{\color[rgb]{1,0.3,0}}%
      \expandafter\def\csname LT8\endcsname{\color[rgb]{0.5,0.5,0.5}}%
    \else
      \def\colorrgb#1{\color{black}}%
      \def\colorgray#1{\color[gray]{#1}}%
      \expandafter\def\csname LTw\endcsname{\color{white}}%
      \expandafter\def\csname LTb\endcsname{\color{black}}%
      \expandafter\def\csname LTa\endcsname{\color{black}}%
      \expandafter\def\csname LT0\endcsname{\color{black}}%
      \expandafter\def\csname LT1\endcsname{\color{black}}%
      \expandafter\def\csname LT2\endcsname{\color{black}}%
      \expandafter\def\csname LT3\endcsname{\color{black}}%
      \expandafter\def\csname LT4\endcsname{\color{black}}%
      \expandafter\def\csname LT5\endcsname{\color{black}}%
      \expandafter\def\csname LT6\endcsname{\color{black}}%
      \expandafter\def\csname LT7\endcsname{\color{black}}%
      \expandafter\def\csname LT8\endcsname{\color{black}}%
    \fi
  \fi
  \setlength{\unitlength}{0.0200bp}%
  \begin{picture}(11520.00,8640.00)%
    \gplgaddtomacro\gplbacktext{%
      \colorrgb{0.00,0.00,0.00}%
      \put(1406,1216){\makebox(0,0)[r]{\strut{}0}}%
      \colorrgb{0.00,0.00,0.00}%
      \put(1406,1990){\makebox(0,0)[r]{\strut{}0.2}}%
      \colorrgb{0.00,0.00,0.00}%
      \put(1406,2764){\makebox(0,0)[r]{\strut{}0.4}}%
      \colorrgb{0.00,0.00,0.00}%
      \put(1406,3538){\makebox(0,0)[r]{\strut{}0.6}}%
      \colorrgb{0.00,0.00,0.00}%
      \put(1406,4312){\makebox(0,0)[r]{\strut{}0.8}}%
      \colorrgb{0.00,0.00,0.00}%
      \put(1406,5087){\makebox(0,0)[r]{\strut{}1}}%
      \colorrgb{0.00,0.00,0.00}%
      \put(1406,5861){\makebox(0,0)[r]{\strut{}1.2}}%
      \colorrgb{0.00,0.00,0.00}%
      \put(1406,6635){\makebox(0,0)[r]{\strut{}1.4}}%
      \colorrgb{0.00,0.00,0.00}%
      \put(1406,7409){\makebox(0,0)[r]{\strut{}1.6}}%
      \colorrgb{0.00,0.00,0.00}%
      \put(1406,8183){\makebox(0,0)[r]{\strut{}1.8}}%
      \colorrgb{0.00,0.00,0.00}%
      \put(1634,836){\makebox(0,0){\strut{}0}}%
      \colorrgb{0.00,0.00,0.00}%
      \put(2686,836){\makebox(0,0){\strut{}2}}%
      \colorrgb{0.00,0.00,0.00}%
      \put(3737,836){\makebox(0,0){\strut{}4}}%
      \colorrgb{0.00,0.00,0.00}%
      \put(4789,836){\makebox(0,0){\strut{}6}}%
      \colorrgb{0.00,0.00,0.00}%
      \put(5840,836){\makebox(0,0){\strut{}8}}%
      \colorrgb{0.00,0.00,0.00}%
      \put(6892,836){\makebox(0,0){\strut{}10}}%
      \colorrgb{0.00,0.00,0.00}%
      \put(7943,836){\makebox(0,0){\strut{}12}}%
      \colorrgb{0.00,0.00,0.00}%
      \put(8995,836){\makebox(0,0){\strut{}14}}%
      \colorrgb{0.00,0.00,0.00}%
      \put(10046,836){\makebox(0,0){\strut{}16}}%
      \colorrgb{0.00,0.00,0.00}%
      \put(304,4699){\rotatebox{90}{\makebox(0,0){\strut{}scale factor $a$}}}%
      \colorrgb{0.00,0.00,0.00}%
      \put(6234,266){\makebox(0,0){\strut{}$t$ (Gyr)}}%
    }%
    \gplgaddtomacro\gplfronttext{%
      \colorrgb{0.00,0.00,0.00}%
      \put(6257,7930){\makebox(0,0)[r]{\footnotesize EdS   }}%
      \colorrgb{0.00,0.00,0.00}%
      \put(6257,7550){\makebox(0,0)[r]{\footnotesize 1.0 Mpc/$h$ }}%
      \colorrgb{0.00,0.00,0.00}%
      \put(6257,7170){\makebox(0,0)[r]{\footnotesize 2.0 Mpc/$h$ }}%
      \colorrgb{0.00,0.00,0.00}%
      \put(6257,6790){\makebox(0,0)[r]{\footnotesize 4.0 Mpc/$h$ }}%
      \colorrgb{0.00,0.00,0.00}%
      \put(6257,6410){\makebox(0,0)[r]{\footnotesize 8.0 Mpc/$h$ }}%
      \colorrgb{0.00,0.00,0.00}%
      \put(6257,6030){\makebox(0,0)[r]{\footnotesize 16.0 Mpc/$h$ }}%
      \colorrgb{0.00,0.00,0.00}%
      \put(6257,5650){\makebox(0,0)[r]{\footnotesize  {\LCDM}   }}%
    }%
    \gplbacktext
    \put(0,0){\includegraphics[scale=0.4]{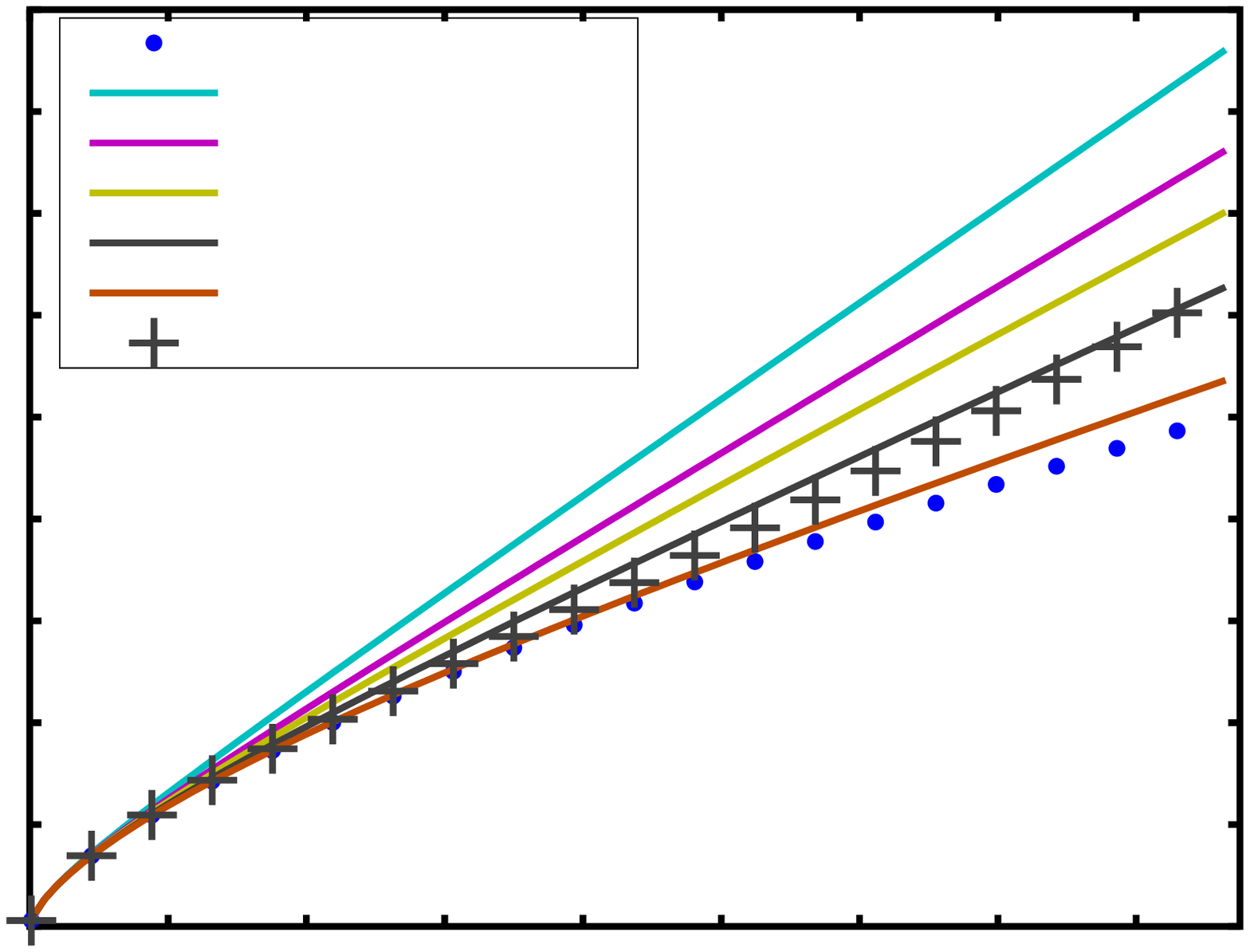}}%
    \gplfronttext
  \end{picture}%
\endgroup

%% file: aRZA_ndtfe128_qdisable1.tex
\begingroup
  \makeatletter
  \providecommand\color[2][]{%
    \GenericError{(gnuplot) \space\space\space\@spaces}{%
      Package color not loaded in conjunction with
      terminal option `colourtext'%
    }{See the gnuplot documentation for explanation.%
    }{Either use 'blacktext' in gnuplot or load the package
      color.sty in LaTeX.}%
    \renewcommand\color[2][]{}%
  }%
  \providecommand\includegraphics[2][]{%
    \GenericError{(gnuplot) \space\space\space\@spaces}{%
      Package graphicx or graphics not loaded%
    }{See the gnuplot documentation for explanation.%
    }{The gnuplot epslatex terminal needs graphicx.sty or graphics.sty.}%
    \renewcommand\includegraphics[2][]{}%
  }%
  \providecommand\rotatebox[2]{#2}%
  \@ifundefined{ifGPcolor}{%
    \newif\ifGPcolor
    \GPcolorfalse
  }{}%
  \@ifundefined{ifGPblacktext}{%
    \newif\ifGPblacktext
    \GPblacktexttrue
  }{}%
  \let\gplgaddtomacro\g@addto@macro
  \gdef\gplbacktext{}%
  \gdef\gplfronttext{}%
  \makeatother
  \ifGPblacktext
    \def\colorrgb#1{}%
    \def\colorgray#1{}%
  \else
    \ifGPcolor
      \def\colorrgb#1{\color[rgb]{#1}}%
      \def\colorgray#1{\color[gray]{#1}}%
      \expandafter\def\csname LTw\endcsname{\color{white}}%
      \expandafter\def\csname LTb\endcsname{\color{black}}%
      \expandafter\def\csname LTa\endcsname{\color{black}}%
      \expandafter\def\csname LT0\endcsname{\color[rgb]{1,0,0}}%
      \expandafter\def\csname LT1\endcsname{\color[rgb]{0,1,0}}%
      \expandafter\def\csname LT2\endcsname{\color[rgb]{0,0,1}}%
      \expandafter\def\csname LT3\endcsname{\color[rgb]{1,0,1}}%
      \expandafter\def\csname LT4\endcsname{\color[rgb]{0,1,1}}%
      \expandafter\def\csname LT5\endcsname{\color[rgb]{1,1,0}}%
      \expandafter\def\csname LT6\endcsname{\color[rgb]{0,0,0}}%
      \expandafter\def\csname LT7\endcsname{\color[rgb]{1,0.3,0}}%
      \expandafter\def\csname LT8\endcsname{\color[rgb]{0.5,0.5,0.5}}%
    \else
      \def\colorrgb#1{\color{black}}%
      \def\colorgray#1{\color[gray]{#1}}%
      \expandafter\def\csname LTw\endcsname{\color{white}}%
      \expandafter\def\csname LTb\endcsname{\color{black}}%
      \expandafter\def\csname LTa\endcsname{\color{black}}%
      \expandafter\def\csname LT0\endcsname{\color{black}}%
      \expandafter\def\csname LT1\endcsname{\color{black}}%
      \expandafter\def\csname LT2\endcsname{\color{black}}%
      \expandafter\def\csname LT3\endcsname{\color{black}}%
      \expandafter\def\csname LT4\endcsname{\color{black}}%
      \expandafter\def\csname LT5\endcsname{\color{black}}%
      \expandafter\def\csname LT6\endcsname{\color{black}}%
      \expandafter\def\csname LT7\endcsname{\color{black}}%
      \expandafter\def\csname LT8\endcsname{\color{black}}%
    \fi
  \fi
  \setlength{\unitlength}{0.0200bp}%
  \begin{picture}(11520.00,8640.00)%
    \gplgaddtomacro\gplbacktext{%
      \colorrgb{0.00,0.00,0.00}%
      \put(1406,1216){\makebox(0,0)[r]{\strut{}0}}%
      \colorrgb{0.00,0.00,0.00}%
      \put(1406,1990){\makebox(0,0)[r]{\strut{}0.2}}%
      \colorrgb{0.00,0.00,0.00}%
      \put(1406,2764){\makebox(0,0)[r]{\strut{}0.4}}%
      \colorrgb{0.00,0.00,0.00}%
      \put(1406,3538){\makebox(0,0)[r]{\strut{}0.6}}%
      \colorrgb{0.00,0.00,0.00}%
      \put(1406,4312){\makebox(0,0)[r]{\strut{}0.8}}%
      \colorrgb{0.00,0.00,0.00}%
      \put(1406,5087){\makebox(0,0)[r]{\strut{}1}}%
      \colorrgb{0.00,0.00,0.00}%
      \put(1406,5861){\makebox(0,0)[r]{\strut{}1.2}}%
      \colorrgb{0.00,0.00,0.00}%
      \put(1406,6635){\makebox(0,0)[r]{\strut{}1.4}}%
      \colorrgb{0.00,0.00,0.00}%
      \put(1406,7409){\makebox(0,0)[r]{\strut{}1.6}}%
      \colorrgb{0.00,0.00,0.00}%
      \put(1406,8183){\makebox(0,0)[r]{\strut{}1.8}}%
      \colorrgb{0.00,0.00,0.00}%
      \put(1634,836){\makebox(0,0){\strut{}0}}%
      \colorrgb{0.00,0.00,0.00}%
      \put(2686,836){\makebox(0,0){\strut{}2}}%
      \colorrgb{0.00,0.00,0.00}%
      \put(3737,836){\makebox(0,0){\strut{}4}}%
      \colorrgb{0.00,0.00,0.00}%
      \put(4789,836){\makebox(0,0){\strut{}6}}%
      \colorrgb{0.00,0.00,0.00}%
      \put(5840,836){\makebox(0,0){\strut{}8}}%
      \colorrgb{0.00,0.00,0.00}%
      \put(6892,836){\makebox(0,0){\strut{}10}}%
      \colorrgb{0.00,0.00,0.00}%
      \put(7943,836){\makebox(0,0){\strut{}12}}%
      \colorrgb{0.00,0.00,0.00}%
      \put(8995,836){\makebox(0,0){\strut{}14}}%
      \colorrgb{0.00,0.00,0.00}%
      \put(10046,836){\makebox(0,0){\strut{}16}}%
      \colorrgb{0.00,0.00,0.00}%
      \put(304,4699){\rotatebox{90}{\makebox(0,0){\strut{}scale factor $a$}}}%
      \colorrgb{0.00,0.00,0.00}%
      \put(6234,266){\makebox(0,0){\strut{}$t$ (Gyr)}}%
    }%
    \gplgaddtomacro\gplfronttext{%
      \colorrgb{0.00,0.00,0.00}%
      \put(6257,7930){\makebox(0,0)[r]{\footnotesize EdS   }}%
      \colorrgb{0.00,0.00,0.00}%
      \put(6257,7550){\makebox(0,0)[r]{\footnotesize 2.0 Mpc/$h$ }}%
      \colorrgb{0.00,0.00,0.00}%
      \put(6257,7170){\makebox(0,0)[r]{\footnotesize 4.0 Mpc/$h$ }}%
      \colorrgb{0.00,0.00,0.00}%
      \put(6257,6790){\makebox(0,0)[r]{\footnotesize 8.0 Mpc/$h$ }}%
      \colorrgb{0.00,0.00,0.00}%
      \put(6257,6410){\makebox(0,0)[r]{\footnotesize 16.0 Mpc/$h$ }}%
      \colorrgb{0.00,0.00,0.00}%
      \put(6257,6030){\makebox(0,0)[r]{\footnotesize 32.0 Mpc/$h$ }}%
      \colorrgb{0.00,0.00,0.00}%
      \put(6257,5650){\makebox(0,0)[r]{\footnotesize  {\LCDM}   }}%
    }%
    \gplbacktext
    \put(0,0){\includegraphics[scale=0.4]{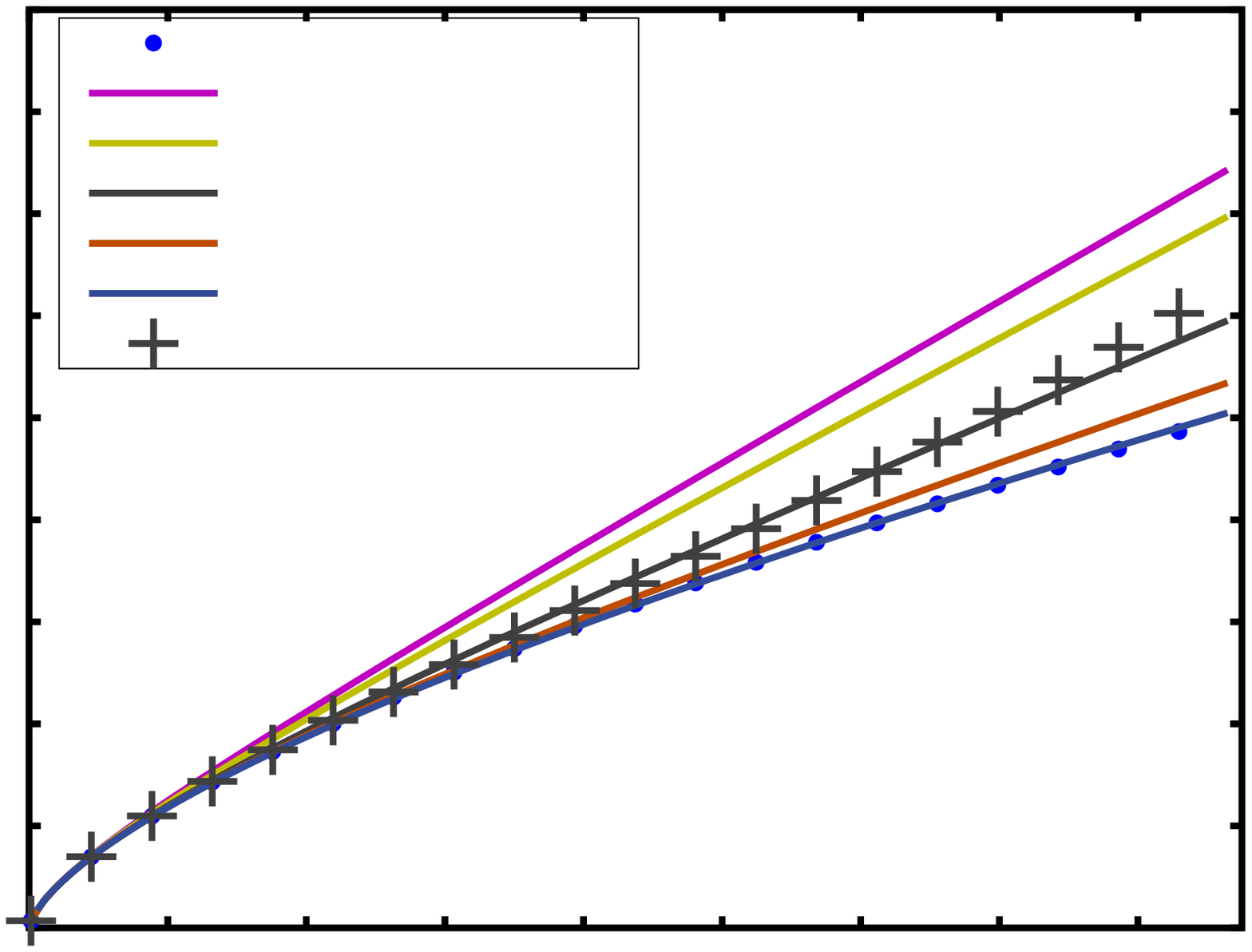}}%
    \gplfronttext
  \end{picture}%
\endgroup

%% file: aRZA_nsc64_ndtfe128_qdisable0.tex
\begingroup
  \makeatletter
  \providecommand\color[2][]{%
    \GenericError{(gnuplot) \space\space\space\@spaces}{%
      Package color not loaded in conjunction with
      terminal option `colourtext'%
    }{See the gnuplot documentation for explanation.%
    }{Either use 'blacktext' in gnuplot or load the package
      color.sty in LaTeX.}%
    \renewcommand\color[2][]{}%
  }%
  \providecommand\includegraphics[2][]{%
    \GenericError{(gnuplot) \space\space\space\@spaces}{%
      Package graphicx or graphics not loaded%
    }{See the gnuplot documentation for explanation.%
    }{The gnuplot epslatex terminal needs graphicx.sty or graphics.sty.}%
    \renewcommand\includegraphics[2][]{}%
  }%
  \providecommand\rotatebox[2]{#2}%
  \@ifundefined{ifGPcolor}{%
    \newif\ifGPcolor
    \GPcolorfalse
  }{}%
  \@ifundefined{ifGPblacktext}{%
    \newif\ifGPblacktext
    \GPblacktexttrue
  }{}%
  \let\gplgaddtomacro\g@addto@macro
  \gdef\gplbacktext{}%
  \gdef\gplfronttext{}%
  \makeatother
  \ifGPblacktext
    \def\colorrgb#1{}%
    \def\colorgray#1{}%
  \else
    \ifGPcolor
      \def\colorrgb#1{\color[rgb]{#1}}%
      \def\colorgray#1{\color[gray]{#1}}%
      \expandafter\def\csname LTw\endcsname{\color{white}}%
      \expandafter\def\csname LTb\endcsname{\color{black}}%
      \expandafter\def\csname LTa\endcsname{\color{black}}%
      \expandafter\def\csname LT0\endcsname{\color[rgb]{1,0,0}}%
      \expandafter\def\csname LT1\endcsname{\color[rgb]{0,1,0}}%
      \expandafter\def\csname LT2\endcsname{\color[rgb]{0,0,1}}%
      \expandafter\def\csname LT3\endcsname{\color[rgb]{1,0,1}}%
      \expandafter\def\csname LT4\endcsname{\color[rgb]{0,1,1}}%
      \expandafter\def\csname LT5\endcsname{\color[rgb]{1,1,0}}%
      \expandafter\def\csname LT6\endcsname{\color[rgb]{0,0,0}}%
      \expandafter\def\csname LT7\endcsname{\color[rgb]{1,0.3,0}}%
      \expandafter\def\csname LT8\endcsname{\color[rgb]{0.5,0.5,0.5}}%
    \else
      \def\colorrgb#1{\color{black}}%
      \def\colorgray#1{\color[gray]{#1}}%
      \expandafter\def\csname LTw\endcsname{\color{white}}%
      \expandafter\def\csname LTb\endcsname{\color{black}}%
      \expandafter\def\csname LTa\endcsname{\color{black}}%
      \expandafter\def\csname LT0\endcsname{\color{black}}%
      \expandafter\def\csname LT1\endcsname{\color{black}}%
      \expandafter\def\csname LT2\endcsname{\color{black}}%
      \expandafter\def\csname LT3\endcsname{\color{black}}%
      \expandafter\def\csname LT4\endcsname{\color{black}}%
      \expandafter\def\csname LT5\endcsname{\color{black}}%
      \expandafter\def\csname LT6\endcsname{\color{black}}%
      \expandafter\def\csname LT7\endcsname{\color{black}}%
      \expandafter\def\csname LT8\endcsname{\color{black}}%
    \fi
  \fi
  \setlength{\unitlength}{0.0200bp}%
  \begin{picture}(11520.00,8640.00)%
    \gplgaddtomacro\gplbacktext{%
      \colorrgb{0.00,0.00,0.00}%
      \put(1406,1216){\makebox(0,0)[r]{\strut{}0}}%
      \colorrgb{0.00,0.00,0.00}%
      \put(1406,1990){\makebox(0,0)[r]{\strut{}0.2}}%
      \colorrgb{0.00,0.00,0.00}%
      \put(1406,2764){\makebox(0,0)[r]{\strut{}0.4}}%
      \colorrgb{0.00,0.00,0.00}%
      \put(1406,3538){\makebox(0,0)[r]{\strut{}0.6}}%
      \colorrgb{0.00,0.00,0.00}%
      \put(1406,4312){\makebox(0,0)[r]{\strut{}0.8}}%
      \colorrgb{0.00,0.00,0.00}%
      \put(1406,5087){\makebox(0,0)[r]{\strut{}1}}%
      \colorrgb{0.00,0.00,0.00}%
      \put(1406,5861){\makebox(0,0)[r]{\strut{}1.2}}%
      \colorrgb{0.00,0.00,0.00}%
      \put(1406,6635){\makebox(0,0)[r]{\strut{}1.4}}%
      \colorrgb{0.00,0.00,0.00}%
      \put(1406,7409){\makebox(0,0)[r]{\strut{}1.6}}%
      \colorrgb{0.00,0.00,0.00}%
      \put(1406,8183){\makebox(0,0)[r]{\strut{}1.8}}%
      \colorrgb{0.00,0.00,0.00}%
      \put(1634,836){\makebox(0,0){\strut{}0}}%
      \colorrgb{0.00,0.00,0.00}%
      \put(2686,836){\makebox(0,0){\strut{}2}}%
      \colorrgb{0.00,0.00,0.00}%
      \put(3737,836){\makebox(0,0){\strut{}4}}%
      \colorrgb{0.00,0.00,0.00}%
      \put(4789,836){\makebox(0,0){\strut{}6}}%
      \colorrgb{0.00,0.00,0.00}%
      \put(5840,836){\makebox(0,0){\strut{}8}}%
      \colorrgb{0.00,0.00,0.00}%
      \put(6892,836){\makebox(0,0){\strut{}10}}%
      \colorrgb{0.00,0.00,0.00}%
      \put(7943,836){\makebox(0,0){\strut{}12}}%
      \colorrgb{0.00,0.00,0.00}%
      \put(8995,836){\makebox(0,0){\strut{}14}}%
      \colorrgb{0.00,0.00,0.00}%
      \put(10046,836){\makebox(0,0){\strut{}16}}%
      \colorrgb{0.00,0.00,0.00}%
      \put(304,4699){\rotatebox{90}{\makebox(0,0){\strut{}scale factor $a$}}}%
      \colorrgb{0.00,0.00,0.00}%
      \put(6234,266){\makebox(0,0){\strut{}$t$ (Gyr)}}%
    }%
    \gplgaddtomacro\gplfronttext{%
      \colorrgb{0.00,0.00,0.00}%
      \put(6029,7930){\makebox(0,0)[r]{\footnotesize EdS   }}%
      \colorrgb{0.00,0.00,0.00}%
      \put(6029,7550){\makebox(0,0)[r]{\footnotesize 0.5 Mpc/$h$ }}%
      \colorrgb{0.00,0.00,0.00}%
      \put(6029,7170){\makebox(0,0)[r]{\footnotesize 1.0 Mpc/$h$ }}%
      \colorrgb{0.00,0.00,0.00}%
      \put(6029,6790){\makebox(0,0)[r]{\footnotesize 2.0 Mpc/$h$ }}%
      \colorrgb{0.00,0.00,0.00}%
      \put(6029,6410){\makebox(0,0)[r]{\footnotesize 4.0 Mpc/$h$ }}%
      \colorrgb{0.00,0.00,0.00}%
      \put(6029,6030){\makebox(0,0)[r]{\footnotesize 8.0 Mpc/$h$ }}%
      \colorrgb{0.00,0.00,0.00}%
      \put(6029,5650){\makebox(0,0)[r]{\footnotesize  {\LCDM}   }}%
    }%
    \gplbacktext
    \put(0,0){\includegraphics[scale=0.4]{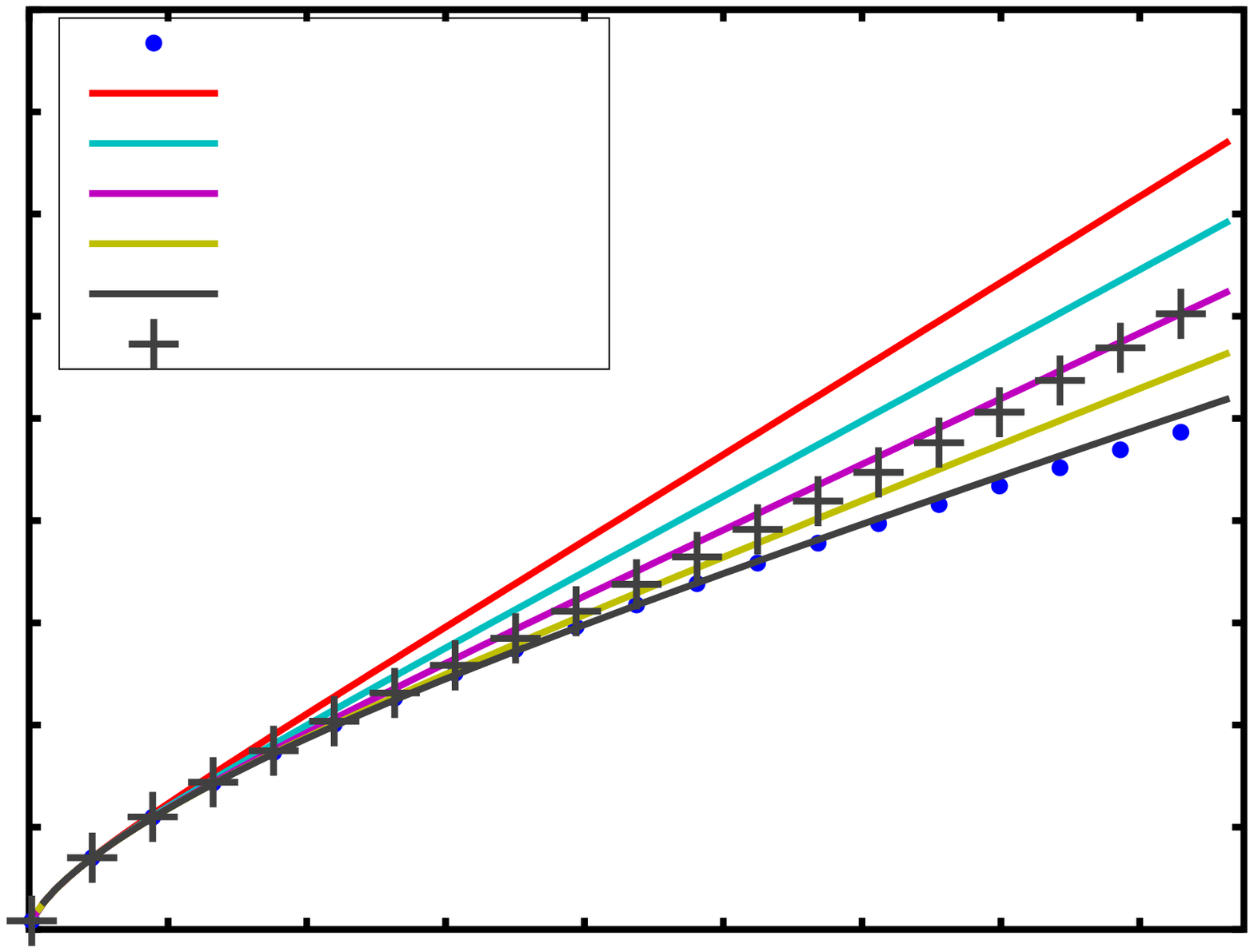}}%
    \gplfronttext
  \end{picture}%
\endgroup

%% file: OmQmean.tex
\begingroup
  \makeatletter
  \providecommand\color[2][]{%
    \GenericError{(gnuplot) \space\space\space\@spaces}{%
      Package color not loaded in conjunction with
      terminal option `colourtext'%
    }{See the gnuplot documentation for explanation.%
    }{Either use 'blacktext' in gnuplot or load the package
      color.sty in LaTeX.}%
    \renewcommand\color[2][]{}%
  }%
  \providecommand\includegraphics[2][]{%
    \GenericError{(gnuplot) \space\space\space\@spaces}{%
      Package graphicx or graphics not loaded%
    }{See the gnuplot documentation for explanation.%
    }{The gnuplot epslatex terminal needs graphicx.sty or graphics.sty.}%
    \renewcommand\includegraphics[2][]{}%
  }%
  \providecommand\rotatebox[2]{#2}%
  \@ifundefined{ifGPcolor}{%
    \newif\ifGPcolor
    \GPcolorfalse
  }{}%
  \@ifundefined{ifGPblacktext}{%
    \newif\ifGPblacktext
    \GPblacktexttrue
  }{}%
  \let\gplgaddtomacro\g@addto@macro
  \gdef\gplbacktext{}%
  \gdef\gplfronttext{}%
  \makeatother
  \ifGPblacktext
    \def\colorrgb#1{}%
    \def\colorgray#1{}%
  \else
    \ifGPcolor
      \def\colorrgb#1{\color[rgb]{#1}}%
      \def\colorgray#1{\color[gray]{#1}}%
      \expandafter\def\csname LTw\endcsname{\color{white}}%
      \expandafter\def\csname LTb\endcsname{\color{black}}%
      \expandafter\def\csname LTa\endcsname{\color{black}}%
      \expandafter\def\csname LT0\endcsname{\color[rgb]{1,0,0}}%
      \expandafter\def\csname LT1\endcsname{\color[rgb]{0,1,0}}%
      \expandafter\def\csname LT2\endcsname{\color[rgb]{0,0,1}}%
      \expandafter\def\csname LT3\endcsname{\color[rgb]{1,0,1}}%
      \expandafter\def\csname LT4\endcsname{\color[rgb]{0,1,1}}%
      \expandafter\def\csname LT5\endcsname{\color[rgb]{1,1,0}}%
      \expandafter\def\csname LT6\endcsname{\color[rgb]{0,0,0}}%
      \expandafter\def\csname LT7\endcsname{\color[rgb]{1,0.3,0}}%
      \expandafter\def\csname LT8\endcsname{\color[rgb]{0.5,0.5,0.5}}%
    \else
      \def\colorrgb#1{\color{black}}%
      \def\colorgray#1{\color[gray]{#1}}%
      \expandafter\def\csname LTw\endcsname{\color{white}}%
      \expandafter\def\csname LTb\endcsname{\color{black}}%
      \expandafter\def\csname LTa\endcsname{\color{black}}%
      \expandafter\def\csname LT0\endcsname{\color{black}}%
      \expandafter\def\csname LT1\endcsname{\color{black}}%
      \expandafter\def\csname LT2\endcsname{\color{black}}%
      \expandafter\def\csname LT3\endcsname{\color{black}}%
      \expandafter\def\csname LT4\endcsname{\color{black}}%
      \expandafter\def\csname LT5\endcsname{\color{black}}%
      \expandafter\def\csname LT6\endcsname{\color{black}}%
      \expandafter\def\csname LT7\endcsname{\color{black}}%
      \expandafter\def\csname LT8\endcsname{\color{black}}%
    \fi
  \fi
  \setlength{\unitlength}{0.0200bp}%
  \begin{picture}(11520.00,8640.00)%
    \gplgaddtomacro\gplbacktext{%
      \colorrgb{0.00,0.00,0.00}%
      \put(1634,1216){\makebox(0,0)[r]{\strut{}$10^{-4}$}}%
      \colorrgb{0.00,0.00,0.00}%
      \put(1634,2377){\makebox(0,0)[r]{\strut{}$10^{-3}$}}%
      \colorrgb{0.00,0.00,0.00}%
      \put(1634,3538){\makebox(0,0)[r]{\strut{}$10^{-2}$}}%
      \colorrgb{0.00,0.00,0.00}%
      \put(1634,4699){\makebox(0,0)[r]{\strut{}$10^{-1}$}}%
      \colorrgb{0.00,0.00,0.00}%
      \put(1634,5861){\makebox(0,0)[r]{\strut{}$10^{0}$}}%
      \colorrgb{0.00,0.00,0.00}%
      \put(1634,7022){\makebox(0,0)[r]{\strut{}$10^{1}$}}%
      \colorrgb{0.00,0.00,0.00}%
      \put(1634,8183){\makebox(0,0)[r]{\strut{}$10^{2}$}}%
      \colorrgb{0.00,0.00,0.00}%
      \put(1862,836){\makebox(0,0){\strut{}0}}%
      \colorrgb{0.00,0.00,0.00}%
      \put(2887,836){\makebox(0,0){\strut{}2}}%
      \colorrgb{0.00,0.00,0.00}%
      \put(3913,836){\makebox(0,0){\strut{}4}}%
      \colorrgb{0.00,0.00,0.00}%
      \put(4938,836){\makebox(0,0){\strut{}6}}%
      \colorrgb{0.00,0.00,0.00}%
      \put(5964,836){\makebox(0,0){\strut{}8}}%
      \colorrgb{0.00,0.00,0.00}%
      \put(6989,836){\makebox(0,0){\strut{}10}}%
      \colorrgb{0.00,0.00,0.00}%
      \put(8015,836){\makebox(0,0){\strut{}12}}%
      \colorrgb{0.00,0.00,0.00}%
      \put(9040,836){\makebox(0,0){\strut{}14}}%
      \colorrgb{0.00,0.00,0.00}%
      \put(10066,836){\makebox(0,0){\strut{}16}}%
      \colorrgb{0.00,0.00,0.00}%
      \put(304,4699){\rotatebox{90}{\makebox(0,0){\strut{}$\left<\OmQD\right>\uncollapsed$}}}%
      \colorrgb{0.00,0.00,0.00}%
      \put(6348,266){\makebox(0,0){\strut{}$t$ (Gyr)}}%
    }%
    \gplgaddtomacro\gplfronttext{%
      \colorrgb{0.00,0.00,0.00}%
      \put(10607,2989){\makebox(0,0)[r]{\footnotesize 0.5 Mpc/$h$ }}%
      \colorrgb{0.00,0.00,0.00}%
      \put(10607,2609){\makebox(0,0)[r]{\footnotesize 1.0 Mpc/$h$ }}%
      \colorrgb{0.00,0.00,0.00}%
      \put(10607,2229){\makebox(0,0)[r]{\footnotesize 2.0 Mpc/$h$ }}%
      \colorrgb{0.00,0.00,0.00}%
      \put(10607,1849){\makebox(0,0)[r]{\footnotesize 4.0 Mpc/$h$ }}%
      \colorrgb{0.00,0.00,0.00}%
      \put(10607,1469){\makebox(0,0)[r]{\footnotesize 8.0 Mpc/$h$ }}%
    }%
    \gplbacktext
    \put(0,0){\includegraphics[scale=0.4]{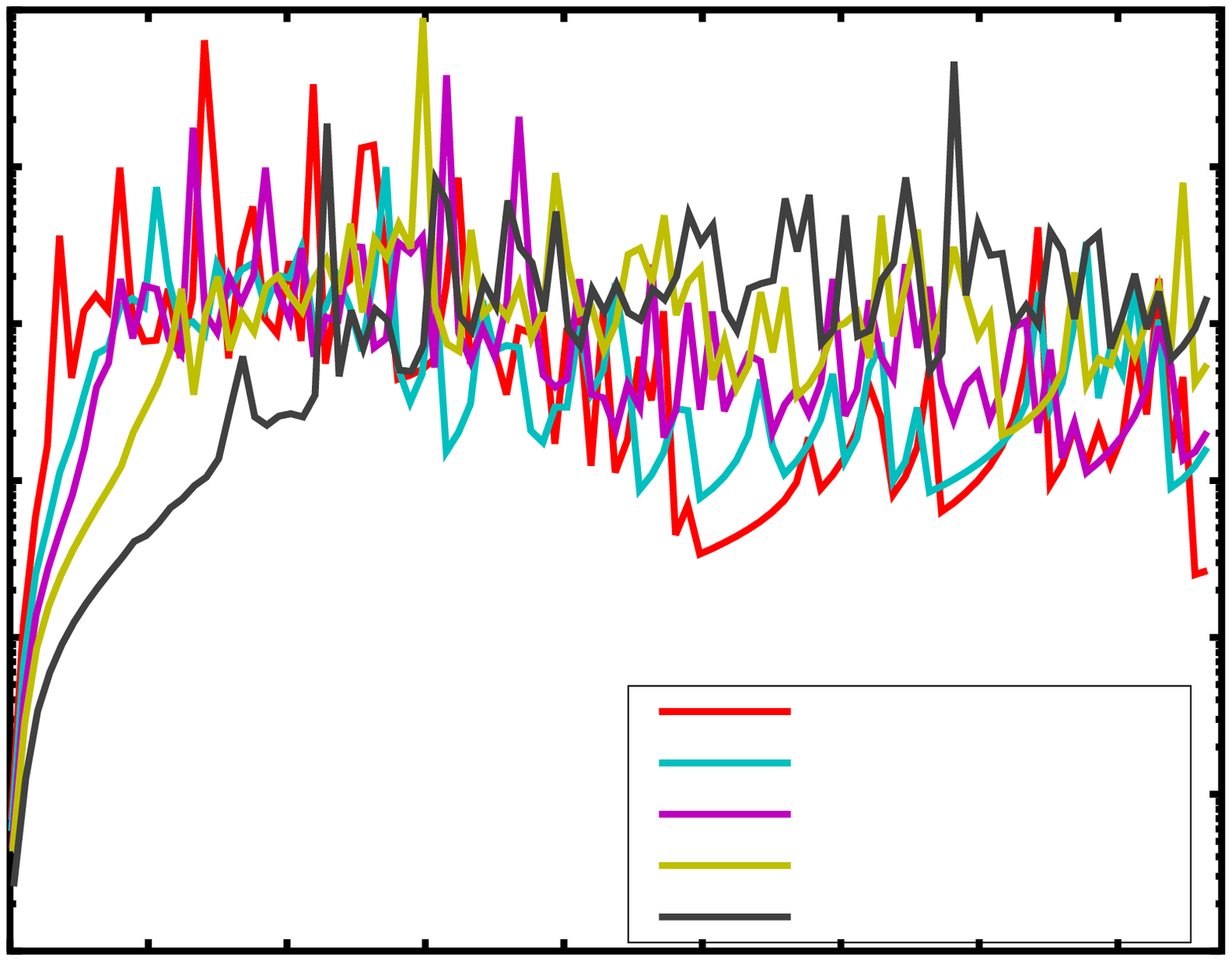}}%
    \gplfronttext
  \end{picture}%
\endgroup

%% file: OmQmedian.tex
\begingroup
  \makeatletter
  \providecommand\color[2][]{%
    \GenericError{(gnuplot) \space\space\space\@spaces}{%
      Package color not loaded in conjunction with
      terminal option `colourtext'%
    }{See the gnuplot documentation for explanation.%
    }{Either use 'blacktext' in gnuplot or load the package
      color.sty in LaTeX.}%
    \renewcommand\color[2][]{}%
  }%
  \providecommand\includegraphics[2][]{%
    \GenericError{(gnuplot) \space\space\space\@spaces}{%
      Package graphicx or graphics not loaded%
    }{See the gnuplot documentation for explanation.%
    }{The gnuplot epslatex terminal needs graphicx.sty or graphics.sty.}%
    \renewcommand\includegraphics[2][]{}%
  }%
  \providecommand\rotatebox[2]{#2}%
  \@ifundefined{ifGPcolor}{%
    \newif\ifGPcolor
    \GPcolorfalse
  }{}%
  \@ifundefined{ifGPblacktext}{%
    \newif\ifGPblacktext
    \GPblacktexttrue
  }{}%
  \let\gplgaddtomacro\g@addto@macro
  \gdef\gplbacktext{}%
  \gdef\gplfronttext{}%
  \makeatother
  \ifGPblacktext
    \def\colorrgb#1{}%
    \def\colorgray#1{}%
  \else
    \ifGPcolor
      \def\colorrgb#1{\color[rgb]{#1}}%
      \def\colorgray#1{\color[gray]{#1}}%
      \expandafter\def\csname LTw\endcsname{\color{white}}%
      \expandafter\def\csname LTb\endcsname{\color{black}}%
      \expandafter\def\csname LTa\endcsname{\color{black}}%
      \expandafter\def\csname LT0\endcsname{\color[rgb]{1,0,0}}%
      \expandafter\def\csname LT1\endcsname{\color[rgb]{0,1,0}}%
      \expandafter\def\csname LT2\endcsname{\color[rgb]{0,0,1}}%
      \expandafter\def\csname LT3\endcsname{\color[rgb]{1,0,1}}%
      \expandafter\def\csname LT4\endcsname{\color[rgb]{0,1,1}}%
      \expandafter\def\csname LT5\endcsname{\color[rgb]{1,1,0}}%
      \expandafter\def\csname LT6\endcsname{\color[rgb]{0,0,0}}%
      \expandafter\def\csname LT7\endcsname{\color[rgb]{1,0.3,0}}%
      \expandafter\def\csname LT8\endcsname{\color[rgb]{0.5,0.5,0.5}}%
    \else
      \def\colorrgb#1{\color{black}}%
      \def\colorgray#1{\color[gray]{#1}}%
      \expandafter\def\csname LTw\endcsname{\color{white}}%
      \expandafter\def\csname LTb\endcsname{\color{black}}%
      \expandafter\def\csname LTa\endcsname{\color{black}}%
      \expandafter\def\csname LT0\endcsname{\color{black}}%
      \expandafter\def\csname LT1\endcsname{\color{black}}%
      \expandafter\def\csname LT2\endcsname{\color{black}}%
      \expandafter\def\csname LT3\endcsname{\color{black}}%
      \expandafter\def\csname LT4\endcsname{\color{black}}%
      \expandafter\def\csname LT5\endcsname{\color{black}}%
      \expandafter\def\csname LT6\endcsname{\color{black}}%
      \expandafter\def\csname LT7\endcsname{\color{black}}%
      \expandafter\def\csname LT8\endcsname{\color{black}}%
    \fi
  \fi
  \setlength{\unitlength}{0.0200bp}%
  \begin{picture}(11520.00,8640.00)%
    \gplgaddtomacro\gplbacktext{%
      \colorrgb{0.00,0.00,0.00}%
      \put(1634,1216){\makebox(0,0)[r]{\strut{}0}}%
      \colorrgb{0.00,0.00,0.00}%
      \put(1634,2609){\makebox(0,0)[r]{\strut{}0.02}}%
      \colorrgb{0.00,0.00,0.00}%
      \put(1634,4003){\makebox(0,0)[r]{\strut{}0.04}}%
      \colorrgb{0.00,0.00,0.00}%
      \put(1634,5396){\makebox(0,0)[r]{\strut{}0.06}}%
      \colorrgb{0.00,0.00,0.00}%
      \put(1634,6790){\makebox(0,0)[r]{\strut{}0.08}}%
      \colorrgb{0.00,0.00,0.00}%
      \put(1634,8183){\makebox(0,0)[r]{\strut{}0.1}}%
      \colorrgb{0.00,0.00,0.00}%
      \put(1862,836){\makebox(0,0){\strut{}0}}%
      \colorrgb{0.00,0.00,0.00}%
      \put(2887,836){\makebox(0,0){\strut{}2}}%
      \colorrgb{0.00,0.00,0.00}%
      \put(3913,836){\makebox(0,0){\strut{}4}}%
      \colorrgb{0.00,0.00,0.00}%
      \put(4938,836){\makebox(0,0){\strut{}6}}%
      \colorrgb{0.00,0.00,0.00}%
      \put(5964,836){\makebox(0,0){\strut{}8}}%
      \colorrgb{0.00,0.00,0.00}%
      \put(6989,836){\makebox(0,0){\strut{}10}}%
      \colorrgb{0.00,0.00,0.00}%
      \put(8015,836){\makebox(0,0){\strut{}12}}%
      \colorrgb{0.00,0.00,0.00}%
      \put(9040,836){\makebox(0,0){\strut{}14}}%
      \colorrgb{0.00,0.00,0.00}%
      \put(10066,836){\makebox(0,0){\strut{}16}}%
      \colorrgb{0.00,0.00,0.00}%
      \put(304,4699){\rotatebox{90}{\makebox(0,0){\strut{}$\mu\left(\OmQD\right)\uncollapsed$}}}%
      \colorrgb{0.00,0.00,0.00}%
      \put(6348,266){\makebox(0,0){\strut{}$t$ (Gyr)}}%
    }%
    \gplgaddtomacro\gplfronttext{%
      \colorrgb{0.00,0.00,0.00}%
      \put(10607,7930){\makebox(0,0)[r]{\footnotesize 0.5 Mpc/$h$ }}%
      \colorrgb{0.00,0.00,0.00}%
      \put(10607,7550){\makebox(0,0)[r]{\footnotesize 1.0 Mpc/$h$ }}%
      \colorrgb{0.00,0.00,0.00}%
      \put(10607,7170){\makebox(0,0)[r]{\footnotesize 2.0 Mpc/$h$ }}%
      \colorrgb{0.00,0.00,0.00}%
      \put(10607,6790){\makebox(0,0)[r]{\footnotesize 4.0 Mpc/$h$ }}%
      \colorrgb{0.00,0.00,0.00}%
      \put(10607,6410){\makebox(0,0)[r]{\footnotesize 8.0 Mpc/$h$ }}%
    }%
    \gplbacktext
    \put(0,0){\includegraphics[scale=0.4]{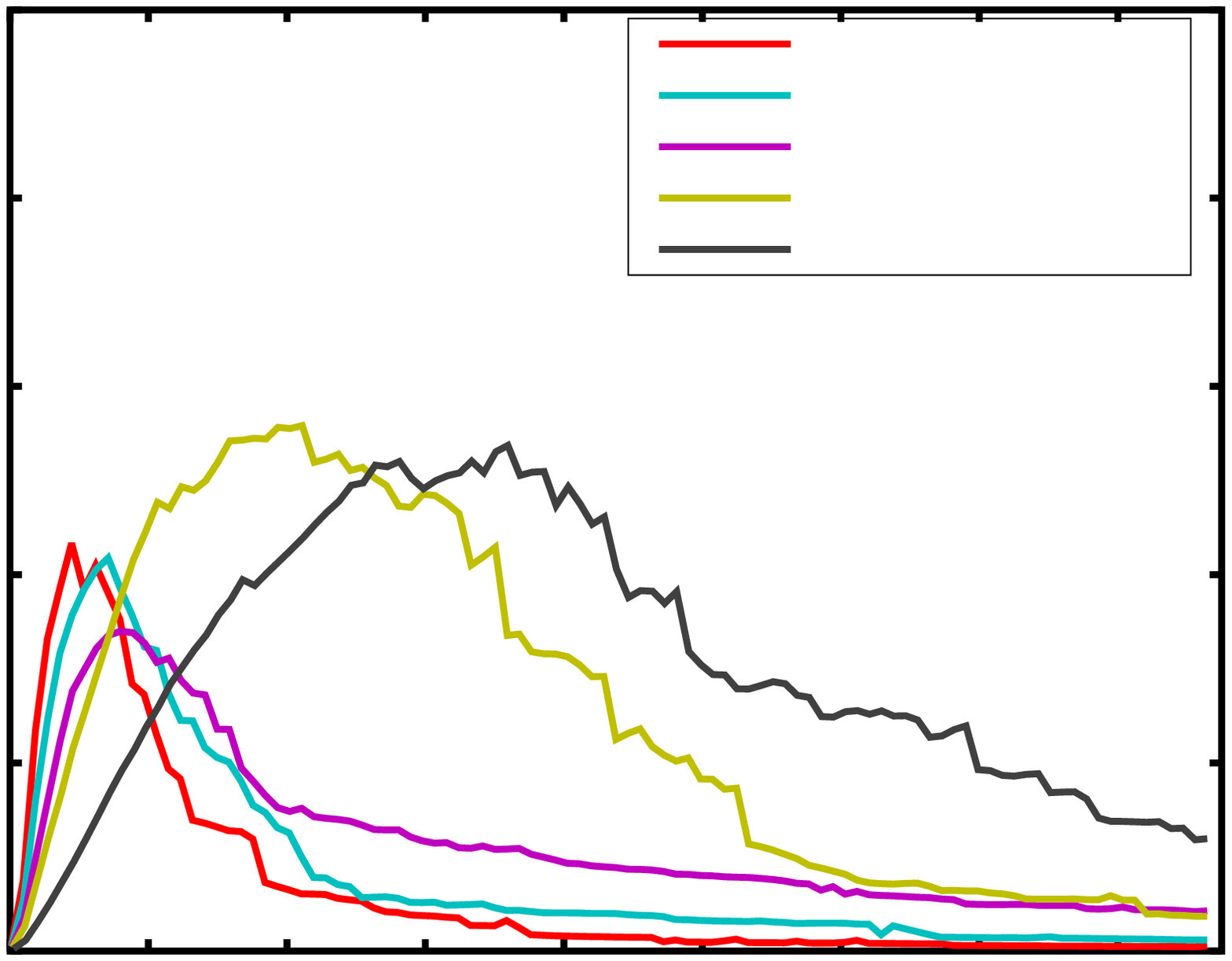}}%
    \gplfronttext
  \end{picture}%
\endgroup

%% file: OmRmean.tex
\begingroup
  \makeatletter
  \providecommand\color[2][]{%
    \GenericError{(gnuplot) \space\space\space\@spaces}{%
      Package color not loaded in conjunction with
      terminal option `colourtext'%
    }{See the gnuplot documentation for explanation.%
    }{Either use 'blacktext' in gnuplot or load the package
      color.sty in LaTeX.}%
    \renewcommand\color[2][]{}%
  }%
  \providecommand\includegraphics[2][]{%
    \GenericError{(gnuplot) \space\space\space\@spaces}{%
      Package graphicx or graphics not loaded%
    }{See the gnuplot documentation for explanation.%
    }{The gnuplot epslatex terminal needs graphicx.sty or graphics.sty.}%
    \renewcommand\includegraphics[2][]{}%
  }%
  \providecommand\rotatebox[2]{#2}%
  \@ifundefined{ifGPcolor}{%
    \newif\ifGPcolor
    \GPcolorfalse
  }{}%
  \@ifundefined{ifGPblacktext}{%
    \newif\ifGPblacktext
    \GPblacktexttrue
  }{}%
  \let\gplgaddtomacro\g@addto@macro
  \gdef\gplbacktext{}%
  \gdef\gplfronttext{}%
  \makeatother
  \ifGPblacktext
    \def\colorrgb#1{}%
    \def\colorgray#1{}%
  \else
    \ifGPcolor
      \def\colorrgb#1{\color[rgb]{#1}}%
      \def\colorgray#1{\color[gray]{#1}}%
      \expandafter\def\csname LTw\endcsname{\color{white}}%
      \expandafter\def\csname LTb\endcsname{\color{black}}%
      \expandafter\def\csname LTa\endcsname{\color{black}}%
      \expandafter\def\csname LT0\endcsname{\color[rgb]{1,0,0}}%
      \expandafter\def\csname LT1\endcsname{\color[rgb]{0,1,0}}%
      \expandafter\def\csname LT2\endcsname{\color[rgb]{0,0,1}}%
      \expandafter\def\csname LT3\endcsname{\color[rgb]{1,0,1}}%
      \expandafter\def\csname LT4\endcsname{\color[rgb]{0,1,1}}%
      \expandafter\def\csname LT5\endcsname{\color[rgb]{1,1,0}}%
      \expandafter\def\csname LT6\endcsname{\color[rgb]{0,0,0}}%
      \expandafter\def\csname LT7\endcsname{\color[rgb]{1,0.3,0}}%
      \expandafter\def\csname LT8\endcsname{\color[rgb]{0.5,0.5,0.5}}%
    \else
      \def\colorrgb#1{\color{black}}%
      \def\colorgray#1{\color[gray]{#1}}%
      \expandafter\def\csname LTw\endcsname{\color{white}}%
      \expandafter\def\csname LTb\endcsname{\color{black}}%
      \expandafter\def\csname LTa\endcsname{\color{black}}%
      \expandafter\def\csname LT0\endcsname{\color{black}}%
      \expandafter\def\csname LT1\endcsname{\color{black}}%
      \expandafter\def\csname LT2\endcsname{\color{black}}%
      \expandafter\def\csname LT3\endcsname{\color{black}}%
      \expandafter\def\csname LT4\endcsname{\color{black}}%
      \expandafter\def\csname LT5\endcsname{\color{black}}%
      \expandafter\def\csname LT6\endcsname{\color{black}}%
      \expandafter\def\csname LT7\endcsname{\color{black}}%
      \expandafter\def\csname LT8\endcsname{\color{black}}%
    \fi
  \fi
  \setlength{\unitlength}{0.0200bp}%
  \begin{picture}(11520.00,8640.00)%
    \gplgaddtomacro\gplbacktext{%
      \colorrgb{0.00,0.00,0.00}%
      \put(1634,1216){\makebox(0,0)[r]{\strut{}-2.5}}%
      \colorrgb{0.00,0.00,0.00}%
      \put(1634,1990){\makebox(0,0)[r]{\strut{}-2}}%
      \colorrgb{0.00,0.00,0.00}%
      \put(1634,2764){\makebox(0,0)[r]{\strut{}-1.5}}%
      \colorrgb{0.00,0.00,0.00}%
      \put(1634,3538){\makebox(0,0)[r]{\strut{}-1}}%
      \colorrgb{0.00,0.00,0.00}%
      \put(1634,4312){\makebox(0,0)[r]{\strut{}-0.5}}%
      \colorrgb{0.00,0.00,0.00}%
      \put(1634,5087){\makebox(0,0)[r]{\strut{}0}}%
      \colorrgb{0.00,0.00,0.00}%
      \put(1634,5861){\makebox(0,0)[r]{\strut{}0.5}}%
      \colorrgb{0.00,0.00,0.00}%
      \put(1634,6635){\makebox(0,0)[r]{\strut{}1}}%
      \colorrgb{0.00,0.00,0.00}%
      \put(1634,7409){\makebox(0,0)[r]{\strut{}1.5}}%
      \colorrgb{0.00,0.00,0.00}%
      \put(1634,8183){\makebox(0,0)[r]{\strut{}2}}%
      \colorrgb{0.00,0.00,0.00}%
      \put(1862,836){\makebox(0,0){\strut{}0}}%
      \colorrgb{0.00,0.00,0.00}%
      \put(2887,836){\makebox(0,0){\strut{}2}}%
      \colorrgb{0.00,0.00,0.00}%
      \put(3913,836){\makebox(0,0){\strut{}4}}%
      \colorrgb{0.00,0.00,0.00}%
      \put(4938,836){\makebox(0,0){\strut{}6}}%
      \colorrgb{0.00,0.00,0.00}%
      \put(5964,836){\makebox(0,0){\strut{}8}}%
      \colorrgb{0.00,0.00,0.00}%
      \put(6989,836){\makebox(0,0){\strut{}10}}%
      \colorrgb{0.00,0.00,0.00}%
      \put(8015,836){\makebox(0,0){\strut{}12}}%
      \colorrgb{0.00,0.00,0.00}%
      \put(9040,836){\makebox(0,0){\strut{}14}}%
      \colorrgb{0.00,0.00,0.00}%
      \put(10066,836){\makebox(0,0){\strut{}16}}%
      \colorrgb{0.00,0.00,0.00}%
      \put(304,4699){\rotatebox{90}{\makebox(0,0){\strut{}$\left<\OmRD\right>\uncollapsed$}}}%
      \colorrgb{0.00,0.00,0.00}%
      \put(6348,266){\makebox(0,0){\strut{}$t$ (Gyr)}}%
    }%
    \gplgaddtomacro\gplfronttext{%
      \colorrgb{0.00,0.00,0.00}%
      \put(10607,2989){\makebox(0,0)[r]{\footnotesize 0.5 Mpc/$h$ }}%
      \colorrgb{0.00,0.00,0.00}%
      \put(10607,2609){\makebox(0,0)[r]{\footnotesize 1.0 Mpc/$h$ }}%
      \colorrgb{0.00,0.00,0.00}%
      \put(10607,2229){\makebox(0,0)[r]{\footnotesize 2.0 Mpc/$h$ }}%
      \colorrgb{0.00,0.00,0.00}%
      \put(10607,1849){\makebox(0,0)[r]{\footnotesize 4.0 Mpc/$h$ }}%
      \colorrgb{0.00,0.00,0.00}%
      \put(10607,1469){\makebox(0,0)[r]{\footnotesize 8.0 Mpc/$h$ }}%
    }%
    \gplbacktext
    \put(0,0){\includegraphics[scale=0.4]{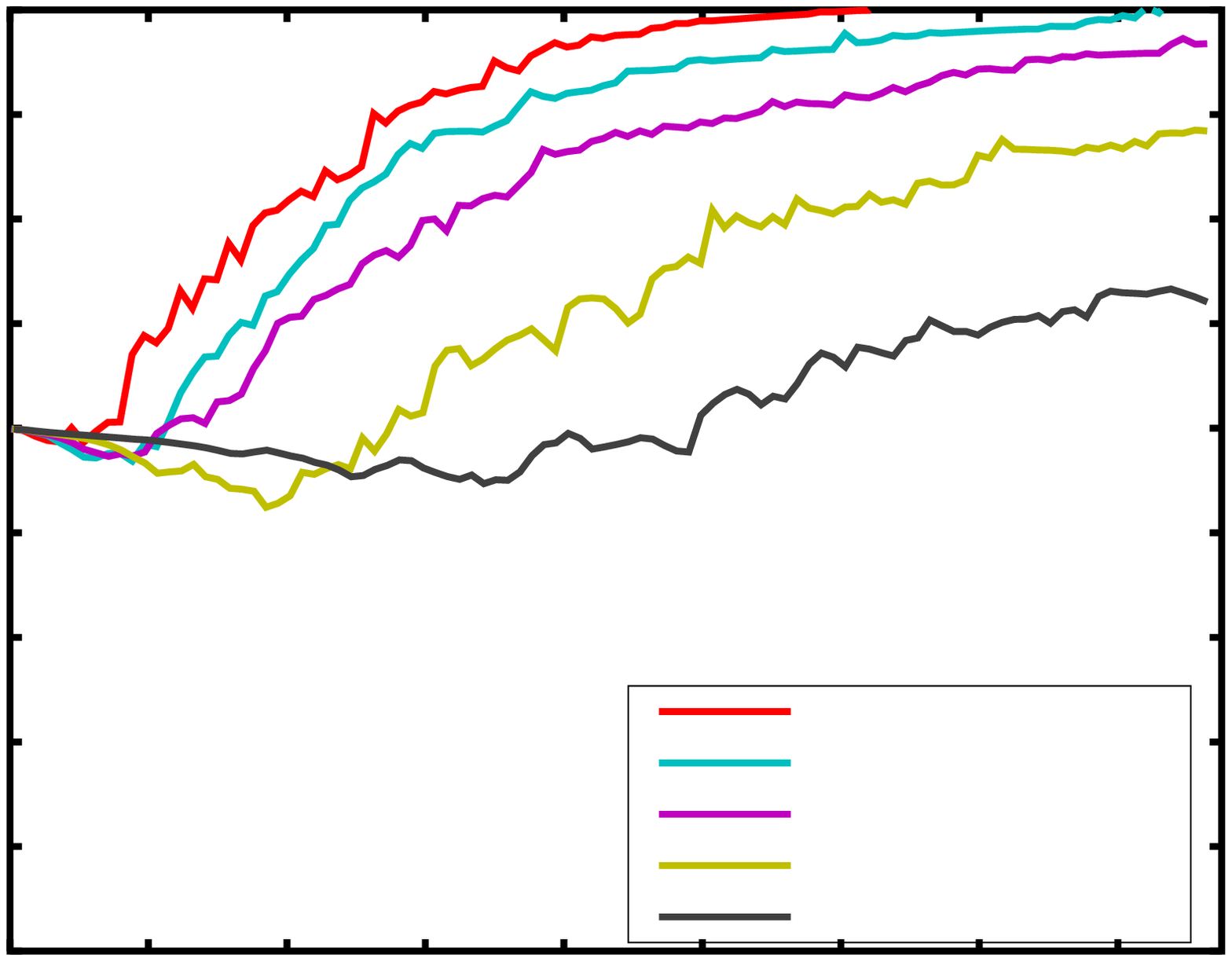}}%
    \gplfronttext
  \end{picture}%
\endgroup

%% file: OmRmedian.tex
\begingroup
  \makeatletter
  \providecommand\color[2][]{%
    \GenericError{(gnuplot) \space\space\space\@spaces}{%
      Package color not loaded in conjunction with
      terminal option `colourtext'%
    }{See the gnuplot documentation for explanation.%
    }{Either use 'blacktext' in gnuplot or load the package
      color.sty in LaTeX.}%
    \renewcommand\color[2][]{}%
  }%
  \providecommand\includegraphics[2][]{%
    \GenericError{(gnuplot) \space\space\space\@spaces}{%
      Package graphicx or graphics not loaded%
    }{See the gnuplot documentation for explanation.%
    }{The gnuplot epslatex terminal needs graphicx.sty or graphics.sty.}%
    \renewcommand\includegraphics[2][]{}%
  }%
  \providecommand\rotatebox[2]{#2}%
  \@ifundefined{ifGPcolor}{%
    \newif\ifGPcolor
    \GPcolorfalse
  }{}%
  \@ifundefined{ifGPblacktext}{%
    \newif\ifGPblacktext
    \GPblacktexttrue
  }{}%
  \let\gplgaddtomacro\g@addto@macro
  \gdef\gplbacktext{}%
  \gdef\gplfronttext{}%
  \makeatother
  \ifGPblacktext
    \def\colorrgb#1{}%
    \def\colorgray#1{}%
  \else
    \ifGPcolor
      \def\colorrgb#1{\color[rgb]{#1}}%
      \def\colorgray#1{\color[gray]{#1}}%
      \expandafter\def\csname LTw\endcsname{\color{white}}%
      \expandafter\def\csname LTb\endcsname{\color{black}}%
      \expandafter\def\csname LTa\endcsname{\color{black}}%
      \expandafter\def\csname LT0\endcsname{\color[rgb]{1,0,0}}%
      \expandafter\def\csname LT1\endcsname{\color[rgb]{0,1,0}}%
      \expandafter\def\csname LT2\endcsname{\color[rgb]{0,0,1}}%
      \expandafter\def\csname LT3\endcsname{\color[rgb]{1,0,1}}%
      \expandafter\def\csname LT4\endcsname{\color[rgb]{0,1,1}}%
      \expandafter\def\csname LT5\endcsname{\color[rgb]{1,1,0}}%
      \expandafter\def\csname LT6\endcsname{\color[rgb]{0,0,0}}%
      \expandafter\def\csname LT7\endcsname{\color[rgb]{1,0.3,0}}%
      \expandafter\def\csname LT8\endcsname{\color[rgb]{0.5,0.5,0.5}}%
    \else
      \def\colorrgb#1{\color{black}}%
      \def\colorgray#1{\color[gray]{#1}}%
      \expandafter\def\csname LTw\endcsname{\color{white}}%
      \expandafter\def\csname LTb\endcsname{\color{black}}%
      \expandafter\def\csname LTa\endcsname{\color{black}}%
      \expandafter\def\csname LT0\endcsname{\color{black}}%
      \expandafter\def\csname LT1\endcsname{\color{black}}%
      \expandafter\def\csname LT2\endcsname{\color{black}}%
      \expandafter\def\csname LT3\endcsname{\color{black}}%
      \expandafter\def\csname LT4\endcsname{\color{black}}%
      \expandafter\def\csname LT5\endcsname{\color{black}}%
      \expandafter\def\csname LT6\endcsname{\color{black}}%
      \expandafter\def\csname LT7\endcsname{\color{black}}%
      \expandafter\def\csname LT8\endcsname{\color{black}}%
    \fi
  \fi
  \setlength{\unitlength}{0.0200bp}%
  \begin{picture}(11520.00,8640.00)%
    \gplgaddtomacro\gplbacktext{%
      \colorrgb{0.00,0.00,0.00}%
      \put(1634,1216){\makebox(0,0)[r]{\strut{}-1}}%
      \colorrgb{0.00,0.00,0.00}%
      \put(1634,2211){\makebox(0,0)[r]{\strut{}-0.5}}%
      \colorrgb{0.00,0.00,0.00}%
      \put(1634,3207){\makebox(0,0)[r]{\strut{}0}}%
      \colorrgb{0.00,0.00,0.00}%
      \put(1634,4202){\makebox(0,0)[r]{\strut{}0.5}}%
      \colorrgb{0.00,0.00,0.00}%
      \put(1634,5197){\makebox(0,0)[r]{\strut{}1}}%
      \colorrgb{0.00,0.00,0.00}%
      \put(1634,6192){\makebox(0,0)[r]{\strut{}1.5}}%
      \colorrgb{0.00,0.00,0.00}%
      \put(1634,7188){\makebox(0,0)[r]{\strut{}2}}%
      \colorrgb{0.00,0.00,0.00}%
      \put(1634,8183){\makebox(0,0)[r]{\strut{}2.5}}%
      \colorrgb{0.00,0.00,0.00}%
      \put(1862,836){\makebox(0,0){\strut{}0}}%
      \colorrgb{0.00,0.00,0.00}%
      \put(2887,836){\makebox(0,0){\strut{}2}}%
      \colorrgb{0.00,0.00,0.00}%
      \put(3913,836){\makebox(0,0){\strut{}4}}%
      \colorrgb{0.00,0.00,0.00}%
      \put(4938,836){\makebox(0,0){\strut{}6}}%
      \colorrgb{0.00,0.00,0.00}%
      \put(5964,836){\makebox(0,0){\strut{}8}}%
      \colorrgb{0.00,0.00,0.00}%
      \put(6989,836){\makebox(0,0){\strut{}10}}%
      \colorrgb{0.00,0.00,0.00}%
      \put(8015,836){\makebox(0,0){\strut{}12}}%
      \colorrgb{0.00,0.00,0.00}%
      \put(9040,836){\makebox(0,0){\strut{}14}}%
      \colorrgb{0.00,0.00,0.00}%
      \put(10066,836){\makebox(0,0){\strut{}16}}%
      \colorrgb{0.00,0.00,0.00}%
      \put(304,4699){\rotatebox{90}{\makebox(0,0){\strut{}$\mu\left(\OmRD\right)\uncollapsed$}}}%
      \colorrgb{0.00,0.00,0.00}%
      \put(6348,266){\makebox(0,0){\strut{}$t$ (Gyr)}}%
    }%
    \gplgaddtomacro\gplfronttext{%
      \colorrgb{0.00,0.00,0.00}%
      \put(10607,2989){\makebox(0,0)[r]{\footnotesize 0.5 Mpc/$h$ }}%
      \colorrgb{0.00,0.00,0.00}%
      \put(10607,2609){\makebox(0,0)[r]{\footnotesize 1.0 Mpc/$h$ }}%
      \colorrgb{0.00,0.00,0.00}%
      \put(10607,2229){\makebox(0,0)[r]{\footnotesize 2.0 Mpc/$h$ }}%
      \colorrgb{0.00,0.00,0.00}%
      \put(10607,1849){\makebox(0,0)[r]{\footnotesize 4.0 Mpc/$h$ }}%
      \colorrgb{0.00,0.00,0.00}%
      \put(10607,1469){\makebox(0,0)[r]{\footnotesize 8.0 Mpc/$h$ }}%
    }%
    \gplbacktext
    \put(0,0){\includegraphics[scale=0.4]{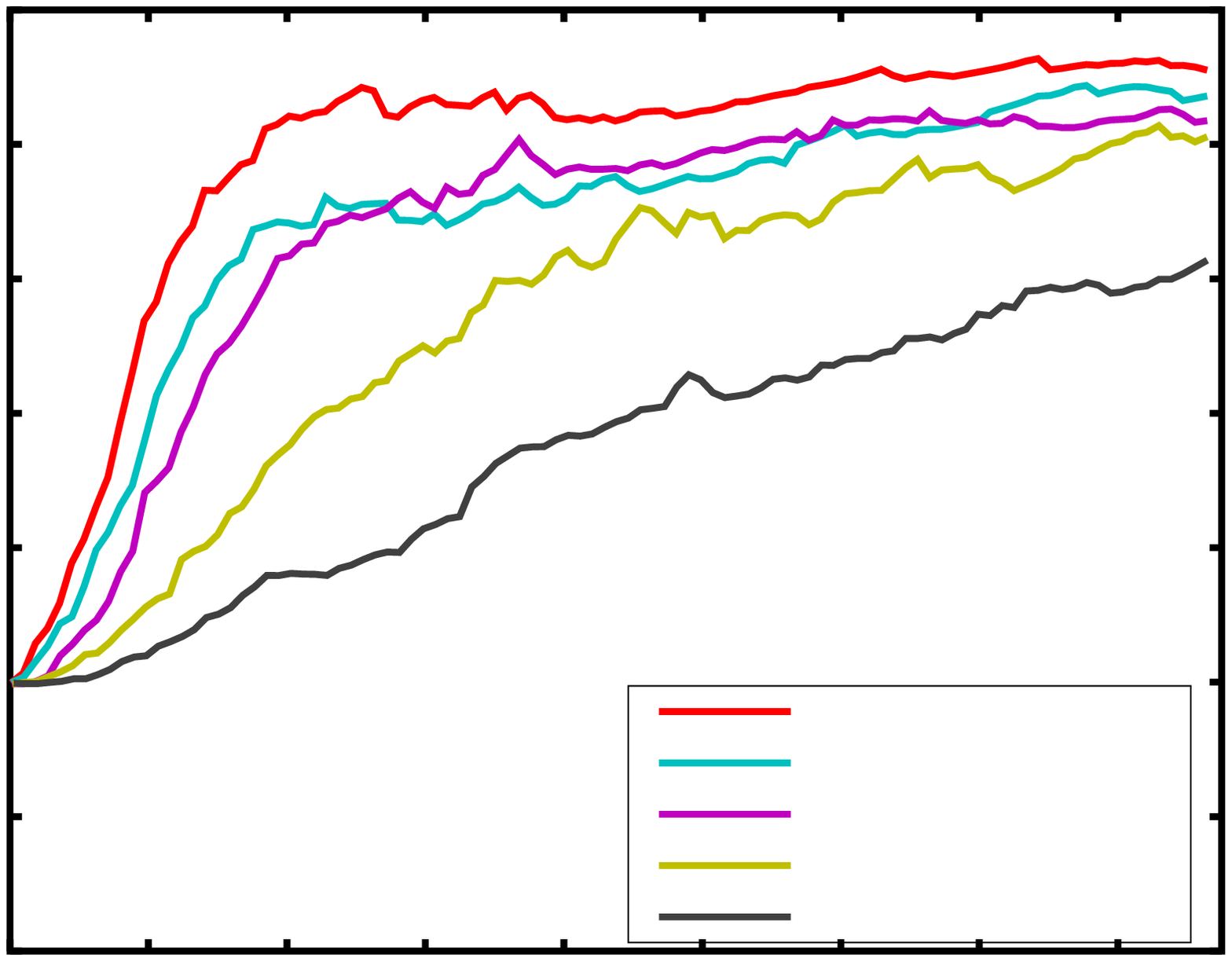}}%
    \gplfronttext
  \end{picture}%
\endgroup

%% file: superEdS_fvir_nsc8_ndtfe32_qdisable0.tex
\begingroup
  \makeatletter
  \providecommand\color[2][]{%
    \GenericError{(gnuplot) \space\space\space\@spaces}{%
      Package color not loaded in conjunction with
      terminal option `colourtext'%
    }{See the gnuplot documentation for explanation.%
    }{Either use 'blacktext' in gnuplot or load the package
      color.sty in LaTeX.}%
    \renewcommand\color[2][]{}%
  }%
  \providecommand\includegraphics[2][]{%
    \GenericError{(gnuplot) \space\space\space\@spaces}{%
      Package graphicx or graphics not loaded%
    }{See the gnuplot documentation for explanation.%
    }{The gnuplot epslatex terminal needs graphicx.sty or graphics.sty.}%
    \renewcommand\includegraphics[2][]{}%
  }%
  \providecommand\rotatebox[2]{#2}%
  \@ifundefined{ifGPcolor}{%
    \newif\ifGPcolor
    \GPcolorfalse
  }{}%
  \@ifundefined{ifGPblacktext}{%
    \newif\ifGPblacktext
    \GPblacktexttrue
  }{}%
  \let\gplgaddtomacro\g@addto@macro
  \gdef\gplbacktext{}%
  \gdef\gplfronttext{}%
  \makeatother
  \ifGPblacktext
    \def\colorrgb#1{}%
    \def\colorgray#1{}%
  \else
    \ifGPcolor
      \def\colorrgb#1{\color[rgb]{#1}}%
      \def\colorgray#1{\color[gray]{#1}}%
      \expandafter\def\csname LTw\endcsname{\color{white}}%
      \expandafter\def\csname LTb\endcsname{\color{black}}%
      \expandafter\def\csname LTa\endcsname{\color{black}}%
      \expandafter\def\csname LT0\endcsname{\color[rgb]{1,0,0}}%
      \expandafter\def\csname LT1\endcsname{\color[rgb]{0,1,0}}%
      \expandafter\def\csname LT2\endcsname{\color[rgb]{0,0,1}}%
      \expandafter\def\csname LT3\endcsname{\color[rgb]{1,0,1}}%
      \expandafter\def\csname LT4\endcsname{\color[rgb]{0,1,1}}%
      \expandafter\def\csname LT5\endcsname{\color[rgb]{1,1,0}}%
      \expandafter\def\csname LT6\endcsname{\color[rgb]{0,0,0}}%
      \expandafter\def\csname LT7\endcsname{\color[rgb]{1,0.3,0}}%
      \expandafter\def\csname LT8\endcsname{\color[rgb]{0.5,0.5,0.5}}%
    \else
      \def\colorrgb#1{\color{black}}%
      \def\colorgray#1{\color[gray]{#1}}%
      \expandafter\def\csname LTw\endcsname{\color{white}}%
      \expandafter\def\csname LTb\endcsname{\color{black}}%
      \expandafter\def\csname LTa\endcsname{\color{black}}%
      \expandafter\def\csname LT0\endcsname{\color{black}}%
      \expandafter\def\csname LT1\endcsname{\color{black}}%
      \expandafter\def\csname LT2\endcsname{\color{black}}%
      \expandafter\def\csname LT3\endcsname{\color{black}}%
      \expandafter\def\csname LT4\endcsname{\color{black}}%
      \expandafter\def\csname LT5\endcsname{\color{black}}%
      \expandafter\def\csname LT6\endcsname{\color{black}}%
      \expandafter\def\csname LT7\endcsname{\color{black}}%
      \expandafter\def\csname LT8\endcsname{\color{black}}%
    \fi
  \fi
  \setlength{\unitlength}{0.0200bp}%
  \begin{picture}(11520.00,8640.00)%
    \gplgaddtomacro\gplbacktext{%
      \colorrgb{0.00,0.00,0.00}%
      \put(1634,1216){\makebox(0,0)[r]{\strut{}0.95}}%
      \colorrgb{0.00,0.00,0.00}%
      \put(1634,2377){\makebox(0,0)[r]{\strut{}1}}%
      \colorrgb{0.00,0.00,0.00}%
      \put(1634,3538){\makebox(0,0)[r]{\strut{}1.05}}%
      \colorrgb{0.00,0.00,0.00}%
      \put(1634,4700){\makebox(0,0)[r]{\strut{}1.1}}%
      \colorrgb{0.00,0.00,0.00}%
      \put(1634,5861){\makebox(0,0)[r]{\strut{}1.15}}%
      \colorrgb{0.00,0.00,0.00}%
      \put(1634,7022){\makebox(0,0)[r]{\strut{}1.2}}%
      \colorrgb{0.00,0.00,0.00}%
      \put(1634,8183){\makebox(0,0)[r]{\strut{}1.25}}%
      \colorrgb{0.00,0.00,0.00}%
      \put(1862,836){\makebox(0,0){\strut{}0.2}}%
      \colorrgb{0.00,0.00,0.00}%
      \put(3358,836){\makebox(0,0){\strut{}0.3}}%
      \colorrgb{0.00,0.00,0.00}%
      \put(4853,836){\makebox(0,0){\strut{}0.4}}%
      \colorrgb{0.00,0.00,0.00}%
      \put(6348,836){\makebox(0,0){\strut{}0.5}}%
      \colorrgb{0.00,0.00,0.00}%
      \put(7844,836){\makebox(0,0){\strut{}0.6}}%
      \colorrgb{0.00,0.00,0.00}%
      \put(9339,836){\makebox(0,0){\strut{}0.7}}%
      \colorrgb{0.00,0.00,0.00}%
      \put(10835,836){\makebox(0,0){\strut{}0.8}}%
      \colorrgb{0.00,0.00,0.00}%
      \put(304,4699){\rotatebox{90}{\makebox(0,0){\strut{}$\aeff/\abg$}}}%
      \colorrgb{0.00,0.00,0.00}%
      \put(6348,266){\makebox(0,0){\strut{}$\fvir$}}%
    }%
    \gplgaddtomacro\gplfronttext{%
      \colorrgb{0.00,0.00,0.00}%
      \put(6385,7930){\makebox(0,0)[r]{{\footnotesize 4.2~Gyr}}}%
      \colorrgb{0.00,0.00,0.00}%
      \put(6385,7550){\makebox(0,0)[r]{{\footnotesize 8.6~Gyr}}}%
      \colorrgb{0.00,0.00,0.00}%
      \put(6385,7170){\makebox(0,0)[r]{{\footnotesize 12.9~Gyr}}}%
    }%
    \gplbacktext
    \put(0,0){\includegraphics[scale=0.4]{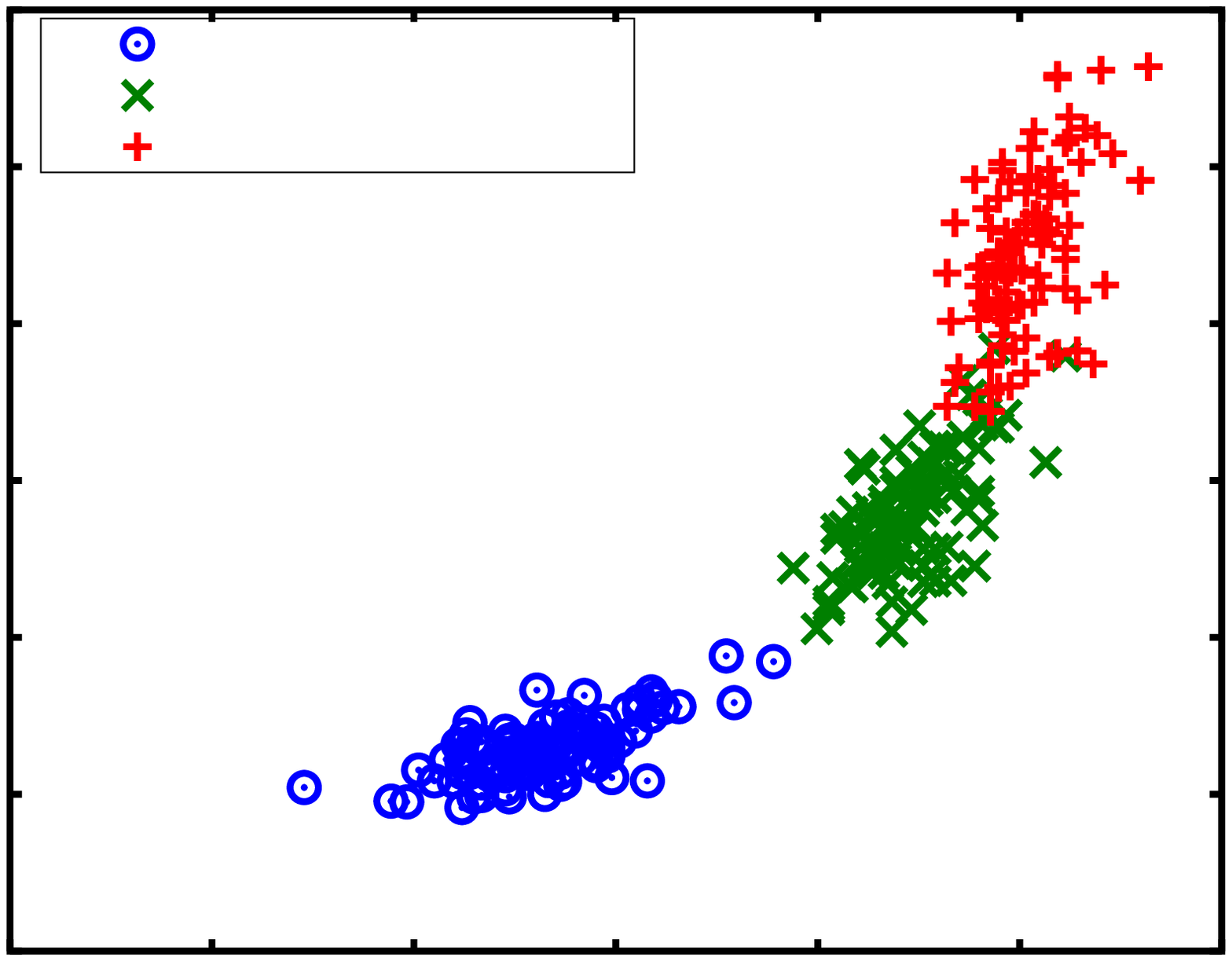}}%
    \gplfronttext
  \end{picture}%
\endgroup

%% file: a_nbody_Lag_128_F.tex
\begingroup
  \makeatletter
  \providecommand\color[2][]{%
    \GenericError{(gnuplot) \space\space\space\@spaces}{%
      Package color not loaded in conjunction with
      terminal option `colourtext'%
    }{See the gnuplot documentation for explanation.%
    }{Either use 'blacktext' in gnuplot or load the package
      color.sty in LaTeX.}%
    \renewcommand\color[2][]{}%
  }%
  \providecommand\includegraphics[2][]{%
    \GenericError{(gnuplot) \space\space\space\@spaces}{%
      Package graphicx or graphics not loaded%
    }{See the gnuplot documentation for explanation.%
    }{The gnuplot epslatex terminal needs graphicx.sty or graphics.sty.}%
    \renewcommand\includegraphics[2][]{}%
  }%
  \providecommand\rotatebox[2]{#2}%
  \@ifundefined{ifGPcolor}{%
    \newif\ifGPcolor
    \GPcolorfalse
  }{}%
  \@ifundefined{ifGPblacktext}{%
    \newif\ifGPblacktext
    \GPblacktexttrue
  }{}%
  \let\gplgaddtomacro\g@addto@macro
  \gdef\gplbacktext{}%
  \gdef\gplfronttext{}%
  \makeatother
  \ifGPblacktext
    \def\colorrgb#1{}%
    \def\colorgray#1{}%
  \else
    \ifGPcolor
      \def\colorrgb#1{\color[rgb]{#1}}%
      \def\colorgray#1{\color[gray]{#1}}%
      \expandafter\def\csname LTw\endcsname{\color{white}}%
      \expandafter\def\csname LTb\endcsname{\color{black}}%
      \expandafter\def\csname LTa\endcsname{\color{black}}%
      \expandafter\def\csname LT0\endcsname{\color[rgb]{1,0,0}}%
      \expandafter\def\csname LT1\endcsname{\color[rgb]{0,1,0}}%
      \expandafter\def\csname LT2\endcsname{\color[rgb]{0,0,1}}%
      \expandafter\def\csname LT3\endcsname{\color[rgb]{1,0,1}}%
      \expandafter\def\csname LT4\endcsname{\color[rgb]{0,1,1}}%
      \expandafter\def\csname LT5\endcsname{\color[rgb]{1,1,0}}%
      \expandafter\def\csname LT6\endcsname{\color[rgb]{0,0,0}}%
      \expandafter\def\csname LT7\endcsname{\color[rgb]{1,0.3,0}}%
      \expandafter\def\csname LT8\endcsname{\color[rgb]{0.5,0.5,0.5}}%
    \else
      \def\colorrgb#1{\color{black}}%
      \def\colorgray#1{\color[gray]{#1}}%
      \expandafter\def\csname LTw\endcsname{\color{white}}%
      \expandafter\def\csname LTb\endcsname{\color{black}}%
      \expandafter\def\csname LTa\endcsname{\color{black}}%
      \expandafter\def\csname LT0\endcsname{\color{black}}%
      \expandafter\def\csname LT1\endcsname{\color{black}}%
      \expandafter\def\csname LT2\endcsname{\color{black}}%
      \expandafter\def\csname LT3\endcsname{\color{black}}%
      \expandafter\def\csname LT4\endcsname{\color{black}}%
      \expandafter\def\csname LT5\endcsname{\color{black}}%
      \expandafter\def\csname LT6\endcsname{\color{black}}%
      \expandafter\def\csname LT7\endcsname{\color{black}}%
      \expandafter\def\csname LT8\endcsname{\color{black}}%
    \fi
  \fi
  \setlength{\unitlength}{0.0200bp}%
  \begin{picture}(11520.00,8640.00)%
    \gplgaddtomacro\gplbacktext{%
      \colorrgb{0.00,0.00,0.00}%
      \put(1406,1216){\makebox(0,0)[r]{\strut{}0}}%
      \colorrgb{0.00,0.00,0.00}%
      \put(1406,1990){\makebox(0,0)[r]{\strut{}0.2}}%
      \colorrgb{0.00,0.00,0.00}%
      \put(1406,2764){\makebox(0,0)[r]{\strut{}0.4}}%
      \colorrgb{0.00,0.00,0.00}%
      \put(1406,3538){\makebox(0,0)[r]{\strut{}0.6}}%
      \colorrgb{0.00,0.00,0.00}%
      \put(1406,4312){\makebox(0,0)[r]{\strut{}0.8}}%
      \colorrgb{0.00,0.00,0.00}%
      \put(1406,5087){\makebox(0,0)[r]{\strut{}1}}%
      \colorrgb{0.00,0.00,0.00}%
      \put(1406,5861){\makebox(0,0)[r]{\strut{}1.2}}%
      \colorrgb{0.00,0.00,0.00}%
      \put(1406,6635){\makebox(0,0)[r]{\strut{}1.4}}%
      \colorrgb{0.00,0.00,0.00}%
      \put(1406,7409){\makebox(0,0)[r]{\strut{}1.6}}%
      \colorrgb{0.00,0.00,0.00}%
      \put(1406,8183){\makebox(0,0)[r]{\strut{}1.8}}%
      \colorrgb{0.00,0.00,0.00}%
      \put(1634,836){\makebox(0,0){\strut{}0}}%
      \colorrgb{0.00,0.00,0.00}%
      \put(2686,836){\makebox(0,0){\strut{}2}}%
      \colorrgb{0.00,0.00,0.00}%
      \put(3737,836){\makebox(0,0){\strut{}4}}%
      \colorrgb{0.00,0.00,0.00}%
      \put(4789,836){\makebox(0,0){\strut{}6}}%
      \colorrgb{0.00,0.00,0.00}%
      \put(5840,836){\makebox(0,0){\strut{}8}}%
      \colorrgb{0.00,0.00,0.00}%
      \put(6892,836){\makebox(0,0){\strut{}10}}%
      \colorrgb{0.00,0.00,0.00}%
      \put(7943,836){\makebox(0,0){\strut{}12}}%
      \colorrgb{0.00,0.00,0.00}%
      \put(8995,836){\makebox(0,0){\strut{}14}}%
      \colorrgb{0.00,0.00,0.00}%
      \put(10046,836){\makebox(0,0){\strut{}16}}%
      \colorrgb{0.00,0.00,0.00}%
      \put(304,4699){\rotatebox{90}{\makebox(0,0){\strut{}scale factor $a$}}}%
      \colorrgb{0.00,0.00,0.00}%
      \put(6234,266){\makebox(0,0){\strut{}$t$ (Gyr)}}%
    }%
    \gplgaddtomacro\gplfronttext{%
      \colorrgb{0.00,0.00,0.00}%
      \put(6029,7930){\makebox(0,0)[r]{\footnotesize EdS   }}%
      \colorrgb{0.00,0.00,0.00}%
      \put(6029,7550){\makebox(0,0)[r]{\footnotesize 0.5 Mpc/$h$ }}%
      \colorrgb{0.00,0.00,0.00}%
      \put(6029,7170){\makebox(0,0)[r]{\footnotesize 1.0 Mpc/$h$ }}%
      \colorrgb{0.00,0.00,0.00}%
      \put(6029,6790){\makebox(0,0)[r]{\footnotesize 2.0 Mpc/$h$ }}%
      \colorrgb{0.00,0.00,0.00}%
      \put(6029,6410){\makebox(0,0)[r]{\footnotesize 4.0 Mpc/$h$ }}%
      \colorrgb{0.00,0.00,0.00}%
      \put(6029,6030){\makebox(0,0)[r]{\footnotesize 8.0 Mpc/$h$ }}%
      \colorrgb{0.00,0.00,0.00}%
      \put(6029,5650){\makebox(0,0)[r]{\footnotesize  {\LCDM}   }}%
    }%
    \gplbacktext
    \put(0,0){\includegraphics[scale=0.4]{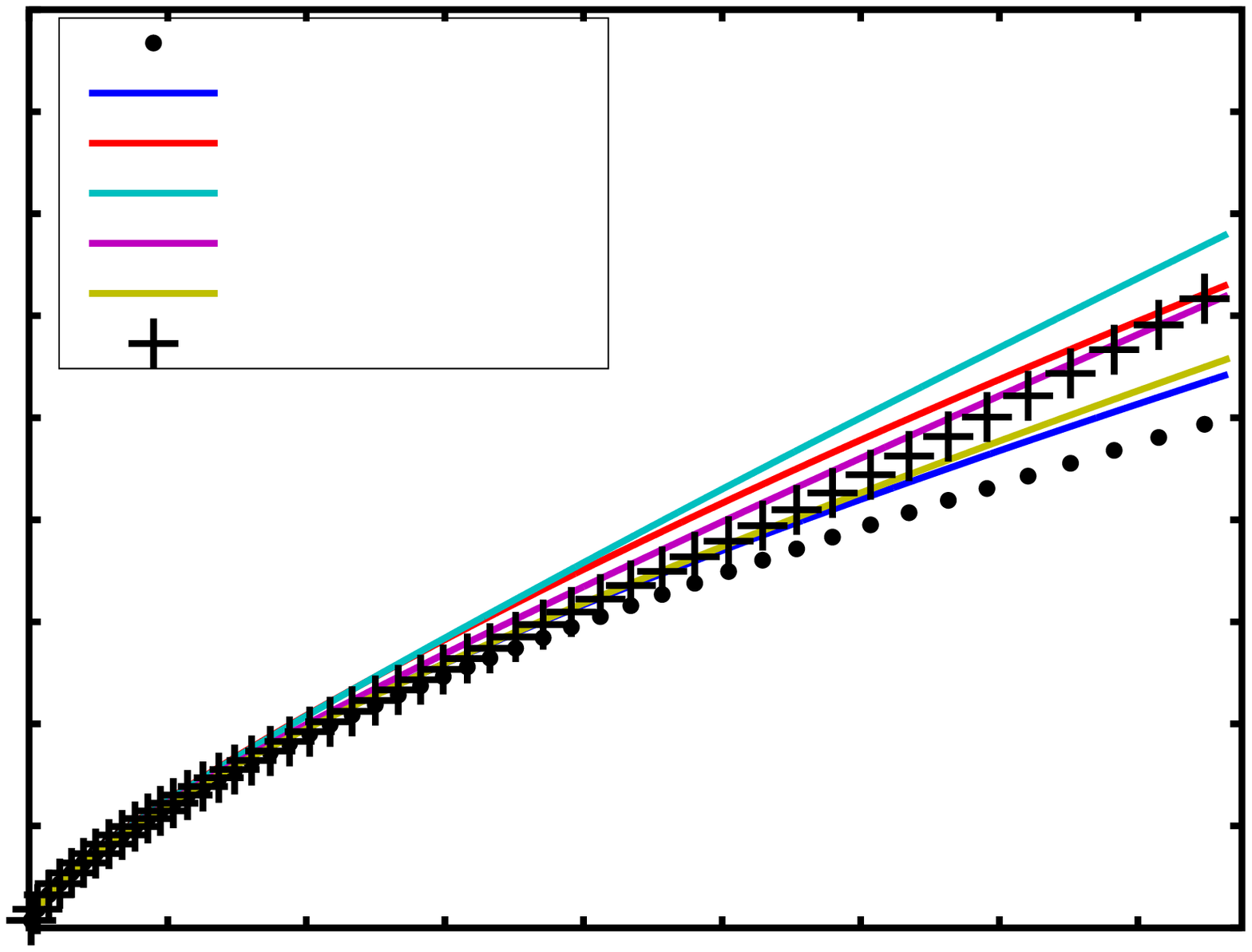}}%
    \gplfronttext
  \end{picture}%
\endgroup

%% file: a_nbody_Lag_128_T.tex
\begingroup
  \makeatletter
  \providecommand\color[2][]{%
    \GenericError{(gnuplot) \space\space\space\@spaces}{%
      Package color not loaded in conjunction with
      terminal option `colourtext'%
    }{See the gnuplot documentation for explanation.%
    }{Either use 'blacktext' in gnuplot or load the package
      color.sty in LaTeX.}%
    \renewcommand\color[2][]{}%
  }%
  \providecommand\includegraphics[2][]{%
    \GenericError{(gnuplot) \space\space\space\@spaces}{%
      Package graphicx or graphics not loaded%
    }{See the gnuplot documentation for explanation.%
    }{The gnuplot epslatex terminal needs graphicx.sty or graphics.sty.}%
    \renewcommand\includegraphics[2][]{}%
  }%
  \providecommand\rotatebox[2]{#2}%
  \@ifundefined{ifGPcolor}{%
    \newif\ifGPcolor
    \GPcolorfalse
  }{}%
  \@ifundefined{ifGPblacktext}{%
    \newif\ifGPblacktext
    \GPblacktexttrue
  }{}%
  \let\gplgaddtomacro\g@addto@macro
  \gdef\gplbacktext{}%
  \gdef\gplfronttext{}%
  \makeatother
  \ifGPblacktext
    \def\colorrgb#1{}%
    \def\colorgray#1{}%
  \else
    \ifGPcolor
      \def\colorrgb#1{\color[rgb]{#1}}%
      \def\colorgray#1{\color[gray]{#1}}%
      \expandafter\def\csname LTw\endcsname{\color{white}}%
      \expandafter\def\csname LTb\endcsname{\color{black}}%
      \expandafter\def\csname LTa\endcsname{\color{black}}%
      \expandafter\def\csname LT0\endcsname{\color[rgb]{1,0,0}}%
      \expandafter\def\csname LT1\endcsname{\color[rgb]{0,1,0}}%
      \expandafter\def\csname LT2\endcsname{\color[rgb]{0,0,1}}%
      \expandafter\def\csname LT3\endcsname{\color[rgb]{1,0,1}}%
      \expandafter\def\csname LT4\endcsname{\color[rgb]{0,1,1}}%
      \expandafter\def\csname LT5\endcsname{\color[rgb]{1,1,0}}%
      \expandafter\def\csname LT6\endcsname{\color[rgb]{0,0,0}}%
      \expandafter\def\csname LT7\endcsname{\color[rgb]{1,0.3,0}}%
      \expandafter\def\csname LT8\endcsname{\color[rgb]{0.5,0.5,0.5}}%
    \else
      \def\colorrgb#1{\color{black}}%
      \def\colorgray#1{\color[gray]{#1}}%
      \expandafter\def\csname LTw\endcsname{\color{white}}%
      \expandafter\def\csname LTb\endcsname{\color{black}}%
      \expandafter\def\csname LTa\endcsname{\color{black}}%
      \expandafter\def\csname LT0\endcsname{\color{black}}%
      \expandafter\def\csname LT1\endcsname{\color{black}}%
      \expandafter\def\csname LT2\endcsname{\color{black}}%
      \expandafter\def\csname LT3\endcsname{\color{black}}%
      \expandafter\def\csname LT4\endcsname{\color{black}}%
      \expandafter\def\csname LT5\endcsname{\color{black}}%
      \expandafter\def\csname LT6\endcsname{\color{black}}%
      \expandafter\def\csname LT7\endcsname{\color{black}}%
      \expandafter\def\csname LT8\endcsname{\color{black}}%
    \fi
  \fi
  \setlength{\unitlength}{0.0200bp}%
  \begin{picture}(11520.00,8640.00)%
    \gplgaddtomacro\gplbacktext{%
      \colorrgb{0.00,0.00,0.00}%
      \put(1406,1216){\makebox(0,0)[r]{\strut{}0}}%
      \colorrgb{0.00,0.00,0.00}%
      \put(1406,1990){\makebox(0,0)[r]{\strut{}0.2}}%
      \colorrgb{0.00,0.00,0.00}%
      \put(1406,2764){\makebox(0,0)[r]{\strut{}0.4}}%
      \colorrgb{0.00,0.00,0.00}%
      \put(1406,3538){\makebox(0,0)[r]{\strut{}0.6}}%
      \colorrgb{0.00,0.00,0.00}%
      \put(1406,4312){\makebox(0,0)[r]{\strut{}0.8}}%
      \colorrgb{0.00,0.00,0.00}%
      \put(1406,5087){\makebox(0,0)[r]{\strut{}1}}%
      \colorrgb{0.00,0.00,0.00}%
      \put(1406,5861){\makebox(0,0)[r]{\strut{}1.2}}%
      \colorrgb{0.00,0.00,0.00}%
      \put(1406,6635){\makebox(0,0)[r]{\strut{}1.4}}%
      \colorrgb{0.00,0.00,0.00}%
      \put(1406,7409){\makebox(0,0)[r]{\strut{}1.6}}%
      \colorrgb{0.00,0.00,0.00}%
      \put(1406,8183){\makebox(0,0)[r]{\strut{}1.8}}%
      \colorrgb{0.00,0.00,0.00}%
      \put(1634,836){\makebox(0,0){\strut{}0}}%
      \colorrgb{0.00,0.00,0.00}%
      \put(2686,836){\makebox(0,0){\strut{}2}}%
      \colorrgb{0.00,0.00,0.00}%
      \put(3737,836){\makebox(0,0){\strut{}4}}%
      \colorrgb{0.00,0.00,0.00}%
      \put(4789,836){\makebox(0,0){\strut{}6}}%
      \colorrgb{0.00,0.00,0.00}%
      \put(5840,836){\makebox(0,0){\strut{}8}}%
      \colorrgb{0.00,0.00,0.00}%
      \put(6892,836){\makebox(0,0){\strut{}10}}%
      \colorrgb{0.00,0.00,0.00}%
      \put(7943,836){\makebox(0,0){\strut{}12}}%
      \colorrgb{0.00,0.00,0.00}%
      \put(8995,836){\makebox(0,0){\strut{}14}}%
      \colorrgb{0.00,0.00,0.00}%
      \put(10046,836){\makebox(0,0){\strut{}16}}%
      \colorrgb{0.00,0.00,0.00}%
      \put(304,4699){\rotatebox{90}{\makebox(0,0){\strut{}scale factor $a$}}}%
      \colorrgb{0.00,0.00,0.00}%
      \put(6234,266){\makebox(0,0){\strut{}$t$ (Gyr)}}%
    }%
    \gplgaddtomacro\gplfronttext{%
      \colorrgb{0.00,0.00,0.00}%
      \put(6029,7930){\makebox(0,0)[r]{\footnotesize EdS   }}%
      \colorrgb{0.00,0.00,0.00}%
      \put(6029,7550){\makebox(0,0)[r]{\footnotesize 0.5 Mpc/$h$ }}%
      \colorrgb{0.00,0.00,0.00}%
      \put(6029,7170){\makebox(0,0)[r]{\footnotesize 1.0 Mpc/$h$ }}%
      \colorrgb{0.00,0.00,0.00}%
      \put(6029,6790){\makebox(0,0)[r]{\footnotesize 2.0 Mpc/$h$ }}%
      \colorrgb{0.00,0.00,0.00}%
      \put(6029,6410){\makebox(0,0)[r]{\footnotesize 4.0 Mpc/$h$ }}%
      \colorrgb{0.00,0.00,0.00}%
      \put(6029,6030){\makebox(0,0)[r]{\footnotesize 8.0 Mpc/$h$ }}%
      \colorrgb{0.00,0.00,0.00}%
      \put(6029,5650){\makebox(0,0)[r]{\footnotesize  {\LCDM}   }}%
    }%
    \gplbacktext
    \put(0,0){\includegraphics[scale=0.4]{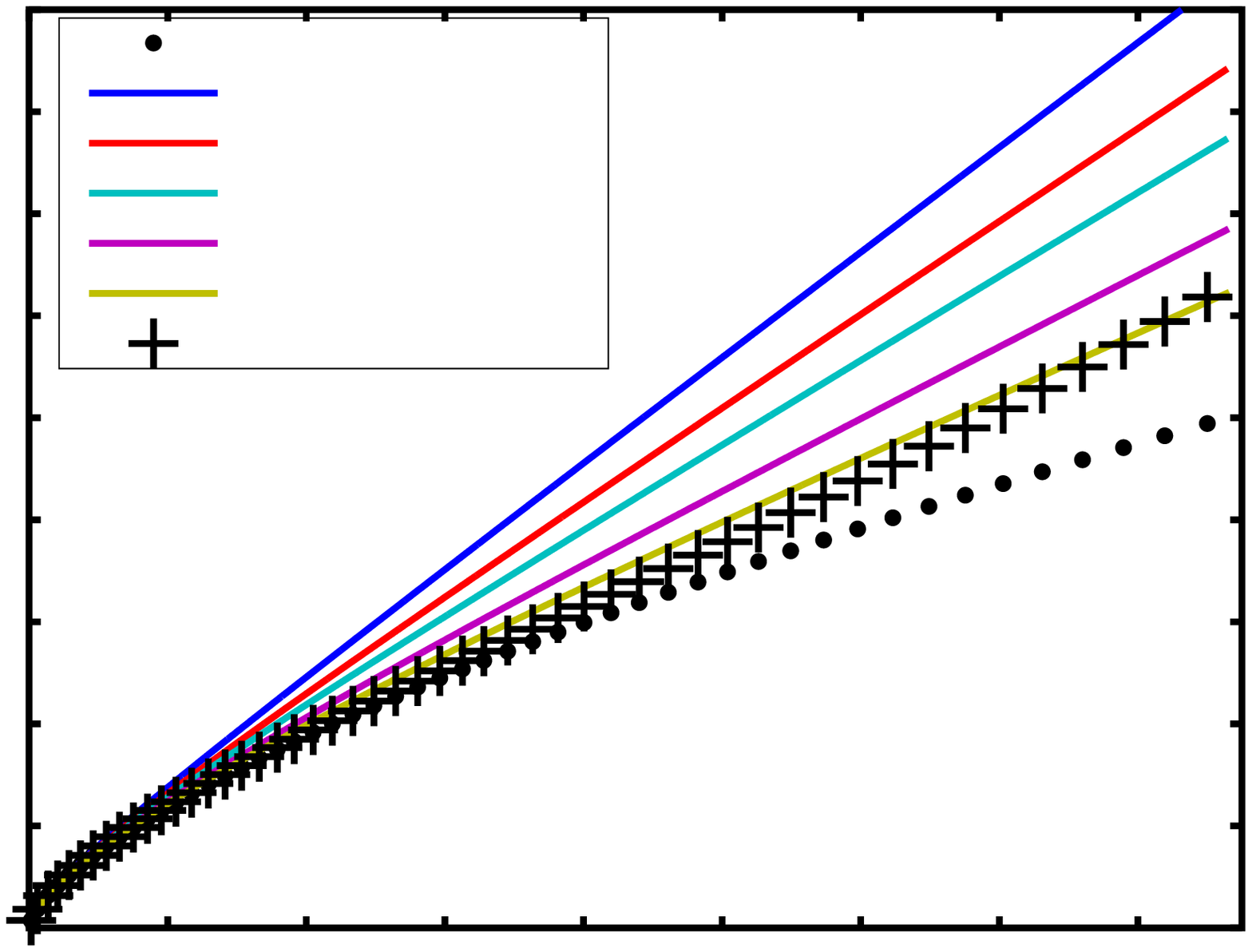}}%
    \gplfronttext
  \end{picture}%
\endgroup